\documentclass[floatfix,aps,rmp,reprint,amsmath,amssymb,graphicx,showkeys]{revtex4-1}

\usepackage{natbib}
\usepackage{graphicx}% Include figure files
\usepackage{dcolumn}% Align table columns on decimal point
\usepackage{bm}% bold math
\usepackage{verbatim}

\newcommand{\calm}{\mbox{${\cal M}$}}

\newcommand{\bftheta}{\mbox{\boldmath $\theta$}}

\newcommand{\bfpsi}{\mbox{\boldmath $\psi$}}

\newcommand{\BY}{\mbox{\boldmath $Y$}}

\newcommand{\simkl}{\stackrel{{<}}{{\scriptstyle\sim}}}
\newcommand{\simgr}{\stackrel{{>}}{{\scriptstyle\sim}}}

\def\boldxih{\boldxi_{\rm h}}

\def\xih{\xi_{\rm h}}
\def\xirnl{\xi_{r,nl}}
\def\xihnl{\xi_{{\rm h},nl}}
\def\boldxi{\xi\kern-0.45em\xi\kern-0.45em\xi}
\def\dd{{\rm d}}
\def\boldF{\mbox{\boldmath$F$}}
\def\boldf{\mbox{\boldmath$f$}}
\def\boldv{\mbox{\boldmath$v$}}
\def\boldr{\mbox{\boldmath$r$}}
\def\bolda{\mbox{\boldmath$a$}}
\def\bolddelr{\mbox{\boldmath$\delta r$}}
\def\div{{\rm div}\,}

\begin{document}

\title[Asteroseismology]{Probing the interior physics of stars through asteroseismology}

\author{C. Aerts (ORCID 0000-0003-1822-7126)}
%\email{Conny.Aerts@kuleuven.be}
\homepage{https://fys.kuleuven.be/ster/staff/conny-aerts}
\email{Conny.Aerts@kuleuven.be}
\affiliation{Institute of Astronomy, Department of Physics \& Astronomy, KU\,Leuven,
Celestijnenlaan 200\,D, 3001 Leuven, Belgium}
\altaffiliation{Also at 
Department of Astrophysics, IMAPP, Radboud University Nijmegen, 
Heyendaalseweg
135, 6525 AJ Nijmegen, the Netherlands and
Max Planck Institute for Astronomy, 
Koenigstuhl 17, 69117 Heidelberg, Germany}

\date{Received: 13 December 2019; Accepted: 5 October 2020; To be published in
  Vol. 93, 2021}

\begin{abstract}
  Yearslong time series of high-precision brightness measurements have been
  assembled for thousands of stars with telescopes operating in space. Such data
  have allowed astronomers to measure the physics of stellar interiors via
  nonradial oscillations, opening a new avenue to study the stars in the
  Universe. Asteroseismology, the interpretation of the characteristics of
  oscillation modes in terms of the physical properties of the stellar interior,
  brought entirely new insights in how stars rotate and how they build up their
  chemistry throughout their evolution. Data-driven space asteroseismology
  delivered a drastic increase in the reliability of computer models mimicking
  the evolution of stars born with a variety of masses and metallicities. Such
  models are critical ingredients for modern physics as a whole, because they
  are used throughout various contemporary and multidisciplinary research fields
  in space science, including the search for life outside the solar system,
  archaeological studies of the Milky Way, and the study of single and binary
  supernova progenitors, among which are future gravitational wave sources.  The
  specific role and potential of asteroseismology for those modern research
  fields are illustrated. The review concludes with current limitations of
  asteroseismology and highlights how they can be overcome with ongoing and
  future large infrastructures for survey astronomy combined with new
  theoretical research in the era of high-performance computing. This review
  presents results obtained through major community efforts over the past
  decade. These breakthroughs were achieved in a collaborative and inclusive
  spirit that is characteristic of the asteroseismology community. The review's
  aim is to make this research field accessible to graduate students and readers
  coming from other fields of physics, with incentives to enjoy and join future
  applications in this glorious domain of astrophysics.
\end{abstract}


\maketitle

\tableofcontents{}

%%%%%%%%%%%%%%%%
%%%%%%%%% Section 1
%%%%%%%%%%%%%%%%

\section{\label{section-intro}Looking deep into stars}

In his News and Views published in 1985 in the journal Nature, Douglas Gough
\citep{Gough1985a} announced the ``{\it Beginnings of
  asteroseismology\/}''\footnote{For the etymology of this specific term in
  astrophysics, see 
  \citet{Gough1996} in response to the anecdote on the terminology raised by
  \citet{Trimble1996}.}.  He
explained that this is
\begin{quote}
``{\it The science of determining the internal structure of stars from
the properties of dynamical oscillations\/}'',
\end{quote}
following the earlier introduction of the term in the
scientific community by J\o rgen
Christensen-Dalsgaard during a conference in Meudon in 1984 \citep{JCD1984}. 
Gough ends his Nature article with the exciting prospect of ``{\it getting direct
  information about the stratification of the energy-generating core of a
  distant star}'', which sounded like a revolutionary idea at that time. A decade later, a
less optimistic view was expressed by \citet{Brown1994} in the introduction of
their review paper entitled ``{\it Asteroseismology}":
\begin{quote}
``{\it
The Sun is (and will likely remain) the outstanding example of the progress that
can be made using seismological methods\/}''.
\end{quote}
It is remarkable that a cosmologist was more optimistic on the matter than the
experts, as expressed by Malcolm Longair \citep{Longair2001} in his invited reflection
entitled ``{\it Facing the Millennium\/}'':
\begin{quote}
``{\it \ldots 
At the same time, we need to understand the internal structures of the stars. In
1915, the breakthrough came with the plotting of the Hertzsprung‐Russell diagram
for a few hundred stars for which distances had been measured. The counterpart
for the 21st century will be asteroseismology, the direct measurement of the
internal structure of the stars by measuring their normal modes of
oscillation. It is salutary to note that helioseismology has revolutionized our
understanding of the interior of the Sun in ways which could not necessarily
have been predicted. The precise location of the boundary between the radiation-
and convection-dominated zones and their three-dimensional structures are
spectacular advances -- a major goal of the astronomy of the future must be to
perform the same studies on the stars present in the Hipparcos
Hertzsprung-Russell diagram\/}''.
\end{quote}
Meanwhile, we have amazing Gaia Hertzsprung-Russell diagrams (HRDs) based on
space astrometry with microarcsecond precision for more than a billion
stars in the Milky Way and beyond \citep{GaiaHRD2018} and asteroseismology for
tens of thousands of those. Indeed, the past decade has seen the assembly of
long-duration (up to four uninterrupted years) photometric data thanks to
dedicated space missions, leading to the time-variable properties of stars
derived with precisions of micromagnitude ($\mu$mag). This corresponds to flux
variations at levels of parts-per-million (ppm).  The primary research goal
of several of those missions was the search for exoplanets around distant stars,
but this is
fine: the machinery delivered the data appropriate for asteroseismology (quite
often, derogatorily, called ``stellar noise'' by exoplanetologists, while it
actually concerns beautiful stellar signal; see the later discussion of
Fig.\,\ref{spbs}.
Asteroseismology based on these space photometric light curves
meanwhile delivered interior rotation rates, stratification properties, and ages
of thousands of distant stars, with impressive relative precisions unachievable
by other methods \citep[][Table\,1]{Aerts2019}.  The past few years, we have
even reached the status of being able to derive the core rotation
frequency for more than a thousand stars and the near-core mixing for many of
them. The Sun is no longer the outstanding example to assess the properties of the
deep interiors of stars.  Those properties have become accessible thanks to the
detection and interpretation of hundreds of nonradial oscillations with probing
capacity of the deepest internal layers of the stars. Such modes
have now been identified in stars across almost the entire stellar mass range and
have turned stellar interiors into observational territory.

History has shown the optimistic daring views to be visionary and at the same
time entirely justified. The study of stellar interiors is nowadays a data-driven modern
topic. As we highlight in this review, asteroseismology not only reveals
the need for better models of stellar structure and evolution but is paving
the way toward them. It brings us
into the {\it renaissance\/} of stellar evolution theory and
does so from
a multidisciplinary approach relying on space science and technology coupled with
mathematical modeling. More particularly, asteroseismology involves
data analysis methods such as time-series analysis, pattern recognition, 
and
statistical modeling, while relying on various fields of physics and
chemistry such as thermodynamics, nuclear and atomic physics, and quantum
mechanics. The bridging of these scientific fields, starting from the
appropriate observational input, allows us to achieve the long-awaited
calibration of the physical properties of stellar interiors.  We 
anticipate that the asteroseismically calibrated stellar models will be highly
beneficial for various fields of research in astronomy and in computational
physics in general.

We now provide concise discussions of some key observational and data analysis
aspects of asteroseismology, omitting many of the details,  which can be found
in the referenced literature. The bulk of the review then focuses on calibrating
and improving the physics of stellar interiors, much in line with the previous quotations
and with the purpose of {\it Reviews of Modern Physics}.

\subsection{\label{vibrations}Stars and their ``good'' vibrations}
 
\begin{figure*}
\begin{center} 
\rotatebox{0}{\resizebox{12.cm}{!}{\includegraphics{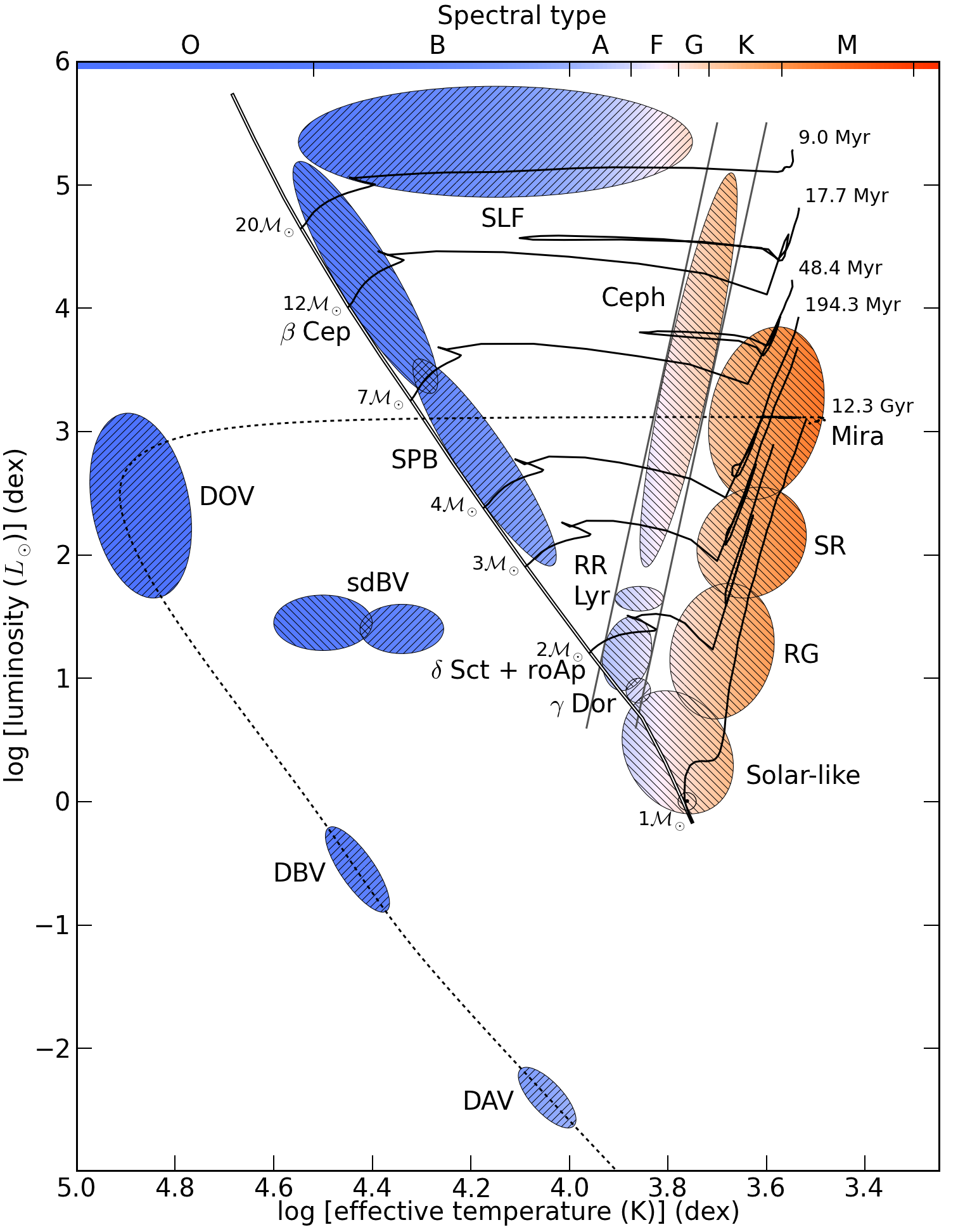}}}
\end{center}
\caption{\label{hrd} Hertzsprung-Russel diagram (HRD) showing the position of
  different classes of pulsating stars.  The abbreviation of the classes follows
  the nomenclature used by \citet[][Chapter\,2]{Aerts2010}, to which we refer for
  extensive discussions of all  indicated classes in terms of the excitation
  mechanisms, along with the typical periods and amplitudes of the oscillations.
  The hatching linestyle used inside each of the ellipses marks the dominant type of
  oscillation mode in each class: $/\!\!/$ for gravity modes and
  $\backslash\!\!\backslash$ for pressure modes.  The recently discovered
  stochastic low-frequency (SLF) variability in O-type stars and blue supergiants is
  discussed in the text and has been added as a comparison with previous versions of this
  plot.  The solid black lines and the black dotted line represent standard
  evolutionary model tracks, with birth masses and evolutionary timescales as
  indicated.  The borders of the classical instability strip are plotted with gray
  lines, while the double line represents the zero-age main sequence.
Early versions of this figure were made by J\o rgen
  Christensen-Dalsgaard (Aarhus University) and by Pieter Degroote (KU\,Leuven).
  This updated version was produced by P\'eter P\'apics based on the version in
  his PhD\,Thesis \citep{Papics2013-PhD}.}
\end{figure*}

Stellar variability is omnipresent in the HRD, which is a key diagnostic diagram
used to evaluate stellar evolution theory. Such evaluations are often done by
comparing the position of observed stars in this diagram with evolutionary
tracks, such as the ones indicated by the full lines in Fig.\,\ref{hrd}. These
tracks are based on particular versions of stellar evolution theories, of which
there are many variants as further outlined in Sec.\,\ref{models}. Those types
of comparisons between observations and theory are merely a crude
evaluation because the HRD relies on only two quantities: the effective
temperature of the star $T_{\rm eff}$ and its luminosity $L$ (usually
expressed in solar luminosity $L_\odot$).  As detailed in
Sec.\,\ref{models}, stellar models contain a multitude of free parameters and
rely on input physics suffering from uncertainties. The evaluation of these
models therefore requires additional observational diagnostics to accompany the
position of an observed star in the HRD. Surface abundances derived from
high-precision spectroscopy offer important constraints in this respect, among
various other observational diagnostics of the stellar atmosphere. These
typically reach relative precisions of 1\% to 5\% for the best cases \citep[see
Table\,1 of][]{Aerts2019}.

A new view on stellar variability in the HRD is offered by data from
the European Space Agency (ESA) Gaia satellite \citep{Eyer2019}. Using 22 months of calibrated
photometric, spectro-photometric, and astrometric Gaia data, this study showed
how the large-amplitude radial modes of classical variables, such as Cepheids,
RR\,Lyr stars, and Miras (indicated in Fig.\,\ref{hrd}),  makes them ``move'' in the
observational analog of the HRD, i.e., a color-absolute magnitude diagram,
during their pulsation cycle.  This introduces a new ``time'' dimension in the
evaluation of stellar evolution theory. These radial pulsators remain of vast
interest and importance for observational cosmology, e.g., \citet{Soszynski2016}
and \citet{Anderson2018}, but are not considered in this review. Our attention is
directed entirely to stars exhibiting multiple
nonradial oscillation modes, which in the context of asteroseismology deserve to be
called ``good vibrations'' after the
eponymous 1966 song by the Beach Boys.

\begin{figure}[t!]
\begin{center} 
\rotatebox{270}{\resizebox{2.8cm}{!}{\includegraphics{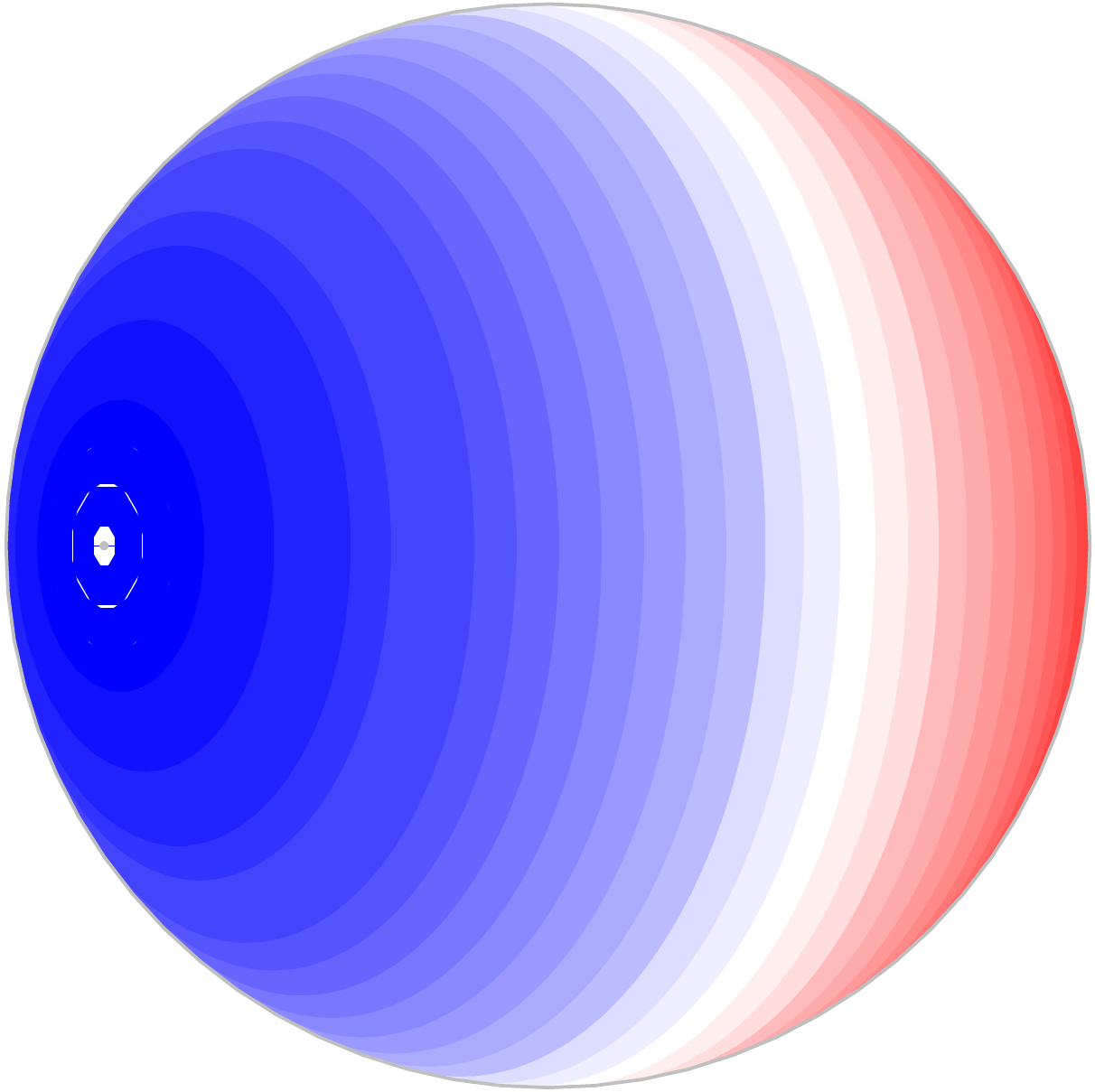}}}
\rotatebox{270}{\resizebox{2.8cm}{!}{\includegraphics{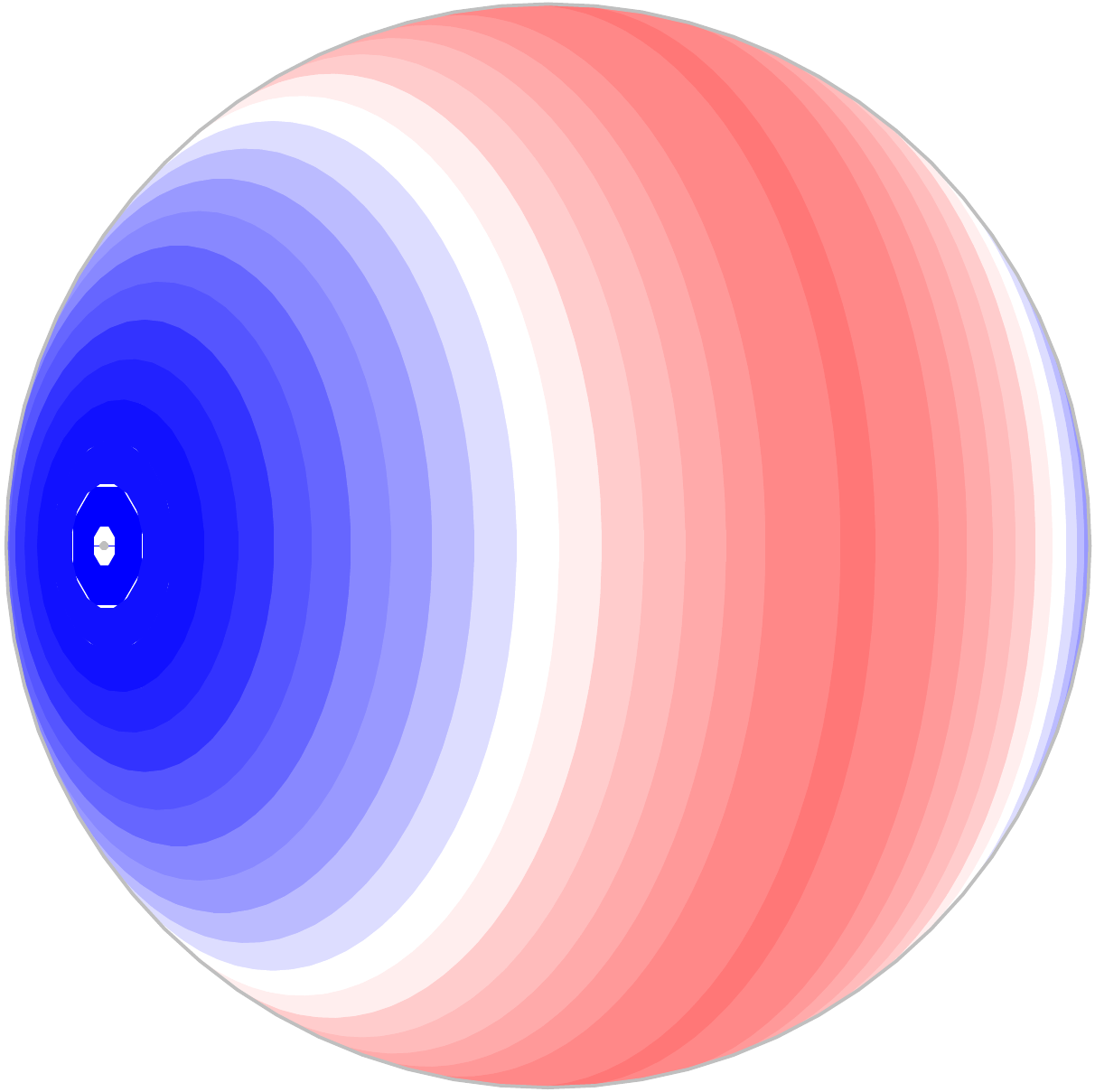}}}
\rotatebox{270}{\resizebox{2.8cm}{!}{\includegraphics{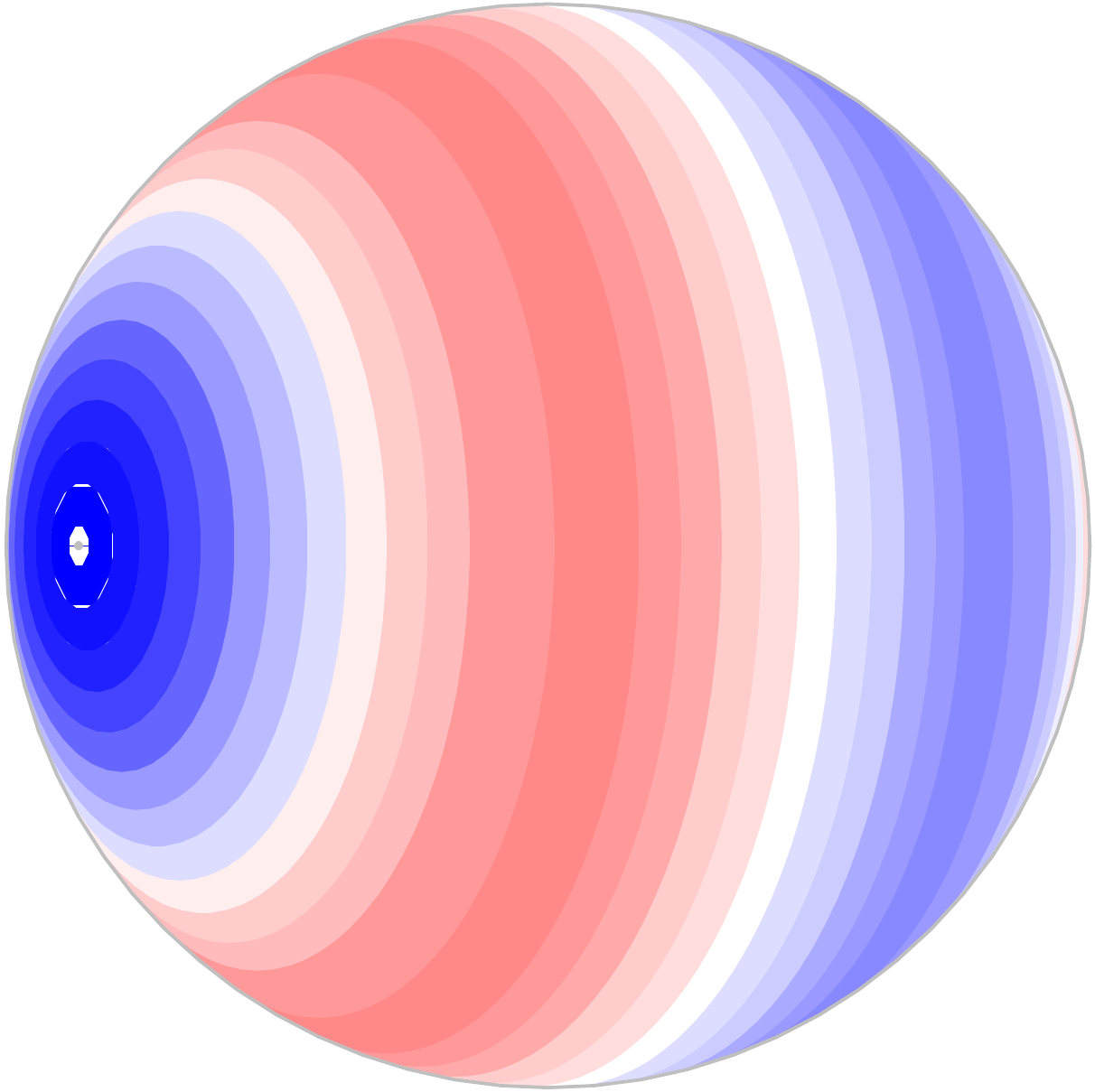}}}%
\\[0.2cm]
\rotatebox{270}{\resizebox{2.8cm}{!}{\includegraphics{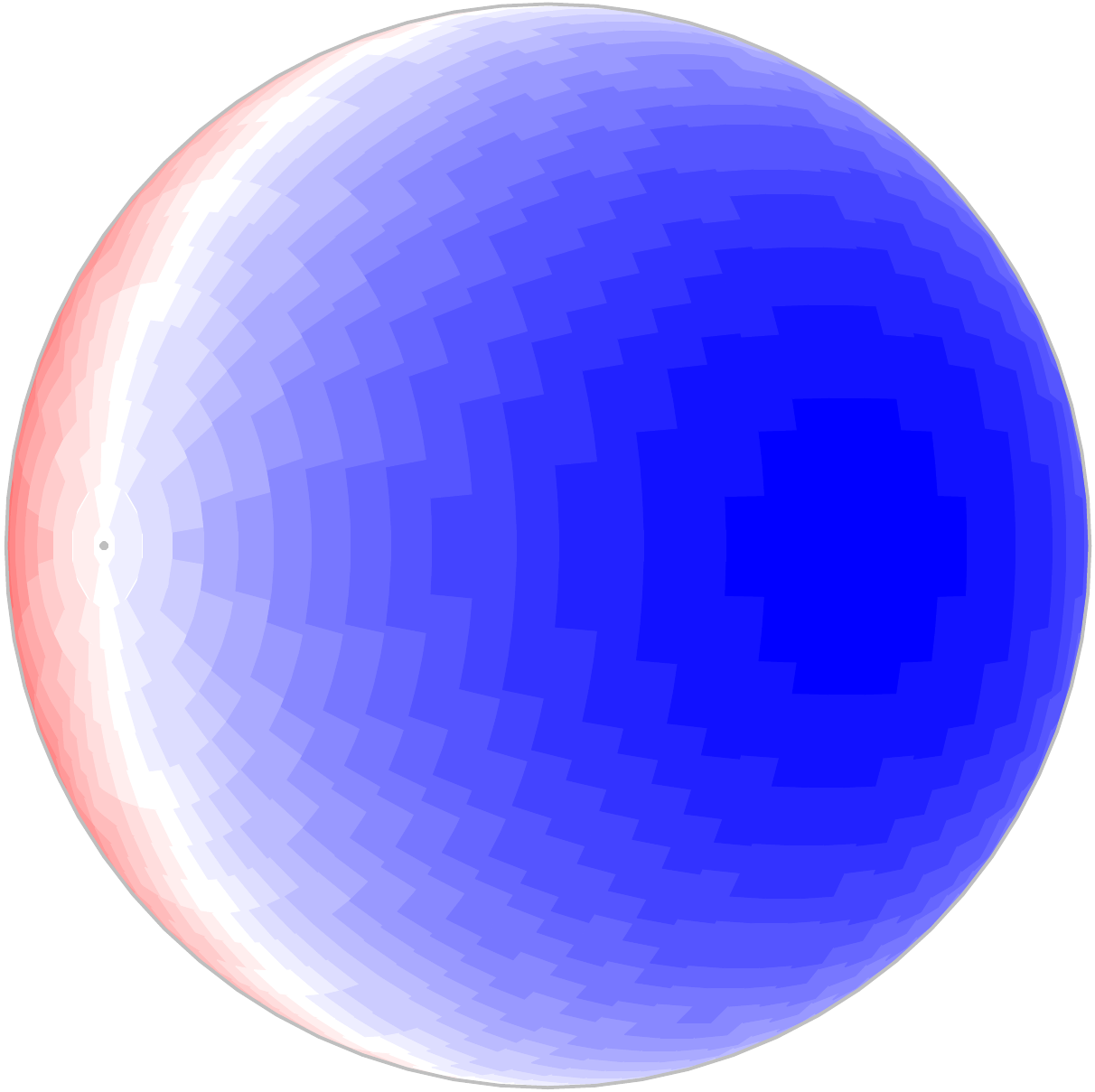}}}
\rotatebox{270}{\resizebox{2.8cm}{!}{\includegraphics{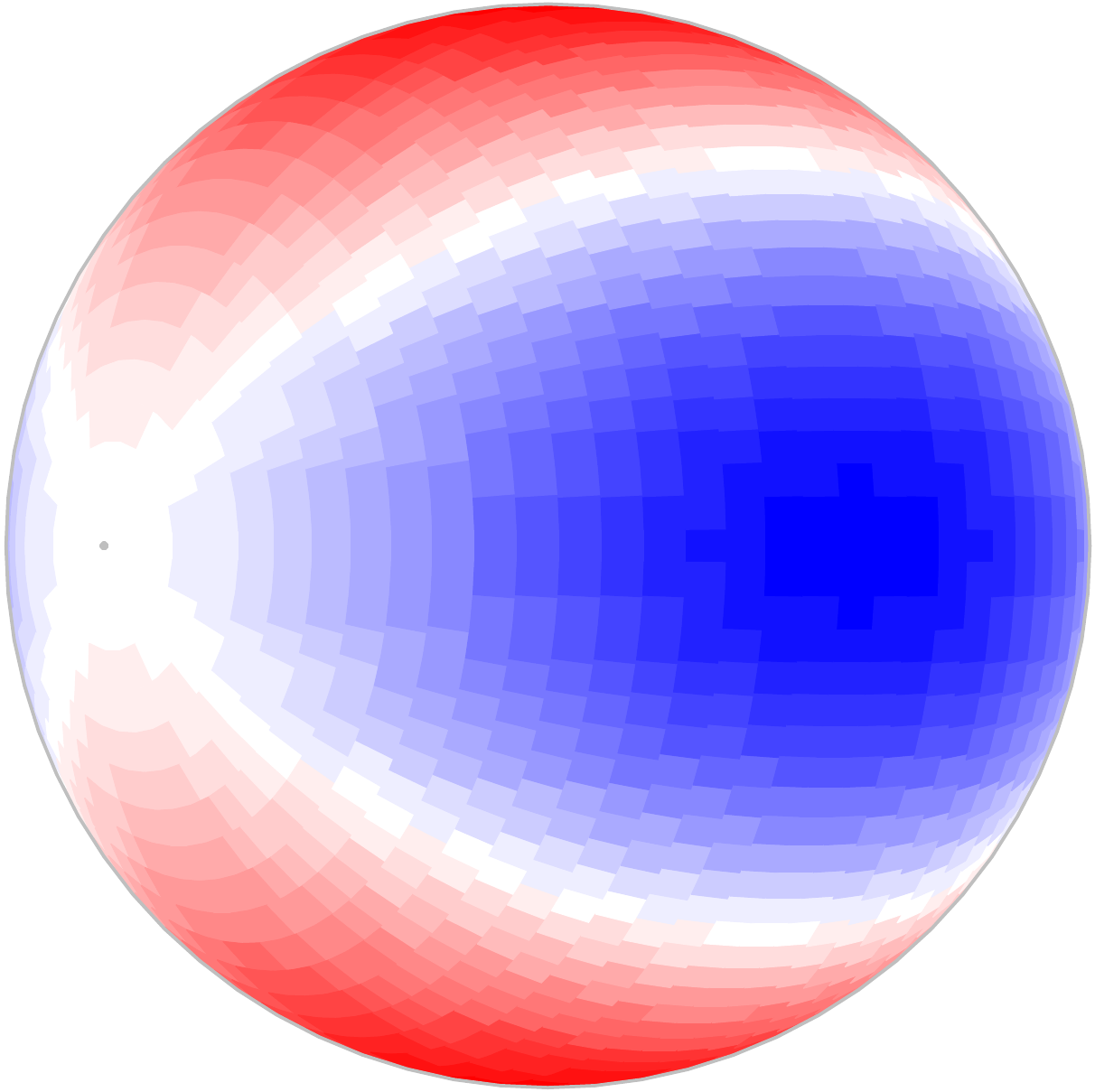}}}
\rotatebox{270}{\resizebox{2.8cm}{!}{\includegraphics{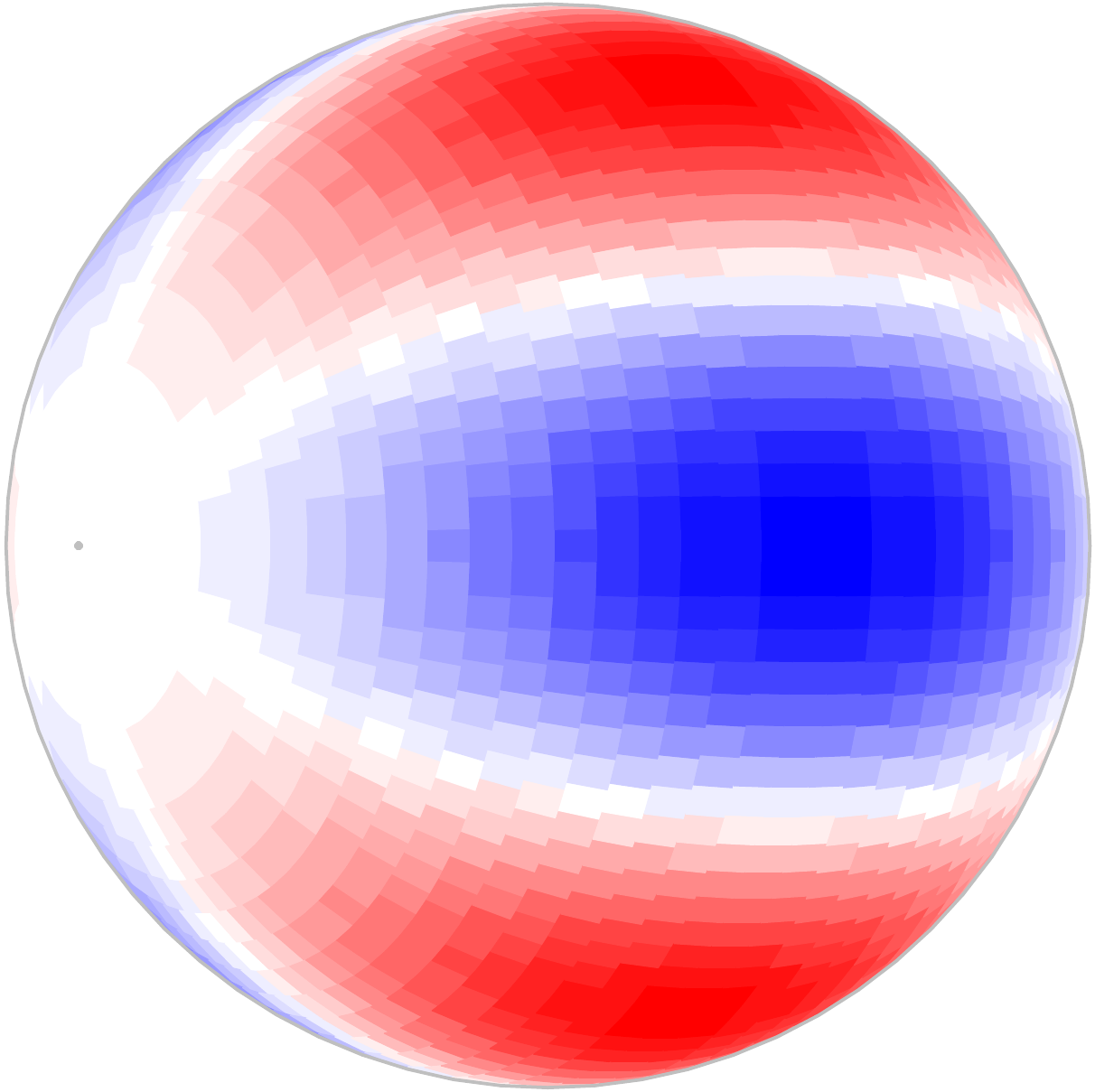}}}%
\\[0.2cm]
\rotatebox{270}{\resizebox{2.8cm}{!}{\includegraphics{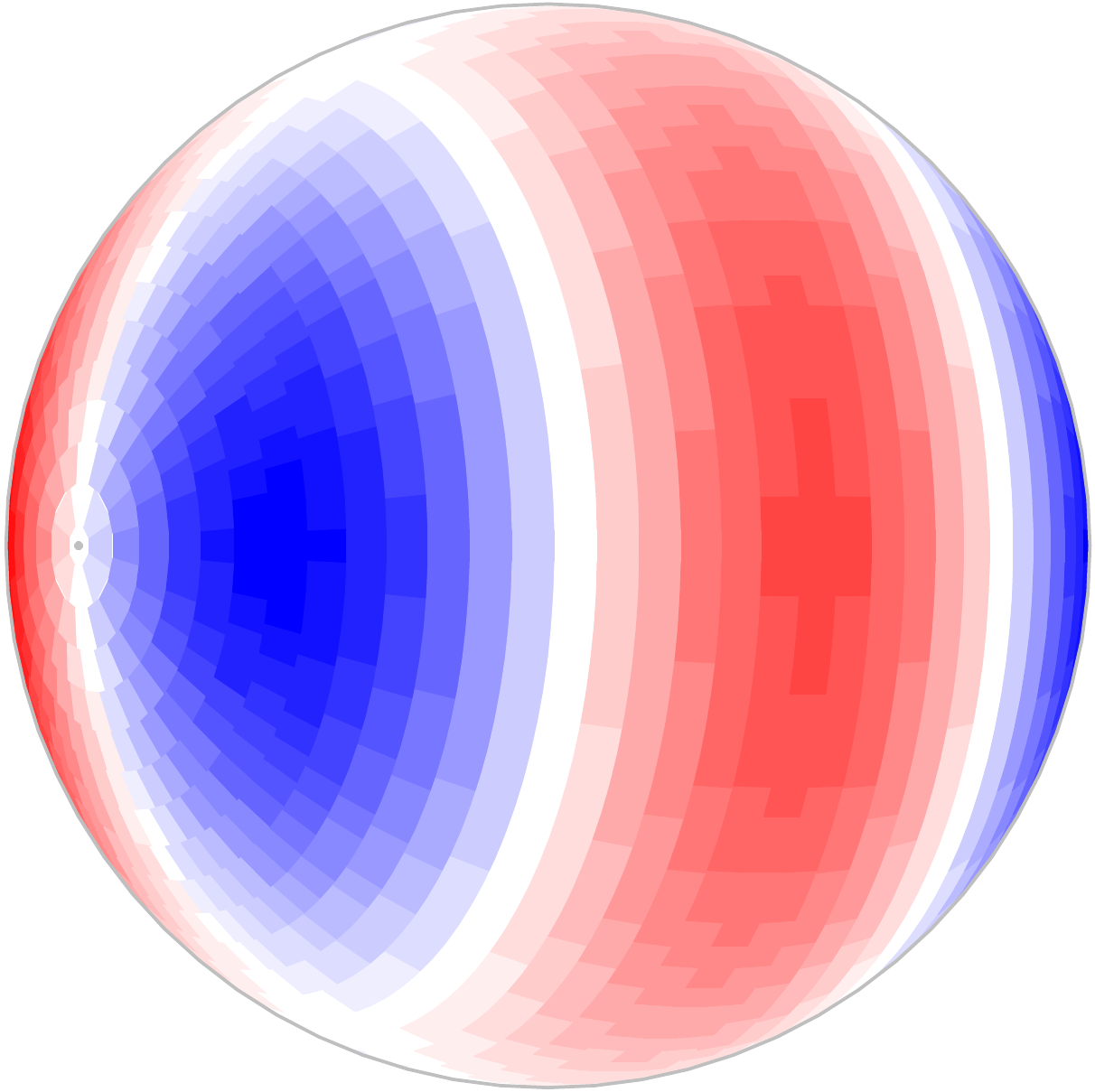}}}
\rotatebox{270}{\resizebox{2.8cm}{!}{\includegraphics{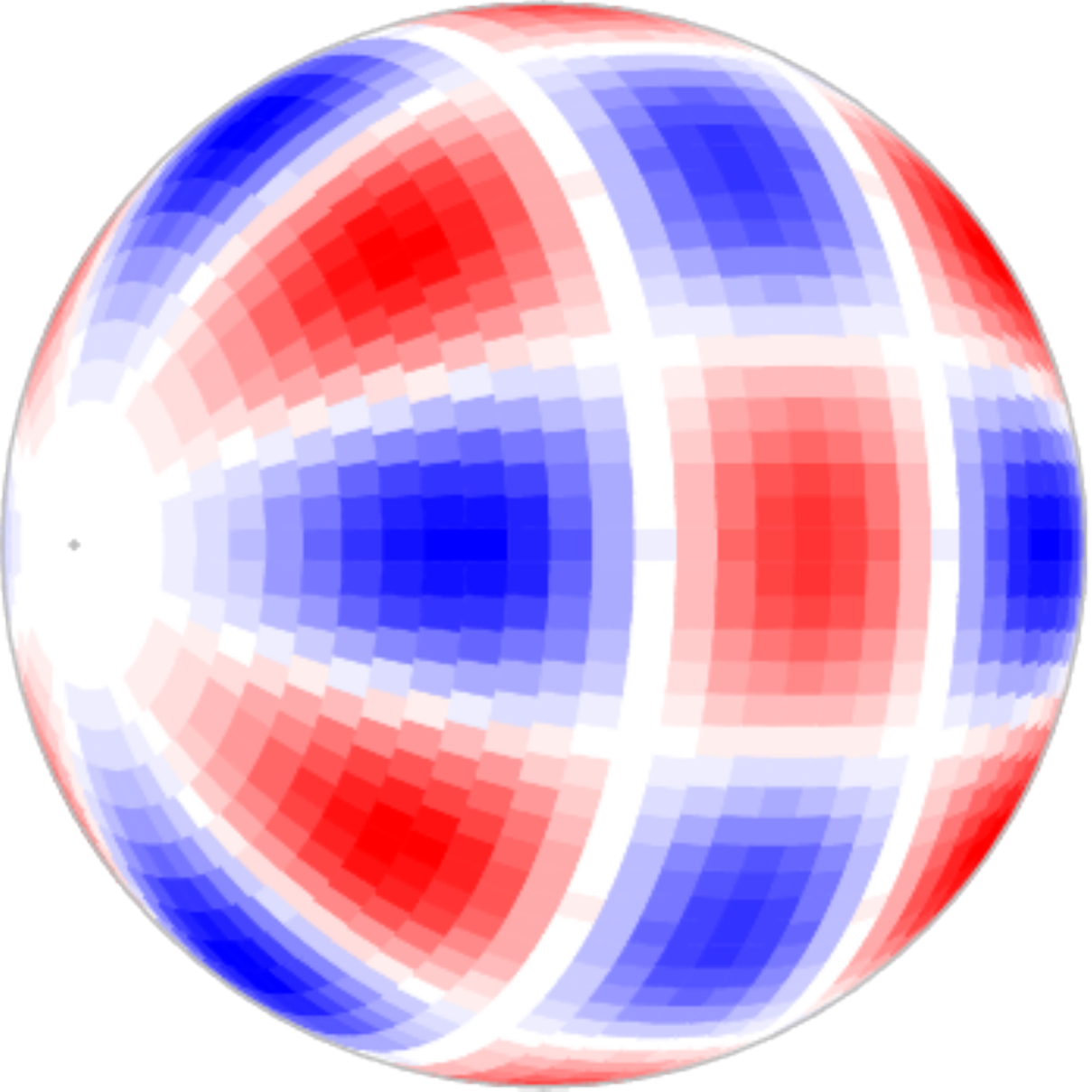}}}
\rotatebox{270}{\resizebox{2.8cm}{!}{\includegraphics{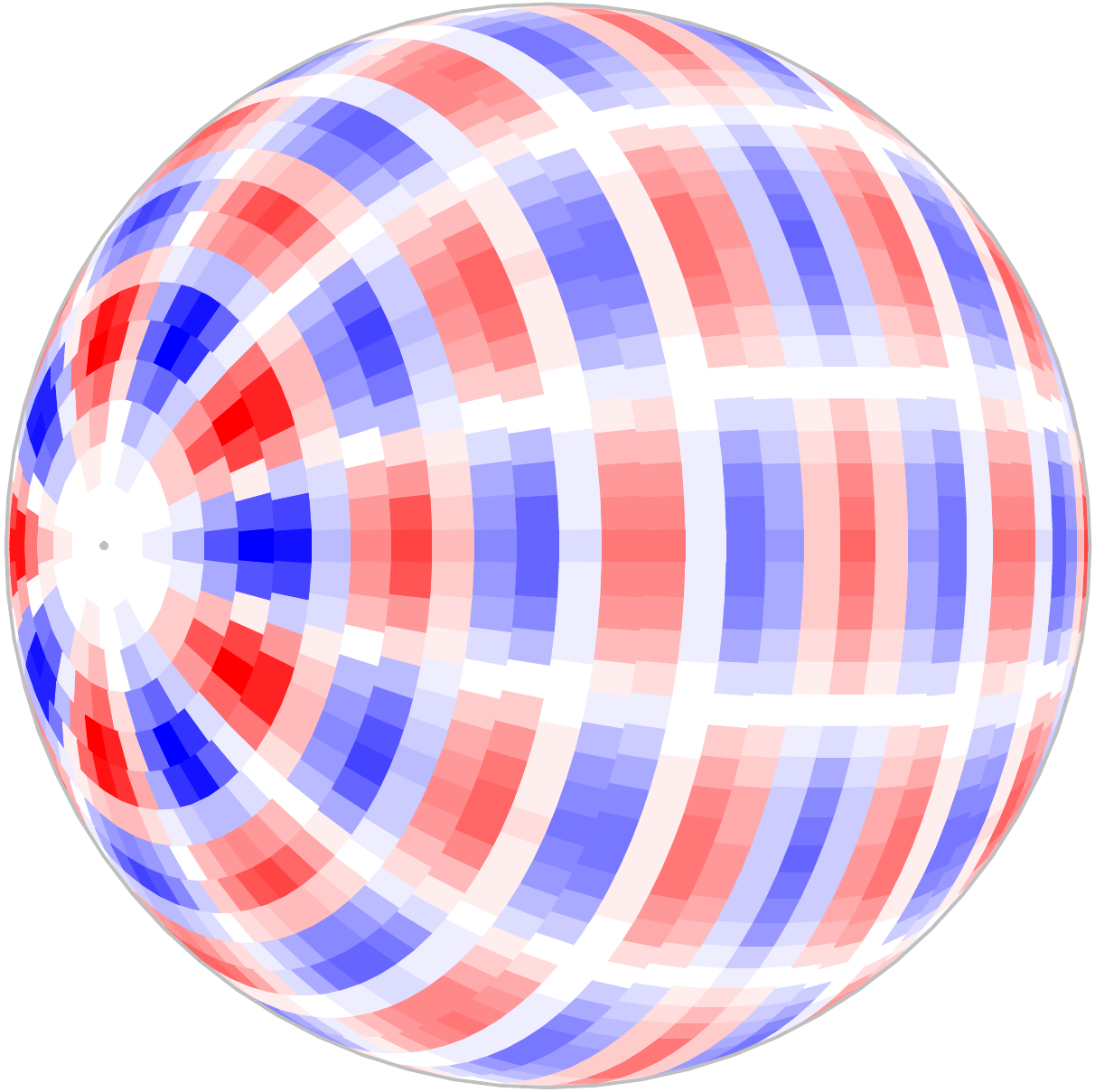}}}
\end{center}
\caption{\label{ylm} Snapshot of the angular dependence of the radial component
  of the displacement vector $\xi_r$ at one point
  in the oscillation cycle for various nonradial modes, seen under an
  inclination angle of $60^\circ$.  White bands indicate the
  positions where $\xi_r=0$; red and blue represent areas at the stellar surface
  moving in (out) at the chosen time. Shown from left to right are first 
  row, axisymmetric ($m=0$) modes with $l=1, 2, 3$; second row, sectoral ($l=|m|$)
modes with
  $l=1, 2, 3$; third row, tesseral ($l\neq |m|$) modes with $(l,|m|)=(3,1), (6,4), (15,5)$.
High-degree
  modes as the two in the third row are usually not detected in space photometry due
  to cancellation effects when integrating the flux variations
  across the visible stellar disk. }
\end{figure}

From a physical viewpoint, nonradial oscillation modes are solutions to the
equation of motion of a star that gets perturbed from its equilibrium.  The
modes are classified into two main groups according to which of the two forces,
the pressure force or the buoyancy force of Archimedes, is dominant in restoring
the equilibrium.  Modes dominantly restored by the pressure force are called
pressure modes, or `p~modes' for short. These mainly have large amplitude in the
envelopes of stars and are characterized by dominant radial motions. Gravity
modes, or `g~modes', are dominantly restored by the buoyancy force of Archimedes
and attain large
amplitudes in the deep interior of the star; they are characterised by dominant
horizontal motions. As stars evolve, a powerful type of modes having a
pressure-mode character in the envelope and a gravity-mode character in the deep
interior emerges. These so-called mixed modes have excellent probing power
throughout the entire star.

A formal mathematical definition of nonradial oscillation modes is given in
Sec.\,\ref{section-nrp}. However, it is instructive to
already know how the modes `look like'. One can consider
nonradial modes of a 3D spherical star as the analogy of the vibration modes of
a 1D string.  Each vibration mode of a string makes it deviate from its
equilibrium position and is characterized by three numbers: its frequency, its
amplitude, and its number of nodes $n$. The nodes are points where the string
does not move during the vibration cycle. One adopts the terminology that $n=0$
corresponds to the fundamental vibration mode of the string, $n=1$ to the
first overtone, $n=2$ to the second overtone, etc.  Each nonradial mode of a
3D star makes the gas particles in this star deviate from their equilibrium
position and is also characterized by a frequency and an amplitude, but now
three integer numbers are needed to indicate the positions of the nodes of the
displacement vector with respect to a symmetry axis of the star.  Given that it
concerns a 3D spherically symmetric body whose fluid elements get displaced from
their equilibrium position by a vector
$\boldxi=(\xi_r,\xi_\theta,\xi_\phi)$, the angular geometry of this vector
is described in terms of a spherical harmonic
function, containing a Legendre polynomial $P_l^m$ as a function of colatitude
$\theta$ and a harmonic function in terms of azimuth $\phi$.  The rotation
axis is usually taken as the symmetry axis of the modes. Thus for each nonradial
oscillation mode, three labels $(l,m,n)$ are used to indicate the nodes of the
mode, where $l$ is the total number of nodal lines on the stellar surface and
$|m|$ of those nodal lines pass through the symmetry axis.  The $n$ value again
indicates the overtone of the mode, which now concerns the number of nodal
shells situated inside the star that do not move during the oscillation cycle.
The special case of a radial mode has $l=m=0$ and displaces the fluid elements
inside the star in the radial direction only.

The symmetry axis of the oscillations is ``inclined'' with the line of sight of
a distant observer by an unknown angle called the inclination angle $i$.
Figure\,\ref{ylm} gives a visual representation of the radial component of the
displacement vector $\xi_r$  for some typical nonradial modes ``observed'' 
under an inclination angle of $60^\circ$.  The term {\it observed\/} is a bit
misleading here because stellar surfaces cannot be resolved well enough to
study the majority of nonradial oscillations of stars, except for the
Sun. Rather, the signatures of the oscillation modes are ``detected'' in
observables that are stellar quantities integrated over the part of the stellar
disk that is visible for an observer. The nonradial oscillations make some parts
of the star move up (indicated in blue in Fig.\,\ref{ylm}), while others are
going down (red patches in Fig.\,\ref{ylm}) periodically according to the
eigenfrequency of the mode. Such motions imply small local changes in the
velocity, temperature, and radius of the stellar gas, creating local flux
variations. These flux variations change periodically in time during the
oscillation cycle, i.e., half a cycle further the red patches in Fig.\,\ref{ylm}
will have become blue and vice versa.  The surface-integrated effect due to each
nonradial mode measured in flux or velocity variations by an observer 
depends on the inclination angle because it is determined by the position of
the surface nodal lines in the line of sight. This interplay between the
geometry of the mode and the value of $i$ gives rise to so-called partial
cancellation due to integration over the visible stellar disk, which increases
as the degree of the mode increases \citep[see Figs.\,1.4 and 1.5
in][]{Aerts2010}.  In particular, when nonradial modes are seen under their
angle of complete cancellation, they do not lead to variability, while the
latter is maximal when seen under their optimal angle of least cancellation. For
the values of these special mode angles, we refer to Table\,B.1 in Appendix\,B
of \citet{Aerts2010}.  It is also noteworthy that partial cancellation works
differently in photometric versus spectroscopic data, because the integrated
flux is highly sensitive to limb darkening but the integrated velocity is less
sensitive to it. In
addition, the effect is different for p~ and g~modes. The sensitivity to limb
darkening is smaller for p~modes because their $\xi_r$ is dominant in the line of sight,
while g~modes have dominant $\xi_\theta$ and $\xi_\phi$ and are hence much 
more prone to limb-darkening effects for an observer.

Measuring the small flux or velocity variations during an oscillation-mode cycle
allows us to derive the mode's period without having to resolve the stellar
surface. This is how the time-variability aspect of asteroseismology works.  It
is in principle an easy aspect of the research provided that one has data with
a high duty cycle, which is defined as the fraction of the mode period covered
with data expressed as a percentage. In practice, the overall beating cycle
encapsulating the global pattern due to all active modes of the star
has to be covered with a high duty cycle. Moreover, the data need to have noise
levels below the amplitudes of the modes in the appropriate frequency
regime. These scientific requirements become easier to meet the longer the
time series and the more data points that one has available.  Detecting oscillation
mode frequencies and estimating their uncertainty is also much easier to do from
uninterrupted data with high duty cycle than from gapped time-series data with
a low duty cycle.

My fellow countryman, Paul Ledoux, proposed the occurrence of two nonradial
p~modes in a rotating star as the explanation for the detected
variable velocity behavior of the star $\beta\,$Canis Majoris
\citep{Ledoux1951}.  His landmark paper provided the
first correct interpretation and understanding of the observed biperiodic
variability (i.e., caused by two simultaneously active p~modes) of a rotating star
in terms of the physics of nonradial oscillations. As a member of the class of
$\beta\,$Cep stars, $\beta\,$Canis Majoris was thus the first star with
confirmed nonradial modes occupying the proper ellipse in Fig.\,\ref{hrd}. It
took another 41 years until the excitation mechanism of those nonradial
oscillations was understood in terms of a heat mechanism, also known as the
opacity mechanism \citep{Moskalik1992}. We return to mode excitation
mechanisms in Sec.\,\ref{excitation}.

During the half century following Ledoux's insightful 1951 paper,
the search for and identification of nonradial oscillations in time-series
observations became an active research field.  Inventories of
nonradial modes and their identification in terms of the spherical wave numbers
$l$ and $m$ (see Sec.\,\ref{section-nrp} and Fig.\,\ref{ylm}) grew
steadily. Nevertheless, asteroseismology in the spirit of Gough and Longair, 
i.e., with the aim to improve the interior physics of stars undergoing
nuclear fusion, was nowhere near the horizon. Major successes were, however,
booked for g~modes of white dwarfs along their cooling track in the HRD
(see Fig.\,\ref{hrd}).  In their review, \citet{Brown1994} indeed discussed the
category of the stars ``unlike the Sun''. They illustrated that white-dwarf
asteroseismology based on weeks-long ground-based multisite monitoring of
nonradial oscillations was furthest advanced, and that the next best cases of
the rapidly-oscillating Ap (roAp) and $\delta\,$Sct stars (see Fig.\,\ref{hrd})
were still limited in terms of physical interpretation. For none of the
other classes of nonradial pulsators in Fig.\,\ref{hrd} did one come anywhere
near making inferences on how to improve the physics of their interiors from
exploitation of the available detected nonradial oscillations.  

Although major achievements were obtained in the decade after this first
review on asteroseismology by \citet{Brown1994}, mainly from ground-based multisite network
campaigns for pulsating white dwarfs \citep[e.g.,][]{Winget1991}, hot subdwarfs
\citep[e.g.][]{Kilkenny1999,Brassard2001}, roAp stars
\citep[e.g.,][]{Kurtz2005}, $\delta\,$Sct stars \citep[e.g.,][]{Breger2005}, and
$\beta\,$Cep stars \citep[e.g.][]{Handler2006}, the plea by \citet{Brown1994} to
replace photometric 
ground-based network observations with data to be taken with spaceborne
telescopes was fully justified. Space data not only would provide much lower
noise by avoiding disturbances due to Earth's
atmospheric variability but also would allow one to increase the duty
cycles of the data significantly, without large daily interruptions of the
time series that plague data from ground-based observatories. Indeed, even
successful multisite campaigns remained below 50\% duty cycle, meaning
that the oscillation cycles were never covered appropriately, except for white
dwarfs, subdwarfs, and roAp stars, all of whose oscillations have periods of
only a few to tens of minutes and dominant mode amplitudes of the order of 
millimagnitudes, corresponding with levels of
parts per thousand (ppt) when considering the star's flux variability rather
than its change in brightness expressed in magnitude. In
retrospect, the gain from space photometry was illustrated by
\citet{Zwintz2000}, who analyzed ten years of Fine Guidance Sensors photometry
of tens of thousands of supposedly ``constant'' guide stars observed with the
Hubble Space Telescope to stabilize the satellite. They found variability
in about 20 stars, among them four K giants revealing periods of a
few hours. They reported this to be incompatible with rotational variability
but did not interpret it in terms of oscillations, which we now know are the cause.

While awaiting space photometry, the hunt for solarlike
oscillations in solar twins from radial-velocity time series grew fast after the
predictions published in the seminal paper by \citet{KjeldsenBedding1995}.  The
ever increasing precision reached by spectrographs led to the first firm discoveries
of individual solarlike oscillation modes in the nearby Sun-like stars
$\eta\,$Boo \citep{Kjeldsen1995}, $\beta\,$Hyi \citep{Bedding2001}, and
$\alpha\,$Cen\,A \citep{Bouchy2001}, after earlier unconfirmed attempts
to find them in Procyon \citep{Brown1991}. Oscillation modes were also detected in
radial-velocity variations of the red giant $\xi\,$Hya \citep{Frandsen2002}. By
the time of space asteroseismology, detections of solarlike
oscillations had been achieved for about 25 bright stars \citep[see Figs.\,2.2 and
2.3 in][for summary plots]{Aerts2010}. All those spectroscopic data revealed
mode frequencies as expected from scaling those of the Sun, which was in line with
the predictions made by \citet{KjeldsenBedding1995}.

\subsection{The beginnings of space asteroseismology}

A half century of intense monitoring of pulsators with nonradial oscillation
modes from ground-based observatories since Ledoux's 1951 inspiring analysis
took place. Despite heroic achievements in terms of number of detected nonradial
mode frequencies in $\delta\,$Sct stars 
\citep[as summarized by Michel Breger,][its pioneer]{Breger2000}, 
the struggle with daily alias frequencies due to
periodic gaps in the data and the lack of unambiguous identification of their
($l,m$) could hardly be overcome. This disappointing situation got
placed in a new light thanks to an opportunity that occurred by accident, and
that is to be taken literally.  The NASA Wide Field Infra Red Explorer (WIRE)
lost its coolant after launch and could not perform the science it was designed
for. Derek Buzasi convinced NASA to reorient the WIRE satellite
project into a proof-of-concept asteroseismology mission, by using its onboard
5\,cm tracker telescope and camera to monitor the variability of various kinds
of bright stars uninterruptedly and with high cadence during several
weeks. Despite major instrumental effects due to telescope jitter (the
machinery was absolutely not built to do what it was used for), this blessing
in disguise immediately showed the potential gain that could be achieved should
a dedicated specifically designed asteroseismology space mission become
available: \citet{Buzasi2000} discovered oscillation modes in the red giant
$\alpha\,$UMa, as anticipated by \citet{Brown1994}. Despite having been an
unplanned pioneer, WIRE achieved ppt-level amplitude
detections and above all illustrated the great improvement of being able to
observe uninterruptedly from space. Among other results, it led to detections of
oscillations in K giants \citep{Stello2008}, showed nonradial modes to be
present in the bright $\delta\,$Sct star Altair \citep{Buzasi2005} where
ground-based monitoring had failed to find any, and drastically improved light
curves of eclipsing binaries \citep{Southworth2007}.

Canada's first space mission, Microvariablity and Oscillations of STars
\citep[MOST, launched in 2003,][]{Walker2003}, was also the first space mission
dedicated to space asteroseismology, although it observed all sorts of stellar
variability. Given its modest aperture of 15\,cm, it is known in the
asteroseismology community as the HST (``Humble Space Telescope'') baptized as
such by its Principle Investigator (PI), Jaymie Matthews. MOST data revealed
numerous oscillation modes in stars belonging to almost all classes indicated in
Fig.\,\ref{hrd}, such as red giants \citep{Barban2007}, 
rapidly oscillation A peculiar (roAp) stars
\citep{Huber2008}, $\delta\,$Sct and $\gamma\,$Dor stars
\citep{Rowe2006,Sodor2014}, emission-line OB stars
\citep{Walker2005a,Walker2005b,Saio2007}, isolated and cluster slowly pulsating
B stars \citep[SPBs,][]{Aerts2006,Cameron2008,Gruber2012}, pre-main sequence
stars -- pre-MS, see \citet{Zwintz2008} and \citet{Zwintz2009}, and many more
\citep[see ][for an early status report]{Matthews2007}. Even though it could
monitor stars for a period of only about six weeks maximally (limiting the
precision of the oscillation frequencies) and its photometric precision in the
time domain was of order ppt, it revealed many more oscillation modes than what
had been achieved from ground-based campaigns.  In many respects MOST was a
highly successful (and inexpensive) planned pioneering mission.

The first ``major'' space mission dedicated to the monitoring of numerous
nonradial pulsators with the aim of space asteroseismology (along
with exoplanet hunting) was the French-led CoRoT mission. It was launched in
2006 into a low-Earth orbit and was operational until 2012
\citep{Baglin2009,Auvergne2009}.  Its original acronym stood for ``Convection
et Rotation'' (CoRot) but later in the project the exoplanet hunting was
added to get the mission funded and hence the ``t'' got upgraded to ``T'' so as
to match ``Convection, Rotation, and exoplanetary Transits''. CoRoT carried a
27\,cm telescope and was dedicated to asteroseismology of tens of
bright stars ($V$ magnitude between 5 and 9) monitored with a cadence of
32\,s and exoplanet hunting around
thousands of faint stars ($V$ magnitude between 11 and 16) measured every
15\,min during each of its pointings \citep[see ][for the technical and
operational details pertaining to the mission]{Auvergne2009}.  
Because of its construction and low-Earth
orbit, CoRoT was able to point in the center or anticenter direction of the
Milky Way over five uninterrupted months (its so-called long runs), 
between which it did short runs of about a month in duration. This meant that
the target selection and the choice of the Fields-of-View (FOVs) to point at were
critical and had to be optimized to meet the wishes of two until then hardly
collaborating communities, the asteroseismologists and the exoplanet hunters,
from the numerous countries that funded the mission. This ``astrosociological''
aspect of the mission led to heated debates (in various languages) during the
so-called ``CoRoT weeks'', which were preparatory workshops held twice a year
to optimize the mission planning and exploitation. Had it not been for the heroic
leadership of the mission PI Annie Baglin, we would have kept on
changing our minds about the pointings until the day of the
launch.

CoRoT was a major success on various fronts. It properly allowed
asteroseismology of Sun-like stars as done by \citet{Michel2008},
\citet{Appourchaux2008}, \citet{Garcia2009}, \citet{Benomar2009},
\citet{Deheuvels2010}, \citet{Mathur2010}, and \citet{Ballot2011}, where the
last study treated an exoplanet host star. It also led to the discovery of
nonradial oscillations in red giants \citep{DeRidder2009}, opening up the major
unexploited parameter space of so-called solarlike oscillations in evolved
stars as studied by \citet{Hekker2009}, \citet{Miglio2009}, \citet{Barban2009},
\citet{Mosser2010}, and \citet{Kallinger2010}.  Applications to other types of
nonradial pulsators are too numerous to mention, but a few breakthroughs were
the discovery of outbursts with accompanying mass loss in Be stars due to the
nonlinear interaction between nonradial modes observed in real time 
\citep{Huat2009}, the occurrence of stochastic nonradial oscillations
in B-type stars \citep[][the latter in the gravito-inertial regime -- see below
for an explanation of these types of
modes]{Belkacem2009,Degroote2010-slo,Neiner2012}, asteroseismic modeling of an
O9 star \citep{Briquet2011}, the discovery of low-frequency variability in
O-type stars that remained unexplained at that time \citep[][see below for
interpretations]{Blomme2011}, and several eclipsing binaries with tidally
induced or tidally affected nonradial oscillations, such as in
\citet{Maceroni2009} and \citet{Maceroni2013} and \citet{daSilva2014}.  Many other results
remain unmentioned here. The special volume 506 (2009) of the journal {\it
  Astronomy and Astrophysics\/}\footnote{Journal volume available in open access
  at {\tt https://www.aanda.org/component/toc/?task=topic\&id=9}.} was
dedicated to 55 CoRoT papers and offers the reader an extensive review
on the mission's instrument performance and scientific results.

\begin{figure*}[t!]
\begin{center} 
\rotatebox{0}{\resizebox{17.cm}{!}{\includegraphics{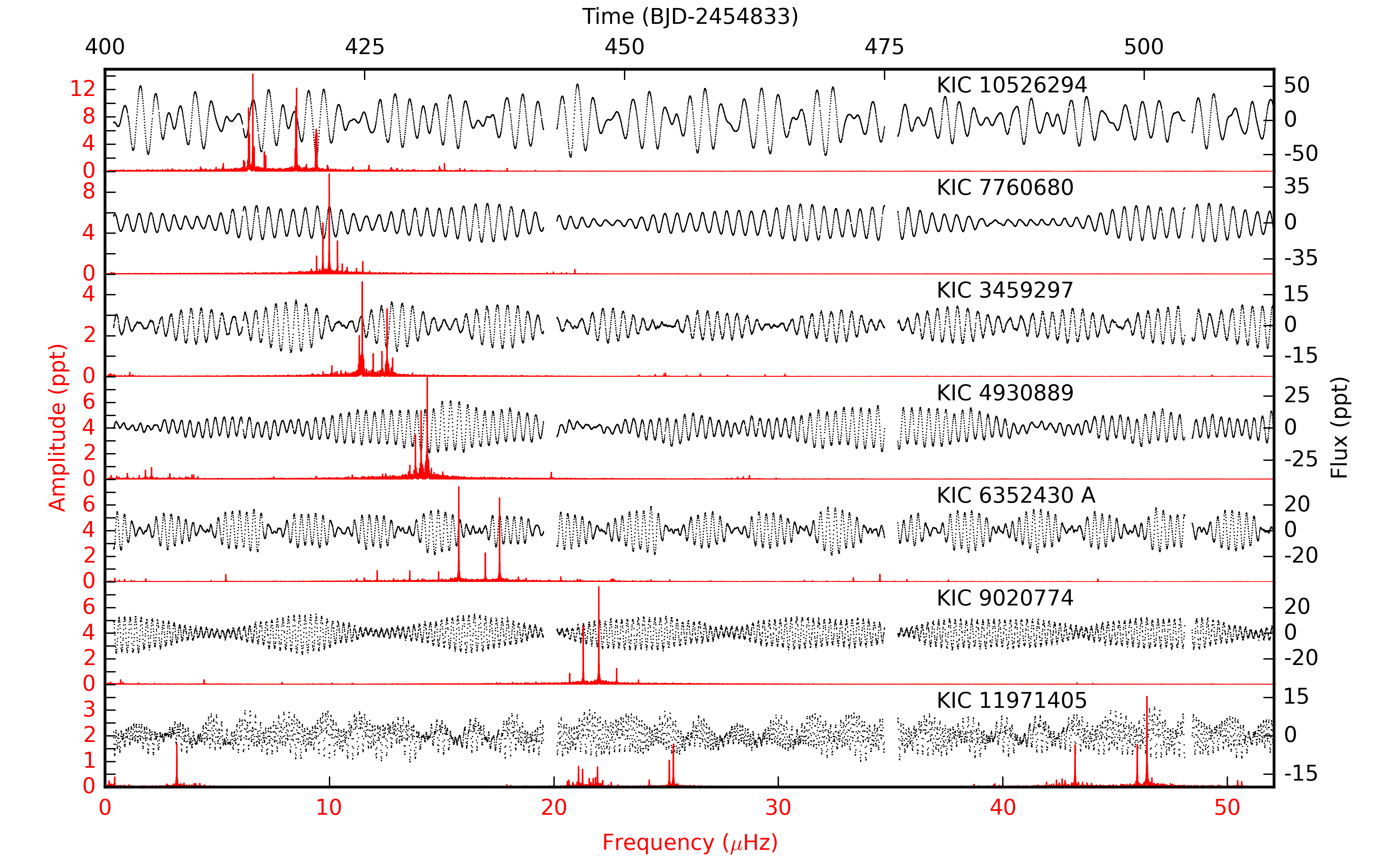}}}
\end{center}
\caption{\label{spbs} Excerpts of 110\,d duration (of the total $\sim\!1500\,$d)
  extracted from the {\it Kepler\/} long-cadence ($\sim\!30$\,min per point) light
  curves (in black dots) of seven slowly pulsating B stars (see
  Fig.\,\ref{hrd}) indicated with their {\it Kepler\/} input catalog
  identification \citep[KIC,][]{Brown2011}.
The amplitude spectrum obtained from a Fourier transform of the full
  {\it Kepler\/} light curves is overplotted in red. These stars exhibit nonradial gravity
  modes with individual mode periods of the order of 1\,d. The light curves
  reveal a gallery of diverse beating patterns among the modes and a gradual
  shift in maximum amplitude from low to high frequency as their rotation
  frequency changes from low in the top panel to high in the bottom
  panel. Despite their large amplitudes of several to $\sim\!12$\,ppt, none of
  these stars were known to have nonradial oscillations prior to the {\it
    Kepler\/} mission; it is notoriously difficult to detect modes with such
  periodicities from ground-based data given their similarity with the rotation
  period of Earth.  Figure based on data from \citet{Papics2017}, by
  courtesy of P\'eter P\'apics, KU\,Leuven.}
\end{figure*}

And then came the amazingly successful NASA {\it Kepler\/} mission
\citep{Koch2010}, launched in 2009, delivering light curves of unprecedented
quality, as shown in Fig.\,\ref{spbs} for a few stars observed in its long
cadence mode (29.43 min). {\it Kepler\/} delivered light curves with a
duration of four years, for about 200000 low- and intermediate-mass stars in
one FOV in the northern sky. These data have a 10 times longer time base and
deliver a factor $\sim\!100$ better precision for the oscillation frequencies
than the CoRoT data, thanks to the larger aperture of the telescope (0.95\,m),
the longer pointing, and the more stable Earth-trailing orbit.  The nominal {\it
  Kepler\/} mission lasted four years and had a dedicated asteroseismology
program \citep{Gilliland2010}, monitoring several hundred low-mass stars at
a short cadence of 58.85\,s.  After the nominal four-year mission, the {\it
  Kepler\/} spacecraft lost two of its four working reaction wheels. The mission
was then repurposed as a space project monitoring fields in the ecliptic,
making clever use of the solar radiation pressure to stabilize the
satellite. This mission operated under the name K2 and monitored 19 fields, each
of which during maximally $\sim\! 80\,$d between February 2014 and October
2018, adopting the same cadence types as {\it Kepler\/}
\citep{Howell2014}.  Because of its superior quality, most of the results
discussed in this review are based on {\it Kepler\/} (or K2) data,
so we do not summarize results here as we did for the other space
missions.

The BRIght Target Explorer constellation (BRITE; launched in 2014 and currently
operational) is a set of Austrian, Canadian, and Polish nanosatellites
assembling multicolor photometry of the brightest stars in the sky for
variability studies and asteroseismology \citep{Weiss2014}. BRITE is unique in
that it offers two-band photometry based on a narrow blue and a broad red
filter.  While the data reduction was initially a challenge, given the limited
weight and pointing stability of the small satellites, its photometric precision
currently reaches ppt per data point. It can monitor selected stars during
about half a year \citep{Pablo2016}.  BRITE is monitoring a variety of
bright variables.  BRITE's data of high-mass stars is complementary to the {\it
  Kepler\/} data in terms of targets. BRITE revealed several more nonradial
oscillation modes than what has been found in ground-based data for OB- and 
Be-type
pulsators \citep[e.g.][]{Baade2016,Pigulski2016,
  Handler2017,Kallinger2017,Ramiaramanantsoa2018}. Combined BRITE data and
archival data assembled from extensive ground-based (multisite) campaigns or
data that are currently being assembled by the NASA Transiting Exoplanet Survey
Satellite \citep[TESS,][]{Ricker2016} hold good potential for asteroseismology
of the highest-mass nearby stars with high-amplitude oscillation modes to
perform modeling of their interior properties.

\citet{Handler2019} illustrated the TESS potential in
revisiting bright B-type pulsators discovered from the ground but lacking
sufficient identified pulsation modes.  The nominal TESS mission is scanning
almost the full sky, delivering high-precision space photometry for millions of
stars with time bases between 27 and 352\,d for each of the hemispheres, with
similar cadences as {\it Kepler}. This difference in duration of the monitoring
is due to its operational scheme of observing in sectors (13 per hemisphere),
each monitored over 27\,d. Increasing partial overlap between the
consecutive sectors occurs for areas in the sky closer to the ecliptic pole,
with a maximum of continuous observation over 352\,d for stars in the TESS
Continuous Viewing Zone (CVZ).  TESS samplings changed to 10\,min and 20\,s
for the long and short cadence modes, respectively, in the extended
mission (started in mid 2020). Owing to the more limited time base it delivers far
less precise oscillation frequencies for asteroseismology than {\it Kepler\/},
but it opens up the entire sky to provide large samples of pulsators. Some of
these samples were not yet treated by CoRoT and {\it Kepler}, including
metal-poor high-mass stars in the Large Magellanic Cloud (LMC).  Thus far large
asteroseismology samples have essentially been limited to red-giant pulsators.  In
that sense, TESS will bring major advances for a variety of stars
across the HRD in Fig.\,\ref{hrd}, from low-mass unevolved Sun-like stars
\citep{Schofield2019}, including exoplanet hosts \citep{Campante2016a}, all the
way up to the most massive stars.

The past decade brought us to a golden era for asteroseismology,
with the BRITE and TESS missions ongoing, the PLAnetary Transits and
Oscillations of stars mission \citep[PLATO,][]{Rauer2014} on the horizon, and an
immense amount of {\it Kepler\/} data yet to be interpreted in the true meaning
of asteroseismology, i.e., with the aim of improving the physics of stellar
interiors. How to achieve that, is discussed in the rest of the review.

\subsection{Asteroseismology to improve stellar evolution}

The HRD in Fig.\,\ref{hrd} reveals that pulsational variables occupy many phases
of stellar evolution. The periods of nonradial oscillations covered by stars and
stellar remnants range from seconds to months or even years and their amplitudes
of brightness variations cover the range of a magnitude to the current detection
treshold of $\mu$mag (corresponding to hundreds of ppt to ppm in flux
variability).  We refer to Table\,A.1 given by \citet{Aerts2010} for a summary of the
pulsation characteristics.  The basic properties and the excitation mechanisms
of all known classes of nonradial pulsators indicated in Fig.\,\ref{hrd}
were discussed in great detail in Chap.\,2 given by \citet{Aerts2010}.  
Noteworthy discoveries of nonradial oscillations not yet firmly
established from ground-based data were made for red giants from CoRoT
\citep{DeRidder2009} and for blue supergiants from MOST \citep{Saio2006},
Hipparcos \citep{Lefever2007}, and {\it Kepler\/} \citep{Aerts2017b}.  While
oscillations were already discovered in red giants from ground-based
spectroscopy, it was still heavily debated whether or not it concerned radial or
nonradial oscillations \citep[e.g.,][]{Frandsen2002}.  Both red giants and blue
supergiants are meanwhile established as nonradial pulsators from early {\it
  Kepler\/} \citep{Bedding2010}, K2 \citep{Bowman2019b}, and TESS data
\citep{Pedersen2019}.  Hence they received their own ellipse and Fig.\,\ref{hrd}
was adapted accordingly.

The {\it Kepler\/} spacecraft led to the discovery of more than 20000 red giants
with nonradial oscillations \citep{Hon2019}, and TESS will undoubtedly provide a
factor of 10 more. Numbers for blue supergiants discovered from K2 and TESS are
much lower, in the order of a few hundred, because these are rare
objects. Moreover, they were omitted from the nominal {\it Kepler\/} FOV so as
not to ``disturb'' the exoplanet hunting. The variability of O-type dwarfs
and blue supergiants is caused by a complex interplay between various phenomena,
which may involve internal gravity waves \citep[IGWs,][]{Rogers2013}, rotational
modulation \citep{Ramiaramanantsoa2018}, sub-surface convection
\citep{Grassitelli2015}, wind variability \citep{Krticka2018}, magnetism
\citep{Sundqvist2013}, nonradial g~modes \citep{Moravveji2012}, and
binarity \citep{Sana2012}.  The discovery of ubiquitous low-frequency power
excess in hundreds of OB dwarfs and supergiants in the upper HRD from K2 and
TESS space photometry by \citet{Pedersen2019} and \citet{Bowman2019b} granted
them an ellipse in Fig.\,\ref{hrd} labeled as ``SLF'' for stochastic
low-frequency variability.

\citet{Gautschy1995,Gautschy1996} announced that the
importance of nonradial oscillation studies would grow as monitoring capacities
to detect ever smaller amplitude variability would improve: a visionary
outlook a few years before space asteroseismology came about with WIRE.
Figure\,\ref{hrd} shows that stars across almost the entire mass range will
encounter nonradial oscillations at particular stages of their evolution. The
characteristics of these nonradial oscillation modes (their periods or
frequencies, amplitudes, and mode lifetimes) offer great diagnostic value for
inferences of the stellar interior. As outlined in detail in
Sec.\,\ref{section-nrp}, the mode frequencies or periods allow high-precision
views of the physical properties inside stars that are not accessible by
classical ``snapshot-type'' data assessing only the surface or atmospheric
properties at a particular time of the variability cycle (such as single-epoch
spectroscopy, color indices, or interferometry).  The combination of
high-cadence time-series data covering the overall pulsational variability
cycle, along with observables representing the position in the HRD
$(T_{\rm eff},L/L_\odot)$ and surface abundances from spectroscopy, constitute
an optimal starting point for asteroseismic modeling.  The power of combining
such a classical and asteroseismic approach for stellar modeling was 
illustrated by \citet{Lebreton2014-EAS} in their Fig.\,13, with major improvement for
stellar aging. In practice, the addition of asteroseismology to stellar
modeling permits to reach levels of $\sim\! 10\%$ precision in stellar ages.
Precise values for the stellar luminosity $L$ or the stellar radius
$R_\star$ of low- and intermediate-mass stars can now be obtained from Gaia
astrometry \citep{Gaia-Brown2018} and interferometry,
respectively. Following the extensive review on how to bridge
asteroseismology and interferometry by \citet{Cunha2007}, this synergy turns out
to be highly successful for the brightest pulsators with space asteroseismic data,
as highlighted by \citet{North2007}, \citet{Bazot2011}, \citet{Huber2012},
\citet{White2013}, and \citet{White2018}.

To conclude this section, nonradial oscillations occur all over the HRD.
This offers the exciting opportunity to perform asteroseismology for members of
the classes indicated in Fig.\,\ref{hrd} and to couple the conclusions into a
coherent picture across stellar evolution. Such a meta-study has been done for
asteroseismic estimates of the interior rotation of low- and intermediate-mass
stars, leading to the conclusion that the theory of angular momentum transport
in stellar interiors already needed improvements from the earliest stages of
stellar evolution.  We come back to this in
Sec.\,\ref{section-astero} but stress here that a global perspective of stellar
evolution across all life phases proved to be necessary to assess the weaknesses
of one particularly important aspect of stellar evolution theory: the transport
of angular momentum is much more efficient than predicted by theory \citep[see
Fig.\,4 in][which summarizes a global result based on major efforts by many
scientists in the asteroseismology community]{Aerts2019}. This is just one
example of how asteroseismology paves the way towards better stellar evolution
models. Data are now being assembled by the TESS mission to reach the same impact
for high-mass stars. For the rest of this review, we adopt the same definition
as \citet{Aerts2019} to discriminate between stars of low
($M_\star \simkl 1.3\,$M$_\odot$), intermediate
($1.3 \simkl M_\star \simkl 8\,$M$_\odot$), and high
($M_\star \simgr 8\,$M$_\odot$) mass.

\subsection{Working in the Fourier domain truly helps}

Figure\,\ref{spbs} is a textbook illustration of how stars of intermediate
mass exhibiting g~modes behave in flux as their interior
rotation increases (from top to bottom in the plot). We explain in
Sec.\,\ref{section-nrp} how to arrive at such a conclusion about their interior
rotation, but for now the message is that this result cannot be ``seen'' easily
in the time domain (i.e., in the light curve in gray), while it can be directly 
distilled from the Fourier spectrum overplotted in red. It is amazing how much
information about the stellar interior is contained in a Fourier transform of a
light curve: the art is to get it out, and asteroseismologists are experts in
this regard.

We see in Fig.\,\ref{spbs} that the duty cycle of the {\it Kepler\/}
data is high, but not 100\%, because the satellite had to be turned every
three months to keep its solar panels pointed to the Sun. Moreover, one-month
data downlink and momentum dump desaturation interruptions occurred
approximately 
every 3\,d.  Moreover, the {\it Kepler\/} photometer consisted of 21 CCD
modules \citep{Koch2010}, but one of them (Module\,3) broke down less than a
year into the mission, causing gaps in the data for stars on that module.  The
{\it Kepler\/} spacecraft sampled the stars at a constant cadence, delivering
data at precise time stamps. The data are modulated with 1-yr periodicity
when transformed to the barycenter of the Solar System \citep{Murphy2013a}.
Given that stellar oscillations are periodic because they occur at the
``eigenfrequencies'' of the star, Fourier analysis offers an optimal frequency
extraction method. It is extensively described in Chap.\,5 given by
\citet{Aerts2010} in the general case of nonequidistant gapped time series of
ground-based data. An excellent crash course on the
topic was given by \citet{Appourchaux2014}, who tuned it to the modern era of space
asteroseismology, while the effect of interruptions in the {\it Kepler\/} light
curves on the frequency analysis for asteroseismology was thoroughly assessed by
\citet{Garcia2014-data}.  \citet{BasuChaplin2017} provided even more detailed information and 
focused on data analysis in the case of
solarlike oscillations. Here we limit ourselves to the bare minimum and pay specific
attention to the aspects of time-series analysis from the viewpoint of having
two major categories of nonradial oscillation modes: those that are damped and
have short lifetimes and those that are undamped and, to a good
approximation, have infinite lifetimes. These two cases require different data
analysis approaches.

As explained later, the three components of the Lagrangian
displacement vector due to a nonradial oscillation mode in the absence of
rotation contain a common time-dependent factor $\exp (-{\rm i\,}\omega t)$,
with $\omega=2\pi\nu$ the angular frequency of the mode, $\nu$ its cyclic
frequency and $P=2\pi/\omega=1/\nu$ its period. In general, $\omega$ is an
imaginary quantity $\omega=\omega_{\rm r}+{\rm i}\,\omega_{\rm i}$, but for the
study of the periodic behavior of the mode to be derived from data we consider
its real part (for simplicity denoted as $\omega$ in the rest
of the review).  Imagine that we want to extract the frequencies present in
time-series data representing a continuous and finite function $x(t)$ (the flux
variations in the case of space asteroseismology, indicated with  the black dots in
Fig.\,\ref{spbs}).  The Fourier transform of $x(t)$ is given by
\begin{equation}
F(\nu)\equiv\int_{-\infty}^{+\infty}x(t)\exp (2\pi{\rm i\,}\nu\,t)dt.
\label{FT}
\end{equation}
By performing this transformation, we move from the time domain (black dots in
Fig.\,\ref{spbs}) to the frequency domain (shown in red in Fig.\,\ref{spbs}).
In the
following
case where $x(t)$ is a sum of harmonic functions with frequencies
$\nu_1,\ldots,\nu_M$ and amplitudes $A_1,\ldots,A_M$~:
\begin{equation}
x(t)=\sum_{k=1}^M A_k \exp (2\pi{\rm i\,}\nu_k t),
\end{equation}
we find that
\begin{equation}
\label{noiselessFT}
F(\nu)=\sum_{k=1}^M A_k\delta (\nu-\nu_k),
\end{equation}
where $\delta$ is Dirac's delta function for which $\delta (\nu-\nu_k)\neq 0$
only for the frequencies $\pm \nu_1,\ldots,\pm \nu_M$.  No matter how good the
{\it Kepler\/} data are, the time series (1) contains discrete data points,
(2) has finite duration, and (3) has gaps (even if they are small). This implies that one
cannot compute the integral in Eq.\,(\ref{FT}). 
However, we can rely on 
the so-called window function defined by the data,  
measured at time points $t_j, j=1,\ldots,N$ during the
time interval $[0,T]$ to obtain:
\begin{equation}
w_N(t)\equiv\frac{1}{N}\sum_{j=1}^N\delta (t-t_j).
\end{equation}
This allows us to write the Discrete Fourier Transform (DFT)
of the function $x(t)$ as
\begin{equation}
F_N(\nu) = N \int_{-\infty}^{+\infty}x(t)w_N(t)\exp (2\pi{\rm i\,}\nu t)dt.
\end{equation}
The following DFT of the window function is called the spectral window $W_N(\nu)$~:
\begin{equation}
W_N(\nu)=\frac{1}{N}\sum_{j=1}^N\exp (2\pi {\rm i\,}\nu t_j).
\label{specwin}
\end{equation}
The DFT of the data is hence the convolution of its spectral window and its
Fourier transform
\begin{eqnarray}
\label{convolution}
F_N(\nu) & = & N (F \ast W_N)(\nu) \nonumber \\
& = & \sum_{j=1}^N x(t_j)\exp (2\pi{\rm i\,}\nu t_j) \Delta t_j\; ,
\end{eqnarray}
where $\Delta t_j\equiv t_j-t_{j-1}$ \citep{Deeming1975}.  Frequencies are often
searched from the power density (PD) instead of the Fourier transform. The PD is defined as
the power present in the signal as a function of frequency, per unit
frequency. Relying on Eq.\,(\ref{convolution}), we get
\begin{equation}
\label{PD}
{\rm PD} (\nu) \equiv \frac{1}{T} \left| F_N(\nu) \right|^2.
\end{equation}
The PD is hence expressed in ppm$^2$/$\mu$Hz when the time series concerns flux
measurements expressed in ppm and frequencies are expressed in microhertz.  

For equidistant data, $t_j=j\cdot\Delta t$, where $\Delta t_j=\Delta t$ is a
constant sampling interval between two consecutive measurements.  In that case, we have
\begin{eqnarray}
\label{DFT}
F_N(\nu) & = & \Delta t\cdot \sum_{j=1}^N x(j\Delta t)
\exp \left[2\pi{\rm i\,}\nu (j\Delta t)\right]\\
\label{alias}
W_N(\nu) & = &
\displaystyle{\exp (\pi {\rm i\,}\nu \Delta t(N+1))
\frac{\sin (\pi \nu N\Delta t)}{N\sin (\pi \nu \Delta t)}}.
\end{eqnarray}
The Nyquist frequency of such data is half of the sampling rate
$\nu_{\rm Ny}=1/2\Delta t$. This frequency is defined as the upper limit of the
frequency range over which the Fourier transform is unique.  In principle this
sets an upper limit for the interval of frequencies to search for. However, a
particular frequency detected below $\nu_{\rm Ny}$ may be an alias of the true
frequency that occurs above $\nu_{\rm Ny}$. Alias frequencies are frequency
values given by the difference between the actual frequency of the signal and
integer multiples of the sampling rate.  Thus, aliasing allows the detection of
true frequencies, even though they occur above $\nu_{\rm Ny}$
\citep{Murphy2013a}.

Equations\,(\ref{DFT}) and (\ref{alias}) reveal that $F_N(\nu)$ reaches maxima
for alias frequencies $\nu_j=j/\Delta t$. Ground-based data are not evenly
sampled but they give rise to 1-d and 1-yr alias frequencies whose values of
$F_N(\nu_j)$ may be similar, preventing from unraveling the ``true'' frequency from
the one introduced by the periodic gaps in the data. This has been a major
show-stopper for ground-based asteroseismology, particularly in the case of
stars with ``slow'' g~modes with periods of approximately 1\,d as in Fig.\,\ref{spbs}.
While these g~modes have amplitudes that are easily detectable with ground-based
instruments, the 1-d aliasing problems are so severe that one can deduce
a few mode frequencies at best, even after yearslong (often multisite)
observations; see e.g.\ \citet{Zerbi1999}, \citet{DeCatAerts2002},
\citet{Mathias2004}, and \citet{Cuypers2009}. This is why {\it
  Kepler\/} was so groundbreaking in the field of slow multiperiodic g-mode
pulsators. {\it Kepler\/} data also revealed that a good fraction of dwarf
pulsators with a convective core (indicated as $\gamma\,$Dor, $\delta\,$Sct,
SPB, and $\beta\,$Cep stars in Fig.\,\ref{hrd}) are actually hybrid pulsators;
i.e., they have both short-period (lasting hours) p~modes and long-period
(lasting days) g~modes. Such pulsators have major potential, as their p and
g~modes offer local {\it in situ\/} measurements in different regions in the star:
p~modes probe the envelope, while g~modes probe the radiative near-core regions
and allow us to turn the study of deep stellar interiors into observational
astronomy.

In practice, applications of frequency analysis to space photometric data are
always done after preprocessing and postprocessing of the light curves deduced from the
raw data. Satellite repointings (every three months in the case of {\it
  Kepler\/}, so-called quarters) imply jumps in the time series. Moreover,
satellite drift has to be corrected for each quarter.  Subsequently, outlier
removal and detrending are applied as standard processing to get interpretable
light curves \citep{Garcia2011}. Such corrections and processing were applied to
get the light curves in the version shown in Fig.\,\ref{spbs}.

\subsubsection{Mode damping and mode lifetimes}

Thus far we have considered a multiperiodic harmonic signal.  For stars like those in
Fig.\,\ref{spbs}, this is relatively straightforward because they have modes
with extremely long lifetimes.  However, a distinction has to be
made between two cases: damped modes with lifetimes of the order of or shorter
than the duration of the time series $T$ and undamped (or so-called
self-excited) modes that never die out and are always active with constant phase
throughout the data gathering. The first option occurs for stars such as the Sun
with
oscillation modes triggered by turbulent motions in their outer convective
envelope. Such ``solarlike'' oscillations are expected to
occur for all stars whose convective envelopes contain sufficient mass to be
highly turbulent.  These oscillations are of stochastic nature in the sense that
they are regularly but randomly excited to more or less the same amplitude but
they damp out relatively quickly, in a time span on the order of days to months,
while continuously being reexcited. At each time stamp, the phase within the
oscillation cycle is perturbed stochastically relative to the previous
measurements.

Stars with radiative envelopes can excite oscillations via a heat mechanism,
because some of their partial ionization layers manage to transform radiative
energy, created in their deep interior by nuclear fusion, into mechanical energy.
This can excite modes that resonate inside a mode cavity. This occurs
because of local opacity peaks in partial ionization layers of hydrogen, helium,
or ironlike species in the outer envelope. Such mode excitation works along the
lines of a valve mechanism (also known as thermodynamical Carnot cycle) and
may excite radial and nonradial oscillation modes.  Because of the key role
played by the opacity in getting the modes excited, it is usually called the
$\kappa$ mechanism.  To a good approximation such resonating modes do not die out
as long as radiative energy is delivered to the excitation layer.  Hence these
modes have ``quasi-infinite'' lifetimes (i.e. they are infinite to a good approximation
compared to the duration of the dataset; hence we drop ``quasi'' in the
text). In such a case, it is predictable at what phase in its oscillation cycle
the mode will be throughout time.

We return to mode excitation in Sec.\,\ref{section-astero} but in any case
we take a data-driven approach to asteroseismology;  i.e., we use as many
independent eigenmode frequencies as possible, as long as we can extract them
from the data, irrespective of how the star managed to excite these
oscillations.

\subsubsection{Undamped oscillations with quasi-infinite lifetimes}

We first treat the case of heat-driven undamped modes.
In this case [see Eq.\,(\ref{noiselessFT})] we seek
to extract the optimal sum of harmonic functions with frequencies
$\nu_1,\ldots,\nu_M$ and amplitudes $A_1,\ldots,A_M$, where the number of modes
$M$ is unknown, keeping in mind the presence of instrumental noise in the
data. Under the optimistic
assumption of uncorrelated data with white Gaussian noise, a
convenient approximation 
of the Fourier transform is the Lomb-Scargle (LS)
periodogram, defined as
\begin{eqnarray}
P_{\rm LS}(\nu) & = & \displaystyle{\frac{1}{2}}
\frac{\displaystyle{\left\{\sum_{i=1}^Nx(t_i)\cos [2\pi \nu (t_i-\tau_0)]\right\}^2}}
{\displaystyle{\sum_{i=1}^N\cos^2 [2\pi \nu(t_i-\tau_0)]}} \nonumber \\
& + & \displaystyle{\frac{1}{2}}
\frac{\displaystyle{\left\{\sum_{i=1}^Nx(t_i)\sin [2\pi \nu(t_i-\tau_0)]\right\}^2}}
{\displaystyle{\sum_{i=1}^N\sin^2 [2\pi \nu(t_i-\tau_0)]}}\; ,
\label{scargle}
\end{eqnarray}
where $\tau_0$ is chosen such that $P_{\rm LS}(\nu)$ becomes invariant with
respect to the starting date of the dataset:
\begin{equation}
\displaystyle{\tan(4\pi \nu\tau_0)=
\frac{\displaystyle{\sum_{i=1}^N\sin (4\pi \nu t_i)}}{\displaystyle{\sum_{i=1}^N\cos
(4\pi \nu t_i)}}}
\label{taunul}
\end{equation}
\citep{Scargle1982}. Along with the DFT, the LS periodogram is
widely used in asteroseismology of stars with heat-driven modes. Both formalisms
are suitable to treat gapped nonequidistant time-series data while requiring
only a short computation time \citep{Kurtz1985}. 
\citet{Horne1986} provide guidance for estimation of the number of
independent frequencies as well as a method for detecting the presence of alias
frequencies caused by the interaction between the window function and the
observed data values.  In the limit of $N\rightarrow\infty$, one has
$P_{\rm LS} (\nu_k) \approx A_k^2N/4$ for each of the modes $k=1,\ldots,M$.  The
LS amplitude spectrum $A_{\rm LS}(\nu) \equiv \sqrt{4P_{\rm LS}(\nu)/N}$
thus gives the physically relevant quantities to perform
asteroseismology, i.e., the frequencies and amplitudes of the modes. This is the
amplitude spectrum shown in red for the seven B-type pulsators in
Fig.\,\ref{spbs}.  

\citet{Scargle1982} showed the maxima of $A_{\rm LS}(\nu)$ to lead to amplitudes
$A_k$ that are statistically equivalent to those obtained by performing a
least-squares optimization to the light curve in the time domain in the limit of
large $N$. Even modern datasets may consist of only a limited
number of data points.  One therefore best performs an optimization to estimate
the amplitudes and frequencies once an approximate value of $\nu_k$ is known
from the DFT or LS periodogram, as a good starting value to perform a
least-squares fitting in the time domain.  This gives rise to the method of
``prewhitening.''

We now consider the dominant mode with frequency
$\nu_1$ deduced from the DFT or LS amplitude spectrum. Minimizing 
the sum of squares of the residuals,
\begin{eqnarray}
\phantom{a} & R^2(\nu_1)  \equiv  
\displaystyle{\sum_{i=1}^N\left[x(t_i)-x^c(t_i)\right]^2}\\
\phantom{a} &\ \  = 
\displaystyle{\sum_{i=1}^N\left\{x(t_i) - \left\{A_1\cos\left\{2\pi\left[\nu_1
          (t_i-\tau)+\psi_1\right]\right\} + c\right\}\right\}^2 } \nonumber
\label{likelihood}
\end{eqnarray}
leads to optimized values for  $\nu_1, A_1, \psi_1$, and $c$, and 
provides the residual light curve with an average value of zero
\begin{equation}
x_R(t_i)=x(t_i)-x^c(t_i).
\label{prewh}
\end{equation}
A second frequency is then searched for by computing the LS amplitude spectrum
for the residual light curve $(t_i,x_R(t_i))$ and optimizing its values
$\nu_2, A_2, \psi_2$, etc. This procedure is repeated until the periodogram no
longer leads to frequencies that are significant for a specified
criterion, such as the ones discussed in \citet{Horne1986},
\citet{Breger1993}, and \citet{Degroote2009}. For a {\it Kepler\/} light curve, this procedure is tedious
and time consuming as it leads to hundreds of significant frequencies. 
To finalize the list of frequencies due to independent oscillation modes and
their uncertainties, great care must be taken to properly account for the
occurrence of combination frequencies and harmonics $r\nu_i+s\nu_j$ with
$i,j \in \mathbb{N}; r$ and $s \in \mathbb{Z}$ due to nonlinearities in the light
curve rather than independent mode frequencies, keeping in mind the spectral
window and the introduction of spurious frequencies due to limited 
time resolution
\citep{Loumos1978} during the prewhitening process.  This was thoroughly
discussed by \citet{Papics2012a}, \citet{Balona2014}, \citet{Kurtz2015}, and
\citet{Bowman2017}.  Moreover, one should correct the error estimation of the
frequencies and their amplitudes for the correlated nature of the data
\citep[e.g., as outlined in][for the highly oversampled CoRoT
data]{Degroote2009}.  Although this correction is often omitted, one 
should apply it to the error estimates of the derived amplitudes, phases, and
frequencies. For the case of $N$ data points, one has
\citep[see ][]{Montgomery1999}:
\begin{equation}
\displaystyle{\sigma_\nu=\frac{D \sqrt{6}\sigma_N}{\pi\sqrt{N} A T},\, 
  \sigma_A= D\sqrt{\frac{2}{N}}\ \sigma_{\rm N},\,
  \sigma_\psi=\frac{D \sigma_N}{\pi\sqrt{2N} A}.}
\label{fout}
\end{equation}
In these three expressions, $\sigma_N$ can be approximated by the standard
deviation of the final residual light curve. The correction factor 
due to the correlated nature of the data $D$ depends on the instrument
properties and the sampling rate and can be estimated as the square root of the
average number of consecutive data points of the same sign in the final residual
light curve, as explained in \citet{Schwarzenberg-Czerny1991} and applied to
ground-based photometry by \citet{Schwarzenberg-Czerny1998} and to space
photometry by \citet{Degroote2009}.  Values for $D$ are typically between two
and ten for the {\it Kepler\/} long-cadence and CoRoT asteroseismology data
sets.  

\subsubsection{Damped oscillations with short lifetimes}

\begin{figure*}
\begin{center}
%\begin{tabular}{cc}
\rotatebox{270}{\resizebox{5.2cm}{!}{\includegraphics{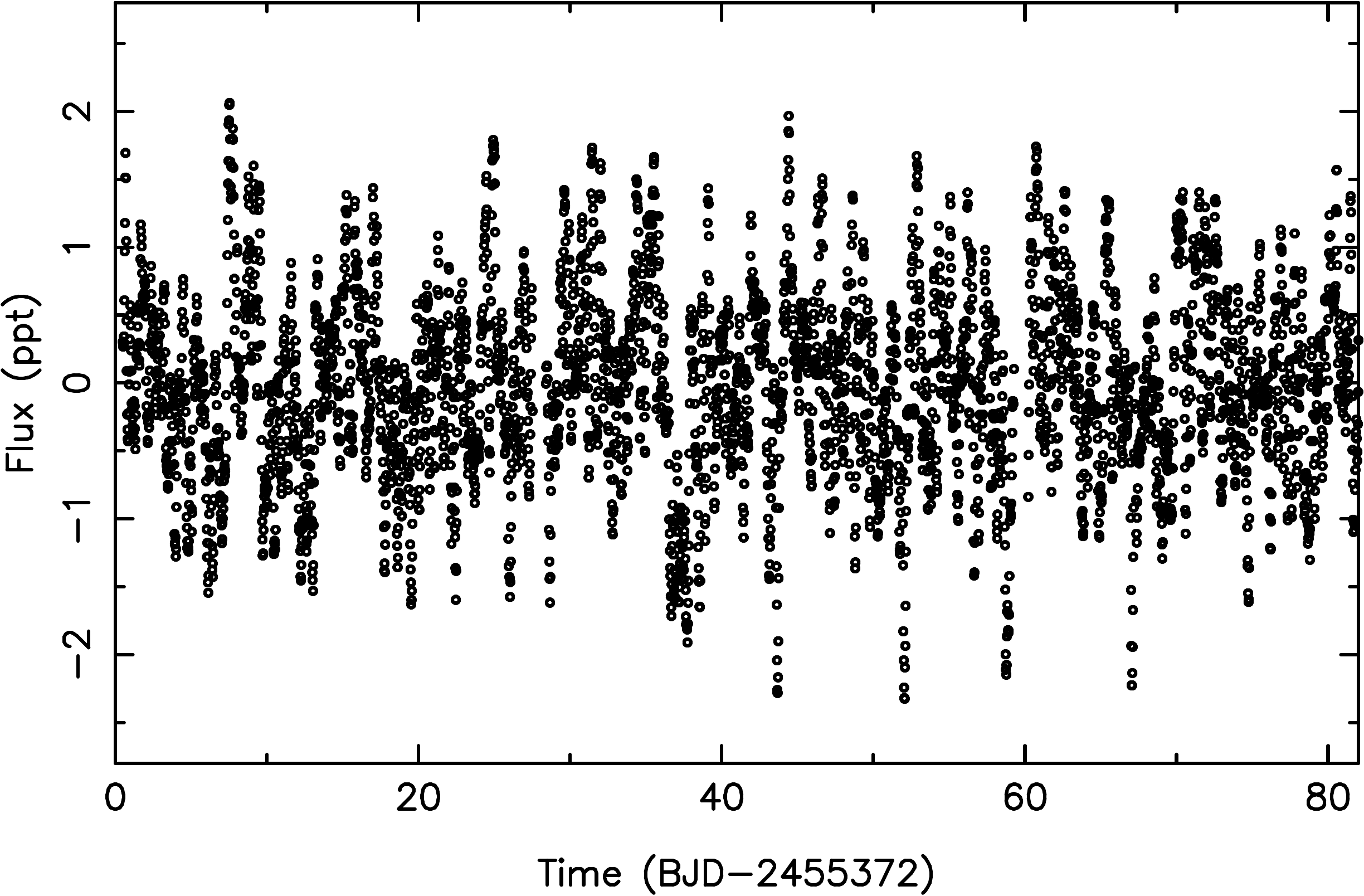}}}\hspace{1.cm}
\rotatebox{270}{\resizebox{5.2cm}{!}{\includegraphics{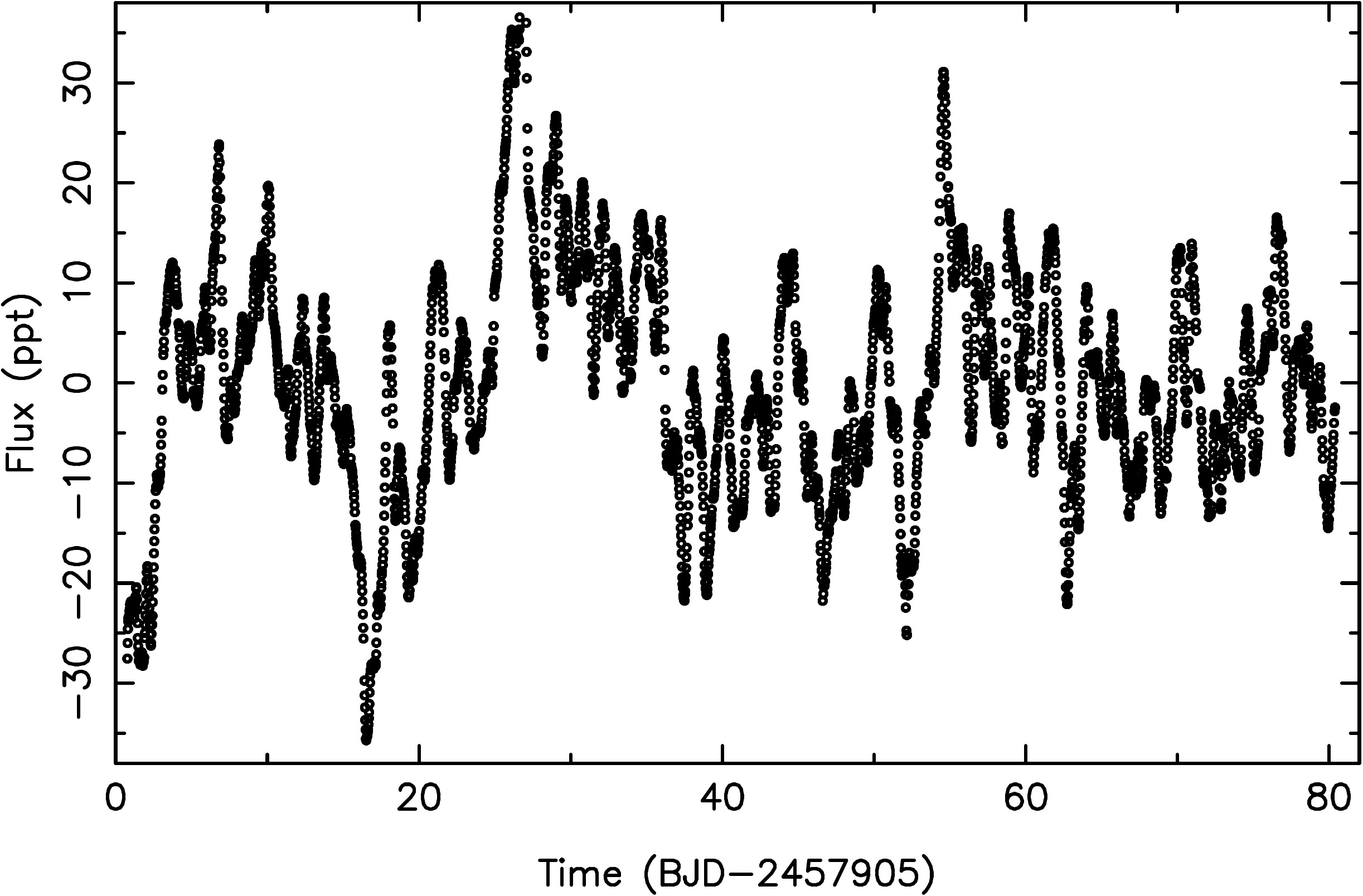}}}\\
\rotatebox{0}{\resizebox{8.9cm}{!}{\includegraphics{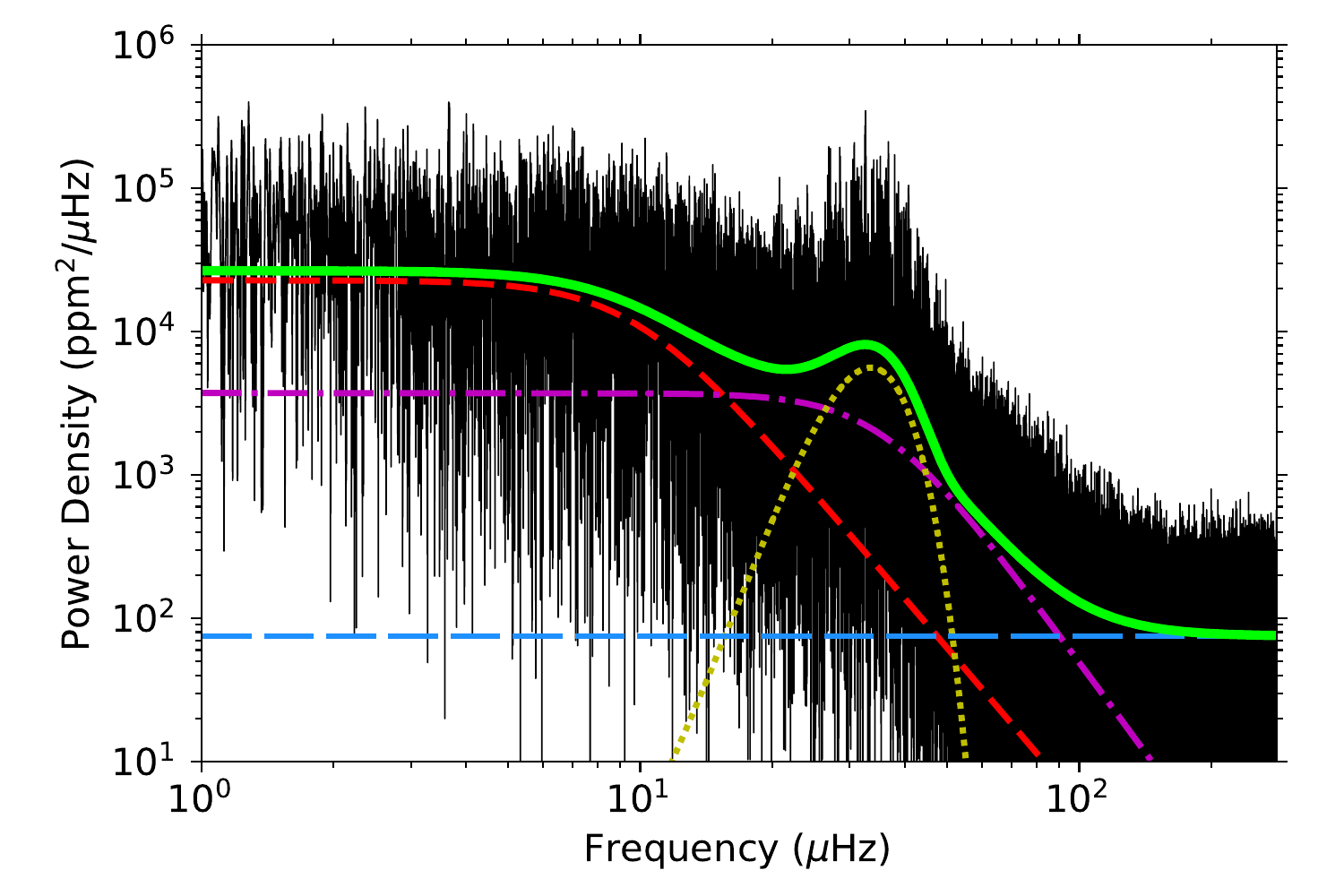}}}
\rotatebox{0}{\resizebox{8.9cm}{!}{\includegraphics{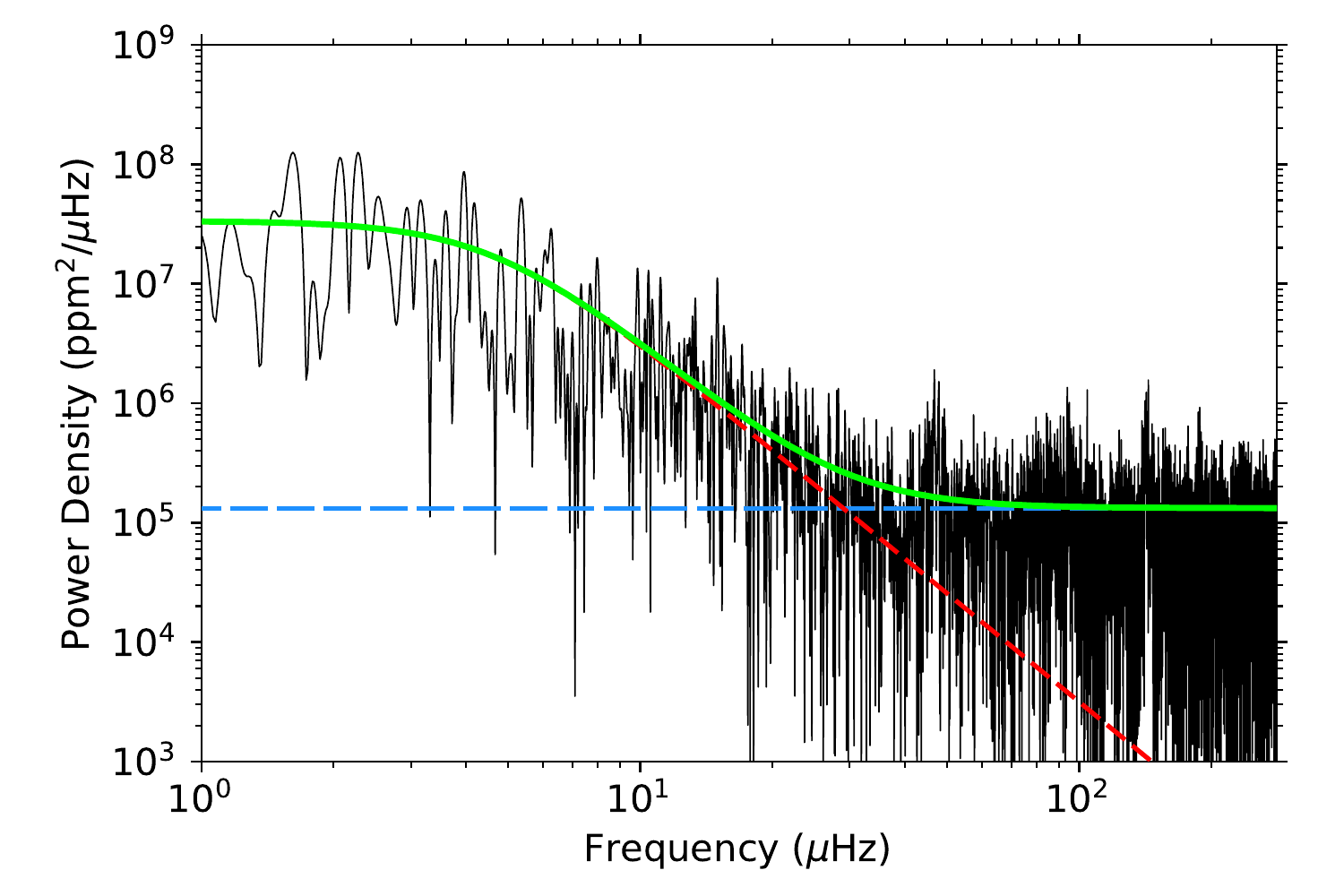}}}
%\includegraphics[width=5cm,angle=270]{RGB-LC.eps} &
%\includegraphics[width=5cm,angle=270]{BSG-LC.eps} \\
%\includegraphics[width=7cm]{RGB_kplr007949599_PSD_fitted_log.eps} &
%\includegraphics[width=7cm]{BSG_EPIC200182931_PSD_fitted_log_PSD.eps}
%\end{tabular}
\end{center}
\caption{\label{RGB-BSG} Part of the {\it Kepler\/} light curve (top left panel) and
  the power density spectrum (bottom left panel, black) of the red-giant star
  KIC\,007949599.  Overplotted in the bottom left panel are the model fits from
  Table\,1 given by \citet{Kallinger2014} based on their Eq.\,(2), representing two
  Lorentzian components due to granulation variability at low frequencies
  (dashed red and dot-dashed purple lines) and a Gaussian component due to stochastic
  p-mode oscillations (dotted yellow line).  K2 light curve (top right panel) and its power
  density spectrum (bottom right panel, black) of the blue supergiant star
  $\rho\,$Leo.  Overplotted in the bottom right is a Lorentzian model fit
  (dashed red line) using the formalism given by \citet{Bowman2019a} to
  represent the power excess due to the low-frequency stochastic variability.  
In both bottom panels, the blue dashed horizontal lines
  indicate the photon noise level and the green lines represent the superposition
  of the individual model components representing the variability.  Figure based
  on data from \citet{Kallinger2014} and from \citet{Aerts2018b}, by courtesy of
  Dominic Bowman, KU\,Leuven.}
\end{figure*}

The frequency analysis for stochastically excited damped solarlike
oscillations requires a different approach.  Because of 
the random excitation and the damping of the modes, the functional form of the light
curve changes.  The simplest case of  one
damped oscillation mode with frequency $\nu_1$ can be described as
\begin{equation}
x(t_i)=A_1\cos\left[2\pi (\nu_1 t_i+\psi_1)\right]\exp(-\eta_1 t_i) + c,
\label{damping}
\end{equation}
with $\eta_1$ the damping rate of the mode, which is the inverse of the mode
lifetime. The latter is unknown and hence must be estimated along with the
frequency. In the hypothetical case of having continuous observations of such a
signal over an infinite amount of time, the power spectrum is given by
\begin{equation}
\displaystyle{P(\nu)=\frac{1}{4}\frac{A_1^2}{\left(4 \pi^2(\nu-\nu_1)^2+(\eta_1)^2\right)}} \; .
\label{powerlorentz}
\end{equation}
In such a simplified (actually, unrealistic) case, the power spectrum thus takes a
Lorentzian profile around the frequency $\nu_1$ with the linewidth of the mode
given by the full width at half maximum $\Gamma\equiv 2\eta_1/2\pi$.
Estimation of $\Gamma$ can be accompanied by major uncertainty, unless long and
high-quality datasets, such as those assembled by {\it Kepler}, are available.
%For elaborate discussions on how the mode linewidths of p~modes were derived
%from {\it Kepler\/} light curves for dwarfs, subgiants, and red giants
%we refer to \citet{Appourchaux2012a} and
%\citet{Vrard2018}. Values
%translate into mode lifetimes from a fraction of a day for dwarfs to several
%tens of days for red giants. \citet{Mosser2018} succeeded in deriving the
%mode lifetimes for the challenging case of mixed dipole modes in red giants,
%with results up to $\sim\! 100\,$d.

Solar-like oscillation frequencies are usually superimposed onto background
power due to low-frequency variability caused by the convective envelope of the
star, such as granulation and/or magnetic activity leading to rotational
modulation.  Stellar granulation occurs in stars with outer convection zones due
to the difference between hotter rising gas and cooler downward moving gas.
Rotational modulation observed in light or velocity curves is attributed to
starspots on the stellar surface, which can have a different temperature,
pressure, or chemistry than their surroundings.  When the spots have properties
that are slowly evolving with time compared to the rotation period, they give
rise to rotational modulation at the frequency of the surface rotation and its
(sub)harmonics. However, the spots may migrate and/or vary in size over time,
leading to low-frequency power excess in the Fourier spectrum.

Both granulation and spots may reach amplitudes that are dominant over those of
the oscillations. Any Fourier transform of a light curve undergoing these
various aspects of variability will be composed of the superposition of all the
harmonic and nonharmonic signals. In such a situation, the method of
prewhitening is not meaningful due to the stochastic nature of the variability.
Rather, one works with the PD and fits this to extract the 
oscillation frequencies.  In Eq.\,(\ref{powerlorentz}) we have assumed
for simplicity that the mode linewidth is independent of frequency.  However, it
may vary with frequency according to some functional form depending on the mode
properties, as discussed for mixed modes in red giants by
\citet{Mosser2018}.

Different models to describe the oscillation modes and the overall
``background'' variability were developed and improved as the {\it Kepler\/}
data got extended. \citet{Kallinger2014} investigated the PD of a large and
homogeneous sample of 1364 stars observed with the {\it Kepler\/} spacecraft,
covering almost all evolutionary stages of stars born with a mass between
approximately 0.7 to 2.5\,M$_\odot$.  All these stars are expected to exhibit
stochastically excited oscillations and granulation triggered by their extended
convective outer envelope. Hence, \citet{Kallinger2014} searched for one global
optimal fitting prescription for the background variability, instead of relying
on the background model used for the Sun and adopted in red-giant
studies based on early {\it Kepler\/} data releases \citep[e.g.,][]{Mathur2011}.
\citet{Kallinger2014} considered various options for the statistical model
formulations of the PD spectra; see their Eq.\,(2) and Table\,1.  In line with
the findings of \citet{Karoff2013} for main-sequence stars, the optimal fits to
the PD spectra of bright red-giant pulsators require more than one Lorentzian
profile to describe the granulation/activity, in addition to a Gaussian power
excess caused by oscillations.  This is illustrated for one of the red giants in
their sample in Fig.\,\ref{RGB-BSG}, where we show the {\it Kepler\/} light
curve in the upper left panel, revealing the stochasticity of the variability
and the PD in the lower left panel.

\citet{Bastien2013} revealed that the scaling of the granulation amplitude
delivers a proper diagnostic for the surface gravity $g$ of the star.  The
scaling was found to be consistent with that of  the pulsation
amplitudes. Moreover, the effective temperature has only a marginal additional
effect on those amplitudes. This brings us to an important quantity in low-mass
star asteroseismology that is further outlined in Sec.\,\ref{section-applics}: the
frequency of maximum power $\nu_{\rm max}$. This quantity was shown to
scale to a good approximation as $g\cdot T_{\rm eff}^{-1/2}$ by
\citet{Brown1994}, \citet{KjeldsenBedding1995}, and \citet{Belkacem2011}. 
It can be deduced from the Gaussian fit represented by the yellow
dotted line in Fig.\,\ref{RGB-BSG}.  

Any asteroseismic modeling requires the oscillation frequencies of the damped
modes to be extracted from the PD. Various methods and implemented pipelines to
do so have been constructed. As said, one does not rely on the method of prewhitening
to achieve this. Instead so-called peak-bagging\footnote{This term became
  standard use in helioseismology and asteroseismology after champion ``peak bagger''
  Jesper Schou made the analogy between the detection of new solar
  oscillation frequencies in data obtained with the SoHO satellite and the
  addition of newly climbed +14\,000 feet summits to his personal bag-pack.}
is done, where the part of interest in the PD is fit with Lorentzian functions
for each of the modes as in Eq.\,(\ref{powerlorentz}), either after subtraction
of or along with the fit to the granulation background. These methods result in
the individual mode frequencies and the mode lifetimes from the linewidths in
the PD. Extensive literature covers the methodologies, where the use of an
Markov Chain Monte Carlo (MCMC)
technique was adopted in this context by \citet{Handberg2011}. Intercomparison
of the results obtained by the various methods is taken into account as part of
the uncertainties of the mode frequencies, amplitudes, and lifetimes; see
\citet{Hekker2011,Hekker2012} and \citet{Appourchaux2012a,Appourchaux2012b}.
Particular care of the correlation structure in the frequency analysis of
exoplanet host stars was taken using a Bayesian unsupervised approach developed by
\citet{Davies2016}.

The {\it Kepler\/} observations led to numerical refinements and testing of
early analytical expressions for uncertainty estimates of the frequencies of
stochastic modes, confirming that they behave as $\sim 1/\sqrt{T}$
\citep[see Eqs.\,(5.57) in][]{Aerts2010} instead of $\sim\,1/T$ as in
Eqs.\,(\ref{fout}).  To get a factor of 2 better frequency precision for
damped oscillations one must hence observe 4 times as long as opposed to 
2 times as long in the case of undamped modes.  For elaborate discussions on how the
mode linewidths of p~modes were derived from {\it Kepler\/} light curves for
dwarfs, subgiants, and red giants, see \citet{Chaplin2009},
\citet{Appourchaux2012a}, \citet{Corsaro2015}, and \citet{Vrard2018}. Values
translate into mode lifetimes from a fraction of a day for dwarfs to several
tens of days for red giants, in good agreement with theoretical predictions by
\citet{Belkacem2012}. \citet{Mosser2018} succeeded in deriving the mode
lifetimes for the challenging case of mixed dipole modes in red giants, with
results up to $\sim\! 100\,$d.  \citet{Chaplin2014} and \citet{Yu2020} contain
summaries of results for about 500 main-sequence stars and $20000$ red giants
observed with the {\it Kepler\/} spacecraft, respectively.

To place the stochastic variability of evolving low-mass stars into a more
global context, stochastic low-frequency (indicated as SLF in Fig.\,\ref{hrd})
variability was also found to occur in high-mass stars, but for different
physical reasons.  While all stars develop a convective outer envelope after
core-hydrogen burning, young high-mass stars are born with a radiative envelope
on top of a convective core. Hence, one does not expect granulation to be
present in their envelope.  As originally discovered from CoRoT data, young hot
O stars reveal SLF in their PD spectra \citep{Blomme2011}. This signature is
different from the frequency spectrum for p-mode oscillations in $\beta\,$Cep
stars (see Fig.\,\ref{hrd}) as found for the O9V CoRoT target HD\,46202
\citep{Briquet2011} in the same pointing of the satellite from which the SLF
signal was found, so an instrumental cause was excluded.  \citet{Tkachenko2014}
and \citet{AertsRogers2015} interpreted the signal detected in the space
photometry of the close massive binary V380\,Cyg and of the O dwarfs as due to
convectively driven IGWs.  These waves are excited stochastically at the
interface between the convective core and the bottom of the radiative envelope
in intermediate- and high-mass stars \citep{Rogers2013}. The simulations by
\citet{Edelmann2019} and \citet{Horst2020} are representative for young
unevolved stars of
high mass and deliver frequency spectra of IGWs and eigenmodes as observed in
dwarfs with CoRoT and {\it Kepler\/} for stars with masses in the range
$[3,25]\,$M$_\odot$.

\citet{Bowman2019a,Bowman2019b,Bowman2020} performed a systematic study 
of OBA-type stars from their 
CoRoT, K2, and TESS data to search for SLF, keeping in mind
that such a signal is hard to find when it occurs beneath the signal of
self-excited modes in the same frequency regime, as for stars in
Fig.\,\ref{spbs}. It was found that SLF variability is ubiquitous in space
photometry of most of these stars, irrespective if they reside in the Milky Way
or in the Large Magellanic Cloud.  The K2 light curve and PD spectrum of one
such case is shown in the right panels of Fig.\,\ref{RGB-BSG}. It concerns the
bright blue supergiant $\rho\,$Leo (spectral type B1Iab) studied by
\citet{Aerts2018b} and \citet{Pope2019} at different levels of sophistication.
The fit to the PD spectrum of $\rho\,$Leo shown in Fig.\,\ref{RGB-BSG} reveals a
characteristic frequency $\nu_{\rm char}=16.624\pm 0.007\mu\,$Hz, corresponding
to a characteristic timescale of 0.7\,d.  The K2 light curve of $\rho\,$Leo also
reveals rotational modulation with a period of 26.8\,d \citep{Aerts2018b},
which corresponds to a frequency of 0.432$\mu\,$Hz.  The astrophysical
interpretation of SLF in intermediate- and high-mass stars may involve a variety
of physical causes given the wide range of evolutionary stages covered by the
sample.

Armed with the knowledge of how to derive the oscillation frequencies,
$\omega_{nlm}^{\rm obs}$, of a pulsating star from high-precision space
photometry, we now move on to their exploitation in terms of the star's interior
physical properties. This requires that we dive into the theory of nonradial
oscillations predicted from equilibrium models of stars. In the next two
sections, we explain how this can be achieved and under which assumptions.
Subsequently we summarize some of the impressive recent results of asteroseismic
modeling.

%%%%%%%%%%%%%%%%
%%%%%%%%% Section 2
%%%%%%%%%%%%%%%%

\section{\label{section-nrp}Nonradial oscillations of stars}

The diagnostic power of nonradial oscillations to probe stellar interiors is
immense, particularly when compared to observations that probe only the stellar
atmosphere.  For example, estimates of $T_{\rm eff}$ of slowly rotating single
stars are dependent on atmosphere models and may reach 1\% precision for the
best cases.  Dynamical masses of binary components are model-independent
observables and may reach 1\% accuracy \citep[see ][for a
summary]{Serenelli2020}.  Oscillation frequencies can be measured directly from
data, without any model dependence, at the level of 0.001\% for p~modes of
low-mass stars and 0.1\% for g~modes of intermediate-mass stars
\citep[][Table\,1]{Aerts2019}.  These precisions of mode frequencies lie at the
heart of the revolution brought by space asteroseismology.

The interpretation of detected oscillation modes requires a good understanding
of the theory of nonradial oscillations and how the modes depend on the
properties of stellar interiors.  This dependence is currently studied from
numerical computations of stellar equilibrium models and their predicted
oscillations, for different sets of input physics and free parameters.  However,
various forms of analytical expressions for the mode properties are highly
insightful for the understanding of the mode behavior. In fact, asymptotic
approximations of the mode frequencies offer an important basis to
interpret the observations, even in current times of large computational
power. This was stressed by \citet{Cunha2015} and illustrated by
\citet{VanReeth2016}, \citet{Ouazzani2017}, \citet{Christophe2018}, and
\citet{Cunha2019b}.

Extensive textbooks on the theory of nonradial oscillations of stars 
were produced by \citet{Unno1989} and \citet{Smeyers2010}, to which we refer for
the historical developments and for mathematical derivations. Here
we limit ourselves to the bare minimum required to understand applications of
asteroseismology. \citet{Aerts2010} provided detailed descriptions of the
general methodology and applications covering all masses and types of nonradial
oscillations, while \citet{BasuChaplin2017} covered applications based on space
photometry but limited to stochastically excited solarlike oscillations. 
\citet{TongGarcia2015} covered synergies between
planetary and stellar seismology.  As outlined in the Introduction, the
asteroseismology revolution of the past decade is so immense that we focus the
rest of this review on applications based on the recent space photometry, even
though this does unjustice to numerous studies and efforts prior to 2010.

\subsection{\label{models}Stars and their hydrodynamics}

The equations describing the oscillations of stars are perturbed versions of the
equations of hydrodynamics applied to a gaseous self-gravitating sphere.  We
introduce the basics of stellar hydrodynamics before moving on to stellar
oscillations. We omit the derivation of these basic equations here, as this is
the topic of various books on fluid dynamics.  A seminal introduction to
hydrodynamics with specific attention to stellar oscillations was given by
\citet{LedouxWalraven1958}.  Here we limit ourselves to the ingredients needed to move on
to asteroseismic modeling while omitting unnecessary details.

\subsubsection{The stellar structure equations} 

The equations to be solved to compute stellar models throughout the evolution of
stars are the general equations of physics, expressing conservation of mass,
momentum, and energy.  In stellar interiors, the circumstances are such that
viscosity can be ignored and the conservation laws can be limited to gaseous
objects.  The derivation of the equations expressing the conservation laws for
stellar structure and evolution was covered by \citet{CoxGiuli1968},
\citet{Hansen2004}, \citet{Maeder2009}, and \citet{Kippenhahn2012}.

Conservation of mass leads to the equation
\begin{equation}
{\partial \rho  \over \partial t} + \nabla ( \rho \boldv ) = 0 \; ,
\label{Eq1}
\end{equation}
where $\rho (  \boldr , t)$  is the local density at position vector $\boldr$
and $ \boldv(\boldr , t)$ the
local velocity vector, both at time $t$.
The equations of motion, expressing conservation of momentum, can be written as
\begin{equation}
\rho\,{\partial \, \boldv \over \partial t} + \rho \boldv \cdot \nabla \boldv =
-\,\nabla p - \rho \nabla\Phi + \rho \, \boldf \; ,
\label{Eq2}
\end{equation}
where $ \boldf$ is body force per unit mass and $\Phi$ is the gravitational
potential satisfying the Poisson equation
\begin{equation}
\nabla^2 \Phi = 4 \pi G \rho \; ,
\end{equation}
and where it is assumed that 
internal friction in the gas can be ignored (i.e., zero viscosity). 
In general, $\boldf$ stands for the electromagnetic and possibly external forces
such as tidal forces in multiple systems.
The energy equation is derived from the thermodynamical properties and the 
energetics of the gas and can be formulated as
\begin{equation}
\rho\, T {\partial \, S \over \partial t} + \rho T \boldv \cdot \nabla S =
\rho\ \varepsilon - \nabla \boldF\; ,
\label{Eq3}
\end{equation}
with $S$ the entropy per unit mass, $\varepsilon$ the energy generation rate per unit mass 
taking into account the energy loss from neutrinos,
and
$\boldF$ the energy flux. 

Further, an equation for the overall energy transport throughout
the star needs to be added.  This is fairly straightforward in radiative zones of the star,
because the mean free path of a photon is ultrashort compared to the length
scales over which the stellar structure changes  ($\sim 2$\,cm in the solar
interior, for instance). In such a case the radiative energy transport is well
described by a diffusion approximation. For stellar interiors, this is given by
\begin{equation}
\boldF  = -{4 \pi  \over 3 \kappa \rho}\,\nabla\,B
= -\,{4 a c T^3 \over 3 \kappa \rho}\,\nabla T \; ,
\label{gradT}
\end{equation}
where $B = (a c / 4 \pi) T^4$ results from integrating Planck's radiation
function, $\kappa$ is the flux-weighted 
opacity, $c$ is the speed of light, and $a$ is the
radiation density constant.  

In convection zones of the stellar interior, the turbulent gas motions transport
the energy in an efficient yet complex manner. In the absence of a proper theory for
the dynamical effect of convection for stellar interiors, the turbulent pressure
is usually ignored and the treatment of convective energy transport in stellar evolution
codes is time-independent.  This approach is a crude approximation; it is based
on pragmatism rather than sophistication. Although various versions exist for
the description of convective instabilities, the most
popular treatment of time-independent convection is the so-called mixing-length
theory \citep[mlt, see ][for a historical
overview]{HoudekDupret2015}. It is characterized by the free parameter
$\alpha_{\rm mlt}$ (expressed in units of the local pressure scale height
$H_p$), which stands for the mean free path over which the convective eddies
travel before dissolving in their environment.  Asteroseismology allows one to infer
the extent of convective regions via estimation of the free parameters of the
convection formulation used for the modeling. This was done for mlt
by \citet{Joyce2018} and \citet{Viani2018}.

Whenever the diffusion of photons is insufficiently efficient as an energy
transport mechanism, convection not only takes over the energy transport, but it
also changes the temperature gradient relative to the radiative one in
Eq.\,(\ref{gradT}). 
From a computational point of view, the calculation of the
energy transport must hence be split up for the radiative and
convective zones inside the stellar model. This is done by testing whether a zone
with temperature gradient 
\begin{equation}
\nabla\equiv\displaystyle{\frac{{\rm d} \ln T}{{\rm d} \ln p}}
\end{equation}
is stable or unstable against convection.  The general condition to test for
convective stability is the so-called Ledoux criterion
\begin{equation}
\nabla_{\rm rad} < \nabla_{\rm ad}+\frac{\varphi}{\delta}\nabla_\mu, 
\label{ledoux}
\end{equation}
where we have introduced
$$\nabla_{\rm rad}= \displaystyle{{3 \over 16 \pi a c G} {\kappa p \over T^4}
  {L(r) \over m(r)}},\  \nabla_{\rm ad}=\displaystyle{\left(\frac{\partial \ln T}{\partial \ln
      p}\right)_S}, $$
$$\nabla_{\mu}=\displaystyle{\frac{{\rm d} \ln \mu}{{\rm d} \ln p}},
\ \delta= - \displaystyle{\left(\frac{\partial \ln \rho}{\partial \ln
      T}\right)_{p,\mu}}, \ \varphi=\displaystyle{\left(\frac{\partial \ln \rho}{\partial \ln
      \mu}\right)_{p,T}},$$ 
with $\mu$ the mean molecular weight of the ionized gas. 
For zones with a homogeneous chemical composition, the Ledoux criterion reduces
to the Schwarzschild criterion
\begin{equation}
\nabla_{\rm rad} < \nabla_{\rm ad}.
\end{equation}
Stars born with a mass above $\sim\!1.7\,$M$_\odot$ have a receding convective
core as they evolve throughout the core hydrogen-burning phase because the
opacity $\kappa$ decreases as the hydrogen
depletes, reducing $\nabla_{\rm rad}$.  The resulting composition gradient gives
rise to $\nabla_{\mu}\neq 0$ and increases stability in that zone.
On the other hand, the change of
$\nabla_{\rm rad}$ for stars born with a mass below $\sim\!1.3\,$M$_\odot$ is
dominated by the factor $L(r)/m(r)$, which increases faster than $\kappa$
decreases. The interplay in importance between $\kappa$ and $L(r)/m(r)$ in the
expression of $\nabla_{\rm rad}$, and along with it the growth or shrinkage of the
convective core, depends on the physical circumstances for masses between 1.3
and 1.7\,M$_\odot$ \citep[e.g.,][]{Mombarg2019}.

In a zone that is stable against convection, a fluid element that gets displaced
by moving up will be pulled back down until it is again situated at its
equilibrium position, thanks to the action of the buoyancy force of
Archimedes. This oscillatory motion of the fluid elements depends on the local
density, pressure, and chemical composition of the gas and happens with the
so-called Brunt-V\"ais\"al\"a frequency, or buoyancy frequency for short, which
can be approximated as
\begin{equation}
N^{2}\simeq\frac{g}{H_p}\left[\delta
\left(\nabla_{\rm ad}-\nabla\right)+\varphi\nabla_\mu\right],
\label{BVformula}
\end{equation}
with $g$ the local gravity.  The $\mu$-gradient affects the local behavior of
$N(r)$ in the radiatively stratified layers of the star. As discussed later,
this will affect stellar oscillations, notably internal gravity
waves. In the case of instability, i.e.\ $N^2<0$, the speed of the fluid element
increases exponentially with time until it breaks up, causing complete and
instantaneous mixing of the chemical species.

Even though Eq.\,(\ref{ledoux}) allows us to derive the zones where
convection takes place inside the star, complications occur in the transition
layers between convective and radiative zones, hereafter termed convective
boundary layers. The fluid elements inside a convection zone experience a
turbulent motion with velocity $\boldv_{\rm conv}$. When they reach the
convective boundary layer, their inertia will prevent them from stopping
abruptly; i.e., they will ``overshoot'' from the convection zone into the
radiative layer over an unknown distance, which we denote as $\alpha_{\rm ov}$
(in analogy to $\alpha_{\rm mlt}$, it is expressed in the unit $H_p$).  The way
in which the fluid elements overshoot the convective boundary depends on the
location of the convection zone inside the star and the physical circumstances
at that position.  For extensive discussions, see \citet{Zahn1991},
\citet{Viallet2015}, \citet{Cristini2016}, \citet{Constantino2017}, and
\citet{Arnett2019}.  Three-dimensional simulation studies have indicated at
least three physical processes that may come into play: penetration by plumes
leading to superadiabatic mixing over a distance $d_{\rm pen}$ \citep{Zahn1991},
subadiabatic thermal diffusion over a distance described by means of an
exponentially decaying mixing profile with parameter $f_{\rm ov}$
\citep{Freytag1996,Herwig2000}, or turbulent entrainment that occurs over a
dissipation length scale expressed as a distance $l_{\rm d}$
\citep{Meakin2007,Viallet2013}. These imply a different and uncalibrated level
and functional form of convective boundary mixing (CBM) and have
a different temperature gradient in the transition layer.  We use the notation
of the free parameter $\alpha_{\rm ov}$ to express the unknown length scale over
which the fluid elements move from inside a convective region into the radiative
adjacent zone, representing any of $d_{\rm pen}$, $f_{\rm ov}$, $l_{\rm d}$, or
other formulations \citep[see ][]{Augustson2019}.

The rate of change of species of type $i$ with relative mass fraction $X_i$ is caused by
various processes, some of which are diffusive but others that are not.  When the
rate of change happens much faster than the nuclear timescale, it is customary
to approach $\partial X_i/\partial t$ by a diffusion equation for computational
convenience.  In the simplest
case of changes due to convective motions, along with nuclear fusion in  
a spherically symmetric star, we can write
\begin{equation}
\label{MLTCBM}
\displaystyle{\frac{\partial X_i}{\partial t}} =  {\cal R}_i  +
\displaystyle{\frac{1}{\rho r^2} \frac{\partial}{\partial r} \left[ \left(D_{\rm
      conv}+D_{\rm ov}\right)
\rho r^2 \frac{\partial X_i}{\partial r} \right],}
\end{equation}
where the rate of change of $X_i$ due to nuclear reactions is denoted
symbolically as ${\cal R}_i$.  
The
diffusion coefficient associated with the convective mixing described by mlt is
given by
\begin{equation}
D_{\rm conv}=\displaystyle{\frac{1}{3}}\,\alpha_{\rm mlt}\,H_{p}\, v_{\rm conv} .
\end{equation}
The unknown profile of CBM due to the overshooting of the fluid elements beyond
the convective boundary is denoted here as $D_{\rm ov}$. Each of the profiles
$D_{\rm conv}(r)$ and $D_{\rm ov}(r)$(expressed in the physical units
cm$^2$\,s$^{-1}$) is in general an unknown function of $r$ and involves at least
one free parameter ($\alpha_{\rm mlt}$ and $\alpha_{\rm ov}$).

For stars with a convective core, the lack of calibration of the physics in the
convective boundary layers implies a serious limitation. Indeed, the CBM
influences the amount of matter that can be brought into the central regions
where nuclear fusion takes place.  The higher the CBM, the more fresh fuel
reaches the nuclear reactor and hence the longer the nuclear fusion can go
on. This has a major impact on the star's core mass and its age. For this
reason, calibration of the amount of matter in the convective core of a star
$M_{\rm cc}$, via an observational estimation of the profile $D_{\rm ov}(r,t)$ and
its feedback throughout the evolution of the star, is a crucial piece of
information to predict a star's life and age.  It was shown that space
asteroseismology has the capacity to deliver such an estimation across a large mass
range by \citet{Deheuvels2016} and \citet{Pedersen2018}, including assessment of
the temperature gradient in the near-core boundary layer
\citep{Michielsen2019}. This potential had already been pointed out by
\citet{Dziembowski1991} but remained without practical application until
recently. Concrete applications to derive $D_{\rm ov}(r)$ based on space
asteroseismology are discussed in Sec.\,\ref{section-applics}.

\subsubsection{Simplification to 1D stellar models}

Because of the immense range in timescales and spatial scales occurring in the interiors
of stars, stellar models must necessarily remain a simplified version of
reality. Indeed, the computation of 3D models across stellar evolution is not
yet feasible. One thus needs to adopt simplifications in the computation of
stellar structure models. With asteroseismic applications in mind, we make two
important approximations: we assume that any equilibrium model, which will be
perturbed to compute a star's oscillations, is spherically symmetric and does
not have a dynamical atmosphere.  The first simplification implies that we can
rely on 1D models in hydrostatic equilibrium computed for stars that do not
rotate close to their so-called critical rotation rate.  The second
simplification allows us to use a static atmosphere model to connect to the
stellar interior as outer boundary, at each time step in a star's evolution.

In practice, stellar evolution codes rely on mass-loss or accretion rates
described by parametrized laws, such that an amount $\dot{M}\cdot\Delta t$ is
peeled off or added to the stellar model after a duration $\Delta t$ of stellar
life has passed. For each particular instance in time, the stellar model is
considered to have a static atmosphere on top of its interior structure. In this
way, the models are built while taking mass loss or accreted mass into
accountand 
ignoring the dynamical properties due to a stellar wind or an accretion
disk. This simplifies the boundary conditions adopted to
close the set of equations to be solved. This basic assumption is justified for
the majority of applications in asteroseismology, because nonradial oscillations
are usually undetectable for stars that have a strong dynamical wind or high
levels of accretion. 

Ignoring fast rotation needs more justification than neglecting the dynamics of
the wind, because rotation is common in stars throughout their lives.  Rotation
acts upon stellar structure in at least three ways: it deforms the star
from spherical symmetry, it leads to higher polar than equatorial flux due to
gravity darkening, and it induces various instabilities and mixing in the
stellar interior. The level of confidence in how to treat these effects is 
different for the three aspects.
Gravity darkening was first discussed by \citet{vonZeipel1924}. It stands for a
reduction in the flux and hence in the effective temperature of the star
resulting from the reduced gravity in the equatorial regions relative to the
polar ones.  The von Zeipel effect is expressed as
\begin{equation}
T_{\rm eff} = T_{\rm eff, p} \left({g_{\rm eff} \over g_{\rm eff, p}} \right)^\beta\; ,
\label{vonZeipel}
\end{equation}
where $T_{\rm eff, p}$ and $g_{\rm eff, p}$ are the effective temperature and
effective gravity at the pole of the star.  For a radiative envelope as
considered by von Zeipel, $\beta \simeq 0.25$. In the presence of a convective
envelope, $\beta$ is usually assumed to be $\beta\!<\!0.1$.  This limited
knowledge of the exponent $\beta$, and along with it a nonsymmetrical stellar
wind, implies a nontrivial treatment in stellar evolution computations in the
presence of rotation.

By definition, the critical (or breakup) velocity is reached when the outwardly
directed centrifugal acceleration is equal to the inward effective gravitational
acceleration at any one point of the stellar surface.  Here we work with the
critical rotation frequency since we are making a comparison to the frequencies of
oscillations.  Usually the Roche approximation is adopted, which assumes that
the mass concentration inside the star is not distorted by the rotation.  In
this case, the polar ($R_{\rm p}$) and equatorial ($R_{\rm e}$) radii of the
star differ by less than a factor of 3/2, where
$R_{\rm e,crit}/R_{\rm p,crit}=3/2$.  This leads to the critical rotation
frequency given by
$\Omega_{\rm crit}=\sqrt{GM_\star/R_{\rm e,crit}^3}= \sqrt{8GM_\star/27R_{\rm
    p,crit}^3}$, with $M_\star$ the mass of the star and
$R_{\rm e,crit}, R_{\rm p,crit}$ its critical equatorial and polar radii.  This is
the solution for the critical rotation frequency when the Eddington parameter
$\Gamma=\kappa L/4\pi cGM_\star<0.639$ \citep{Maeder1999}. The other solution is
not considered here, as almost all applications in asteroseismology thus far occur
for stars without a strong radiation-driven wind.

The prediction $R_{\rm e,crit}/R_{\rm p,crit}<3/2$, along with von Zeipel's
formula (\ref{vonZeipel}), can be evaluated directly from interferometric
measurements of stellar surfaces.  Such observations indeed show that fast
rotators are oblate \citep[e.g.][]{Souza2018} and that their surface properties
and winds are not spherically symmetric, as revealed by, e.g.,
\citet{Kervella2006} and \citet{Souza2014}.  However, fast rotating stars do not
necessarily comply with the Roche approximation.  Interferometry of the Be star
Archernar \citep{Souza2003} led to $R_{\rm e}/R_{\rm p}\simeq\,1.56$.  Moreover,
von Zeipel's law is not adhered to by the stars having gravity darkening
measurements from interferometry \citep{Souza2018}. This led
\citet{Gagnier2019-vcrit} to perform a new evaluation of the critical rotation
of a star from 2D static models, with the conclusion that $\beta$ decreases from
0.25 to 0.13 for rotation rates evolving from slow to critical.

In conclusion, the computation of 2D equilibrium models in the presence of
rotation comes with major uncertainty, even in its simplest aspects of the local
surface and its flux.  For this reason, computations of stellar oscillations
from 2D equilibrium models are often restricted to static polytropic models
\citep{Lignieres2006,Reese2006}.  Although progress is steadily achieved
\citep{Ouazzani2012,Reese2013}, fitting of measured frequencies to perform 2D
asteroseismic modeling is not within reach yet.  We thus do not treat 2D
equilibrium models as input for oscillation-mode computations.
Rather, we focus in this review on tuning the stellar interior
quantitatively by taking into account the
Coriolis acceleration at the level of the mode computations while relying on 
1D equilibrium models. 
\citet{Lignieres2006}, \citet{Ballot2010} and
\citet{Ouazzani2017} provided comparisons between oscillation frequency
predictions of rapid rotators from 1D versus 2D equilibrium models using
higher-order perturbative expressions for the effect of rotation. These studies
revealed that a 1D treatment for p~modes leads to appropriate oscillation
predictions for stars rotating up to $\sim\!15\%$ of the critical rotation
frequency. For high-order g~modes, the 1D treatment is justified up to $\sim\!70\%$ of the
critical rotation frequency \citep{Henneco2021}.  Within these regimes, it is justified to work with
1D equilibrium models as input for the computation of oscillation modes, where
no rotation or 
only the spherically symmetric component of the centrifugal force is included in
Eq.\,(\ref{Eq2}). In the latter case, the simplified equation of hydrostatic
equilibrium reads
\begin{equation}
\label{HE+cent}
{\partial p \over \partial r} = -{G m \rho \over r^2} 
+ {2 \over 3} \rho\,r\,\Omega^2\; .
\end{equation}

\subsubsection{Standard 1D stellar models in hydrostatic equilibrium} 

Thus far we have focused on the internal structure of the star, but we have hardly
considered its chemical evolution.  The chemical composition inside the star
at time $t$ is described by relative mass fractions of species $i$,
$X_i=X_i(r,t)$, where $r\in[0,R_\star(t)]$ with $R_\star(t)$ the radius of the
spherically symmetric star at time $t$ in its evolution.  These profiles are an
important aspect of stellar models, because they determine the
opacity, thermodynamical characteristics, and energy production $\varepsilon$
due to nuclear reactions as in Eq.\,(\ref{Eq3}).  These reactions, in turn,
change the chemical composition and rule the life of the star.

To solve the stellar structure equations (\ref{Eq1}) --
(\ref{Eq3}) along with the energy transport equation(s) and the changes in the
chemical profiles $X_i=X_i(r,t)$, the microscopic properties of the stellar
matter need to be known as a function of $p(r,t)$, $T(r,t)$, etc. This requires
adopting an equation of state, various thermodynamical properties, opacity
tables to compute the Rosseland mean opacity, a network of nuclear reaction
rates, etc.  This is jointly referred to as ``input physics'' when one computes
stellar models.  Further, proper boundary conditions at the center and surface
of the star and initial conditions characterizing the star's properties at birth
($\tau\equiv 0$) when it has arrived at the so-called Zero-Age-Main-Sequence
(ZAMS) have to be chosen \citep[this is not discussed here, see,
e.g.,][]{Kippenhahn2012}. The solution of the equations for chosen boundary
conditions delivers what is called a stellar equilibrium model at age $\tau$
described by $m(r,\tau), p(r,\tau), L(r,\tau), T(r,\tau), X_i(r,\tau)$, and by
all other relevant functions that can be derived from these solutions, with
$m(r,\tau)$ the mass enclosed by the shell positioned at $r\in[0,R_\star(\tau)]$
inside the star.

The ZAMS $\tau\equiv 0$ is defined as the point in time when hydrogen fusion
occurs in full equilibrium in the center of the star.  At the ZAMS, the star has
a specific yet unknown chemical mixture of species $X_i$ in its interior. This
mixture is the result from the initial chemistry it received from its birth
cloud when it started as a fully convective protostar on the so-called Hayashi
track \citep[see, e.g.,][for a definition]{Kippenhahn2012} and from changes in
this mixture due to nuclear reactions and due to mixing taking place during the
contraction phase from the Hayashi track towards the ZAMS.  Often, the
computation of stellar models for low-mass stars adopts the solar mixture
using the Sun's current or initial surface abundances \citep{Asplund2009}.  On
the other hand, the surface abundances of B-type stars in the solar
neighborhood \citep{Przybilla2013} constitute a logical choice for the initial
chemical mixture when computing high-mass stellar models.  With a specified
chemical mixture, the initial composition is an input for the 1D evolutionary
computations.  We denote this initial composition as
$X_{\rm ini}, Y_{\rm ini}, Z_{\rm ini}$, which stand for the initial hydrogen,
helium and metal mass fractions, complying with
$X_{\rm ini}+Y_{\rm ini}+Z_{\rm ini}=1$.  For most of the phases of stellar
evolution, the stars do not change on a dynamical timescale, as it is much
shorter than the contraction timescale and the nuclear timescale.  Whenever
this is the case, the left hand side of Eq.\,(\ref{Eq2}) is zero and the
resulting stellar model is in hydrostatic equilibrium.

In the simplest case of a nonrotating nonmagnetic single star without a stellar
wind, there is no additional body force $\boldf$ in Eq.\,(\ref{Eq2}) such that
the pressure and gravity forces are the actors that compensate for each other.
Such simplifications lead to equilibrium models that resemble
reality well for many of the stars and during large fractions of their
life.  Stellar models computed with those simplifications for the interior and
with a static atmosphere as outer boundary are called standard stellar
models.  Evolutionary tracks representing such standard models are included
as full lines in Fig.\,\ref{hrd}. Extensive comparisons of stellar evolution
models computed with independently developed codes have been done for low-mass
stars in the context of CoRoT \citep{Monteiro2009} and show impressive agreement
when the same input physics is considered. This is in sharp contrast to the
major differences occurring for stellar evolution computations based on similar
input physics for high-mass stars, even for nonrotating models
\citep{MartinsPalacios2013}.

\subsubsection{Nonstandard 1D models with microscopic atomic diffusion}

Composition changes do not occur only in regions where nuclear reactions take
place.  In addition to full and instantaneous mixing in convective regions and full
or partial mixing in convective boundary layers, the chemical profiles in
radiative regions may also change due to microscopic and macroscopic transport
processes \citep[e.g.][for reviews]{Pinsonneault1997,Salaris2017}.  Which
of those is dominant depends on the timescales upon which they act.
Macroscopic mixing may be induced by turbulence, magnetic fields, waves,
rotation, etc. In this section, we focus on the microscopic scale and consider
element transport caused by microscopic atomic diffusion.  The accompanying
local chemical composition changes induced by it are caused by gradients
operating in the radiative layers of the star. These gradients may introduce
lower or higher concentrations of particular chemical species in particular
layers of the radiative envelope.  Section\,II.A.5 treats macroscopic
element transport.

A key aspect of assessing the importance of microscopic diffusion is that the
timescales upon which it acts are significantly different for the atmosphere than for
the interior of the star \citep{Michaud2015}. Diffusion timescales are
typically less than a century for the stellar atmosphere, while millions to
billions of years for the interior regions.  Given that we focus on
asteroseismic applications and on the tuning of stellar interiors, we do not
consider modeling surface abundances affected by atomic diffusion as observed in
some intermediate-mass stars (so-called Ap and Bp stars).  Rather, we limit
ourselves to those aspects of atomic diffusion that act on long timescales in
radiative parts of the stellar interior, keeping in mind the importance of
atomic diffusion for the solar case, as demonstrated from helioseismology by
\citet{JCD1993}.

\begin{figure*}[t!]
\begin{center}
\rotatebox{0}{\resizebox{17cm}{!}{\includegraphics{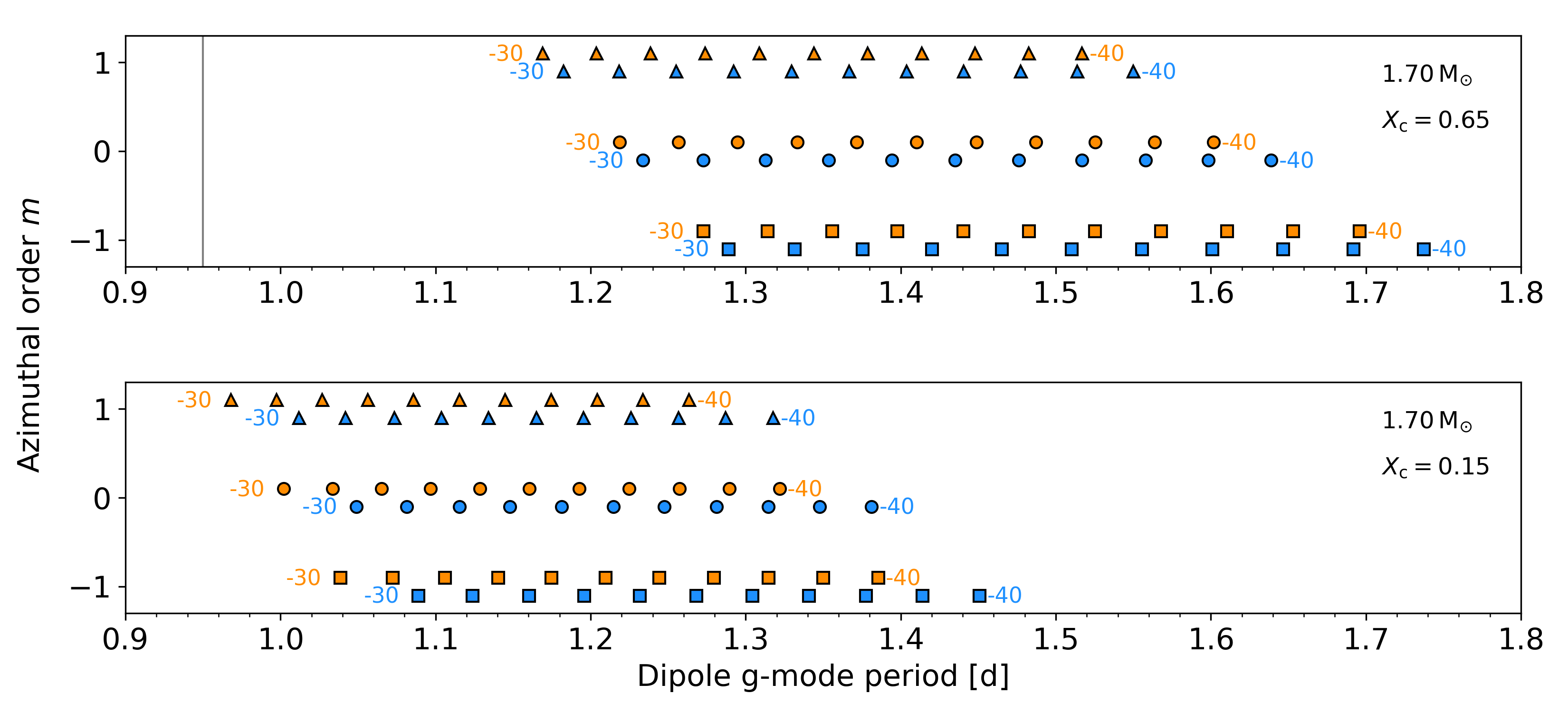}}}
\end{center}
\caption{\label{levitation-gdor} Shifts in the periods of dipole ($l=1$) triplet
  g~modes of two equilibrium models [blue (orange), without (with) atomic diffusion,
  including radiative levitation] with the same input physics, for parameters
  $M=1.7\,$M$_\odot$, $X_{\rm ini}=0.7154$, $Z_{\rm ini}=0.022$, and at two
  different evolutionary stages expressed in terms of the central hydrogen
  fraction $X_c$.  The g~modes were computed assuming rigid rotation with a
  period of 14.4\,d.  The radial orders $n$ are labeled. The vertical line
  indicated on the left side in the upper panel denotes the measurement uncertainty
  for such mode periods from a 4-yr nominal {\it Kepler\/} light curve.
  Figure based on stellar models in \citet{Mombarg2020} by courtesy of Joey
  Mombarg, KU\,Leuven.}
\end{figure*}

Following \citet{Thoul1994}, \citet{Chayer1995},
\citet{Richer2000}, \citet{Vandenberg2002}, \citet{Richard2002a,Richard2002b},
\citet{Michaud2004}, \citet{Hu2011}, \citet{Theado2012}, and \citet{Deal2016},
four different aspects of microscopic atomic diffusion are considered in stellar
models. These occur due to pressure, temperature, and concentration gradients on
the one hand, and radiative forces on the other hand. While pressure and
temperature gradients augment the concentration of more massive species toward
the center of the star, concentration gradients have the opposite effect.  On
the other hand, radiative forces levitate species with an efficiency that
depends on the details of the atomic structure of the involved isotopes.  The
calculation of the appropriate radiative accelerations is therefore challenging
in terms of computation times.  The accelerations can be computed from atomic
data by treating the appropriate multicomponent gas \citep{Burgers1969}. This
requires evaluations of the frequency-dependent absorption coefficients derived
from a screened Coulomb potential \citep{Paquette1986}, taking into account
partial ionization, and this for all the layers inside the star
\citep[see ][]{Thoul1994}.  Once the overall local velocities $w_i$ for each of
the species $i$ involved in the atomic diffusion are computed, they can be
inserted in the equation governing the time evolution of the mass fraction
$X_i$:
\begin{eqnarray}
\label{atomicdiffusion}
\displaystyle{\frac{\partial X_i}{\partial t}} & = & \displaystyle{{\cal R}_i 
- \frac{1}{\rho r^2} \frac{\partial}{\partial r} \left(\rho r^2X_i w_i\right)} \\ \nonumber
& + & 
\displaystyle{\frac{1}{\rho r^2} \frac{\partial}{\partial r} \left[ \left(D_{\rm
      conv} +D_{\rm ov}\right)
\rho r^2 \frac{\partial X_i}{\partial r} \right],}
\end{eqnarray}
where the second term on the right-hand side is the result of the microscopic
atomic diffusion acting upon species $X_i$ and the third term is the result of 
macroscopic transport of the chemical species due to convection and overshooting.

If atomic diffusion can be treated without the radiative effects, which is a
good approximation for cool stars with extended convective envelopes such as the
Sun, its impact on the computation time required for evolutionary model
calculations is modest.  As a consequence, the use of such simplified
microscopic diffusion computations without levitation in evolutionary models is
widespread \citep[e.g.,][]{Chaboyer1995,Pinsonneault1997}.  With levitation
included, the computation of $w_i$ and the solution of the set of
equations\,(\ref{atomicdiffusion}) at each step of the evolution is a major
challenge. Nevertheless, such computations have been done with the specific aim
of asteroseismic applications, adopting various levels of complexity.  
Studies of stellar interiors of A- and F-type stars were given by
\citet{Turcotte1998}, \citet{Deal2018}, and \citet{Verma2019b}, subdwarf B
stars were studied by \citet{Hu2011} and \citet{Bloemen2014}, and  white dwarfs
were studied by
\citet{Romero2017} and \citet{Geronimo2019}, where the last two papers did not
include radiative levitation.  Figure\,\ref{levitation-gdor} shows the influence
of atomic diffusion on g-mode frequencies of intermediate-mass stars whose
rotation period is about 10 times longer than its dipole-mode periods. 
The frequency shifts induced by atomic diffusion are much larger
than the measurement uncertainties, highlighting the fact that asteroseismology has the
capacity to evaluate the need (or not) of radiative levitation in models of such
stars, as illustrated by \citet{Mombarg2020}.

It is instructive, particularly for the later nuclear burning stages, to compare
the asteroseismic results based on evolutionary models with those obtained from
static structure models that are more sophisticated in
some aspects of the structure yet less prone to unknown aspects of the physics
in the models that accumulate throughout the evolution. This approach was
followed by \citet{Charpinet2011} and \citet{VanGrootel2013}, as well as by
\citet{Giammichele2018} and \citet{Charpinet2019} for subdwarfs and white
dwarfs, respectively.  Differences in the stellar structure profiles from such
static models [$m(r), p(r), T(r), L(r)$, and $X_i(r)$] compared to those
obtained from evolutionary models can then be used to improve the input physics
adopted for full evolutionary computations via an iterative loop between
asteroseismology and the equilibrium models as given by \citet{Timmes2018} and
\citet{Geronimo2019}.

Atomic diffusion impacts the concentration of the species in the stellar
interior on timescales that are relevant for stellar evolution. Its effect is
hard to unravel from a star's luminosity and effective temperature, which are
the two quantities that define the evolutionary tracks in an HRD. Models with
and without atomic diffusion (either with or without levitation) usually differ
far less than typical observational errors of $L$ or $\log\,g$ plotted versus
$T_{\rm eff}$ as shown by \citet{Dotter2017} and by \citet{Deal2018}.  As a
confrontation between data and theory in the HRD is commonly the only assessment
to evaluate stellar evolutionary theory, and given the computational
requirements, microscopic atomic diffusion is often ignored in stellar and
galactic astrophysics.  Its inclusion is, however, critical when evaluating
surface abundances for archaelogical chemical tagging \citep{Dotter2017} and to
interpret asteroseismic data as done by \citet{Verma2017},
\citet{Deal2018,Deal2020}, and \citet{Mombarg2020}.

\subsubsection{Nonstandard 1D models with rotation and waves}

Rotation has a major effect on stellar evolution \citep{Maeder2009}. Yet in
this era of space asteroseismology, it has become clear that its treatment in
stellar interiors is up for improvement.  Predictions based on 
local conservation of angular momentum and rotational mixing, both of which
have been used extensively in stellar evolution models the past few decades,
lead to predictions that are incompatible with asteroseismology, as
discussed later.  Even 1D models of slowly rotating stars face
challenges. Improving them is a major aim of asteroseismology and of stellar
astrophysics in general.  We have now reached the stage where asteroseismic
inferences based on high-quality space photometry can be used to derive
$\Omega(r)$ for stars and to provide a calibration of the poorly known physical
ingredients of rotating stellar models, which is in line with Gough's quote in
the Introduction: asteroseismology in action.

\begin{figure*}[t!]
\begin{center} 
\rotatebox{270}{\resizebox{6.4cm}{!}{\includegraphics{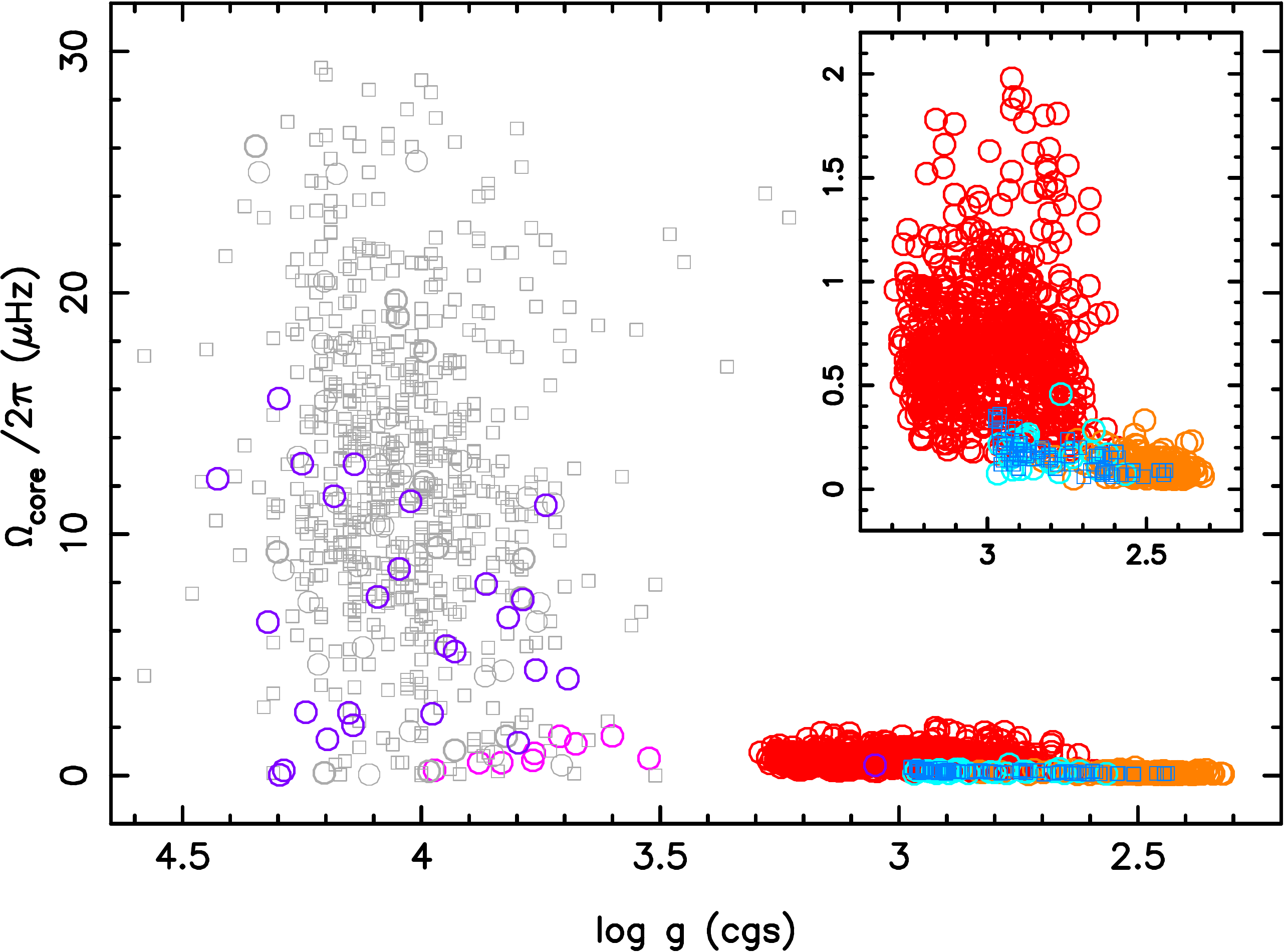}}}\hspace{0.25cm}
\rotatebox{270}{\resizebox{6.4cm}{!}{\includegraphics{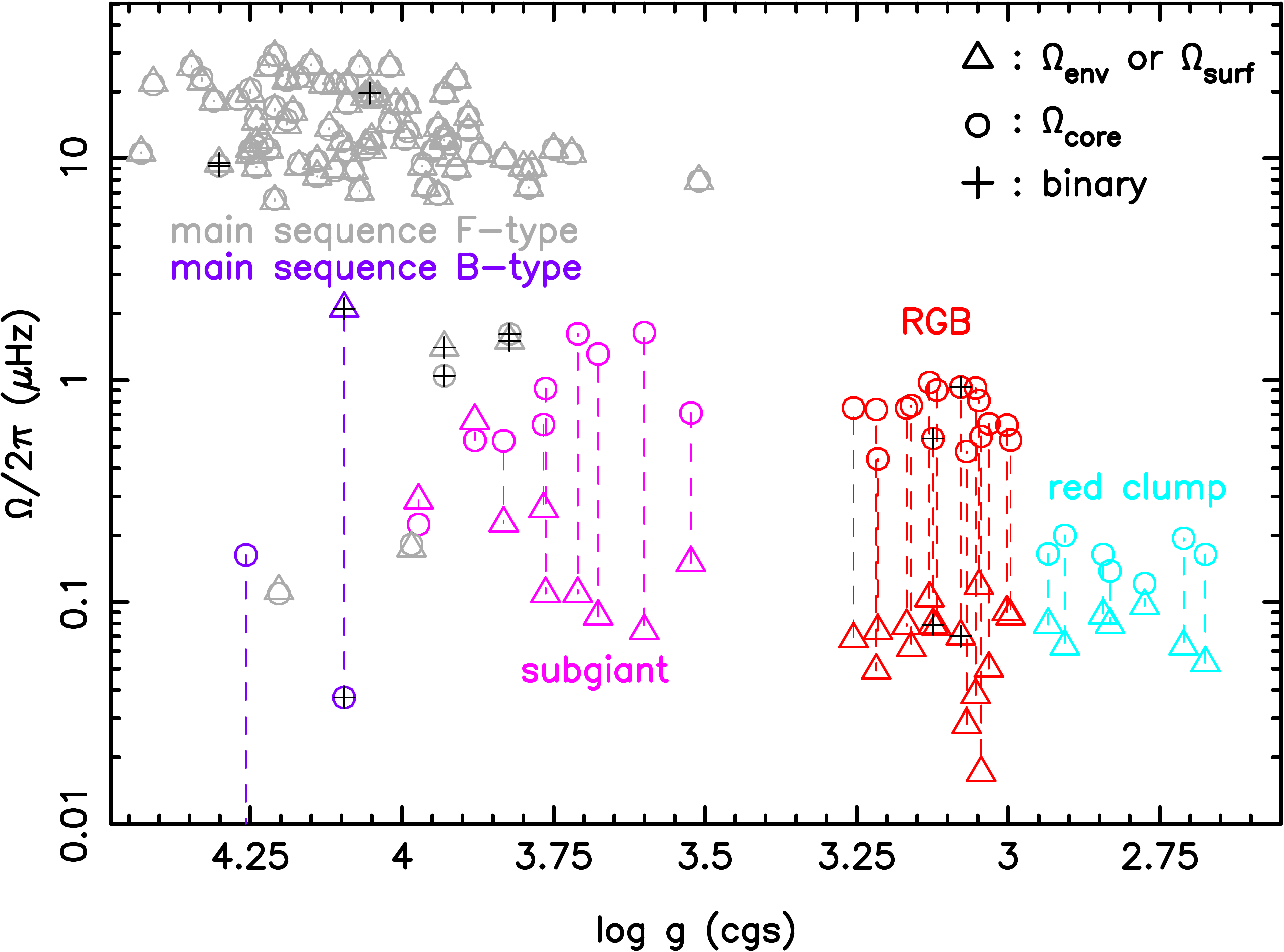}}}
\end{center}
\caption{\label{araa-update} Left panel: core (near-core) rotation rates derived from
  mixed (gravity) modes for stars in core-hydrogen burning (purple and gray),
  hydrogen-shell burning (pink and red), and core-helium burning (after the
  helium flash in orange and avoiding the helium flash in cyan). The circles
  indicate stars with asteroseismic estimates of $\Omega_{\rm core}$ and
  $\log\,g$ taken from \citet{Aerts2019}; their errors are smaller than the
  symbol size. The squares are additional stars with
  asteroseismic determinations for $\Omega_{\rm core}$ but with less reliable
  values for $\log\,g$ from spectroscopy and/or stellar models from \citet[][622
  $\gamma\,$Dor stars, indicated in gray]{GangLi2020} and from \citet[][72
  core-helium-burning red giants in blue]{Tayar2019}.  Their uncertainties for
  $\log\,g$ range from 0.2 (blue squares) to 0.5\,dex (gray squares) and are omitted
for clarity.  Right panel: all stars with an additional
  measurement of the envelope (from p~modes) or surface (from rotational
  modulation) rotation frequency.  This figure is an update of Fig.\,4 in
  \citet{Aerts2019} and is based on data kindly provided by Gang Li, Sydney
  University and Jamie Tayar, University of Hawaii.}
\end{figure*}

To compute equilibrium models including rotation,  
we need to know $\Omega(r)$ and how it changes throughout
stellar evolution.  Asteroseismology is a major and thus far unique game changer
on this front. While we discuss methods to deduce $\Omega (r)$ in
Sec.\,IV.D., we provide the current status of the rotation frequencies in the region
just outside the convective core (denoted as $\Omega_{\rm core}$ throughout the
review) in Fig.\,\ref{araa-update}. We also highlight the envelope
($\Omega_{\rm env}$) or surface ($\Omega_{\rm surf}$) rotation frequency for the
stars with this information.  Figure\,\ref{araa-update} updates the work of 
\citet{Aerts2019}, who presented these asteroseismic measurements for
low- and intermediate-mass stars distributed across all evolutionary stages from
{\it Kepler\/} photometry. We discuss these results extensively in
Sec.\,\ref{section-applics}, but we point out here that almost all single stars
in the covered mass range of $[0.8,3.3]$\,M$_\odot$ were found to rotate nearly
uniformly during the core-hydrogen-burning and core-helium-burning phases and that the
angular momentum of the helium-burning core of these stars is in agreement with
the angular momentum of white dwarfs.  Figure\,\ref{araa-update} implies a
strong decrease of core angular momentum in the phase when stars have a convective
core. Current stellar evolution theory of rotating stars 
cannot explain these asteroseismic
results. This calls for a reevaluation of 1D models with rotation.

Stellar models with rotation often adopt the approximation of shellular
rotation, following \citet{Zahn1992}. In this approximation, one assumes that
the chemical composition and the angular velocity remain constant on isobars. As
such, the ratio of the rotation frequency of the star $\Omega (r)$ with
respect to $\Omega_{\rm crit}$ [or the accompanying $v(r)/v_{\rm crit}$] is used
as input for the numerical computations of the stellar models. Given the limited
knowledge on angular momentum evolution during the contraction phase, the
input ratio $\Omega(R_\star)/\Omega_{\rm crit}$ is usually taken at the ZAMS,
assuming a rigid rotation profile to start the evolutionary computations.

From a theoretical perspective, rotation is expected to induce a myriad of
processes and instabilities in the stellar interior, leading to transport of
angular momentum and of chemical species. This was extensively discussed 
by \citet{Maeder2009}.  As recently reviewed in the modern context of
asteroseismology, these macroscopic processes can be classified into four main
categories \citep[][Sec.\,3]{Aerts2019}: meridional circulation, hydrodynamical
instabilities, magnetorotational instabilities, and IGWs.  However, the concept
of ``rotational mixing'' in stellar evolution computations and in the literature
often stands for the macroscopic element transport due to the action of
circulation and all instabilities together.  Further, in analogy to rotational
mixing, we will use the term ``pulsational mixing'' for element transport caused
by waves. Because rotational or pulsational mixing is expected to homogenize
the chemical mixture in the layers where they are active on short timescales,
models including these ingredients often ignore the microscopic atomic diffusion
effects leading to concentrations of species.  However, there is no justified
physical reason for this ``computationally convenient'' simplification when the
timescales of these processes are similar \citep[see ][]{Deal2020}.

\begin{figure*}[t!]
\begin{center} 
\rotatebox{0}{\resizebox{15.cm}{!}{\includegraphics{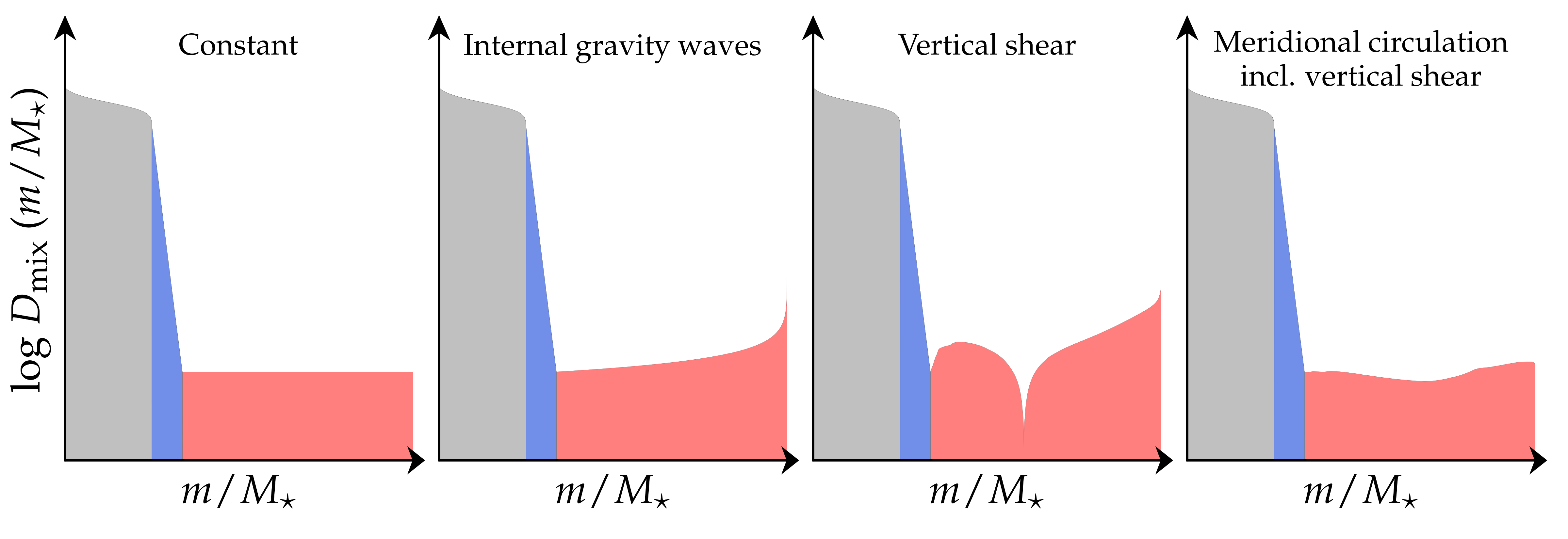}}}
\rotatebox{0}{\resizebox{15.cm}{!}{\includegraphics{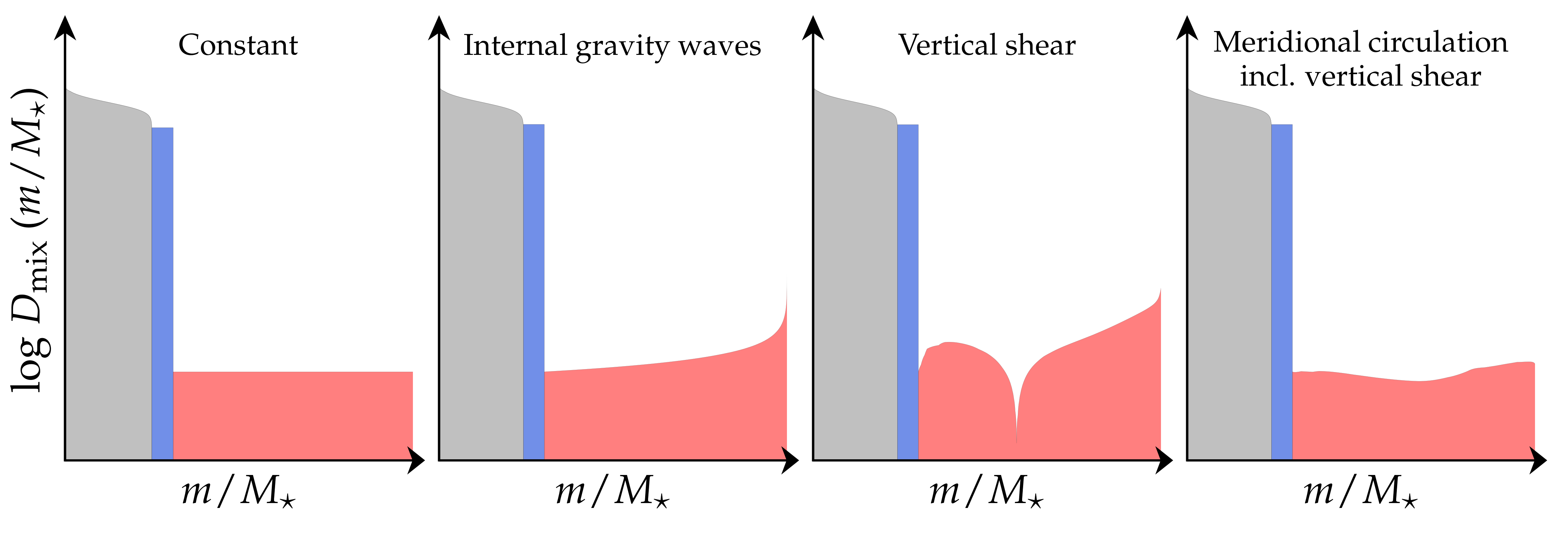}}}
\end{center}
\caption{\label{geneva} Schematic representation of mixing profiles due to
  various transport processes in stars with a convective core (indicated in
  gray) and a radiative envelope for exponentially decaying
  diffusive core overshooting (upper panels) and convective penetration
  (lower panels) as CBM (purple). Four types of envelope mixing based on different
  theoretical frameworks are considered  (pink), as labeled.  Figure courtesy of
  May Gade Pedersen, KU\,Leuven, based upon data for 5\,M$_\odot$ and
  3\,M$_\odot$ ZAMS models kindly made available by Sylvia Ekstr\"om and Tami
  Rogers, from \citet{Georgy2013a} and \citet{RogersMcElwaine2017},
  respectively.}
\end{figure*}

The transport equation controlling the evolution of the angular momentum
$r^2\Omega(r)$ reads
\begin{eqnarray}
\label{angmom}
\displaystyle{\frac{\partial}{\partial t} \left(r^2\Omega\right)} & = & 
\displaystyle{\frac{1}{5\rho r^2} \frac{\partial}{\partial r} 
\left[\rho r^4\Omega U(r)\right]} \\ 
& + & \displaystyle{\frac{1}{\rho r^2} \frac{\partial}{\partial r} 
\left(\rho r^4 D_{\rm shear}  \frac{\partial\Omega}{\partial r}\right)}. \nonumber
\end{eqnarray}
Here $U(r)$ is the radial component of the velocity due to meridional
circulation and the diffusion coefficient $D_{\rm shear}$ represents a variety
of vertical shear instabilities occuring between layers subject to different velocities
\citep{Maeder2009}.  In addition to these instabilities, IGWs also occur in the
radiative zones of stellar interiors.  Given that the dominant restoring force
for an IGW is the buoyancy force of Archimedes, the frequencies of IGWs are below
$N(r)$. These IGWs propagate in the radiative zones of the star, where they
dissipate, depositing angular momentum efficiently in the layers where they
break.  A pioneering study of the excitation and propagation of IGWs in Sun-like
stars was presented by \citet{Charbonnel2005}. It demonstrated convincingly the
capacity of IGWs to transport angular momentum in an efficient way,
explaining the flat rotation profile of the Sun derived from helioseismology.

As for the transport of the chemical species due to rotation,
\citet{Chaboyer1992} showed that it can be approximated as a diffusive process
in the presence of strong horizontal turbulence due to shear instabilities. For
this reason, the diffusive part in the chemical composition equations in
Eq.\,(\ref{MLTCBM}) gets extra terms due to various effects of rotation, each
with its own diffusion coefficient \citep{Maeder2009}. Aside from rotation,
additional causes of element mixing are also considered, particularly in
transition layers that are stable against the Schwarzschild criterion, but
unstable against the Ledoux criterion, for the cases of both $\nabla_\mu>0$
(called semiconvective mixing) and $\nabla_\mu<0$ (called thermohaline mixing).
Magnetism and IGWs may also affect the mixing.  Overall, this brings a multitude
of extra diffusion coefficients that affect the chemical composition profiles of
the star, aside from $D_{\rm conv}(r)$ and $D_{\rm ov}(r)$ included in
Eq\,(\ref{MLTCBM}). For rotation, these have been grouped as $D_{\rm shear}(r)$
and $D_{\rm eff}(r)$ adopting the notation by \citet{Maeder2009}, where the
latter is due to meridional circulation in the approximation of strong horizontal
turbulence and the former stands for the joint effect of vertical shear due to
all sorts of rotational (and possibly magnetic) instabilities. Pulsational
mixing profiles due to IGWs, adopting a diffusion approximation, were
derived from hydrodynamical simulations for a 3\,M$_\odot$ ZAMS star by
\citet{RogersMcElwaine2017}, resulting in a diffusion coefficient depending on
the density as $D_{\rm IGW}(r)\sim D_{\rm IGW} \cdot \rho^{-\gamma}(r)$ with
$\gamma\in [0.5;1]$.

Figure\,\ref{geneva} offers a schematic representation of mixing profiles
adopted in stellar evolution computations, where the envelope mixing profiles
were stitched to the CBM at an arbitrary level.  The two rightmost panels show
profiles for a 5\,M$_\odot$ ZAMS model rotating at 50\% of the critical rate
taken from \citet{Georgy2013a}; the particular shape of the third panel from the
left in Fig.\,\ref{geneva} is due to the drop in $U(r)\simeq 0$ in the envelope
layers near $m/M_\star\simeq 0.5$ \citep[cf.,][]{Maeder2003}.  Similar
sharp-peaked mixing profiles based on independently developed stellar evolution
codes were found by \citet{Heger2000}, \citet{Chieffi2013}, and
\citet{Paxton2013}, among others.  The profile labeled as Internal gravity waves
is from a
3\,M$_\odot$ nonrotating model computed by \citet{RogersMcElwaine2017}.  In
general, the mixing profiles indicated in Fig.\,\ref{geneva} vary strongly
during the evolution of the star, but it is poorly understood how.  In this
sense, none of these profiles are calibrated. Asteroseismology offers a major tool
to infer the overall mixing profiles throughout stars, denoted as of now as
$D_{\rm mix}(r,t)$.

Inferences of the internal mixing in stars received less attention than the
probing of $\Omega(r,t)$ thus far. The reason is simple: estimation of
$D_{\rm mix}(r,t)$ is much harder than of $\Omega(r,t)$. The latter
can be achieved in a quasi-model-independent way and (almost) directly from the
Fourier transform of the data, as explained in Sec.\,\ref{theoryNRP}.
Given that the levels of $D_{\rm mix}(r,t)$ as displayed in Fig.\,\ref{geneva}
differ by orders of magnitude in the literature, it is highly beneficial
to infer asteroseismic levels of mixing (and the accompanying convective core
mass) to bring the models into agreement with measurements of nonradial
oscillations of intermediate- and high-mass stars.  Asteroseismology of
$\gamma\,$Dor, SPB, and $\beta\,$Cep stars has the potential to provide the
answer if proper ensembles of such pulsators are subjected to asteroseismic
inference. We return to this potential and its first applications in
Sec.\,\ref{section-applics}.

\subsubsection{One-dimensional equilibrium models as input for
  asteroseismology} 

For stars with detected nonradial oscillations, space asteroseismology brings an
entirely new way to assess the rotation frequency $\Omega (r,t)$ and the overall
chemical mixing $D_{\rm mix}(r,t)$ in the radiative zones of stars. Indeed, mode
frequencies provide high-precision observational constraints coming directly
from the deep stellar interior. Assembling asteroseismic data for stars in
various evolutionary stages allows one, in principle, to assess the change of
$\Omega (r)$ and $D_{\rm mix}(r)$ as a function of stellar age. Yet
asteroseismic probing capacities for $\Omega (r,t)$ and $D_{\rm mix}(r,t)$ are
different for low-mass stars with a radiative core and a convective envelope
than for high-mass stars with a convective core and a radiative envelope. They
also differ for young stars burning hydrogen in their core and for old stars
close to their final fate as stellar remnant. To understand
why, it is necessary to dive into the nature of nonradial oscillations based on
1D equilibrium models.

As discussed earlier, the simplest versions of 1D stellar equilibrium models are
nonrotating nonmagnetic models having only 6 free parameters for fixed
choices of the input physics: the stellar birth mass $M_\star$, the initial
chemical composition guided by a galactic enrichment law \citep[e.g., as
in][]{Verma2019a} and expressed as relative mass fractions
$(X_{\rm ini},Y_{\rm ini})$ (or equivalently $X_{\rm ini},Z_{\rm ini}$), the
mixing-length value $\alpha_{\rm mlt}$ that gives rise to the mixing profile
$D_{\rm conv}(r)$, the convective overshoot length scale $\alpha_{\rm ov}$ that
leads to the CBM profile $D_{\rm ov}(r)$, and the age $\tau$.  Asteroseismic
modeling will then consist of determining the maximum likelihood estimators
(MLEs) of these six free parameters from measured oscillation mode
frequencies (often accompanied by other observables).  For a rotating star, at
least one additional parameter has to be added ($\Omega$ for the simplest case
of rigid rotation).  Once the most likely 6D or 7D parameter vector
$\bftheta\equiv (M_\star,X_{\rm ini},Y_{\rm ini},\alpha_{\rm mlt},\alpha_{\rm
  ov},\tau)$ has been found, the exercise can be repeated for other choices of
the input physics to come to an overall selection of the best stellar models for an
ensemble of stars. Any residual values between the measured and theoretically
predicted oscillation frequencies of unambiguously identified modes can then be
exploited to assess shortcomings in $\Omega (r,\tau)$ and $D_{\rm mix}(r,\tau)$
for the fixed chosen input physics.  Once a sufficiently large and unbiased (in
terms of rotation, initial chemical composition, etc.)  sample of nonradial
pulsators with suitable modes is available from observations, we can investigate
whether they adhere to the same theory of stellar structure and evolution or
instead 
need different internal mixing profiles as in Fig.\,\ref{geneva}.  We return
to this procedure of ``ensemble asteroseismology'' and will discuss
simplifications and applications of it for various types of stars in
Sec.\,\ref{section-applics}. An overall scheme representing this approach is
graphically visualized in Fig.\,\ref{flowcharts} and is discussed in
Sec.\,\ref{section-astero}.

\subsection{\label{theoryNRP}Linear nonradial oscillation modes}

We now consider small perturbations to 1D spherically symmetric stellar models
in hydrostatic equilibrium, whose quantities we assume to have been derived from
solving the stellar structure equations. We denote the equilibrium solutions
at age $\tau$ as $m_0(r), p_0(r), L_0(r), T_0(r), X_{i,0}(r)$. We assume that
the oscillations cause 3D periodic deviations from equilibrium with
amplitudes that justify a linear approach in the derivation of the pulsation
equations. In practice this implies that we perturb Eqs.\,(\ref{Eq1}),
(\ref{Eq2}), and (\ref{Eq3}) while retaining only the linear terms in the
perturbations. For example, a fluid element at position vector $\boldr_0$ in the
equilibrium model of the star is displaced due to the 3D stellar
oscillations to the vector $ \boldr_0 + \bolddelr$, where $\bolddelr$ is the
Lagrangian perturbation of the position vector. The Lagrangian perturbation to
the pressure then becomes
\begin{equation}
\delta p (\boldr)\!=\!p (\boldr_0+\bolddelr)\!-\!p_0(\boldr_0)\!=\!p(\boldr_0
)\!+\!\bolddelr\!\cdot\!\nabla p_0\!-\!p_0 (\boldr_0)\; .
\label{delp}
\end{equation}
All perturbed quantities that occur in Eqs.\,(\ref{Eq1}), (\ref{Eq2}),
and (\ref{Eq3}) can be deduced in a similar way. The linearized versions of the perturbed
equations are obtained by inserting expressions like Eq.\,(\ref{delp}) into the
full equations, subtracting the version of those equations for the static
equilibrium solutions, and neglecting all terms of order higher than 1 in the
perturbed quantities.  \citet[][Chap.\,3]{Aerts2010} gave 
full derivations; we adopt the notations from that book. Additional
extensive discussions on the theory of nonradial oscillations were given 
by \citet{Cox1980,Unno1989} and \citet{Smeyers2010}, where the last
work includes a particularly extensive historical perspective of the topic.

We argued in Sec.\,\ref{models} that it is meaningful to ignore the
nonradial components of the centrifugal force for stars that rotate up to
$\sim\!70\%$ of their critical rotation frequency and to treat the Coriolis and
Lorentz forces only at the level of the 3D perturbations to computed nonradial
g~modes, but not for the equilibrium models.  For p~modes, this validity
already breaks down above $\sim\!15\%$.  In the following sections, we gradually
build up the complexity of the treatment of the oscillations.  An obvious
simplification occurs when we consider the adiabatic approximation for the
computation of the modes. This means that we can ignore the perturbations of the
entropy $S$ in Eq.\,(\ref{Eq3}). We do so in the rest of this
section. Working in the adiabatic approximation is good and fully justified as
long as we consider modes that are mostly sensitive to the physics in the deep
stellar interior where adiabaticity is well met. This restriction is a point of
attention when dealing with modes that have their dominant energy in the
envelope of the star, close to the stellar surface; cf.\,p~modes in low-mass
stars as in Fig.\,\ref{eigenfunctions}, which is discussed later.

\subsubsection{Pressure and gravity modes}

We simplify the perturbed stellar structure equations maximally by ignoring the
Lorentz and Coriolis forces.  In that case, the only forces at play are the pressure
force and gravity.  These simplifications offer maximal separability in terms of
spherical polar coordinates $(r, \theta , \phi )$ and time $t$, where $r$ is the
distance to the center of the star, $\theta$ is the angle from the polar axis,
which is taken to coincide with the rotation axis of the star, and $\phi$ is the
longitude.  The displacement $\bolddelr$ can then be separated 
into radial and horizontal components as
\begin{equation}
\bolddelr = \xi_r \bolda_r+ \boldxih \; ,
\end{equation}
where $\bolda_r$ is a unit vector directed radially outward.  The solutions to
the resulting perturbed versions of the equations, along with proper boundary
conditions for the center and for the surface of the star \citep[not discussed
here, see, e.g.,][]{Unno1989} lead to nontrivial solutions only for the nonzero 
eigenfrequencies $\omega$ of the stellar equilibrium model. Each of
these eigenfrequencies corresponds to a so-called time-dependent spheroidal mode
of oscillation.  Because the equations are homogeneous, the eigensolutions are
determined only up to a constant factor.

Each of the nonradial eigenmodes of the equilibrium model corresponds to a
displacement vector $\boldxi$ whose components are written in terms of a mode
degree $l$, azimuthal order $m$, and radial order $n$ as
$\boldxi (r, \theta, \phi, t) = [(\xirnl \bolda_{r} + \xihnl \nabla_{\rm h})
Y_l^m(\theta,\phi)] \exp (- {\rm i}\,\omega_{nlm} t )$.  Modes with $m=0$ are
called axisymmetric (or zonal) modes; these reveal $l$ latitudinal surface nodal
lines.  For $|m|=l$, all surface nodal lines are lines of longitude. These
modes are called sectoral modes. Modes with $0\neq |m|<l$ are called tesseral
modes and have $|m|$ longitudinal and $l-|m|$ latitudinal nodal lines.  As a
special case, radial oscillations have $l=0$; i.e., they do not reveal any nodal
lines on the stellar surface.  The angular dependence of the radial eigenvector
component ($\xi_r$) of some nonradial modes was graphically illustrated
in Fig.\,\ref{ylm}.  Space photometry has predominantly given rise to the detection
of low-degree modes, typically with $l<4$. As discussed in
Fig.\,\ref{ylm}, the higher the degree of the mode, the more the detection is
prone to partial cancellation due to the integration of the mode's overall
perturbation over the visible stellar surface in the line of sight. The
cancellation gets more pronounced as $l$ increases, because more and smaller
patches with opposite sign occur in the spherical harmonic $Y_l^m$ that
represents $\xi_r$. The level of cancellation also depends on the angle between
the rotation axis and the line of sight (chosen as $60^\circ$
in Fig.\,\ref{ylm}).  The radial order of the mode, $n$, represents the number
of nodes of $\xi_r$ in the stellar interior, where these nodes are counted
positively for p~modes and negatively for g~modes.  For unevolved stars,
the assignment of $n$ is straightforward in that modes with $n=0$ have no nodes
aside from the stellar center. These modes are called fundamental modes,
abbreviated as f~modes.  For mixed modes in evolved stars, however, one could
obtain $n=0$ from the occurrence of pairs of nodes for $\xi_r$ in the p- and
g-mode cavities. Since the assignment of the radial order $n$ is used to
classify modes, such classification may be subject to quite complex
issues, as explained in Sec.\,3.5.2 of \citet{Aerts2010}, to which we refer for
more details.  Further, we adopt the convention that the sign of $m$
distinguishes prograde ($m>0$) from retrograde ($m<0$) modes, where the former
represent motions along the rotation of the star and the latter represent
motions against it.

The general system of differential equations that lies at the basis of the
eigenvalue problem describing nonradial oscillation modes is of fourth order in
the unknown perturbed quantities, which are $\xi_r$ and the perturbations to the
pressure $\delta p$, gravitational potential $\delta \Phi$ and the derivative of
$\delta \Phi$. These equations have
$Y_l^m(\theta,\phi) \exp (- {\rm i}\ \omega_{nlm} t )$ as a common factor. Hence
this factor can be divided out.  The resulting ordinary differential equations
to solve for the radial component of the unknown eigenfuntions do not depend on
the azimuthal order $m$ due to the assumption of having a spherically symmetric
equilibrium model.  This fourth-order system of equations needs four boundary
conditions to be solved.  However, it is often appropriate to ignore the
perturbation to the gravitational potential because this perturbation is
sufficiently small relative to the perturbation to the density. This is known as
the Cowling approximation \citep{Cowling1941}. It renders the system of
equations to second order and thus requires only two boundary conditions to get
physically meaningful solutions.  These are $\xi_r\simeq l\xi_h\sim r^{l-1}$ for
$r\rightarrow 0$ and $\delta p=0$ for $r\rightarrow R_\star$ \citep{Unno1989}.
This also allows for the derivation of an analytical expression for the ratio of the
horizontal to the radial displacement at the stellar surface, which depends only
on the frequency of the mode:
\begin{equation}
\label{K-value}
\displaystyle{\frac{\xi_h(R_\star)}{\xi_r(R_\star)}\simeq \frac{GM_\star}{\omega_{nl0}^2R_\star^3}}\; .
\end{equation}
This ratio is called the ``K-value'' by observers.  Typical values for this
ratio are below 0.001 for high-order p~modes as in the Sun and 10 -- 1000 for
high-order g~modes of core-hydrogen-burning stars. Mathematically, the Cowling
approximation is valid only for modes of high radial order $n$ and of ``high''
degree $l$.  One should therefore not expect this to be an optimal approximation
for low-order low-degree modes, and in particular not for $l=1, n=0$ f~modes
\citep[Sec.\,3.4.1 in][]{Aerts2010}. Hence, observed stars may reveal frequency
values for their f~modes that do not coincide with those computed in the Cowling
approximation.

The two pulsation equations resulting from adoption of the Cowling approximation 
can be combined into a single approximative second-order differential equation
for $\xi_r$ as follows:
\begin{equation}
\label{Cowling}
{\dd^2 \xi_r  \over \dd  r^2} \simeq {\omega^2  \over {c_{\rm s}^2}}
\left( 1 - {N^2  \over \omega^2} \right) \left( 
{S_l^2  \over \omega^2} - 1 \right) \xi_r \; ,
\end{equation}
with $N(r)$ as given in Eq.\,(\ref{BVformula}) and where we
have introduced the following local characteristic acoustic frequency (also
called the Lamb frequency) for the mode with degree $l$:
\begin{equation}
S^2_l(r)\equiv \frac{l(l+1) c_{\rm s}^2}{r^2} \; ,
\end{equation}
with $c_{\rm s}$ the sound speed in the stellar interior.  While Eq.\,(\ref{Cowling}) is
the simplest form in which nonradial oscillations can be described, it still
leads to a good approximation for the mode
frequencies, and more importantly to insightful interpretations and an elegant way to 
introduce the so-called mode cavities. These are illustrative when plotted in
propagation diagrams. Solutions for $\xi_r$ from solving Eq.\,(\ref{Cowling})
are oscillatory as a function of $r$ when (a) $| \omega | > | N |$ and
$| \omega | > S_l$ or when (b) $ | \omega | < | N | $ and $ | \omega | < S_l$. The
position inside the star where these conditions are met correspond with the
zones in the stellar interior where the modes resonate inside a cavity. In this
sense, the modes correspond to standing waves in their mode cavity
and are said to be trapped there.  The modes that meet
conditions a) are dominantly restored by the pressure force and are therefore
called pressure modes, usually labeled as p~modes. Within their mode cavity,
these modes are resonating sound waves (also called acoustic waves).
By convention, we denote their number of nodes in the stellar interior as $n>0$.
Buoyancy is the dominant restoring force when conditions b) are met and these
modes are therefore called gravity modes, labeled as g~modes.  Their radial
order is denoted by $n<0$, which means that they have $-n>0$ nodes in the interior of
the star.  Within their mode cavity, they behave like low-frequency (i.e.,
slow) internal gravity waves with a dominant horizontal displacement in a gas that
is radially stratified due to gravity.  Finally, solutions for $\xi_r$ when
solving Eq.\,(\ref{Cowling}) have an exponential behavior 
when $ | N | < | \omega | < S_l$ or $ S_l < | \omega | < | N | $. 
The zones in which the modes behave exponentially are called 
evanescent regions and the eigensolutions decrease or increase 
exponentially the farther away they are from the mode cavities.

\begin{figure*}[t]
\begin{center} 
\rotatebox{0}{\resizebox{8.8cm}{!}{\includegraphics{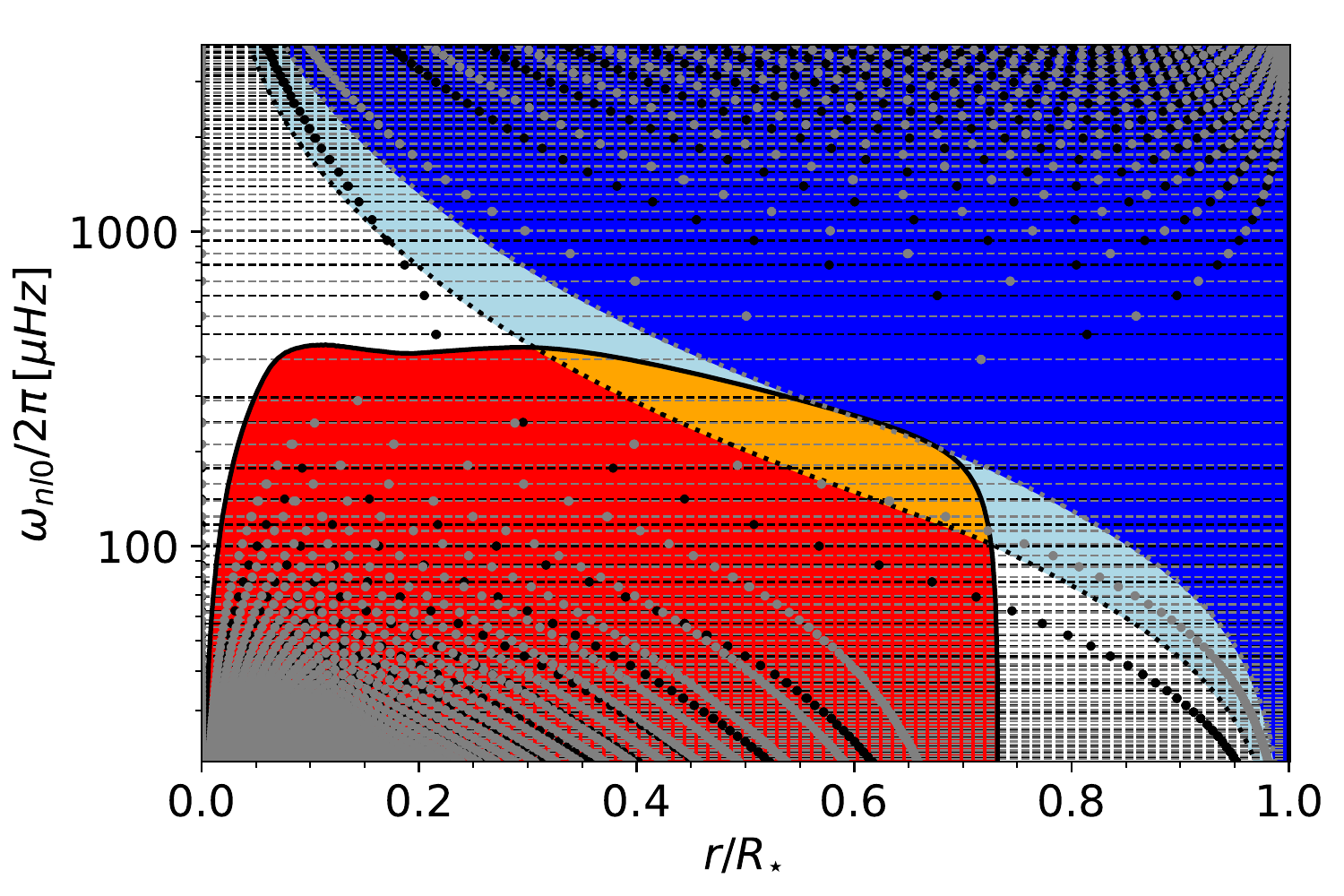}}}
\rotatebox{0}{\resizebox{8.8cm}{!}{\includegraphics{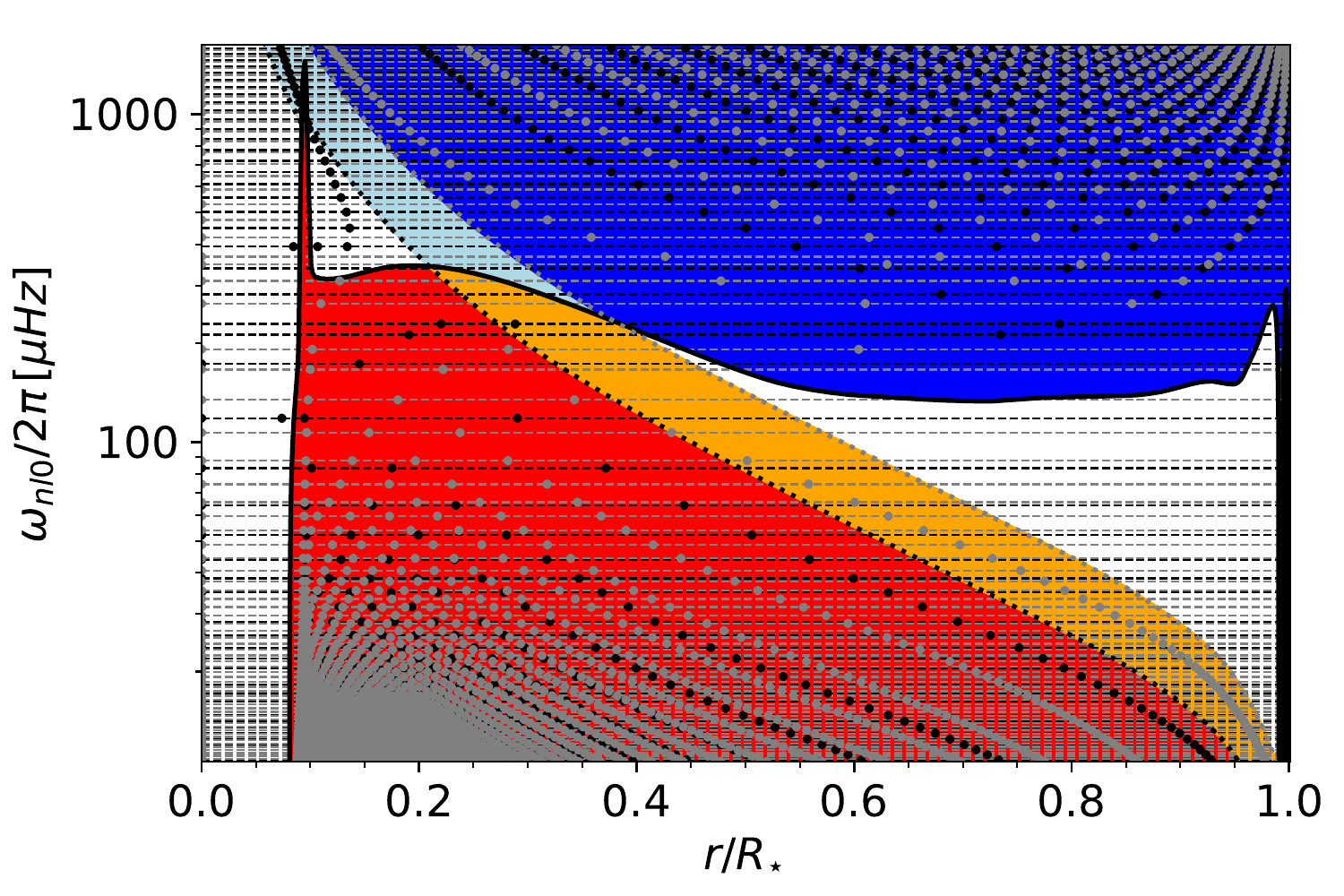}}}\\[0.2cm]
\rotatebox{0}{\resizebox{8.8cm}{!}{\includegraphics{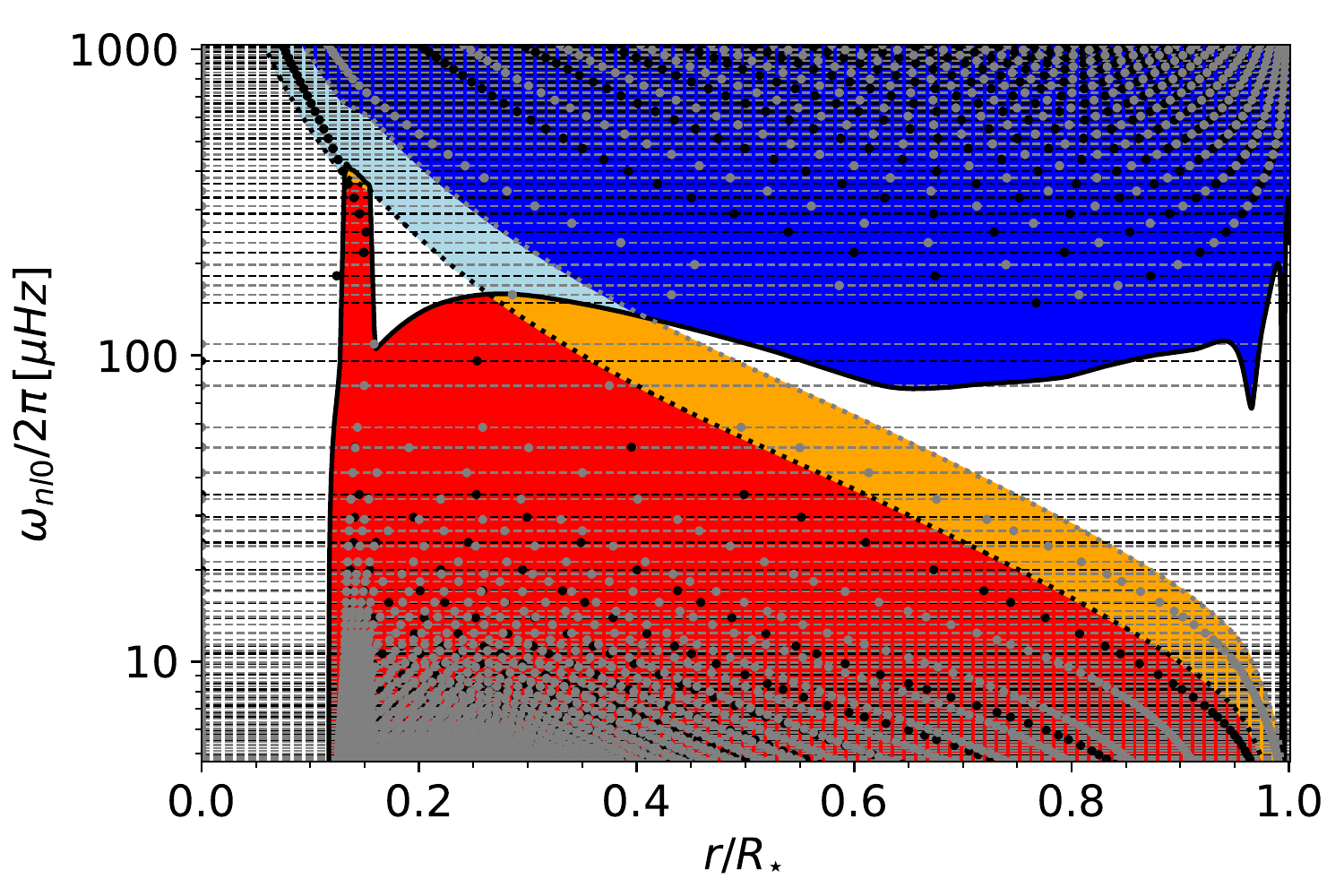}}}
\rotatebox{0}{\resizebox{8.8cm}{!}{\includegraphics{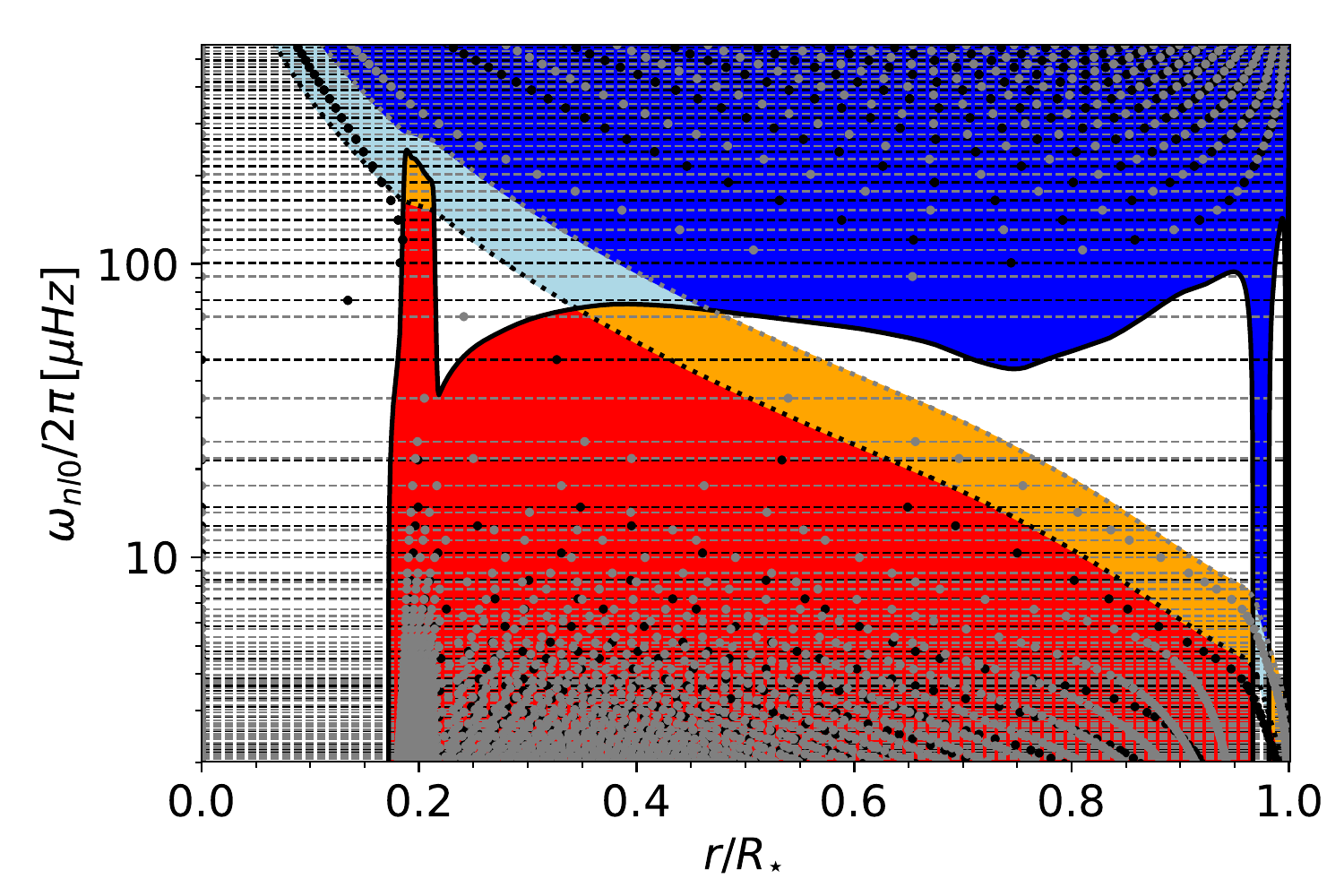}}}
\end{center}
\caption{\label{cavity} Propagation diagrams showing the mode cavities of
  axisymmetric p and g~modes in four stellar models halfway through the
  core-hydrogen-burning stage of evolution. The models have masses of 1, 1.7, 5,
  15\,M$_\odot$ from top left to bottom right. The thick solid black line
  indicates $N(r)$, while the dotted black and gray lines represent $S_1(r)$ and
  $S_2(r)$, respectively. The values of the dipole (quadrupole) mode frequencies
  are indicated as black (gray) horizontal lines. The position of the nodes of
  $\xi_r$ are indicated as thick black and gray dots for $l=1$ and 2,
  respectively.  The red region is the g-mode cavity for dipole modes; it is
  extended by the orange part for quadrupole modes.  The dark blue region is
  the mode cavity of quadrupole ($l=2$) p~modes. It is extended by the light
  blue region for dipole ($l=1$) p~modes. The modes correspond to evanescent
  waves in the white regions in the stellar envelope. The g~modes cannot
  propagate in the convective core of the three most massive stellar models, nor
  in the outer $\sim\!26\%$ of the convective envelope of the 1\,M$_\odot$
  model, where $N^2(r)<0$. Figure courtesy of Joey Mombarg, KU\,Leuven.}
\end{figure*}
Figure\,\ref{cavity} shows propagation diagrams for four stellar models that
represent stars about halfway through their core-hydrogen-burning stage, with
birth masses of 1, 1.7, 5, and 15\,M$_\odot$ and with solar chemical composition
and mixture. The oscillation modes were computed with the open source pulsation
code {\tt GYRE} \citep{Townsend2013,Townsend2018}, coupled to equilibrium
models computed with the open source code Modules for Experiments in Stellar
Astrophysics
\citep[{\tt MESA}][]{Paxton2011,Paxton2013,Paxton2015,Paxton2018,Paxton2019}.  The
results in Fig.\,\ref{cavity} were obtained not by relying on the Cowling
approximation but rather by from solving the fourth-order set of equations, as in
modern applications of asteroseismology. The mode cavities for axisymmetric
dipole ($l=1$) and quadrupole ($l=2$) modes are indicated, as are the mode's
eigenvalues (horizontal lines) and positions of the radial nodes (dots). The
importance of the receding convective core and the accompanying shape of $N(r)$
for the g-mode oscillations in intermediate- and high-mass models is visible in
the lower panels of Fig.\,\ref{cavity}.

\begin{figure*}[t]
\begin{center} 
\rotatebox{0}{\resizebox{17.cm}{!}{\includegraphics{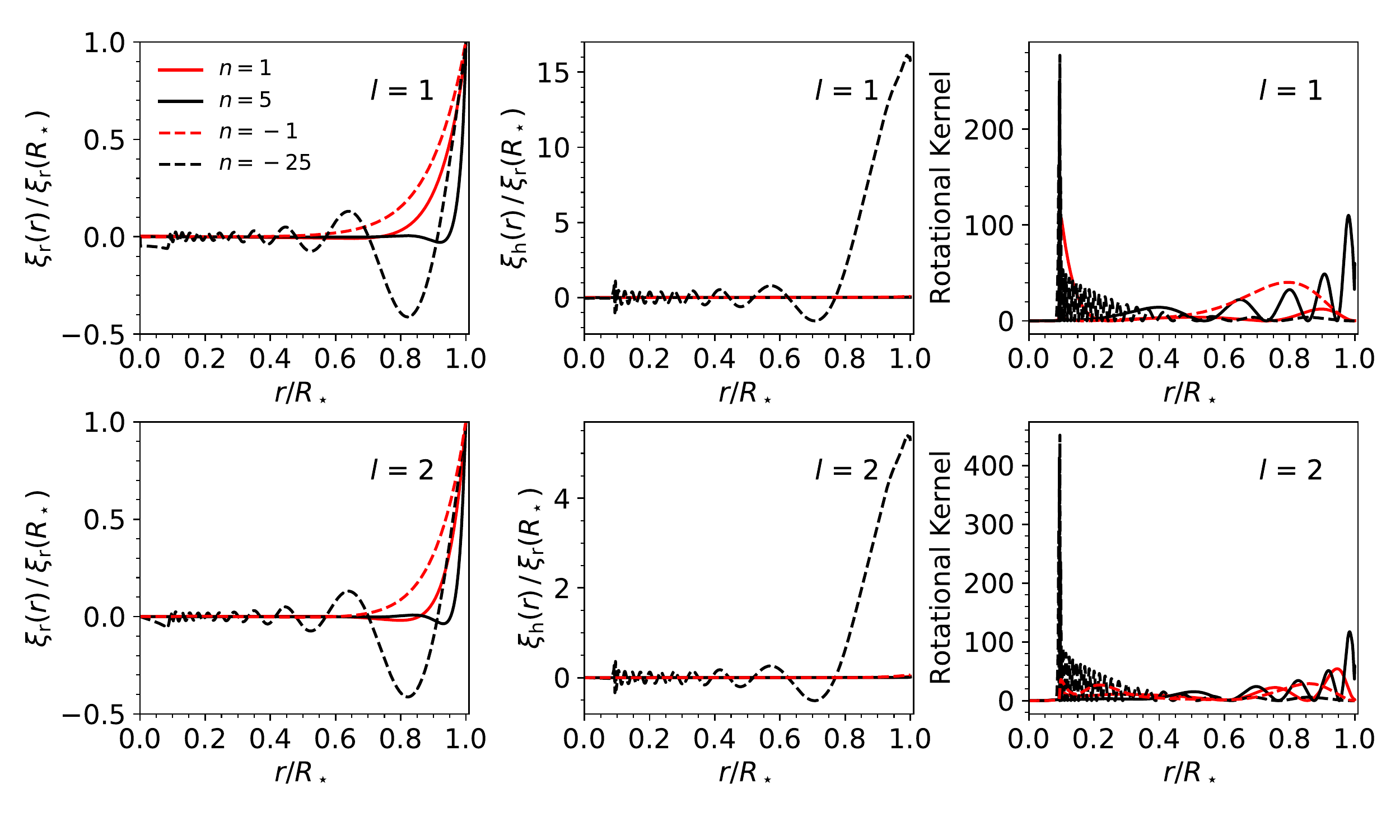}}}
\end{center}
\caption{\label{eigenfunctions} Radial (left panels) and horizontal components
  (middle panels) 
  of the Lagrangian displacement of four indicated axisymmetric ($m=0$)
p and g~modes for $l=1$ (top panels)
  and $l=2$ (bottom panels) of a stellar model with $M_\star=1.7\,$M$_\odot$ halfway through
  its core-hydrogen-burning stage of evolution. The right panels show the
  rotation kernel defined by Eq.\,(\ref{kernel}), which represents the probing
  power of an oscillation mode.  Figure courtesy of Joey Mombarg, KU\,Leuven.}
\end{figure*}
Eigenfunctions for eight modes are shown in the left and middle panels of
Fig.\,\ref{eigenfunctions} for the stellar model whose mode cavities are
displayed in the upper right panel of Fig.\,\ref{cavity}.  In the absence of
predictive power for the mode amplitudes, we normalize the modes such that
$\xi_r(R_\star)=1$ in Fig.\,\ref{eigenfunctions}.  It can be seen by comparing
the left and middle panels that high-order g~modes have dominant horizontal
displacements, while it is the opposite for p~modes. This is in line with the
predictions based on the Cowling approximation. It is a general property of p
and g~modes. Moreover, Fig.\,\ref{eigenfunctions} shows that p~modes have higher
amplitudes in the outer stellar envelope than in the inner regions, 
while g~modes have their highest amplitude in
the regions near the convective core. 

The mode cavities change as a star evolves, reflecting the increased
density contrast in the stellar interior. This drastically changes the profile
of the sound speed $c_{\rm s}(r)$ and hence the profile of $S_l(r)$ as well. As
a result, the p-mode cavities decrease in frequency and the evanescent zones
become narrower. Their exponential decay may hence be limited, allowing them
to reach the g-mode cavity and couple to the eigenfrequencies of the g~modes. Such is the
case for dipole modes in red-giant stars. These modes are therefore called mixed
modes: they have a p-mode character in the outer envelope and a g-mode
character in the inner regions of the star. We refer to Figs.\,2 to 4 in the
Supplemental Material given by \citet{Aerts2019} for propagation
diagrams of mixed modes in red-giant stars and refrain from repeating such diagrams here
for brevity. This mixed character of these dipole modes was predicted
theoretically by \citet{Dziembowski1971} and \citet{Shibahashi1979}.
\citet{Dupret2009} pointed out their probing power
for the center of evolved stars following the discovery of nonradial
oscillations in red giants from CoRoT \citep{DeRidder2009}, 
prior to their actual detection in {\it Kepler\/}
data. We return to this capacity in Sec.\,\ref{section-applics}.

\subsubsection{Asymptotic representations of high-order modes}

\begin{figure}
\begin{center} 
\rotatebox{0}{\resizebox{8.8cm}{!}{\includegraphics{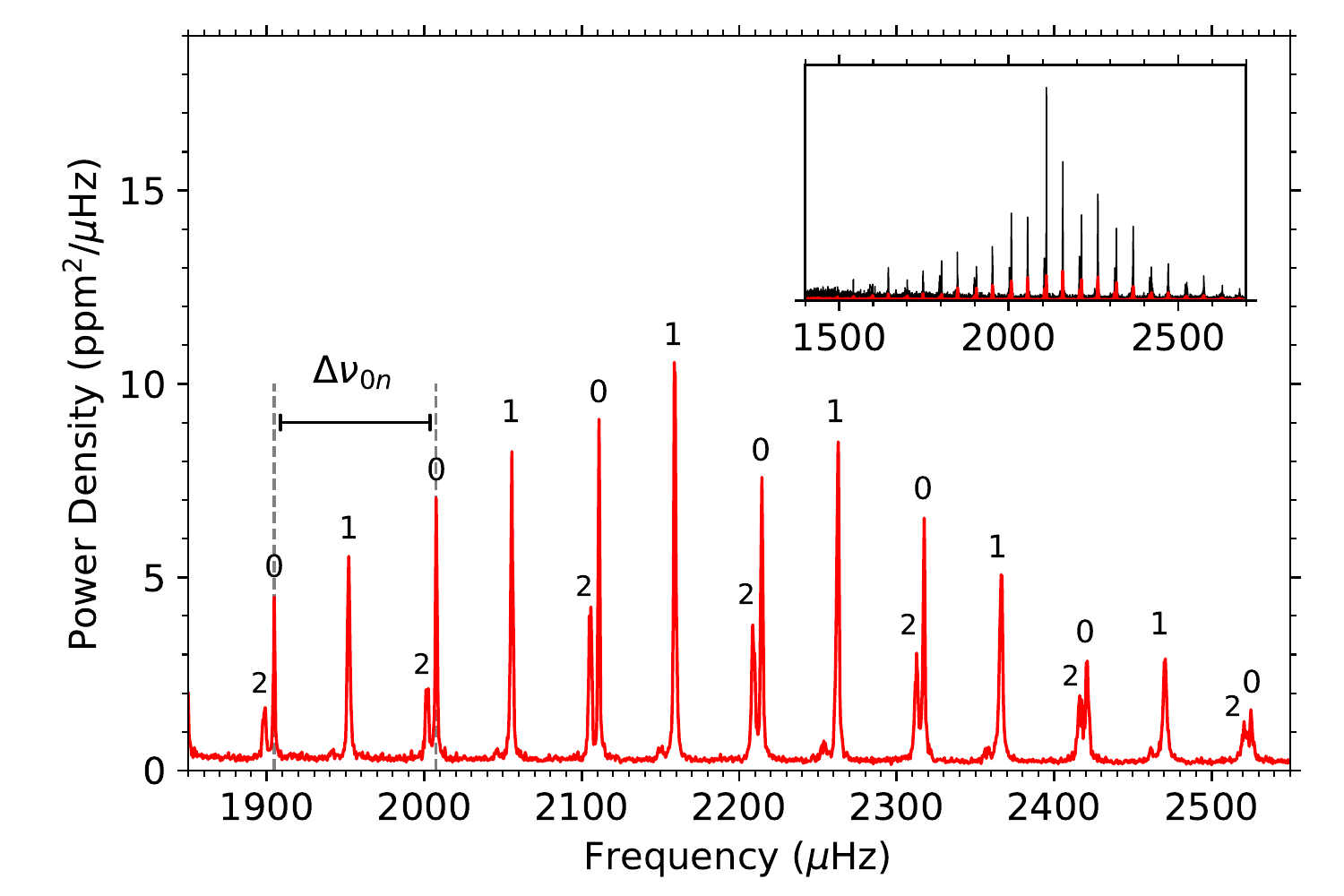}}}
\end{center}
\caption{\label{16cyg} Enlargement (red) of the observed envelope of oscillation signal
  revealed by the power density spectrum (black)
  of the solar analog 16\,Cyg\,A as deduced from data assembled with the {\it
    Kepler\/} satellite. The p~modes are labeled by their degree $l$. The large
  frequency separation based on the detected radial mode frequencies is
  indicated.  Figure based on data in \citet{ChaplinMiglio2013} by courtesy of
  Dominic Bowman, KU\,Leuven.}
\end{figure}

As discussed in Sec.\,\ref{section-astero}, mode identification is a critical step
to be taken before any asteroseismic inference can be made. Indeed, a comparison
between the detected and theoretically computed oscillation mode frequencies,
$\omega_{nlm}$, can be made only after the mode labels $(n,l,m)$ have been
derived. Given that we cannot resolve the surfaces of pulsating stars in
sufficient detail (except for the Sun), we cannot identify the spherical
wave numbers $(l,m)$ of the nonradial modes from maps of the eigenfunctions, as
in the graphical representation in Fig.\,\ref{ylm}. We somehow have to
derive the mode identification from the observables.  To this end, asymptotic
representations of high-order modes help a great deal, although other more
empirical methods
for mode identification of modes exist as well \citep[Chap.\,6
in][]{Aerts2010}.  Here, we limit the discussion to mode identification
based on patterns deduced among the detected oscillation mode frequencies or mode
periods.

The asymptotic theory of nonradial oscillations is based on second-order
differential equations describing the modes, which illustrates again why the
Cowling approximation is so useful for asteroseismology.  The convenience of
asymptotic representations of high-order modes was initially considered for the
case of linear radial modes by \citet{Ledoux1962} (in French). He
recognized that the radial-mode properties can be derived from a second-order
differential equation, which constitutes a Sturm-Liouville eigenvalue problem
with singular endpoints at $r=0$ and $r=R_\star$. The asymptotic properties of
nonradial oscillation modes have been studied more generally ever since and are
well covered in the literature, at various levels of mathematical detail.  
See the extensive papers by \citet{Tassoul1980,Tassoul1990} and see 
Sec.\,3.4 and Appendix\,E given by \citet{Aerts2010} for a general background and
results. \citet{Smeyers2010} provides in their Chaps.\,14--18
thorough mathematical
details and comparisons for the different regimes of validity while considering
different types of modes and various types of equilibrium models.\\[0.05cm]

\noindent\underline{High-order p~modes.\\[0.1cm]}
We first consider the case of low-degree high-order axisymmetric p~modes.
To leading order in the asymptotics, the frequencies of such modes comply with
\begin{equation}
\nu_{nl} \equiv {\omega_{nl}  \over 2 \pi}
 \simeq \left( n + {l \over 2 }+ {1 \over 4 }+ \alpha \right) \Delta \nu \; ,
\label{freqsep}
\end{equation}
where we have dropped the $m=0$ wave number in the notation and where we 
use the cyclic frequencies of the oscillation modes.
In Eq.\,(\ref{freqsep}) 
\begin{equation}
\label{Deltanu}
\Delta \nu = \left( 2 \int_0^R {\dd r \over c_{\rm s} }\right)^{-1} \; 
\end{equation}
is called the large frequency separation.  It is the inverse of twice the sound
travel time between the center and the surface of the star. On the basis of this
theoretical prediction, one expects the frequencies of the p~modes with
sufficiently high $n$ to be equally spaced and modes with the same value of
$n + l /2$ to have almost the same frequency values, since
$\nu_{nl} \simeq \nu_{n-1\,l
  +2}$. Such frequency patterns have indeed been observed for the solar
low-degree p~modes and these observational findings have given rise to the
research field of helioseismology \citep[see ][for an extensive review,
including historical aspects of the development of asteroseismology of ``our own''
star]{JCD2002}.

Given that the excitation and damping of the solar oscillations is due to the
turbulent convection in its outer envelope, we expect similar asymptotic
behavior for the high-order p~modes of all stars with a convective
envelope. This was confirmed almost two decades ago from ground-based
velocity data for $\beta\,$Hydri \citep{Bedding2001} and $\alpha\,$Cen\,A
\citep{Bouchy2001} and prior to space photometry for several tens of stars
\citep{Aerts2010}. Space photometry confirms that stars with a
convective envelope comply with asymptotic theory, from dwarfs to the bottom of
the asymptotic giant branch. One of the best datasets of solarlike oscillations in a star
different from the Sun was assembled for 16\,Cyg\,A; its PD spectrum in
Fig.\,\ref{16cyg} illustrates the validity of the asymptotic theory.

Equation\,(\ref{freqsep}) is based on the dominant term in the asymptotic
representation of low-order p~modes. The second-dominant term in the expansion leads to
the so-called small frequency separation, given by 
\begin{equation}
\delta \nu_{nl} \equiv \nu_{nl} - \nu_{n-1 \, l+2}
\simeq - ( 4 l + 6 ) 
{\Delta \nu  \over 4 \pi^2 \nu_{nl}}
\int_0^R {\dd c_{\rm s}  \over \dd r }{\dd r \over r } \; ,
\label{freqsep2}
\end{equation}
where $c_{\rm s}(R_\star)\simeq 0$ was assumed to arrive at this approximation.
From this expression, it is clear that $\delta \nu_{nl}$ probes the sound-speed
gradient in the deep stellar interior. For stars in the core-hydrogen-burning
stage, $\dd c_{\rm s}/\dd r$ is
highly sensitive to the hydrogen and helium composition profiles, which are
directly impacted by the nuclear fusion. 
It is then readily understood that $\delta \nu_{nl}$ is
of major diagnostic value to estimate the age of the exoplanet host star by
comparing its observed values with predictions of this quantity based on
equilibrium models.  In the case of 16\,Cyg\,A, as can be seen in
Fig.\,\ref{16cyg}, $\delta \nu_{nl}$ can be measured with high precision from
the radial and quadrupole modes. This, and more sophisticated diagnostics for
additional modes, was used by \citet{Bellinger2017} to find an
asteroseismic estimate of $\tau=6.9\pm 0.4$\,Gyr. This is in excellent agreement
with other methods for this well-characterized bright exoplanet host binary
\citep{Maia2019}.

\begin{figure}
\begin{center} 
\rotatebox{0}{\resizebox{8.8cm}{!}{\includegraphics{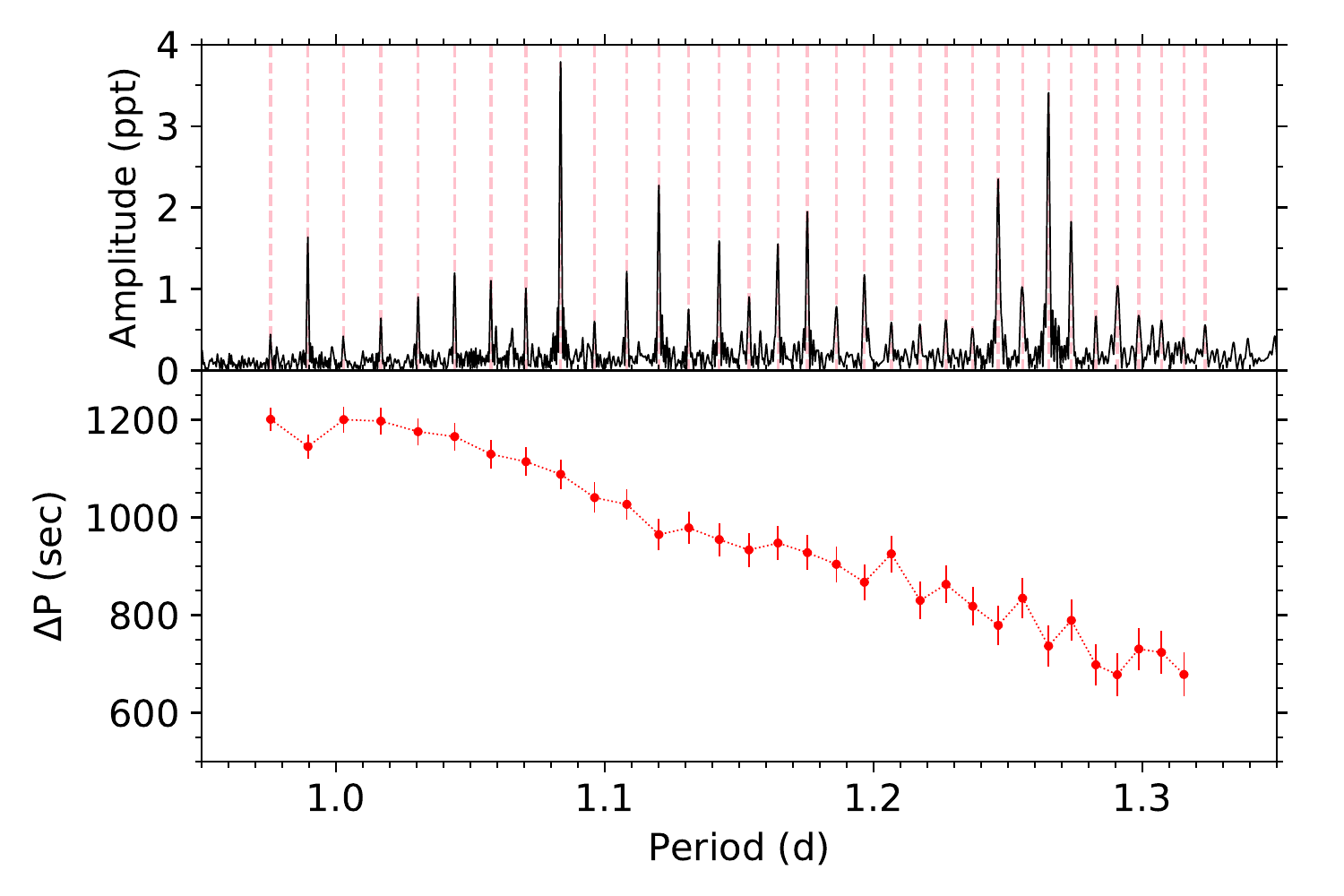}}}
\end{center}
\caption{\label{gdor} Top panel: observed amplitude spectrum (black) in terms of
the   period for the $\gamma\,$Dor star KIC\,11721304 from its light curve observed
  with the {\it Kepler\/} satellite. The mode periods with dominant amplitude
  are indicated with red dashed vertical lines as a guide for the eye. Bottom panel:
 period-spacing pattern deduced from the dipole sectoral prograde modes of
  consecutive radial order $n$ indicated in the top panel.  Figure based on
  data in \citet{VanReeth2015} by courtesy of Timothy Van Reeth, KU\,Leuven.}
\end{figure}

The similarity of the nonradial oscillations of 16\,Cyg\,A to those of the Sun
as illustrated in Fig.\,\ref{16cyg}, is representative of low-mass dwarfs
with convective envelopes. This observational finding is of key diagnostic
importance to estimate stellar masses, radii, and ages of such stars, as we discuss
further in Sec.\,\ref{section-applics}.\\[0.05cm]

\noindent\underline{High-order g~modes.\\[0.1cm]}
For high-order low-degree axisymmetric 
g~modes, $\omega<\!\!<N$ over most of the mode cavity;
cf.\ Fig.\,\ref{cavity}. We denote with $r_1$ and $r_2$ the inner and
outer positions of the g-mode cavity.  In this case, the asymptotic analysis
based on the Cowling approximation by \citet{Tassoul1980} led to
\begin{equation}
\label{periodspacing}
P_{nl} = {\Pi_0 \over \sqrt{l(l+1)}} (|n| + \alpha_{l,\rm g} ) \; ,
\end{equation}
where
\begin{equation}
\label{Pi0}
\Pi_0 \equiv 2 \pi^2 \left( \int_{r_1}^{r_2} N {\dd r \over r} \right)^{-1} \; .
\end{equation}
The quantity $\Pi_0$ stands for the buoyancy travel time and represents a 
characteristic period for the g modes of
the star (as inverse of a frequency, it is expressed in the unit of time). In  
this case, the mode periods are asymptotically equally spaced in the order of
the mode and the period-spacing value decreases with increasing $l$.  The phase
term $\alpha_{l, \rm g}$ depends on whether the star has a radiative or a
convective core. 

Long before space asteroseismology, g-mode period spacing patterns have been
extensively exploited for pulsating white dwarfs, based on photometric data
assembled with the Whole Earth Telescope \citep{Winget1991,Winget1994}. The
short periods of their g~modes  (a few to tens of minutes) imply
beating patterns in the light curves of only a few days, while the modes may have
high amplitudes of percentage level. This led to the detection of tens of dipole and
quadrupole modes that are subject to strong mode trapping in the outer thin H
and He layers of these objects, where $N(r)$ experiences spikes due to strong
changes in $\mu$.  In the context of white dwarfs, polytropes can be taken as
equilibrium models, leading to analytical expressions for
$\alpha_{l,\rm g}$ that allow detailed interpretation of the mode trapping in
terms of the chemical composition and mass of the outer layers of such pulsators
\citep[][for a seminal paper]{Brassard1992}. Ground-based asteroseismology was
therefore already highly successful for white dwarfs in the early 1990s.

For applications to SPB stars and $\gamma\,$Dor stars, which are both
core-hydrogen-burning g-mode pulsators with a convective core
(cf. Figs.\,\ref{hrd} and \ref{cavity}), $\alpha_{l,\rm g} = \alpha_{\rm g}$
turns out to be independent of the mode degree $l$ and one gets
$\Delta P\equiv P_{nl} - P_{n-1 l} = \Pi_0/\sqrt{l(l+1)}$.  \citet{Smeyers2007}
provided more sophisticated asymptotic analyses based on the full fourth-order
theory of nonradial oscillations; i.e., they omitted the Cowling approximation. They
developed the asymptotic approximations for both the cases of a radiative (SPB
stars) and a convective ($\gamma\,$Dor stars) envelope. Although they derived
more complicated expressions, the patterns to be expected from observations are
well captured by Eq.\,(\ref{periodspacing}). However, the number of nodes may
differ by 1 compared with the simpler treatment from \citet{Tassoul1980} based
on the Cowling approximation. This implies that one should consider an
uncertainty in the assignment of the radial order $n$ by at least 1 in any
practical asteroseismic modeling based on observed g-mode period spacings.  In
general, modern pulsation codes usually adopt the \citet{Takata2012}
classification scheme to assign the radial order $n$ to modes.

The period-spacing pattern of high-order g~modes offers a direct probe of the
physical conditions in the region near the convective core of main-sequence 
stars. This offers interesting applications to assess the mixing at the
bottom of the radiative envelope of core-hydrogen-burning stars, which is one
of the major uncertainties in the theory of stellar evolution, as discussed in
Sec.\,\ref{models}.  A seminal paper on this probing capacity was written by
\citet{Miglio2008}. In retrospect, this paper offered a remarkable sneak-preview
of the major insights to come from g-mode space asteroseismology when turned
into practice.  Aside from the somewhat controversial detection and
interpretation of g~modes in the Sun \citep{Garcia2007}, the 
first g-mode period-spacing pattern
detection for an intermediate-mass dwarf came from CoRoT data of the
$\sim\!7\,$M$_\odot$ B3V star HD\,50230. Eight axisymmetric g~modes with
consecutive radial order could be deduced from a 137-d-long light curve
by \citet{Degroote2010}. This star revealed periodic deviations from a uniform
spacing, which is in line with the theoretical predictions by
\citet{Miglio2008}. This detection allowed researchers to assess the level of $D_{\rm ov}$
and to derive an upper limit for $D_{\rm mix}(r)$ in the radiative
envelope. These results were confirmed by
independent asteroseismic modeling by \citet{WuLi2019-HD50230}. 

Given the immense asteroseismic potential of g-mode period-spacing
patterns, the CoRoT discovery opened the floodgates in the hunt for such
patterns in SPB and $\gamma\,$Dor stars, once the 4-yr light curves of the
{\it Kepler\/} spacecraft became available.  Meanwhile $\Pi_0$ has been measured
for hundreds of stars, one of which is shown in Fig.\,\ref{gdor}. It can be seen
that a clear pattern emerges from the data but that $\Delta P$ is not constant
as predicted by Eq.\,(\ref{periodspacing}). Rather, it decreases for increasing
mode period and reveals substructures. Such ``tilted'' $\Delta P$ patterns turn
out to be common in {\it Kepler\/} data of SPB stars as found by
\citet{Papics2015,Papics2017} and \citet{Szewczuk2018} and of $\gamma\,$Dor 
stars as revealed by  
\citet{VanReeth2015}, \citet{Bedding2015},
\citet{Keen2015}, \citet{Ouazzani2017}, and 
\citet{GangLi2019a,GangLi2019b,GangLi2020}.
The slope in these observed g-mode $\Delta P$ patterns is caused by the rotation
frequency of the star as deduced by 
\citet{VanReeth2016}, \citet{Ouazzani2017}, \citet{Christophe2018}, and
\citet{GangLi2020}. This rotation frequency turns out to
be of a similar order as the g-mode frequencies and puts these modes into the
gravitoinertial regime \citep{Aerts2017}. This  implies 
the need to include
the Coriolis force into the theory at the level of the pulsation equations for a
proper asteroseismic interpretation. We do so in the next two sections,
following the recent review on angular momentum transport by
\citet{Aerts2019} which includes more details and illustrations.

\subsubsection{Rotational splitting in a perturbative approach}

Thus far we have simplified the equations to compute the stellar oscillations by
ignoring the stellar rotation. Rotation affects the observed oscillation
frequencies in several ways.  We choose a reference frame with the polar axis
along the rotation axis of the star and corotating with the star under the
assumption of a constant rotation frequency $\Omega$. This leads to
a purely geometric shift of
\begin{equation}
\omega_{nlm}  = \omega_{nl}  + \ m\Omega \;  
\label{geometry}
\end{equation}
for a mode with frequency $ \omega_{nl}$ in the nonrotating case.  Further,
both the Coriolis and centrifugal forces come into play in the stellar structure
equations; see Eq.\,(\ref{HE+cent}).  The Coriolis force lifts the
degeneracy of the modes in the star with respect to the azimuthal order $m$.
Each mode frequency $\omega_{nl}$ of the eigenvector $\xi_{n l} $
$=( \xirnl , \xihnl )$ for the nonrotating case gets split into $2l+1$
frequency multiplet components due to the influence of the Coriolis force.
Hence, each mode degree $l$ can occur with $2l+1$ different values for $m$,
namely $-l, -l+1, \ldots, -1, 0, 1, \ldots, l-1, l$. Moreover, each of these
multiplet components gets shifted over $m\Omega$ as in Eq.\,(\ref{geometry}) in
the inertial coordinate system of the observer.

We now consider the case where the so-called spin parameter
$s=2\Omega/\omega<\!\!< 1$ for all involved mode frequencies, with $\omega$ an
abbreviated global notation for the oscillation frequencies in the corotating
frame.  This allows us to treat the Coriolis force as a small perturbation in
the pulsation equations. This condition is usually met for p~modes in low-mass
stars with convective envelopes, for p and mixed modes in red giants, and for
g~modes in subdwarfs and white dwarfs, all of which are slow rotators. This
simplification is not justified for the g~modes observed in the majority of
intermediate- and high-mass dwarfs, as these modes occur in the gravito-inertial
regime and require the Coriolis force to be treated nonperturbatively
\citep{Aerts2017}.  We return to the case of gravitoinertial modes later
but first treat the easier case of a perturbative treatment
of the Coriolis force.

Following the same arguments as for the shellular rotation, we
simplify the problem to be solved by assuming that the rotation profile
depends only on the radial coordinate $\Omega(r)$.  In this case, multiplet
components in the inertial coordinate system are, up to first order in
$\Omega(r)$, given by \citep[see Chapter\,6
and Sec.\,3.8 in][for the derivations]{Unno1989,Aerts2010}
\begin{equation}
\omega_{nlm}  = \omega_{nl}  + \ m\ (1-C_{nl}) \int_0^R K_{nl} (r) 
\Omega (r) \dd r  \; , 
\label{LedouxSplitting}
\end{equation}
where
\begin{equation}
K_{nl} (r) =
{ \left( {\xi_r}^2 +  [l(l+1)] \xih^2
 - 2 {\xi_r} \xih - \xih^2 \right) r^2 \rho   \over \int_0^R \left( {\xi_r}^2 + 
[l(l+1)] \xih^2 - 2 {\xi_r} \xih - \xih^2 \right) r^2 \rho \dd r} \; ,
\label{kernel}
\end{equation}
is the rotational kernel and
\begin{equation}
C_{nl}  = 
{\int_0^R \left( 2 {\xi_r} \xih + \xih^2 \right) r^2 \rho \dd r  \over 
\int_0^R \left( {\xi_r}^2  + [l(l+1)] \xih^2 \right)  r^2 \rho \dd r} \; 
\label{Cnl}
\end{equation}
is the Ledoux constant \citep{Ledoux1951}.  Rotational kernels for dipole and
quadrupole modes of four radial orders are plotted in the right panels of
Fig.\,\ref{eigenfunctions} for a 1.7\,M$_\odot$ star halfway through its
core-hydrogen-burning stage. It can be seen that the
high-order g~modes have far better probing potential for the core regions of the
star than the low-order modes. It is then understood from the profile shape
of $K_{nl}(r)$, which acts as a weighting function to the rotation
profile, why it is far easier to estimate the near-core values of $\Omega(r)$
than the envelope values for g~modes in stars with a convective core and a radiative
envelope, once rotational splitting has been detected from data.
Finally, we see from Eqs.\,(\ref{kernel}) and (\ref{Cnl}) that they depend on the
equilibrium model via its density profile $\rho(r)$ and its influence
on the 
eigenfunctions. 

In the limit of high-order or high-degree p~modes, one can show that
$C_{nl}\simeq 0$ and  ${\xi_r}\!>\!\xi_h$ 
\citep[][Sec.\,3.8]{Aerts2010}. 
On the other hand, for
high-order high-degree g~modes, one has ${\xi_r}\!<\!\xi_h$ as shown in
Fig.\,\ref{eigenfunctions} and one may neglect the terms with ${\xi_r}$ in
Eqs.\,(\ref{kernel}) and (\ref{Cnl}). In this way, the simplification
\begin{equation}
C_{nl} \simeq {1 \over [l(l+1)]} \; 
\end{equation}
emerges. For uniform rotation, this implies that the measured rotational
splitting provides a good measure of the average of $\Omega (r)$, weighted with
the squared eigenfunction.
A further simplification is useful, as  \citet{VanReeth2018} showed
that intermediate-mass stars with a convective core are quasiuniform rotators.
In the case of constant $\Omega$ one has
\begin{equation}
\omega_{nlm} = \omega_{nl} + m (1 -\  C_{nl}) \Omega \; 
\end{equation}
and $C_{nl}$ fully determines the shifts of the frequencies due to the Coriolis
force, i.e., those shifts do not depend on the rotational kernels.  This means
that the adjacent frequencies in a high-order p-mode multiplet belonging to
$m=-l,\ldots,+l$ give a direct measure of the average rotation frequency in the
stellar envelope, without depending on the equilibrium model (because
$C_{nl} \simeq 0$). In the case of high-order dipole g~modes, $\Omega$ is found
 to be twice the splitting value in a triplet since $C_{nl} \simeq 1/2$.

More complicated perturbative approaches treating the Coriolis force up to
second and third order at the level of the pulsation computations, while still
relying on 1D equilibrium models, have been developed. We refrain from including
the results here for conciseness and refer the interested reader to
\citet{Saio1981}, \citet{Dziembowski1992}, \citet{LeeBaraffe1995},
\citet{Soufi1998}, \citet{Daszynska2002}, \citet{Suarez2006}, and
\citet{Suarez2010}. Few of those theories have been applied to modern space
photometric data because the stars for which they are most appropriate are
rapidly rotating p-mode pulsators, such as $\delta\,$Sct stars and $\beta\,$Cep
stars; cf.\ Fig.\,\ref{hrd}. In the case of the $\delta\,$Sct stars, lack of
mode identification prevents applications, although \citet{Bedding2020} managed
to overcome this hurdle for a limited sample of such stars.  The few observed
$\beta\,$Cep stars with precise space photometry either lack mode identification
\citep{Burssens2019} or rotate slow enough to stick to the first-order
perturbative approach. \citet{Suarez2010} made a careful analysis of
second-order effects in $\Omega$ for stochastic p~modes and found those to
become important for equatorial rotation velocities above some
15\,km\,s$^{-1}$. This is also the limiting value for the treatment of g~modes
in intermediate-mass stars derived by \citet{SchmidAerts2016}. For rotation
speeds above this value, a perturbative analysis should be abandoned as
illustrated from their asteroseismic modeling of the high-order g~modes
in the two F-type p- and g-mode hybrid pulsators in the eccentric binary
KIC\,10080943. For faster rotation, the g~modes enter into the gravitoinertial
regime, where one can no longer treat the Coriolis force perturbatively
($s>1$). As outlined by \citet{Aerts2017} and shown in Fig.\,\ref{araa-update},
this is the case for the observed g~modes in almost all intermediate- and
high-mass stars. We thus conclude that the treatment of g~modes in stars with a
convective core requires a nonperturbative treatment of the Coriolis force.

\subsubsection{Gravito-inertial modes in the Traditional Approximation}

A major achievement resulting from the 4-yr light curves assembled with
the {\it Kepler\/} satellite is the discovery of g~modes with period-spacing
patterns such as the one illustrated in Fig.\,\ref{gdor} in hundreds of stars
covering spectral types early-F to early-B along the main sequence;  cf.\
Fig\,\ref{hrd}. Except for the few (less than 10\%) stars for which a surface
magnetic field was detected in this
range of spectral type \citep{Wade2016}, such stars are in
general moderate to fast rotators. They are indeed not subject to
braking due to the lack of a magnetic field, which does occur in low-mass stars with
an appreciable convective envelope. The high-order g~modes in these stars of
intermediate mass have periods similar to their rotation period 
so the oscillations are gravitoinertial modes \citep[see Fig.\,5 given by
][]{Aerts2019}. 
\citet{VanReeth2016} and \citet{Aerts2017} computed the spin
parameters for more than 1650 g~modes in 37 $\gamma\,$Dor stars and found the
majority to have subinertial values, defined as the regime for which $s>1$. In practice, the spin
parameter covered values $s=2\Omega/\omega_{nlm}^{\rm co}\in
[1,15]$, where $\omega_{nlm}^{\rm co}$ is the mode frequency in the corotating frame.

Taking full account of the Coriolis force in the equation of momentum
conservation, even in the adiabatic and Cowling approximations while ignoring
the centrifugal force, does not lead to separability of the pulsation equations
in terms of the coordinates $(r, \theta , \phi )$.  This is why
\citet{LeeSaio1987a,LeeSaio1987b,LeeSaio1989} considered the
so-called ``Traditional Approximation of Rotation'' (TAR) 
in their theoretical studies of low-frequency g~modes. In the TAR, one
ignores the horizontal component of the rotation vector such that the equations
can be separated in each of the coordinates. This approximation leads us to the
Laplace tidal equations \citep{Laplace1799}, which are commonly used in geophysics
\citep{Eckart1960}.  The TAR is a particularly good approximation for the
g~modes in intermediate- and high-mass main-sequence stars \citep[as well as in
neutron stars, cf.][]{Bildsten1996}, given that their Lagrangian displacement
vector is dominantly horizontal; see Fig.\,\ref{eigenfunctions}.  For
derivations of the pulsation equations in the TAR and their asymptotic analysis
in a modern numerical context, see
\citet{LeeSaio1997}, \citet{Townsend2003a,Townsend2003b}, and 
\citet{Mathis2013LNP}. Here we provide
the outcome in concise notation that allows for easy comparison with
Eq.\,(\ref{periodspacing}). For uniform rotation, the TAR leads to the following
g-mode period-spacing pattern in the corotating frame of reference:
\begin{equation}
\Delta P_{l,m,s}^{\rm co} = {\Pi_0 \over \sqrt{\lambda_{lms}}} \; ,
\label{spacing-TAR}
\end{equation}
with $\lambda_{lms}$ the eigenvalue of the Laplace tidal equation for the g~mode
with quantum numbers $(l,m)$ in a star with spin parameter $s$. In the limit of 
 $s \rightarrow 0$, $\lambda \rightarrow l(l+1)$ is recovered.
Numerical
computation of the eigenvalues $\lambda_{lms}$
for a chosen 1D equilibrium model of
the star then allows for the identification of $(l,m)$, as well as estimation of
the spin parameter along with $\Omega$ from an observed period-spacing pattern
as in Fig.\,\ref{gdor}. This opportunity was developed theoretically
by \citet{Bouabid2013} and was put into practice for the past five years
after careful frequency analysis based on the 4-yr light curves assembled
with the {\it Kepler\/} spacecraft. We highlight some of the recent
achievements on asteroseismic derivations of $\Omega(r)$ along with 
opportunities to estimate $D_{\rm mix}(r)$ from the period-spacing diagnostics
in Sec.\,\ref{section-applics}.

\citet{Mathis2009} generalized the TAR to take into account differential
rotation with a profile $\Omega (r,\theta)$, while \citet{Mathis2019} included
the centrifugal force for slightly deformed stars in the case of
close-to-uniform rotation, deriving an analytical expression for the period
spacing patterns in the Cowling and other justified approximations.
In addition, \citet{Prat2017} derived an asymptotic period
spacing for axisymmetric gravitoinertial waves taking into account all the
components of the rotation vector; i.e., they went beyond the treatment of the TAR.
This work was further generalized by \citet{Prat2018} into an asymptotic theory
for gravitoinertial waves for a differential rotation profile
$\Omega (r,\theta)$. Finally, \citet{Prat2019} derived a period-spacing
expression in the presence of uniform rotation on top of an axisymmetric fossil
magnetic field with poloidal and toroidal components.  None of these recent new
theoretical developments have yet been applied to measured g-mode frequencies.
This obviously constitutes several future paths for improved asteroseismic
modeling compared to the current state of the art. The
{\it Kepler\/} data of gravitoinertial pulsators are currently under study with
this purpose.

As a noteworthy side step, we point out that only one intermediate-mass g-mode
pulsator with a detected surface magnetic field has been the subject of magnetogravito
asteroseismology thus far \citep{Buysschaert2018}.  This led to the conclusion
that the frequency shifts for g~modes due to the Lorentz force are far smaller
than those due to the Coriolis force for meaningful values of the interior
magnetic field strength \citep{Prat2019}. This is quite different from the case
of high-frequency magnetoacoustic modes, which occur on the other side of the
frequency spectrum in terms of the validity (or lack thereof) of a perturbative approach to
treat the Coriolis and Lorentz forces \citep[see Fig.\,5
in][]{Aerts2019}. Inspired by the solar oscillations, \citet{Gough1990} derived
expressions for the pertubation to the eigenfunctions caused by rotation and a
magnetic field in the stellar interior.  Their asymptotic analysis and numerical
results for high-order solar acoustic modes for various magnetic field
configurations, including a localized magnetic field at the base of the
convection zone, provide estimates for the frequency splitting when the
magnetic field and rotation vary smoothly. This work is a convenient guide to compare
with observations.  

Aside from Sun-like stars, the best known magnetic pulsators are the roAp stars, 
which were discovered in 1978 by Don Kurtz \citep[][for his
review on these stars]{Kurtz1990} and later studied in great detail
\citep[][for a more recent review]{Saio2014}. 
These core-hydrogen-burning stars oscillate in high-$n$
low-$l$ p~modes according to an axis that may be misaligned with respect to both the magnetic
and rotation axes, although it is usually close to the magnetic axis.  Their
magnetic field strengths are up to a few thousand gauss, while they are slow
rotators and this implies that the Lorentz force is more important than the
Coriolis force. This is thus a case where the symmetry axis for the oscillations
is inclined with respect to the rotation axis. Although their magneto-acoustic modes have
sufficiently high amplitudes and periods of only a few minutes, making them easily
accessible from ground-based asteroseismology, their recent studies have
benefited greatly from modern space photometry. The oblique pulsator model of
roAp stars has constantly been in need of improvement as more data become
available, as shown by \citet{Shibahashi1993} and \citet{Bigot2002}. This
model was again challenged and refined thanks to the high-frequency precision
obtained from space asteroseismology, which revealed that some roAp stars seem
to have multiple pulsation axes \citep{Kurtz2011} and others oscillate in
distorted pulsation modes \citep{Holdsworth2016}.  Recent TESS data have been
used to find the shortest period roAp star, with a pulsation period of only
4.7\,min 
\citep{Cunha2019}. A growing number of roAp stars have been found to pulsate
above their acoustic cutoff frequency, which presents another challenge to
current pulsation theory.

\subsubsection{Rossby modes}

\begin{figure*}
\begin{center} 
\rotatebox{0}{\resizebox{17.cm}{!}{\includegraphics{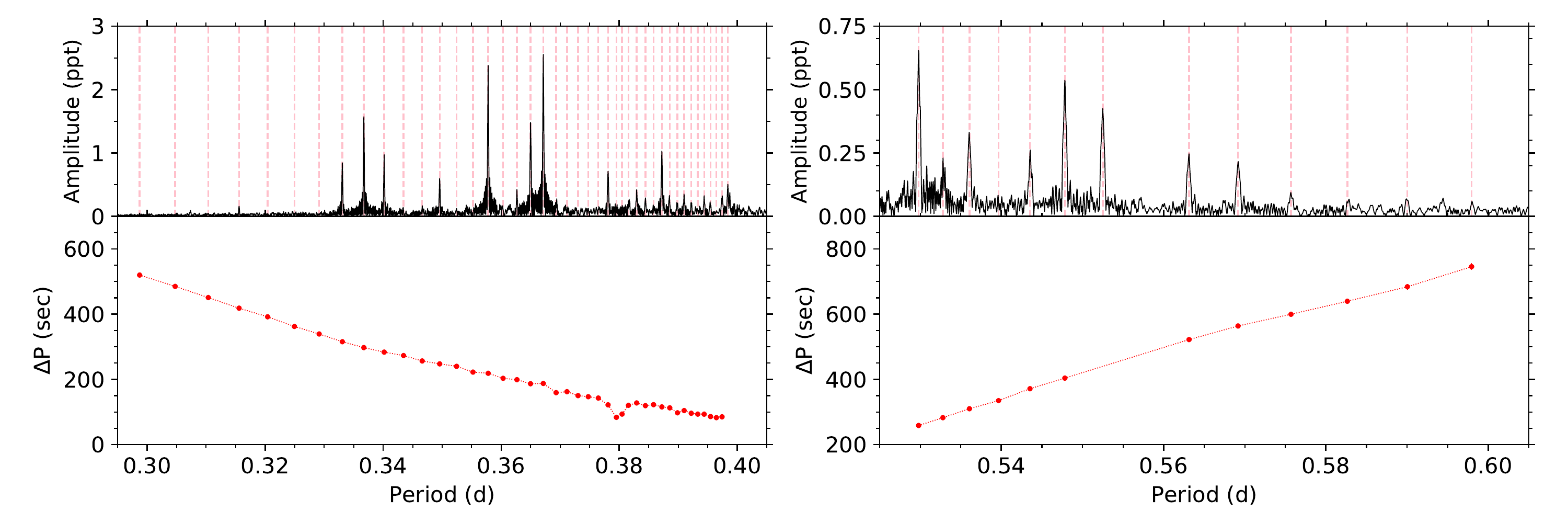}}}
\end{center}
\caption{\label{rossby} Same as Fig.\,\ref{gdor}, but for the $\gamma\,$Dor star
  KIC\,12066947 exhibiting both prograde dipole gravitoinertial modes with
  $(k,m)=(0,+1)$ and retrograde Rossby modes with $(k,m)=(-2,-1)$.  In contrast
  to the case of KIC\,11721304 shown in Fig.\,\ref{gdor}, the errors in the
  period-spacing pattern are smaller than the symbol size.  Figure based on data
  in \citet{VanReeth2015} by courtesy of Timothy Van Reeth, KU\,Leuven.}
\end{figure*}

\begin{figure}
\begin{center} 
\rotatebox{0}{\resizebox{8.5cm}{!}{\includegraphics{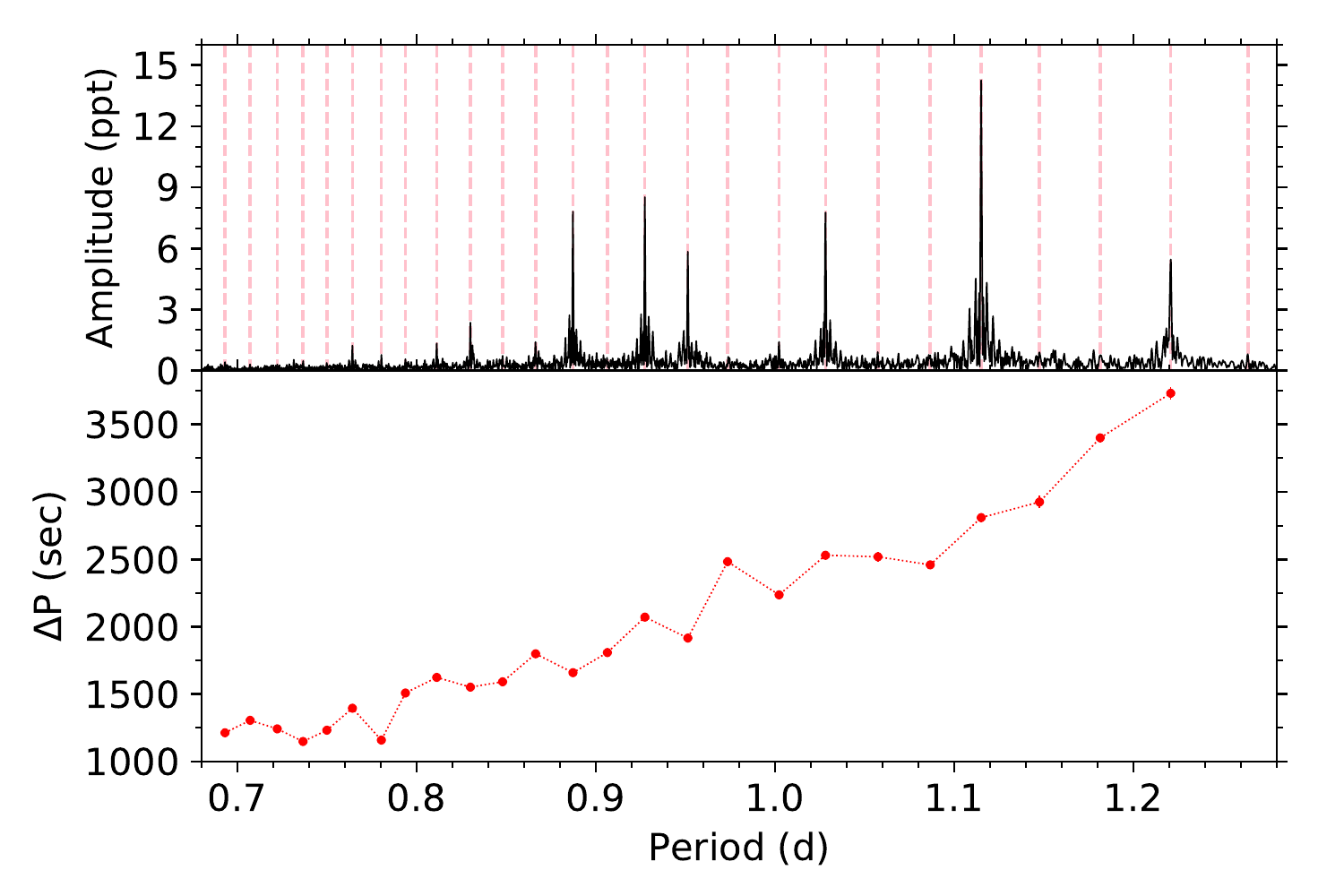}}}
\end{center}
\caption{\label{yanai} Same as Fig.\,\ref{gdor}, but for the $\gamma\,$Dor star
  KIC\,6425437 exhibiting retrograde Yanai modes with $(k,m)=(-1,-1)$.  In
  contrast to the case of KIC\,11721304 shown in Fig.\,\ref{gdor}, the errors in
  the period-spacing pattern are smaller than the symbol size.  Figure based on
  data in \citet{VanReeth2015} by courtesy of Timothy Van Reeth, KU\,Leuven.}
\end{figure}

We now return to the maximally simplified version of the stellar pulsation
equations deduced from perturbing Eq.\,(\ref{Eq2}) in the absence of rotation
and magnetism. This approach allowed us to introduce the time-dependent
spheroidal modes of oscillation known as p and g~modes. However, if we keep the
Coriolis force in Eq.\,(\ref{Eq2}) and perturb that version of the equation,
then two families of eigenvalue problems actually result, each with nonzero
eigenvalues. The first family is the one we have been discussing, leading
to spheroidal normal modes of a star. We now pick up the second family of
eigensolutions, termed toroidal normal modes. In particular, we consider the
Rossby modes, also termed and abbreviated as ``r~modes'' by \citet{PapPringle1978}.
This is a family of toroidal normal modes that become time-dependent (and
hence nonzero) only in a rotating star. The dominant restoring force of these modes
is the Coriolis force.  This is why they cannot be deduced from Eq.\,(\ref{Eq2})
unless a nonzero rotation vector is considered.

Toroidal modes comply with $\div{\bf \xi}=0$ and $\xi_r=0$. Therefore, just as
with the gravitoinertial modes discussed in the previous section, the
eigenvalues of Rossby modes can be deduced with excellent precision by adopting
the TAR and solving the Laplace tidal equations.  For the eigenfrequencies of
the spheroidal g~modes we had the limiting case of $\lambda \rightarrow l(l+1)$
as $s \rightarrow 0$. For the Rossby modes, one obtains $\lambda \rightarrow 0$ as
$s \rightarrow [l(l+1)]/m$ \citep{PapPringle1978}.  For this reason,
\citet{LeeSaio1997} adopted an ordering of the eigenvalues by introducing a
labelling scheme that allows one to treat the cases of gravitoinertial g~modes and
pure inertial modes with one set of indices $(k,m)$, with $k=l-|m|\geq 0$ for
gravitoinertial modes and $k<0$ for purely inertial modes \citep[see the
seminal paper by][for the various types of low-frequency modes in rotating
pulsators]{Townsend2003a}. 
Rossby modes have frequencies below the rotation
frequency in the corotating frame and are therefore always retrograde modes in
the inertial frame of the observer \citep{Saio1982}. 
The occurrence of the temperature variations at the
stellar surface due to Rossby modes and for various values of the spin parameter
is shown in Fig.\,2 given by \citet{Saio2018a} and omitted here for conciseness.

With the labeling scheme introduced by \citet{LeeSaio1997}, the period-spacing
pattern of Rossby modes becomes
\begin{equation}
\Delta P_{k m s}^{\rm co} = {\Pi_0 \over \sqrt{\lambda_{k m s}}} \; ,
\label{spacing-Rossby}
\end{equation}
with $\lambda_{k m s}$ again the eigenvalues of the Laplace tidal equation. It
was shown by \citet{Townsend2003b} that the eigenvalues for Rossby modes comply
with $\lambda_{k m s}\approx m^2(2|k|-1)^{-2}$ for $s>\!\!>1$ and $k\leq -2$.
From this, it is found that the period-spacing value of Rossby modes of
consecutive radial order as seen by an observer increases with increasing mode
period. This is illustrated for the $\gamma\,$Dor star KIC\,12066947 observed by
the {\it Kepler\/} spacecraft in Fig.\,\ref{rossby}.  This star has both prograde
gravitoinertial dipole modes with $k=0$ and retrograde Rossby modes with
$k=-2$.  For its sectoral gravitoinertial g~modes, just as for the ones
observed for KIC\,11721304 shown in Fig.\,\ref{gdor}, the label is $k=0$ and we
recover the treatment of the period-spacing pattern represented by
Eq.\,(\ref{spacing-TAR}).

\citet{VanReeth2016} made the first discovery of Rossby modes in {\it Kepler\/}
data. This was achieved for ten $\gamma\,$Dor stars, all of which were found to
have spin parameters $s\in [14,30]$ \citep{Aerts2017}. Meanwhile, Rossby modes
were found to be common in F-, B-, Be-, and A-type stars, as well as in eccentric
binaries, all of which were observed with the {\it Kepler\/} spacecraft and studied by
\citet{Saio2018a} and \citet{GangLi2019b}.  These discoveries offer the
opportunity to assess whether the interior rotation is constant or differential from combined
g- and r-mode asteroseismology. We also note in passing that retrograde Yanai
modes were discovered in {\it Kepler\/} data of seven $\gamma\,$Dor stars so far
by \citet{VanReeth2018} and \citet{GangLi2019b}. As explained by
\citet{Townsend2003b}, this family of modes behaves like gravitoinertial modes
when they are prograde, while the retrograde Yanai modes behave as Rossby modes
but have $k=-1$. We show the results for KIC\,6425437, which is one such star
revealing a period-spacing pattern of Yanai modes, in Fig.\,\ref{yanai}
\citep{VanReeth2018}.  Just as with the Rossby modes, the full potential of Yanai-mode
frequencies in terms of asteroseismic probing of the properties of the interior
physics has yet to be investigated and exploited, given their recent
discoveries.

%%%%%%%%%%%%%%%%
%%%%%%%%% Section 3
%%%%%%%%%%%%%%%%

\section{\label{section-astero}Principles of asteroseismic modeling}

The theory of nonradial oscillations outlined in Sec.\,II rests on
the assumption of linearity.  Although the amplitudes of stochastically excited
modes of the Sun can now be predicted from the damping rate and the
stochastic energy from 3D convection simulations \citep{Zhou2019}, this is not
the case for the amplitudes of the modes excited by other mechanisms.  Thus,
most interpretations in asteroseismology rely on the properties of the detected
mode frequencies and not on their amplitudes. Asteroseismic modeling is
therefore usually done in a linear adiabatic framework.  After having derived
the frequencies of the modes $\omega$ and their uncertainty 
$\sigma_{\omega}$ from data, interpretations in terms of the theory of nonradial
oscillations computed from perturbing stellar equilibrium models
(see Sec.\,\ref{section-nrp}) can be done only if the modes' identifications
have been achieved.  This means that we must be able to label the radial order,
the degree, and the azimuthal order ($n,l,m$) for each of the modes
corresponding to the measured oscillation frequencies $\omega_{nlm}^{\rm obs}$.

Mode identification is usually based on patterns recognized from the list of
adiabatic frequencies $\omega_{nlm}^{\rm theo}$, predicted from the
perturbation of 1D equilibrium models as outlined in Sec.\,\ref{section-nrp}.
Easily interpretable patterns concern those due to rotational splitting or
corresponding to the predictions from the asymptotic theory, as outlined in
Sec.\,\ref{section-nrp}.  Comparison between predicted and detected patterns
such as
those shown in Figs.\,\ref{16cyg} -- \ref{rossby} can then be fed
with the ``wisdom'' of the applicant to identify ($n,l,m$).  This wording
already indicates why asteroseismologists tend to be ``Bayesianminded'' when
identifying modes and performing asteroseismic modeling (see
\citet{Gruberbauer2012}, \citet{Bazot2012}, \citet{Appourchaux2014}, and
\citet{Aerts2018a} for thorough discussions).  Nevertheless, MLE and model
selection with so-called noninformative (flat) priors is often enlightening and
sometimes necessary to avoid too much prejudice in the prior, particularly on
the appropriateness of the equilibrium models used to compute the mode
predictions.

Despite our inability to predict reliably which of the eigenmodes should get
excited to observable amplitudes, the mode excitation mechanisms are understood
in general terms for the classes in Fig.\,\ref{hrd}.  Therefore, even though the
nonadiabatic treatment of the oscillations is not sufficiently established to
derive perfect mode excitation and amplitude predictions, it is still
instructive to consider the regimes of mode excitation before tackling the task
of mode identification and asteroseismic modeling.

\subsection{\label{excitation} Excitation mechanisms}

\subsubsection{Heat mechanisms and stochastic driving}

Thus far we have ignored the perturbation of the entropy in Eq.\,(\ref{Eq3}),
which greatly simplifies the theory of nonradial oscillations. However, to get
an understanding of mode excitation, nonadiabatic theory needs to be
considered. This is extensively discussed in Chaps.\,IV and V given by
\citet{Unno1989} and also in \citet[][Sec.\,3.7]{Aerts2010}, which addressed 
the general problem and the development of the so-called
quasiadiabatic approximation.  The prediction for a mode to get excited relies
on the computation of its growth rate. This quantity is positive for modes that
get excited (or modes that are unstable, as is often used as terminology), while
it is negative for modes that are overdamped.  Derivation of the growth rate of
a mode requires the computation of the imaginary part of its eigenfrequency
\citep[see Eq.\,(3.282) given by ][]{Aerts2010}. When considering the theoretical
expression, one finds that excitation occurs whenever the compression of the gas
and its heating happen in phase with each other. This is completely in line with
the operation of a thermodynamical heat engine.

The perturbations to both the flux (radiative + convective,
$\boldF=\boldF_{\rm rad}+\boldF_{\rm conv}$) and the energy generation stemming
from Eq.\,(\ref{Eq3}) go into the overall expression for the heat. For each of
these three contributions, one adopts a specific terminology.  When the
perturbation of the energy generation is dominant in the heat term that sets the
imaginary part of the mode frequency [see Eq.\,(\ref{Eq3})] the
$\varepsilon\,$mechanism may operate. This can obviously happen only in the deep
stellar interior. In the case in which the radiative flux delivers most of the
heat, it is often due to the increased opacity that acts as the heat engine in
the thin partial ionization layers in the envelope of the star.  This driving of
oscillations is therefore often called the $\kappa\,$mechanism, giving rise to
self-excited modes with infinite lifetimes.  Theoretical predictions of
nonradial mode excitation via the $\kappa\,$mechanism in intermediate-mass stars
along the main sequence are generally good. \citet{Pamyatnykh1999},
\citet{Bouabid2013}, and \citet{Szewczuk2017} conducted extensive studies. Yet
they are not perfect: we observe more modes than predicted for OB-type
pulsators, particularly in the g-mode regime. Bringing theory and observations
into agreement requires either higher-than-standard opacities in the partial
ionization zones of ironlike species situated in the layers with temperatures
$\sim\,2\times 10^5$\,K, as shown by \citet{Moravveji2016-opacity} and
\citet{Daszynska2017}, or higher metal abundances in the local region of the
excitation (for instance, as a consequence of atomic diffusion).  Similarly,
iron and nickel opacity enhancements are needed to explain the g~modes observed
in cool pulsating subdwarf B stars, as emphasized by \citet{Fontaine2003},
\citet{Jeffery2006}, and \citet{Bloemen2014}. Their hotter counterparts were
predicted theoretically by \citet{Charpinet1997} in terms of p~modes excited by
the $\kappa\,$mechanism at about the same time as, but independently of their
observational discovery by \citet{Kilkenny1997}. Opacity bumps due to carbon-
and oxygen in layers of $\sim\,10^6$\,K result in heat-driven mode excitation of
helium-rich subdwarfs \citep{SaioJeffery2019} and GW\,Vir variables (also known
as DO white dwarfs).  \citet{Corsico2019} provided a recent summary of pulsating
white dwarfs, and \citet{Montgomery2020} inferred nonstatic convection zones in
white dwarfs from observational limits on their mode coherence.

The perturbation to the convective flux and its contribution to the heat presents
a much larger challenge than the case of radiative flux, because it is
coupled to the properties of the turbulent pressure. For the deep stellar
interior, one may assume that this is time independent and well described by
mlt. However, for convective outer envelopes, the perturbations to the
convective flux and the turbulent pressure render the modes stable, such that
the heat-engine mechanism does not drive oscillations. Instead, excitation
occurs through stochastic forcing, where the energy in the acoustic noise in the
outer convection zone triggers some of the global eigenmodes. This stochastic
forcing happens in stars with an outer convective envelope,
leading to the excitation of damped and continuously reexcited oscillation
modes. In this case, predictions of the excitation and properties of the
modes are challenging due to the limited knowledge of the time-dependent
properties of $\boldF_{\rm conv}$ in the equilibrium models. This propogates
into theoretical uncertainty for the perturbation of the
time-variable convective envelope of a pulsating star.

In the limit of extremely long convective timescales relative to the periods of
the oscillations, the convective flux does not react to the pulsations and
convective flux blocking becomes an efficient excitation mechanism. This excites
g~modes in the thin convective envelopes of the $\gamma\,$Dor stars as shown by
\citet{Guzik2000} and further elaborated upon by \citet{Dupret2005}.
Gravitoinertial modes in the radiative envelope of such stars may couple
resonantly to inertial modes in the convective core of rapid rotators
\citep{Ouazzani2020}. On the other hand, \citet{LeeSaio2020} found mode
excitation due to resonant coupling between convective $g^-$ modes active in the
core of rapidly rotating 2\,M$_\odot$ stellar models and g~modes in the
radiative envelope, for frequencies $|m|\Omega_{\rm core}$.  For the
intermediate-mass $\delta\,$Sct and $\gamma\,$Dor stars, the time-dependence in
the pulsation-convection interaction is known as the problem of the red edge of
instability strip; see Fig.\,\ref{hrd}.  The convective timescales in the
thin outer convection zones of DA and DB white dwarfs are much shorter than
those of their g-mode pulsation periods, leading to mode excitation
\citep[e.g.,][]{GoldreichWu1999}.

The general case where the convective and mode timescales are similar is much
more challenging to treat in terms of stochastic mode excitation by the
turbulent pressure perturbation. This was developed by \citet{Houdek1999} and
\citet{Dziembowski2001}. Major improvements in excitation theory were
achieved for low-mass stars across stellar evolution by
\citet{Belkacem2008}, \citet{Dupret2009}, \citet{Belkacem2011,Belkacem2012}, and
\citet{Grosjean2014}; \citet{HoudekDupret2015} provided a good
summary. Despite this progress, considerable uncertainty in the predictions for
mode excitation and damping properties, as well as for the amplitudes, remain
due to uncertainties in the equilibrium structure of the superadiabatic outer
envelope and its coupling to the atmosphere.  Convection and nonadiabaticity
also affect the oscillation frequencies, and despite major efforts and progress
on this front for the Sun \citep[e.g.,][]{Houdek2017}, current theoretically
predicted frequency values are not yet at a level such that they can be fitted to
the observed frequencies.  This is known as the problem of the ``surface
effects'' in asteroseismology of stars with solarlike oscillations.

A rough global summary of the observed mode periods and amplitudes in pulsators
excited by heat mechanisms and stochastic driving for the classes indicated in
Fig.\,\ref{hrd} is provided in Table\,A.1 given by \citet{Aerts2010}. The periods
range from minutes to months. While this is already a broad range, at least
three more additional cases of mode excitation are in order.

\subsubsection{Nonlinear resonant mode excitation}

Many of the CoRoT and {\it Kepler\/} light curves reveal nonlinear
effects. Combination frequencies are omnipresent in the oscillation spectra of
$\kappa-$driven pulsators along the main sequence, as discussed by
\citet{Degroote2009}, \citet{Papics2012a}, \citet{Kurtz2015}, and
\citet{Bowman2016}.  Combination frequencies got lost in ground-based data as
they often have amplitudes below ppt.  They may be due to nonlinearities in the
light curves due to deviations from sinusoidal variations, because the modes
have amplitudes beyond the linear regime. However, given the density of g-mode
eigenfrequency spectra, combination frequencies may also occur at actual
eigenmode frequencies of the star that get excited by nonlinear resonant mode
coupling. A distinction between these two cases is not evident when
dealing with hundreds of frequencies deduced from a long-duration light curve.

Excitation of nonradial ``daughter'' modes via nonlinear ``parent'' mode
coupling is expected for particular low-order combination frequencies from
theoretical considerations based on the method of amplitude equations, as
developed by \citet{Buchler1984}, \citet{Goupil1994}, \citet{VanHoolst1994}, and
\citet{Buchler1997}. This mode excitation may give rise to time-variable mode
amplitudes, as commonly observed in space photometry of the higher-amplitude
nonradial pulsators.  Mainly due to a lack of proper data, theoretical predictions
on nonlinear mode excitation remained largely unexploited prior to space
asteroseismology, with the notable exception of g~modes in white dwarfs. For
those, the phenomenon of nonlinear mode coupling was already accessible from
ground-based data, thanks to their short pulsation periods. This allowed one to
assess the depth of the outer convection zones following \citet{Wu2001} and
\citet{Montgomery2005}. Meanwhile, nonlinear mode interactions were detected in
the short-cadence {\it Kepler\/} data in the cool pulsating DA white dwarfs
KIC\,4552982 \citep{Bell2015} and PG\,1149+057 \citep{Hermes2015}. Both of these DA
pulsators revealed large-amplitude regular outbursts with timescales of days,
which is much longer than the individual pulsation-mode periods.  This nonlinear
behavior in pulsating white dwarfs remained unknown prior to space
asteroseismology, even though large flux variations up to $\sim\!20\%$
recur. Nonlinear asteroseismology was also performed for the DB white-dwarf star
KIC\,8626021 \citep{Zong2016b} and for the pulsating subdwarf B star
KIC\,10139564 \citep{Zong2016a}, where rotation was found to be a key actor in
the detected resonances of the latter.

Nonlinear mode coupling behaves in diverse ways as predicted by theory and it is
a phenomenon occurring across the entire HRD. \citet{Weinberg2019} invoked
cascades of daughter modes resulting from nonlinear mixed-mode parents as an
explanation for the suppression of mixed-mode amplitudes observed in about a
quarter of the pulsating red giants. This study was triggered by the suggestion
made by \citet{Mosser2017a} that not all suppressed dipole modes found in
red-giant 
pulsators with this phenomenon can be explained by the magnetic greenhouse
effect originally proposed by \citet{Stello2016}. Indeed, some of the
stochastic dipole modes with depressed amplitudes are mixed modes with a g-mode
character in the stellar interior rather than p~modes, as assumed in 
theoretical developments by \citet{Fuller2015} and
\citet{Cantiello2016}. Although progress in interpretations based on a strong
internal magnetic field was achieved by \citep{Loi2020}, \citet{Loi2020-DeltaP},
and \citet{Bugnet2021}, the interpretation by \citet{Weinberg2019} does not
require core magnetism. This nonlinear theory is an alternative and
complementary explanation, which is in line with {\it Kepler\/} data of
intermediate-mass dwarfs revealing g~modes (SPB and $\gamma\,$Dor stars in
Fig.\,\ref{hrd}). Their period-spacing patterns would be affected by a strong
magnetic field, following the theory by \citet{Prat2019} and \citet{Prat2020}
and accompanying predictions by \citet{VanBeeck2020}. To date there has been no
observational evidence of internal magnetic fields from observed g-mode period
spacings of intermediate-mass stars revealed by \citet{VanReeth2015},
\citet{Papics2017}, \citet{GangLi2019a},
\citet{GangLi2019b}, and \citet{GangLi2020}.

This brings us to nonlinear nonradial asteroseismology for $\kappa$-driven
main-sequence stars from {\it Kepler\/} data. This underdeveloped research field
within asteroseismology holds great potential, but the theory is still to be
refined up to the level of the {\it Kepler\/} data. Unraveling nonlinear
effects in the light curves from nonlinear mode coupling via resonant excitation
is possible in principle for modes with an infinite lifetime, as the two lead to
distinguishable properties. A distinction between the two cases can be made
via the phase behavior of the combination frequencies of parent and daughter
modes. One expects phase locking to take place whenever low-order combination
frequencies occur exactly at another eigenfrequency of the star, such that the
latter gets excited by energy exchange between the two or more parent and
daughter modes involved in the resonance. Such phase locking was observed in the
CoRoT data of the large-amplitude $\beta\,$Cep star HD\,180642
\citep{Degroote2009}. Energy exchange due to resonantly coupled nonradial
g~modes was invoked as the cause of outbursts in pulsating Be stars, a
phenomenon first observed in a Be star by the CoRoT satellite \citep{Huat2009}
and later for several Be pulsators with the BRITE constellation
\citep[e.g.,][]{Baade2018}.  Detailed mode coupling studies are currently being
undertaken from {\it Kepler\/} long-cadence data of p- and g-mode pulsators
along the main sequence \citep[e.g.,][for a $\gamma\,$Dor pulsator]{Saio2018b}.
Given the observed amplitude and frequency modulations in numerous $\delta\,$Sct
and $\gamma\,$Dor stars \citep{Bowman2017}, the prospects for data-driven
nonlinear asteroseismology to be put into practice are excellent.  Even if mode
excitation by nonlinear resonances is not completely understood, one can take
the pragmatic approach of exploiting the detected combination frequencies
involved in resonance locking and, once identified, test to see whether adding them to the
list of identified pulsation modes used for asteroseismic inferences improves in
terms of precision of the inferred internal physics. A good target to test this
is the {\it Kepler\/} eclipsing binary KIC\,3230227 \citep{Guo2020}.

\subsubsection{Convectively driven internal gravity waves}

Traveling damped IGWs can be generated at the interfaces between convective and
radiative zones by turbulent convective flux forcing, as studied for the Sun by
\citep{Rogers2005} and for solar-type stars by \citet{Dintrans2005}.  The
pioneering study by \citet{Charbonnel2005} showed that inward traveling IGWs in
low-mass stars result in retrograde waves with the capacity to impose near-rigid
rotation on timescales much shorter than the evolutionary timescale.  For
intermediate- and high-mass stars, IGWs travel outward from the convective core
with a similar capacity \citep{Rogers2013}, which explains observed
asteroseismic rotation properties \citep{Rogers2015}.

\begin{figure}
\begin{center} 
\rotatebox{0}{\resizebox{7.cm}{!}{\includegraphics{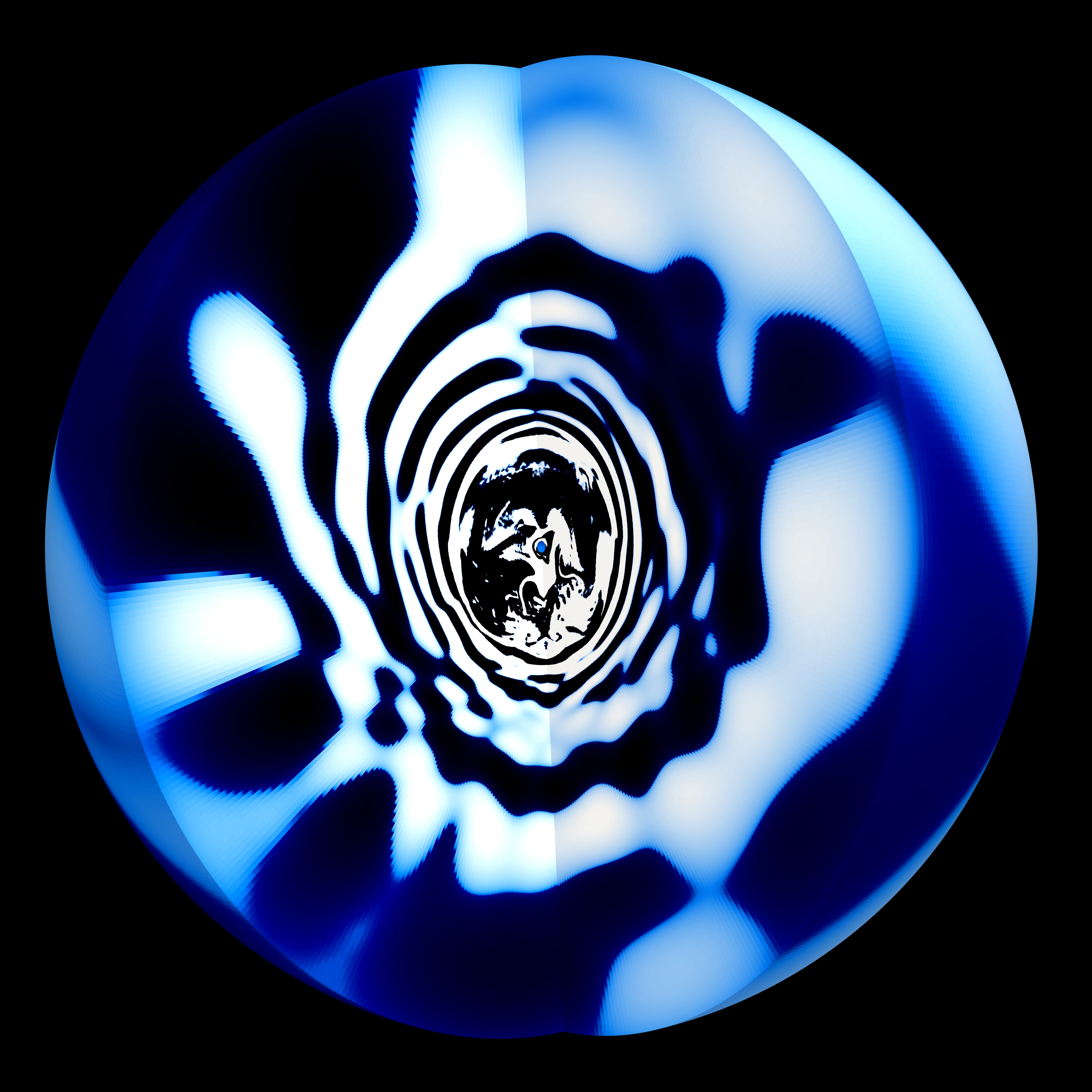}}}
\end{center}
\caption{\label{Philipp} Snapshot of 3D hydrodynamical simulations
  representing the temperature fluctuations induced by IGWs excited by stochastic
  forcing at the transition layer between the convective core and the bottom of
  the radiative envelope of a 3\,M$_\odot$ ZAMS star.  The color coding
represents fluctuations up to $10^5$\,K with respect to an
  equilibrium model.  Figure based on data in \citet{Edelmann2019}
  by courtesy of Philipp Edelmann, Newcastle University.}
\end{figure}

Although the 3D simulations of solarlike stars by \citet{Alvan2014,Alvan2015}
and of intermediate-mass stars by \citet{Edelmann2019} and \citet{Horst2020}
showed clear modal structure of internal g~modes, it remains unclear if and
which of the internal waves become resonant modes. This depends on the profile
of the wave's eigenfunction, on its propagation and dissipation properties, on
the efficiency of the radiative damping, and on the onset of nonlinearity
\citep[e.g.,][]{Rathish2019}. A snapshot of the temperature variations
accompanying the driving of IGWs and their propagation from the 3D hydrodynamical
simulations by \citet{Edelmann2019} is shown in Fig.\,\ref{Philipp}. This gives
the reader a grasp of the large-scale fluctuations induced in the stellar
interior.  The theoretical predictions coupled to the setup of the 3D
hydrodynamical simulations is subject of intense debate among various research
teams, because it is hard to drive IGWs from convective flows by heating. 
For this reason, most numerical simulations adopt artificial luminosity boosting in the
convective region to get the flows going into the radiative region at velocities
compliant with mlt.  The dependence of  IGW behavior on the level of
boosting remains to be studied in detail.  The fully compressible 2D simulations
by \citet{Horst2020} have lower numerical viscosity and a factor of 1000 lower
luminosity boosting than the 3D simulations by \citet{Edelmann2019}, yet
lead to similar results in terms of IGW properties. Moreover, these simulations
lead to appropriate predictions for p~modes and for SLF variability
in agreement with space observations of high-mass stars
\citep{Bowman2019a,Bowman2019b,Bowman2020}.

The overall spectra of IGWs can be triggered by convective cores, convective
envelopes or thin convection zones due to shell burning or opacity bumps in
radiative envelopes. These various cases of IGWs generation were extensively
discussed by \citet{Rogers2013}, \citet{Talon2008}, \citet{Fuller2014}, and
\citet{Cantiello2009}, respectively. Because of an inability to predict which of
the waves within the entire generated spectrum of IGWs could get excited as
resonant g~modes with observable amplitude, we are still far from pinpointing
their $\lambda_{lms}$ values from SLF detected in space photometry, as shown in
the right panels of Fig.\,\ref{RGB-BSG}.  Just as with the observed g~modes,
convectively triggered waves will occur mostly in the gravitoinertial regime
for the majority of main-sequence intermediate- and high-mass stars, because
these waves have spin parameters $s>1$ for the measured rotation rates of such
stars.  In view of this, theoretical and numerical studies should consider the
driving, propagation, and dissipation of stochastic gravitoinertial waves (GIWs)
in rotating stars \citep{Augustson2020}, rather than IGWs in nonrotating stars.
Synergies between GIWs predicted from 3D simulations and nonradial nonadiabatic
oscillation modes computed in the TAR from 1D stellar equilibrium models are yet
to be explored, starting with the observational constraints on the detected
frequency regimes, in the spirit of Fig.\,1 given by \citet{Aerts2019}. In that way,
one may develop asteroseismology based on the observed spectra of the GIWs.
\citet{Neiner2020} offer a step in this direction with their application
of GIW asteroseismology to the rapidly rotating pulsating Be star HD\,49330
observed by CoRoT.

\subsubsection{Tidal excitation of nonradial modes}

\begin{figure*}
\begin{center} 
\rotatebox{0}{\resizebox{17.cm}{!}{\includegraphics{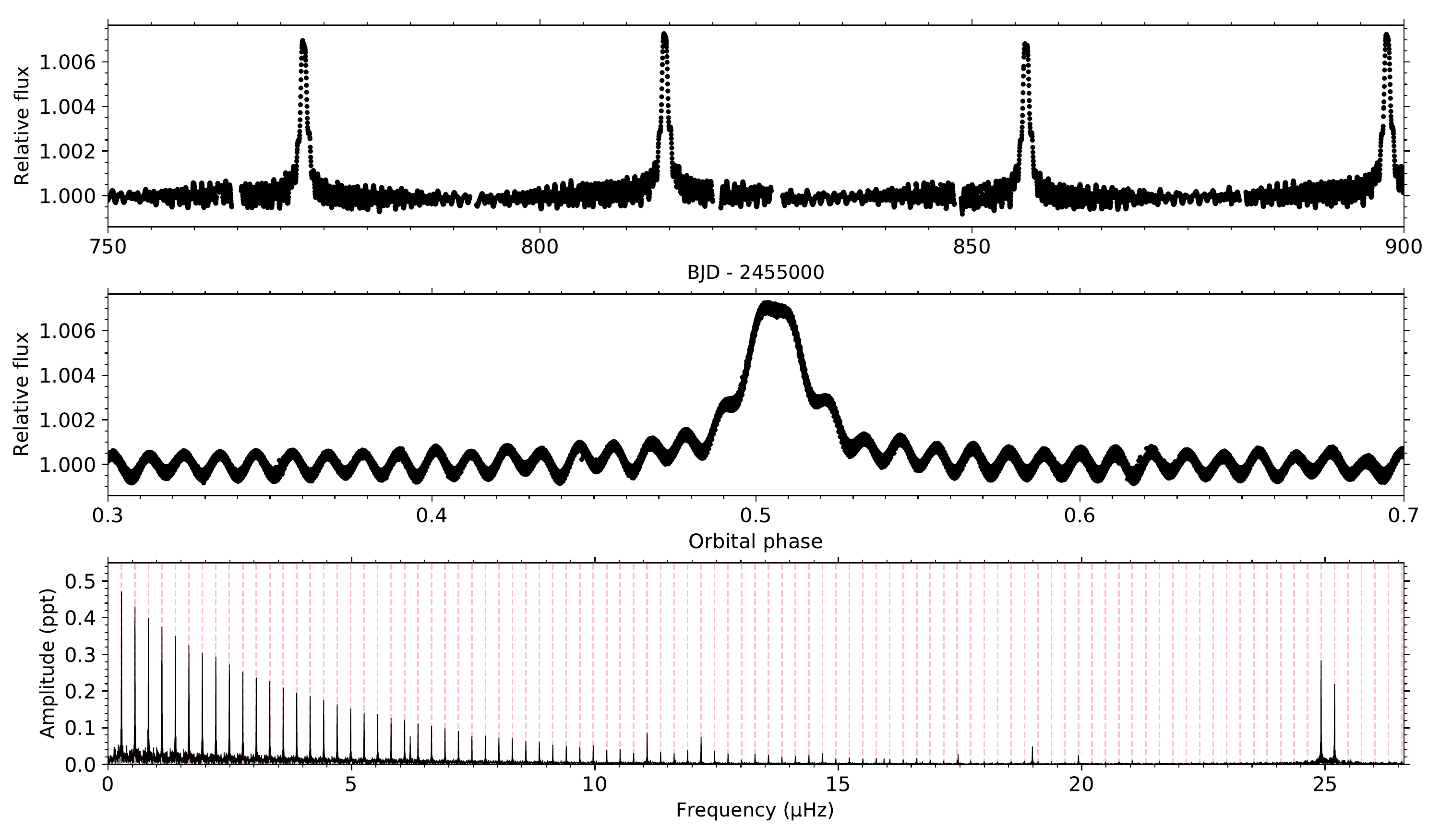}}}
\end{center}
\caption{\label{KOI-54} Excerpt of the {\it Kepler\/} light curve (top panel),
  phase folded according to the orbital frequency (middle panel), and LS amplitude
  spectrum (bottom panel) of the eccentric eclipsing binary KIC\,8112039. All
  but two frequencies (not indicated by a thin red vertical line) in the bottom
  panel are caused by tidal excitation.  Figure based on data in
  \citet{Welsh2011} by courtesy of Timothy Van Reeth, KU\,Leuven.}
\end{figure*}

The tidal action of a companion in a close binary is yet another way to excite
nonradial oscillation modes. This was realized by \citet{Cowling1941}
when he introduced ``his'' Cowling approximation. Tidally excited nonradial
oscillations and their effect on stellar evolution have been studied extensively
in the literature from a theoretical viewpoint, for various types of close
binaries; see \citet{Zahn1975}, \citet{Pap1997}, \citet{Savonije1997},
\citet{Terquem1998}, \citet{Witte1999}, \citet{Willems2003}, and
\citet{Fuller2011} for studies across the stellar mass range prior to the {\it
  Kepler\/} mission. \citet{Fuller2017a} covered the
case of eccentric binaries discovered from space photometry.  For this
excitation mechanism to work, the properties of the nonradial eigenmodes must be
``suitable'' compared to the period and the eccentricity of the binary
orbit. The component masses and radii must also be in the proper regime to
trigger nonradial modes by the tidal forces.  The tide-generating potential
within an eccentric binary is related to an infinite number of partial dynamic
tides with forcing frequencies and it is dominated by spherical harmonics of
degree $l=2$.  Whenever one of those forcing frequencies gets into resonance
with an eigenfrequency of a free oscillation mode of one of the components, the
tidal action exerted by the companion may excite this mode.

Tidally excited oscillations are expected to
occur at exact multiples of the orbital frequency. This makes them easy to spot
in the Fourier transforms of the light curves, particularly when they occur in
eclipsing binaries for which the orbital period is directly accessible from the
space photometry.  The occurrence of resonances between partial dynamic tides
and free oscillation modes is particularly relevant for the excitation of
g~modes, because of their similarity in period compared to orbital periods of
close binaries. Just as for the g~modes in single stars, CoRoT already allowed us
to discover tidally excited modes in binaries, but the 4-yr-long time base of
the {\it Kepler\/} mission implied the true beginnings of tidal
asteroseismology.  Numerous cases  have meanwhile been discovered and analyzed,
the most spectacular one shown in Fig.\,\ref{KOI-54} (another one is 
discussed in Sec.\,\ref{section-tides} and is shown in Fig.\,\ref{Zhao}).  The
stunning light curve of KIC\,8112039, also known as Kepler-Object-of-Interest
number 54 \citep[KOI-54,][]{Welsh2011} shown in Fig.\,\ref{KOI-54} was the first
object of a new class of high-eccentricity pulsating binaries whose {\it
  Kepler\/} light curve resembles the signal of a human heartbeat measured in a
cardiogram, hence this class was named ``heartbeat stars'' by
\citet{Thompson2012}. However, we prefer a naming based on the physical
properties, and hence refer to the class as high-eccentric binaries. KOI-54 has
more than 100 tidally excited g~modes, indicated by the red vertical lines
in Fig.\,\ref{KOI-54}.  Those with dominant amplitude occur at 90 and 91 times
the orbital frequency and are interpreted in terms of prograde sectoral
quadrupole modes excited by dynamical tides in a system where the rotation axis
of the primary star is almost aligned with the orbital axis. These two
dominant modes are locked in resonance with the orbit. Such an interpretation
leads to the other g~modes being near-resonant quadrupole zonal modes, while the
two modes whose frequency is not a multiple of the orbital frequency are due to
three-mode nonlinear mode coupling \citep{Fuller2012}.
With the discovery of numerous close binaries revealing multiple oscillation
modes, tides are an important excitation mechanism for 
nonradial modes in binaries, while they also affect free oscillation modes that would
occur if the star were single \citep[e.g.,][]{Guo2020-Excitation}. The
interpretation of tidally excited or tidally affected nonradial modes is a
tedious job, because the tidal forces imply deformations of the mode cavities
relative to the cavities for single stars (cf.\ Fig.\,\ref{cavity}). This bare
fact has yet to be exploited in detail.

The opposite situation of having fewer modes than expected also occurs and may be
connected to binarity.  Evidence for suppression or even absence of solarlike
oscillations in low-mass stars with convective envelopes by stellar companions
was found \citep[e.g.][]{Derekas2011,Gaulme2016,Schonhut2020}. Systematic
large-scale observational studies with spectroscopy, interferometry, or adaptive
optics are required to deduce the cause(s) of the absence of expected
oscillations in (unknown spectroscopic) binaries, e.g., to discriminate between
tidal changes of the mode cavities versus dilution of oscillation amplitudes due
to contaminating flux from visual companions.  Evidence for damping of solarlike
oscillations or amplitude suppression due to magnetic activity has also piled up
for both single and binary low-mass stars \citep{Mathur2019}.

As a general conclusion, one cannot rely on current theory to deliver a complete list of
unstable mode frequencies $\omega_{nlm}^{\rm theo}$ to be compared with the
observed ones to identify the mode wave numbers belonging to each detected
frequency.  We observe more  eigenmodes than predicted by the current
nonadiabatic nonradial oscillation theory, particularly in the g-mode
regime. This observational fact points to current limitations in nonadiabatic
mode excitation and damping computations, due to missing opacity, 
to overinterpretation of radiative damping,  or to yet other unknown physical
phenomena in the outer envelopes of stars. It is therefore premature to rely on
nonadiabatic predictions of mode excitation when performing asteroseismic
modeling.  Instead, the mode excitation predictions should be used as a good but
not perfect guideline of the modes to be expected from the current knowledge
of input physics in equilibrium models while being aware that modes predicted
not to be excited do occur in real stars and the other way around. Once the best
stellar model has been found from adiabatic asteroseismic modeling, one can
check its nonadiabatic mode excitation and damping predictions and use those
modes that are predicted to be stable yet observed as an excellent guide to
improve the input physics of the models and the excitation theory. 

\subsection{\label{MI}Mode identification}

Inferences on stellar interiors from asteroseismology provide tighter
constraints the more oscillation modes are involved in the modeling. For this
reason, asteroseismic modeling takes a pragmatic data-driven approach: we
thankfully use all detected frequencies offered by the stars to work with as
long as we can label their wave numbers $(l,m,n)$ or $(k,m,n)$ from adiabatic
eigenfrequency predictions. When uncertainty in the labeling occurs, the
frequency can still be used but the best equilibrium model selection should be
done in a Bayesian way, encapsulating the uncertain mode identification in the
prior(s).  Thanks to space photometry, identification of $(l,m)$ can often be
achieved from pattern recognition, notably when rotational multiplets as in
Eq.\,(\ref{LedouxSplitting}) are detected. We now highlight a few of the
current methods to deduce the mode identification. These depend on the kind of
pulsator and type of mode(s).

It is noteworthy that asteroseismic modeling to estimate stellar properties
other than rotation is usually done from axisymmetric ($m=0$) modes.  These tend
to be available in low-mass stars with stochastically-excited modes. Such modes
are often not detected in the case for p-, g-, or r-modes in heat-driven
pulsators.  Moderate to fast rotators tend to reveal mostly prograde or
retrograde modes with $m\neq 0$ in the data. For such cases, identification of
$m$ is also needed, aside from labeling $n$ and $l$ (or $k$ in the case of
Rossby modes).

\subsubsection{Mode identification from \'echelle diagrams}

For low-mass stars with stochastically-excited modes, as those shown in
Fig.\,\ref{16cyg}, one uses so-called \'echelle diagrams to identify the $l$-
and $n$-values of the modes.  An \'echelle diagram is a plot of the detected
mode frequencies as a function of the frequencies modulo the large frequency
separation as given in Eqs.\,(\ref{freqsep}) and (\ref{Deltanu}) and readily
accessible from the PD as illustrated in Fig.\,\ref{16cyg}.  In practice, an
observed PD spectrum is cut into segments of length $\Delta\nu$ and these
segments are subsequently stacked on top of each other to make a 2D map of $\nu$
vs $\nu$mod$(\Delta\nu)$. When doing that, modes of the same degree $l$
``line up'' along quasivertical ridges. This was found to be a convenient way
to represent and identify the solar oscillation frequencies by
\citet{Grec1980}, who introduced the terminology of \'echelle (French for
``ladder'') diagram. \'Echelle diagrams are commonly used ever since to identify
$l$ and $n$ in low-mass pulsators with stochastically excited modes \citep[we
refer to Fig.\,2 given by ][for colorful examples from {\it Kepler\/}
data]{ChaplinMiglio2013} and recently also for p~modes of young $\delta\,$Sct
stars \citep{Bedding2020}.

A computationally convenient way to identify modes after derivation of the large
and small frequency separation in the case of noisy data was developed by
\citet{Roxburgh2006}.  This method relies on the autocorrelation function (ACF)
and allows one to deduce the diagnostics $\Delta\nu$ and $\delta\nu$ without
being capable to derive the individual mode frequencies. The ACF is defined as
the Fourier spectrum of a filtered Fourier transform of the time series, where
the choice of the filter function can be optimized according to the envelope of
the observed signal in the PD spectrum as shown in Fig.\,\ref{16cyg}. 
\citet{MosserAppourchaux2009} provided a formal definition of the ACF
and additional details. The ACF method to derive the large and small
frequency separations as a way to achieve the mode identification is
efficient, relies on the physical properties of the wave behavior in the
mode cavities, and allows one to suppress disturbing effects of the noise in the PD
spectrum. This is why the ACF is currently being used in frequency analysis
pipelines, although \'echelle diagrams remain visually attractive and
insightful.

Any departure from the asymptotic relation given by Eq.\,(\ref{freqsep}),such as
considering the lowest frequency regime of p~modes, will introduce
curvature in the \'echelle diagrams. This curvature is also found as the star
evolves and mixed modes occur, creating ``bumps'' in the \'echelle
diagram. These phenomena are effectively illustrated in Fig.\,13 given by 
\citet{GarciaBallot2019}.  For an enlightening discussion of a doubtful
identification for the CoRoT F-type pulsator HD\,49333 due to mode bumping, and
how to treat this ``doubt'' in the context of a Bayesian prior, see 
\citet{Appourchaux2014}.

\subsubsection{Mode identification from rotationally split multiplets}

Rotational splitting following Eq.\,(\ref{LedouxSplitting}) gives rise to
multiplet structures in the data: $l=1$ triplets, $l=2$ quintuplets, etc. This
has long been known from ground-based data of self-excited modes and is also
observed from space photometry, e.g.\,\citet{Kurtz2014}, \citet{Papics2014}, and
\citet{Reed2014}. The detection of complete multiplets with an odd ($2l+1$)
number of components as in these examples immediately reveals the $l$ and $m$
values of the modes from the Fourier transform of the data. Rotational splitting
also gives information on $\Omega (r)$, as discussed by \citet{Aerts2019}, which
includes illustrations based on {\it Kepler\/} data.

\subsubsection{Mode identification from period-spacing patterns}

Period spacings $\Delta P$ for low-degree zonal gravitoinertial modes
of main-sequence F-type stars typically range from 2000 to 4000\,s for
dipole modes and from 1000 to 2500\,s for quadrupole modes. These ranges
are obtained when varying the mass (from 1.3 to 2.0\,M$_\odot$), age,
metallicity, and mixing properties of models in appropriate regimes according to
the observed mode frequencies \citep{VanReeth2016}.  Main-sequence B-type g-mode
pulsators, on the other hand, reveal a much broader range covering roughly
$\Delta P \in [1000,15000]$\,s \citep{Papics2017,Szewczuk2017}. These stars
have masses between 3 and 10\,M$_\odot$.

Gravity-mode period-spacing patterns such as those shown in Figs.\,\ref{gdor},
\ref{rossby}, and \ref{yanai} immediately reveal the sign of $m$. Indeed,
prograde modes have a $(P, \Delta P)$ pattern with a downward trend
(Fig.\,\ref{gdor}), while the pattern of retrograde modes reveal an upward trend
(Figs.\,\ref{rossby} and \ref{yanai}). The slope of these patterns allows
estimation of the rotation frequency in the region where the g-mode kernels
$K_{nl}(r)$ are dominant. As shown in Fig.\,\ref{eigenfunctions}, the kernels of
high-order g~modes are sharply peaked near the convective core. Hence, observed
gravitoinertial g~modes, Rossby modes, and Yanai modes allow
$\Omega_{\rm core}$ to be assessed from the slope of the period spacing
patterns by exploiting the relationship between the observed series of
$\lambda_{lms}$ and $\Pi_0$, $l$ (or $k$ for Rossby or Yanai modes), and
$m$. Indeed, $\lambda_{lms}$ depends on the spin parameter and the value of the
asymptotic period spacing, as revealed by Eq.\,(\ref{spacing-TAR}). Slightly
different methods to identify the mode numbers $l$ (or $k$) and $m$, along with
estimation of $\Omega_{\rm core}$, were developed by \citet{VanReeth2016} based
on model grids with varying $M$, $Z$, $X_{\rm ini}$, and $D_{\rm ov}$ and by
\citet{Ouazzani2017} and \citet{Christophe2018} based on stretching the
$\Delta P$ patterns to obtain $\Pi_0$. This stretching is done by searching the
value of $\lambda$ such that $\sqrt{\lambda} P_{l m s}^{\rm co}$ are equally
spaced by $\Pi_0$. Both methods give excellent agreement on the estimation of
$\Omega_{\rm core}$ \citep[][Fig.\,5]{Ouazzani2019}.  \citet{Takata2020} came up
with yet another tool for mode identification. It is based on a diagram in which
the frequency is plotted against the square root of the frequency. This allows one
to identify prograde sectoral modes and deduce at once the average rotation rate
and $\Pi_0$, which is in line with the numerical method by \citet{VanReeth2016}
that delivers mode identification along with estimation of $\Omega_{\rm core}$ and
$\Pi_0$.

As discussed earlier, Rossby modes are always retrograde in the inertial
reference frame of an observer. These modes occur at similar radial order
\citep[$n$ typically between -10 and -80,][]{GangLi2020} but have higher spins
\citep[][values between 15 and 30]{Aerts2017} than gravitoinertial g~modes
($n$ roughly between -10 and -100 and spins between 1 and 15). \citet{Saio2018a}
studied the observational appearances of even and odd Rossby
modes by computing their visibilities. It was found that the amplitude
distributions of odd ($k=-1$) modes are located at lower frequencies than those
of even ($k=-2$) modes for any given $m$ and that the amplitudes decrease
strongly as $m$ increases; see their Fig.\,4. These theoretical predictions
offer a good way to identify the wave numbers $(k,m)$ for these
modes. 

\subsection{\label{modeling}Asteroseismic modeling using mode 
frequencies}

\subsubsection{Some modeling preliminaries}
Seeking agreement between the detected identified oscillation mode frequencies
$\omega_{nlm}^{{\rm obs},i}\pm\sigma_{\omega_{nlm}^{\rm obs,i}}$ and those
predicted by equilibrium models $\omega_{nlm}^{{\rm theo},i}$ for
$i=1,\ldots,N_\omega$, with $N_\omega$ the number of detected identified
oscillation frequencies, constitutes a multivariate (nonlinear) regression
problem.  Fitting these identified frequencies can generally be done with or
without the addition of other seismic diagnostics (such as mean frequency
separations, frequency ratios, or other combinations for particular modes) or
by adding other observables into the
fitting process [$T_{\rm eff}$, $\log\,g$,
$\log\,(L/L_\odot)$, an interferometrically deduced $R_\star$, a dynamical
binary component mass $M_\star$, etc.]. 
In general, we consider an observed vector $\BY^{\rm obs}$
consisting of $i=1,\ldots,N_\omega+M$ components $Y^{\rm obs}_i$ derived from 
$N_\omega$
observed and identified oscillation frequencies and $M\geq\,0$ additional observational
constraints.  Comparison of $\BY^{\rm obs}$ with the corresponding
$\BY^{\rm theo}$ predicted by model computations is an extremely powerful method
to determine the interior and global properties of stars, including their
rotation, mixing, and composition profiles as well as their mass, radius, bulk
metallicity, and age. Nonradial oscillations occur in different types of
stars in almost all phases of stellar evolution; see Fig.\,\ref{hrd}. This,
along with the availability of long-duration high-precision space photometry,
has turned the potential of an asteroseismic calibration of the theory of
stellar structure and evolution into a reality. The level of sophistication
adopted for asteroseismic modeling is highly variable. Here, we summarize
methodology that can handle the challenging case of pulsating stars having a
convective core and rotating up to considerable fraction of their critical rate.

In our description of asteroseismic modeling via regression, we follow the
notations and concepts given by \citet{Aerts2018a};  i.e., we denote
equilibrium models generically as $\calm(\bftheta,\bfpsi)$, where $\bftheta$
stands for the vector containing the 
free parameters to be estimated for the fixed choices of the
input physics $\bfpsi$ (i.e., frozen microscopic and macroscopic
input physics).  The goal is to fit as closely as possible the observed and identified
oscillation frequencies and other observables by theoretical values derived from
the 3D perturbation of $\calm(\bftheta,\bfpsi)$. We keep in mind the following important
aspects:
\begin{enumerate}
\item
Theoretically predicted oscillation mode frequencies have uncertainties 
$\sim\!0.001$\,d$^{-1}$ ($\sim\!0.01\mu$Hz) due to limitations in our knowledge of
physics and due to numerical implementations. The observed oscillation
frequencies from space asteroseismology are typically one to several orders of magnitude more precise
than the theoretical predictions \citep[see Table\,1 in][]{Aerts2019}.
\item
The components of $\BY^{\rm theo}$ are strongly correlated.
\item
The components of $\BY^{\rm theo}$ and of $\BY^{\rm obs}$ 
may have very different variances, i.e., 
heteroscedasticity has to be included in the formalism.
\item
The components of $\bftheta$ may also be strongly correlated.
\end{enumerate}
These four properties result in a challenging modeling problem.  
Indeed, the theoretical uncertainties stemming from limitations in $\bfpsi$ and from
numerical implementation to compute the equilibrium models dominate over the
measurement uncertainties. This fact is often ignored in the modeling procedure.
Moreover, the correlated nature of the fitting problem implies that the
uncertainty regions for the parameters in $\bftheta$ tend to be of
multidimensional elongated shape, and therefore hard to interpret. We
provide a mathematical scheme that takes into account these challenges.

The limitations  of the
equilibrium models are due to restrictions to nonrotating 1D models,
missing atomic physics, poor opacities, imperfect numerical schemes to
solve the differential equations, etc.  Some of the distributions of systematic
uncertainties in the theoretical predictions of oscillation frequencies are
shown in Figs.\,2 to 10 given by \citet{Aerts2018a} for the case of g~modes in stars
with a convective core. An assessment of some of the systematic biases in the
case of solarlike oscillations was provided by \citet{Gruberbauer2013}.  From the
viewpoint of improving stellar structure and evolution theory, the aim is to
select the most likely physical model $\calm(\bftheta,\bfpsi)$ from an unbiased
sample of stars without introducing {\it a priori\/} bias by restricting too
narrowly the choice of the input physics $\bfpsi$.

\begin{figure*}
\begin{center} 
\rotatebox{0}{\resizebox{16.cm}{!}{\includegraphics{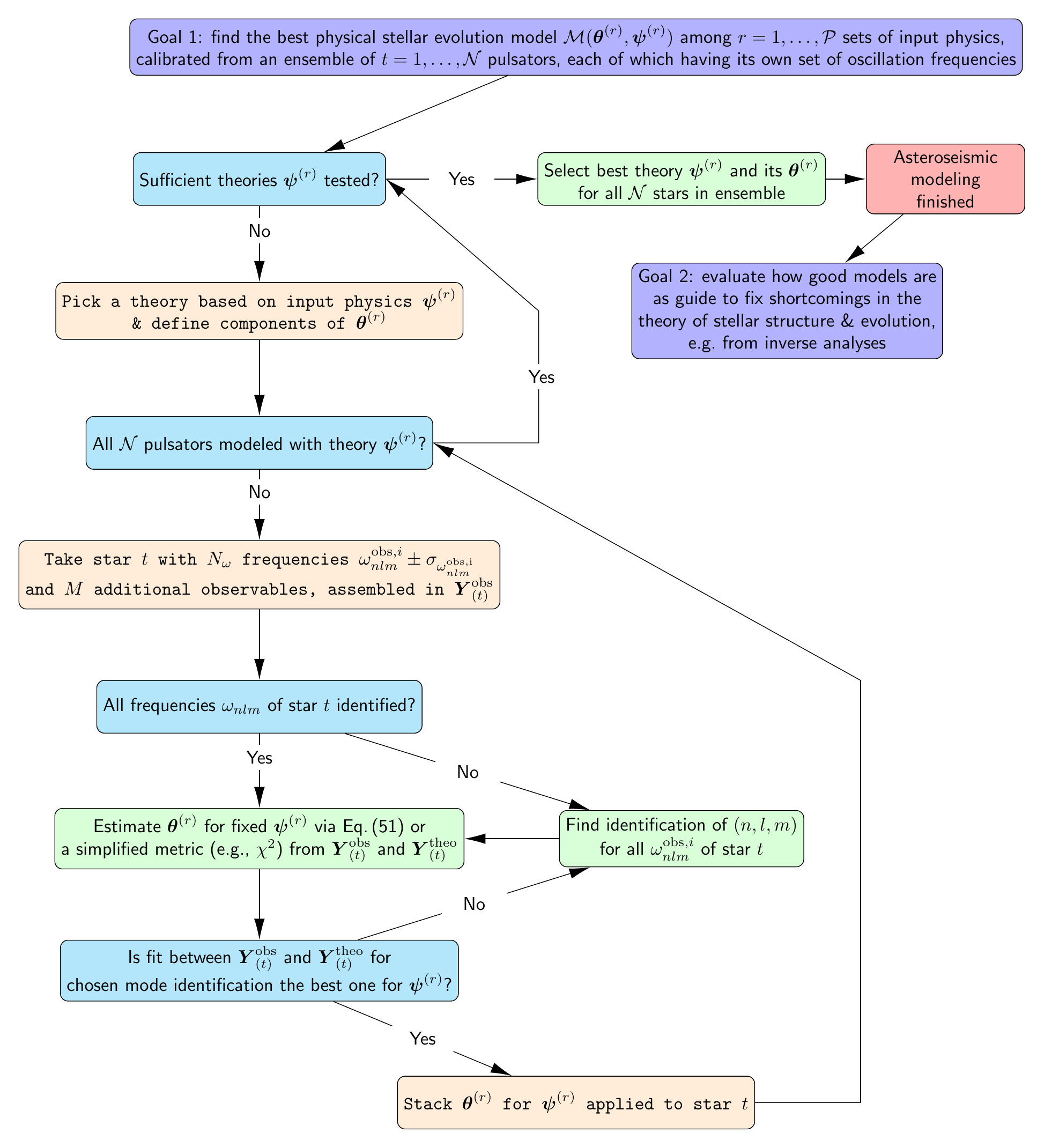}}}
\end{center}
\caption{\label{flowcharts} Schematic representation of the procedure of
  asteroseismic modeling for an ensemble of stars that summarizes the steps
  discussed in Sec.\,\ref{modeling}. See the text 
for the meaning of the notation. The green boxes involve
statistical methods in the topics of maximum likelihood estimation, pattern
recognition, and model selection. }
\end{figure*}

The general procedure of asteroseismic modeling of an ensemble of stars
is graphically depicted in the flowchart in Fig.\,\ref{flowcharts}. We discuss
this framework in the rest of this section, but it is not necessary to
digest the details of this flowchart to understand the
applications treated in Sec.\,\ref{section-applics}. 

\subsubsection{Setup of the modeling approach}

In the stellar modeling problem at hand, the choices of $\bftheta$ and $\bfpsi$
are different for stars with a radiative versus a convective core.  Moreover,
the probing power of p- and g~ modes is different; see 
Fig.\,\ref{eigenfunctions}.  Hence the choice of the parameters in the vector
$\bftheta$ to be estimated and the level of redundancy in the observables to be
used also differ; see \citet{Angelou2017} for an enlightening discussion on
this topic.
While the observed mode frequencies in slowly rotating stars can be
condensed into a few ``observables'' derived from asymptotic approximations as
in Eq.\,(\ref{freqsep}) or (\ref{periodspacing}), this is not the case for
moderate to fast rotators because their patterns depend on the interior rotation
and are different for different $l$ and $m$, as revealed by
Eq.\,(\ref{spacing-TAR}).  Even though a stretching method has been devised to
transform the observed period-spacing patterns due to g~modes into one
observable \citep{Christophe2018}, its value depends strongly on $D_{\rm ov}$
and $Z$ \citep{Mombarg2019}. One can therefore not assume that the
frequencies of g~modes can be transformed into simple ``summary'' diagnostics as
in the case of solarlike oscillations in slow rotators (where $\Delta\nu$ and
$\nu_{\rm max}$ play that role).  To treat asteroseismic modeling in general
terms, including applications to hybrid pulsators with both p and
g~modes, we consider a formulation for the matching of individual mode
properties. We denote it here for the frequencies, but it can be done for any
observable (e.g., period spacings).

Asteroseismic modeling of an individual star is done from an observed vector
$\BY^{\rm obs}$, which includes a set of $N_\omega$ observables
based on the observed frequencies $\omega_{nlm}^{{\rm obs},i}$, where each mode
has its own mode cavity, lifetime, and probing power for the interior physics;
see Fig.\,\ref{eigenfunctions}. Moreover, one often chooses to add $M$
nonseismic observables to the modeling, depending on their capacity to assess
the models and their correlation with respect to other components already
chosen for $\BY^{\rm obs}$.  In setting up the problem to solve, keep
in mind that several observables may be measured independently from each other
and occur without covariance, while they do not provide extra information about
the star. Singular value decomposition (SVD) methods, among them principal
component analysis (PCA), are therefore useful techniques to reduce the
dimensionality or to plan or not plan follow-up data once asteroseismic information
has been deduced. A SVD approach was introduced for helioseismic inversions by
\citet{JCD-Thompson1993}, while PCA applications for solarlike and g-mode
oscillations were done by \citet{Angelou2017} and \citet{Mombarg2019},
respectively.

The length of the vector $\BY^{\rm obs}$ is in general star specific and each of
its components is accompanied by its own error measurement
$\epsilon_i^{\rm obs}$, for which we assume normality, i.e.,
$\epsilon_i^{\rm obs}\sim {\cal N}(0,\sigma_i^2)$, as justified by
\citet{Gruberbauer2013}, \citet{Appourchaux2014}, and \citet{Aerts2018a}.  For
each pulsator in an ensemble of stars, the aim is to find the value for
$\bftheta$ that best predicts the observables $\BY^{\rm obs}$, with
corresponding value $\BY_0\equiv\BY_0(\bftheta_0,\bfpsi)$. We need to select
$\bftheta_0$ such that the distance between $\BY^{\rm obs}$ and $\BY^{\rm theo}$
is minimal for $\bftheta=\bftheta_0$, keeping in mind the correlated nature of
the parameters and the observables, as well as any systematic uncertainties
in the theoretical predictions.  In such a case, a natural merit function to
minimize for the estimation of $\bftheta$ is the Mahalanobis distance
\citep{JohnsonWichern2007}. This merit function represents a
generalized distance. It has been introduced in stellar variability
classification studies for CoRoT data \citep{Debosscher2007}.  In the current
context of asteroseismic modeling, it takes the form
\begin{equation}
\label{maldist}
\bftheta_0\!=\!\arg\!\min_{\bftheta}\!\left\{\!(\BY(\bftheta)\!-\!\BY^{\rm obs})^\top\!(V\!+\!\Sigma)^{-1}\!(\BY(\bftheta)\!-\!\BY^{\rm obs})\!\right\},
\end{equation}
where $V=\mbox{var}(\BY)$ is the variance-covariance matrix of the vector
$\BY (\bftheta,\bfpsi)$ and $\Sigma$ is the matrix with diagonal elements
$\sigma_i^2$ for $i=1, \ldots, N_\omega+M$.  The notation $X^\top$ stands for
the transpose of $X$.  The matrix $V$ can be estimated so as to capture the
variance for each of the components of the theoretically predicted vector
$\BY^{\rm theo}$, keeping in mind that the uncertainties due to the
limitations of the input physics $\bfpsi$ are much larger than the measurement
errors and that correlations among the vector components occur. The components
of $\BY^{\rm theo}$ must cover an appropriate range due to the free parameter
ranges of $\bftheta$. For these reasons, the matrix $V$ can be assessed from
grids of models $\calm(\bftheta,\bfpsi)$ covering a broad range of $\bftheta$
for various $\bfpsi$, as illustrated by \citet{Aerts2018b} for the case of
g~modes.  The Mahalanobis distance defined by Eq.\,(\ref{maldist}) provides a
more sophisticated merit function than the often used $\chi^2$ based on an
Euclidian distance, because it takes into account the variance-covariance
structure connected with $\BY^{\rm theo}$ and uncertainties stemming from the
limitations of $\bfpsi$. It also considers the overall correlated nature of
$\bftheta$, $\BY^{\rm theo}$, and their interconnection.

Asteroseismic modeling has thus far mostly been done from minimizing a $\chi^2$
merit function relying only on the measurement uncertainties
$\epsilon_i^{\rm obs}$.  This was improved upon by \citet{Gruberbauer2012} by
taking into account unknown systematic uncertainties of $\BY^{\rm theo}$ in a
Bayesian framework, using a $\chi^2$ formulation. The advantage of minimization
as in Eq.\,(\ref{maldist}) is that it allows for heteroscedasticity in and
correlation structures among the components of $\BY^{\rm theo}$.  Minimizations
by Eq.\,(\ref{maldist}) and $\chi^2$ were compared
by
\citet[][Table\,3 and Fig.\,12]{Aerts2018a} for a case of g-mode
asteroseismology of a SPB star, leading to a somewhat different best solution
for $\bftheta$.  Care must be taken when estimating $\bftheta_0$ due to
the correlated nature of $\BY^{\rm theo}$ and $\bftheta$, keeping in mind
systematic biases in the theory of stellar interiors and ensuring that $V$ is
of proper rank.

Estimation of the uncertainty regions for the components of $\bftheta_0$ is hard
to achieve if only one star is modeled, even if many identified frequencies and
high-precision classical observables are jointly included in $\BY^{\rm obs}$.
This is due to $\bftheta$ being of high dimension and containing correlated
vector components, as stressed by \citet{Angelou2017} and \citet{Aerts2018a}.
The multi-D error regions are usually elongated.  In such a case, inference on
the errors of $\bftheta_0$ is conveniently achieved from a MCMC
approach \citep[see ][for a popular tool used in
astronomy]{MCMC2013}.  For such applications, clever ways to sample are in
order to avoid getting stuck in too few local minima in the case of strong
covariances \citep{Handberg2011}. For this reason, nested sampling in a
Bayesian setting is often considered \citep{Corsaro2014}.  A practical MCMC
application to asteroseismology of $\alpha\,$Cen\,A was made by
\citet{Bazot2012}.

\subsubsection{Considering individual stars and ensembles}

A major challenge for low-mass stars with a convective outer envelope is to deal
with the 1D treatment of this envelope in the equilibrium models.  The outer
boundary condition adopted to compute the equilibrium model should come from a
proper 3D and time-dependent treatment of convection, while it is usually
simplified with time-independent mlt and a 1D atmosphere model. This
aspect of asteroseismic modeling for stars with damped modes excited by
turbulent convection is known as the problem of the ``unknown surface effects.''
Clever ways to deal with this, with the attitude of getting rid of the problem,
have been developed by using specific combinations of mode frequencies, such as
ratios of frequency separations \citep{Roxburgh2003}. These ratios were shown to
have probing power for the deeper layers of the star while being less sensitive
to the physics in the outer layers.  In the case of a star with a radiative
envelope, the challenge for the modeling is not so much the outer boundary, for
which the simple approximation of an Eddington gray atmosphere is fine, but
rather how to deal with the near-core boundary mixing and for the high-mass
stars also with mass loss due to a radiatively driven and possibly dynamical
wind (for $M>15\,$M$_\odot$).

In the case of solarlike oscillations in low-mass stars, the input physics
$\bfpsi$ of the models $\calm (\bftheta,\bfpsi)$ is often taken to be similar to
that of solar models calibrated from helioseismology. This is fine because
one can rely on the reasonable assumption that such stars adhere to similar
physics as the Sun, given that they are slow rotators with an extended
convective outer envelope. In that case, one can limit the estimation to the
minimal set of free parameters to compute the equilibrium models,
$\bftheta=(M_\star,X_{\rm ini},Y_{\rm ini},\tau)$. More sophisticated
applications based on machine-learning techniques treating higher dimensions,
e.g., by including $\alpha_{\rm mlt}$ and $D_{\rm mix}$, are done as well
\citep{Bellinger2016}.

For intermediate- and high-mass stars, $\bftheta$ is always more than
four dimensional due to non-negligible interior rotation, core overshooting, and
envelope mixing.  Even  when one can ignore rotation in the
computation of the oscillation modes, one deals with a higher-dimensional
problem compared to low-mass stars.  For pulsators with a convective core, the
MLE has to be done minimally with
$\bftheta=(M_\star,X_{\rm ini},Z_{\rm ini},D_{\rm ov},D_{\rm mix},\tau)$ in the
case where rotation can be ignored \citep{Moravveji2015}. When
dealing with gravitoinertial or Rossby modes, i.e., beyond the perturbative
treatment of rotation, an estimation of $\Omega(r)$ has to be included in
$\bftheta$, increasing dimensionality even further \citep[see ][for
examples]{Moravveji2016,VanReeth2016}.

The multidimensional uncertainty regions of $\bftheta_0$ are hard to determine
for such a complex modeling problem \citep[see ][for a detailed
discussion]{Johnston2019a}.  It is also challenging to discriminate among
candidate theories $\bfpsi$ from the modeling of only one or a few stars. This
is why applications optimally consider ensembles of stars. In that case, one has
the opportunity to derive the error regions for the individual members or to
consider one global average error estimate for each of the components of
$\bftheta$ for the entire population. This approach was applied by
\citet{SilvaAguirre2017} and \citet{Mombarg2019} to solarlike and $\gamma\,$Dor
pulsators, respectively.

As graphically shown in Fig.\,\ref{flowcharts}, an important aspect of the
ensemble modeling is to assess the quality of a collection of candidate
theories, i.e., to consider the following:
\begin{itemize}
\item observables $\BY^{\rm obs}_{(t)}$ for $t=1,\dots, {\cal Q}$ members of a
representative sample of pulsators, and
\item theories $\calm(\bftheta^{(r)},\bfpsi^{(r)})$, $r=1,\dots, {\cal P}$, each
  of which delivering predicted values $\BY^{\rm theo}_{(r)}$.
\end{itemize}
In such a setting, the goal is to select the most appropriate theory among the
${\cal P}$ candidate theories after applying Eq.\,(\ref{maldist}) or a
simplified version of it (e.g., $\chi^2$) to every star $t$.  This can be done
using a grid-based approach where extensive grids of models
$\calm(\bftheta^{(r)},\bfpsi^{(r)})$ are computed \citep[e.g.][]{Pedersen2021},
or from optimizations ``on the fly'' via a genetic algorithm approach
\citep{Metcalfe2014}, or via Bayesian methods coupled to MCMC
\citep{Bazot2012}. Akaike or Bayesian information criteria are proper
statistical tools to select the best physical model $\bfpsi$
\citep{Claeskens2008}. Care should always be taken to penalize for higher
degrees of freedom when doing the model selection, keeping in mind the dimension
of $\bftheta$.

The ultimate goal of ensemble asteroseismology is to have a pathway to improve
the input physics of the theoretical models (indicated as second goal in
Fig.\,\ref{flowcharts}).  Hence, once the best of the currently available model
sets $\bfpsi$ is chosen according to the first goal in the scheme in
Fig.\,\ref{flowcharts}, one should evaluate how good or bad it represents the
data in the details of each of the individual stars in the sample and for the
sample as a whole.  Stellar models can subsequently be improved, 
from inversion methods originally developed in the framework of
helioseismology \citep{Gough1985b}. Such methods are usually applied on a
star-by-star basis once the best 1D model for the appropriate
$\bftheta_0$ has been found \citep[see ][for a discussion of the
methodology]{BasuChaplin2017}.  Initial applications of this technique have 
led to the interior rotation profiles of six subgiants and young red
giants \citep{Deheuvels2014} and in core-helium-burning red giants
\citep{Deheuvels2015}. Detailed analyses resulting in profiles $\Omega(r)$ were
obtained for the SPB star KIC\,10526294 from g-mode triplets \citep{Triana2015},
the red-giant star KIC\,4448777 from dipole mixed modes \citep{DiMauro2016},
and the differential envelope rotation of 16\,Cyg\,A and B by
\citet{Bazot2019}. Inversion methods have led to an evaluation of the interior
structure for 16\,Cyg\,A and B, revealing discrepancies between the sound speed
in the cores of these two stars with respect to those in the 1D models at the
level of $\sim\!5\%$ \citep{Bellinger2017}, although this
binary-exoplanet system is the best calibrated solar analog
\citep{Davies2015}. Inversions applied to the exoplanet host star {\it
  Kepler}-444 (also known as KOI-3158) resulted in high-precision mass and
radius estimates of $M_\star=0.75\pm 0.03\,$M$_\odot$,
$R_\star=0.75\pm 0.01\,$R$_\odot$ and revealed that this star must have had a
convective core during the first 8 Gyr of its 11-Gyr lifetime \citep{Buldgen2019}.

%%%%%%%%%%%%%%%%
%%%%%%%%% Section 4 is applications per physical processes to improve
%%%%%%%%%%%%%%%%

\section{\label{section-applics}Applications of asteroseismic modeling}

At least four reviews and a book have recently been published on the topic
of asteroseismic applications based on {\it Kepler\/} or K2 data.  Low- or
intermediate-mass stars with an outer convective envelope reveal p~modes or
mixed modes.  The solarlike oscillations of these stars were reviewed
by \citet{ChaplinMiglio2013}, \citet{HekkerJCD2017},
\citet{GarciaBallot2019}, and a book on the data analysis
methodology was written by \citet{BasuChaplin2017}. We revisit some general results based on
solarlike oscillations, focusing on what asteroseismology of such stars can
deliver to other fields in astrophysics and on opportunities to improve the
theory of stellar interiors.  The evolution and g~modes of white dwarfs were 
recently reviewed by \citet{Corsico2019}, to which we refer for
asteroseismic modeling applications to the stellar remnants of low- and
intermediate-mass stars.  All of these reviews focused on ``fast'' modes,
i.e., high-frequency modes in the sense that their periods are much shorter than
the rotation period of the star. In such a case, the Coriolis force can be
ignored or treated using a perturbative approach.  For this reason, such
applications are relatively easy compared to cases where the rotation and
oscillation-mode periods are comparable, demanding a nonperturbative
treatment.

We start this section with the simplest applications of asteroseismology and
gradually increase the level of complexity, putting more emphasis on
applications that have been less summarized in reviews thus far. We 
focus here on ``convenience of use'' for the nonexpert while
highlighting selected striking results and opportunities to improve stellar
physics. The topics of Secs.\,IV.A--IV.F were chosen without any
attempt to be exhaustive because that would fill up an entire encyclopedia.

\subsection{\label{astero-lowM}
Sizing, weighing, and aging stars with convective envelopes}

The rotation of stars with $M \simkl 1.3\,$M$_\odot$ slows down efficiently,
as first reported by \citet{Skumanich1972} and studied from
space photometry \citep[e.g.,][]{Meibom2015}.  Although the details of their
rotational evolution are not yet fully understood \citep[e.g.][]{vanSaders2016},
this efficient slowdown is interpreted in terms of magnetic braking induced by
the dynamo created in their convective envelope and angular momentum loss via a
thin stellar wind.  These stars, as well as all evolved stars, have extended
convective envelopes that are the seeds of stochastic driving of modes by
turbulent convection.  \citet{Chaplin2014} provided a summary of the
asteroseismic properties of the ensemble of dwarfs and subgiants observed with
the {\it Kepler\/} spacecraft and found the modes occurring near the frequency of
maximum power in the PD spectra to have radial orders ranging from $n=17$ to 19
for dwarfs and from $n=15$ to 19 for subgiants.  For stars born with a radiative
core ($M \simkl 1.1\,$M$_\odot$) and similar metallicity as the Sun, the
transition from hydrogen-core to hydrogen-shell burning occurs near
$\nu_{\rm max}\simeq 2000\,\mu$Hz. This transition gradually shifts to lower
frequencies for larger and more massive stars.  The transition from core to
shell burning occurs at $\nu_{\rm max}\simeq 800\,\mu$Hz for stars near the
upper mass for which solarlike oscillations still occur
($M\simeq\,1.5\,$M$_\odot$).

Basic key ingredients of asteroseismic applications based on solarlike
oscillations are the global seismic scaling relations relying on solar
values. These were derived prior to space asteroseismology by
\citet{KjeldsenBedding1995}.  These scaling relations are based on the large
frequency separation defined in Eq.\,(\ref{Deltanu}) and the frequency of
maximum power $\nu_{\rm max}$ already discussed in Sec.\,\ref{section-intro}:
\begin{eqnarray}
\label{scalings}
\displaystyle{\frac{\Delta\nu}{\Delta\nu_\odot}} & \approx & 
\displaystyle{\left(\frac{M}{M_\odot}\right)^{1/2}\cdot
                                                             \left(\frac{R}{R_\odot}\right)^{-3/2}}\; , \\[2mm]
\displaystyle{\frac{\nu_{\rm max}}{\nu_{{\rm max},\odot}}} & \approx & 
\displaystyle{\left(\frac{M}{M_\odot}\right) \cdot
                                                                       \left(\frac{R}{R_\odot}\right)^{-2} \cdot
\left(\frac{T_{\rm eff}}{T_{{\rm eff},\odot}}\right)^{-1/2}}\; . \nonumber
\end{eqnarray}
In Eqs.\,\ref{scalings}), the solar reference values have to be computed using the
same methodology as for the star(s) under study to achieve meaningful and
consistent results. In their Table\,1, \citet{Pinsonneault2018} listed solar reference
values for various asteroseismic pipelines in use today.  It is seen in
Eqs.\,(\ref{scalings}) that the large frequency separation scales with the
square-root of the mean density of the star.  Because one relies on the mass,
radius, and oscillations of the Sun for a particular choice of input physics,
$\bfpsi_{\rm Sun}$, one has a quick and easy way to deduce the mass and
radius of the star under study. This type of stellar weighing and sizing implies
a major simplification: none of the steps in the procedure shown in
Fig.\,\ref{flowcharts} have to be taken, because one assumes that the input
physics to model the Sun, $\bfpsi_{\rm Sun}$, is also valid for the star(s)
under study and one does not test any models with other choices for the input
physics $\bfpsi$.  This does not allow one to perform model selection
among candidate theories $\bfpsi$, as one freezes the latter to the solar one
calibrated from helioseismology.  Moreover, by relying on the solar values
$\Delta\nu_\odot$ and $\nu_{{\rm max},\odot}$ one implicitly assumes that the
star under study has the same metallicity and chemical mixture as the Sun.  By
using the scaling relations in this way, there is no such thing as asteroseismic
modeling in the sense of Fig.\,\ref{flowcharts}.

\begin{figure}
\begin{center} 
\rotatebox{0}{\resizebox{8.5cm}{!}{\includegraphics{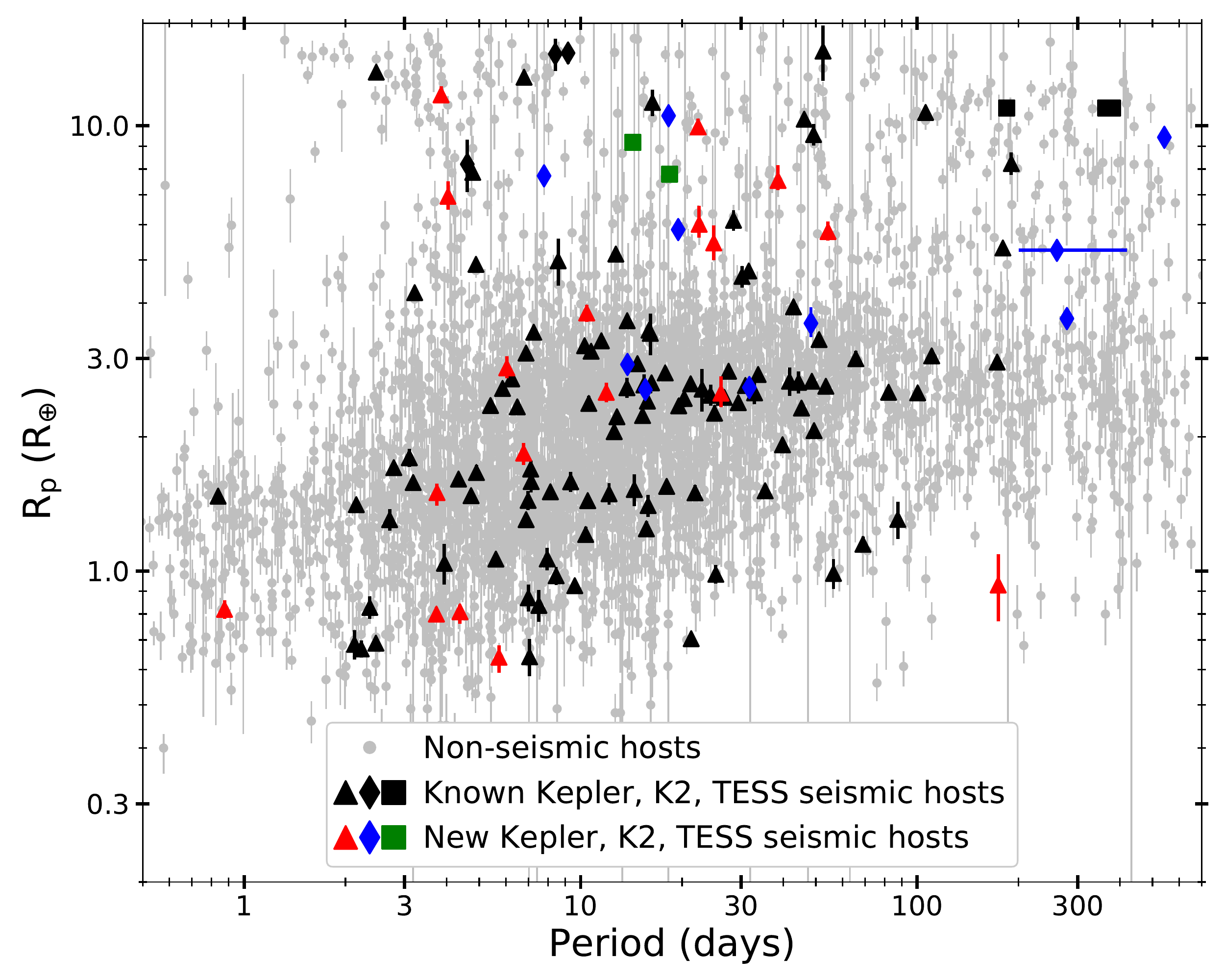}}}
\end{center}
\caption{\label{exoplanets} Planetary radii as a function of orbital period,
  where the properties of exoplanet host stars with and without asteroseismic
  estimation are compared. Asteroseismology of the host star not only provides
  the age of the exoplanetary system but also improves the planetary radii by 
  a factor of $\sim\!2$ compared to the case where such data are not available.
  Figure based on the sample study by \citet{Chontos2019} and Chontos et al.\
  (in preparation) by courtesy of Ashley Chontos, University of Hawaii.}
\end{figure}

For frozen $\bfpsi_{\rm Sun}$, it follows from the scaling relations in
Eq.\,(\ref{scalings}) that a measurement of $T_{\rm eff}$, $\Delta\nu$, and
$\nu_{\rm max}$ suffices to deduce the mass and radius of a star.  For such an
application, it is not even necessary to know the identification of the
individual p~modes, as long as one can estimate $\Delta\nu$ and
$\nu_{\rm max}$ from the PD spectrum, with or without the help from an ACF.
This has major applications, as asteroseismic $M_\star$ and $R_\star$
values can be computed easily for use in various fields of astrophysics, even
from short and/or gapped time series. For this reason, substantial effort has
been put into testing the scaling relations from independent methods,
notably from an interferometric radius, as done by
\citet{Huber2012} and \citet{White2013} or from an astrometric radius, as treated
by \citet{SilvaAguirre2012}, \citet{Huber2017}, and \citet{Zinn2019}. Overall
agreements are excellent for stars with a fairly large range in radii, from
about 0.8\,R$_\odot$ to above 10\,R$_\odot$: this is a tribute to the trio of
asteroseismology, astrometry, and interferometry.

Thanks to their simplicity, the scaling relations in Eq.\,(\ref{scalings}) have
been used extensively to deduce the masses and radii of stars with solarlike
oscillations observed with space photometry. Aside from the recent {\it
  Kepler\/} catalog papers on dwarfs and subgiants \citep{Chaplin2014}
and red giants \citep{Yu2018}, we refer the interested reader to earlier
CoRoT catalogs for red giants by \citet{Kallinger2010} and 
\citet{Mosser2010,Mosser2012}, as well as to the K2 catalog by
\citet{Stello2017}. The scaling relations were even expanded to the extremely low
frequencies of solarlike oscillations in M giants belonging to the class of
Semi-Regular Variables; see Fig.\,\ref{hrd}. This allowed researchers to assess and
interpret the period-luminosity relations derived from ground-based microlensing
surveys in terms of the interior structure of such highly evolved stars
\citep{Mosser2013-SR}.  Analyses of M giants observed with {\it Kepler\/}
led to the discovery of nonradial oscillations of low inertia in such stars,
interpreted as f~modes by \citet{Stello2014-SR}.  This discovery holds major
potential and has yet to be explored further. In particular, it paves the way to
perform extragalactic asteroseismology from observations of M giants in the
Magellanic Clouds.

As another key application, masses and radii for exoplanet hosts based on the
scaling relations have been published in a number of studies, among them
\citet{Huber2013}, \citet{VanEylen2014}, \citet{SilvaAguirre2015},
\citet{Lundkvist2016}, \citet{Campante2016b}, and \citet{VanEylen2018}.  A
radius estimate of an exoplanet host star from the scaling relations propagates
directly into a radius estimate of its exoplanets for which a transit has been
measured. Asteroseismic sizing from scaling relations is therefore particularly
convenient for exoplanet studies.  Once an exoplanet has been detected in space
photometry, the latter can be revisited in an attempt to measure its host
star's values of $\Delta\nu$ and $\nu_{\rm max}$ \citep{Chontos2019}.
Figure\,\ref{exoplanets} shows the planet radii versus orbital periods for an
assembly of {\it Kepler\/}, K2, and TESS exoplanets, comparing those with and
without asteroseismology of the host star.  As highlighted in the figure, more
measurements of $\Delta\nu$ and $\nu_{\rm max}$ from the PD spectra are achieved
as detection methods are refined and spectroscopy (to estimate $T_{\rm eff}$)
are assembled.  The gain in precision of the exoplanetary radius is typically a
factor of $\sim\!2$ when asteroseismic sizing of the host star from scaling
relations can be done relative to the case where no oscillations are detected.

Once the mass and radius of a star have been estimated from the scaling
relations, its age and hence its evolutionary stage can be assessed. This requires
evolutionary models and was originally done for the Sun.  \citet{JCD1988}
introduced the so-called CD diagram to estimate the age of solar-type stars from
their large and small frequency separations.  \citet{Chaplin2014} made a
thorough analysis to estimate the ages of the more than 500 dwarfs and subgiants
observed with {\it Kepler}. This ensemble analysis was based on six different
data analysis pipelines and 11 stellar model grids in order to assess the
combined effect of observational and model uncertainties by taking into account
$V+\Sigma$, as discussed after Eq.\,(\ref{maldist}).  This is an application of
Fig.\,\ref{flowcharts} where one does not use the individual frequencies
$\omega_{nlm}^{{\rm obs},i}\pm\sigma_{\omega_{nlm}^{\rm obs,i}}$ as input, but
rather
$\BY^{\rm obs}=(\nu_{\rm max}, \Delta\nu,\delta\nu, T_{\rm eff}, {\rm [Fe/H]})$
to estimate the parameters $\bftheta=(M,R,\tau)$ and quantities that can be
derived from these three (such as the mean density and gravity). For this
application, Chaplin et al.\ assumed that the components
of $\BY^{\rm theo}$ are not subject to uncertainties and are not correlated with
each other, such that $\chi^2$ can be used as a merit function.  Under these
assumptions, this ensemble modeling led to average relative precisions of
approximately 5.4\% in mass, 2.2\% in radius, and $\sim\!10\%--20\%$ in age.
Such relative precisions are within reach when spectroscopic estimates of
$T_{\rm eff}$ and ${\rm [Fe/H]}$ are available.  If only $\nu_{\rm max}$ and
$\Delta\nu$ are available, the relative precisions are downgraded by about a
factor of 2, which is still excellent and often the only way to get an age
estimate of isolated stars in the Milky Way.

\citet{Bellinger2019a,Bellinger2019b} derived scaling relations for the ages of
dwarfs, subgiants and red giants from
$\BY^{\rm obs}=(\nu_{\rm max}, \Delta\nu,\delta\nu, T_{\rm eff}, {\rm [Fe/H]})$.
For the dwarfs, the relations were deduced from fits to these quantities for 80
stars whose measurement uncertainties for $\delta\nu$ and $\nu_{\rm max}$ are
better than 10\% and 5\%, respectively.  This leads to age precision estimates
of about 10\% for dwarfs. These formulas are easy to use (e.g., for exoplanet
host aging), but users have to keep in mind that the relations explicitly rely
on the solar input physics via homology relations. Thus, the fact that the
interior rotation, mixing, and magnetism of the stars might be different than
those of the Sun is ignored.

A major breakthrough in asteroseismology was achieved upon the detection of
mixed dipole modes in {\it Kepler\/} data of evolved low-mass stars
\citep{Beck2011,Bedding2011} after they had been  theoretically predicted in the
context of CoRoT by \citet{Dupret2009}.  The mixed modes can have a
gravity-dominated or a pressure-dominated character, depending on the extent and
shape of their propagation cavity. Such dipole mixed modes in evolved stars
occur together with radial and quadrupole p~modes, which obey the asymptotic
relation in Eq.\,(\ref{Deltanu}) and probe the convective envelope of the star.
Measurement of $\Delta\nu_{nl}$ and $\Delta\,P_{nl}$ thus allows one to derive the
mass and radius of the star from scaling relations, as well as its evolutionary
stage \citep{Bedding2011}. Indeed, the gravity-dominated mixed modes probe the
deep stellar interior and have different values for hydrogen-shell-burning red
giants than for core-helium-burning red giants. This allows to deduce the
nuclear burning stage of these two types of red giants, while their surface
properties are the same.  Period spacings of dipole mixed modes lead to
higher-precision age estimates than the large frequency separation from
p~modes. Moreover, the period spacings are a sensitive probe for internal mixing
in intermediate-mass dwarfs and allow one to calibrate the core overshooting
efficiency using low-luminosity red-giant stars \citep{Hjorringgaard2017}. We
return to this capacity for g~modes in Sec.\,\ref{sec-mixing}.

Estimation of $R_\star$ from asteroseismology combined with a spectroscopic
measurement of $T_{\rm eff}$ allows one to deduce the luminosity of the star and
hence derive an asteroseismic parallax
\citep{SilvaAguirre2012}. Comparisons between such asteroseismic
parallax with the one from Gaia astrometry reveals excellent agreement for
dwarfs \citep[e.g.,][]{DeRidder2016} and red giants
\citep[e.g.][]{Huber2017}. This allows one to probe the deep end of the Milky Way
with luminous pulsating stars \citep{Mathur2016}.  The capacity of joint
asteroseismic aging, sizing, and distance estimation opened up the field of
galactic archaeology, which had already een jump-started prior to the Gaia era by
\citet{Miglio2009}. Their study of various populations of core-helium-burning
red giants in the galactic disk observed with CoRoT opened a new field of
mapping and dating stellar populations from red-giant asteroseismology
\citep{Miglio2013}. Meanwhile, extensive progress has been made in
archaeological studies for the pointing directions in the Milky Way covered by
CoRoT, {\it Kepler\/}, and K2 coupled with large spectroscopic surveys and/or Gaia
data; see \citet{Stello2015,Stello2017}, \citet{Anders2017a,Anders2017b},
\citet{Serenelli2017}, \citet{Pinsonneault2018}, \citet{SilvaAguirre2018},
\citet{Sahlholdt2018}, \citet{Rendle2019}, \citet{Zinn2019}, and
\citet{Sharma2019}.  Major potential for extending this topic toward all-sky
coverage is being offered by the ongoing TESS mission \citep{Ricker2016} and
the future PLATO mission \citep{Rauer2014}. These surveys should optimally be
coupled with spectroscopic surveys with multiobject spectrographs to target
hundreds of thousands of asteroseismically aged and sized red giants. The
beginnings of such large-scale asteroseismic archaeology have already
revealed abundances and distances that allowed one to separate high- and
low-[$\alpha$/Fe] populations in the Milky Way disk \citep{Chiappini2015}. In
this way, asteroseismology has become a key ingredient in the study of the
multiple populations and of the chemical evolution of our Milky Way.

To circumvent computationally intensive age derivation from evolutionary models,
age scaling relations were derived by \citet{Bellinger2019b}. These rely on the
asteroseismic properties of $\sim\!1000$ red giants and are convenient for
galactic archaeologists.  These age relations assume that stars adhere to
$\psi_{\rm Sun}$.  Their quoted precision of $\sim\!15\%$ does not take into
account systematic uncertainty due to the unknown evolutionary properties on the
main sequence. As highlighted by Fig.\,\ref{geneva}, uncalibrated descriptions
of internal mixing in dwarfs with a convective core are used in the
models. Moreover, asteroseismology revealed the theory of angular momentum
transport to be limited \citep{Aerts2019}; see also Fig.\,\ref{araa-update}.
As long as users of ``recipe-type'' aging recognize this
major culprit stemming from fixing the input physics, the scaling relations are
a convenient tool for initial asteroseismic and comparative stellar aging of
populations, preventing us from  having to go through {\it \`a la carte\/} modeling, a term
introduced by \citet{Lebreton2014} and \citet{Lebreton2014-EAS} that is represented by
Fig.\,\ref{flowcharts}.  To get the maximum precision out of the data
of a particular star, including aging to better than 10\%, detailed modeling
according to Fig.\,\ref{flowcharts} is in order. Treating
populations of ${\cal N}$ stars in this way is much more cumbersome than
applying age scaling relations, but it is the only way to properly take into
account the fact that the internal mixing of stars can be diverse (see Fig.\,\ref{geneva}),
even for a population of stars born with the same metallicity and similar
rotation; see Table\,\ref{coreM}.

\subsection{\label{sec-glitches}Assessing sharp features in stellar structure}

Fitting the frequencies or periods of oscillations for individual stars and
particularly for an ensemble of stars allows one to learn more about the quality of
the input physics $\bfpsi$ of stellar models following Fig.\,\ref{flowcharts}.
This can be done from fitting (some of) the individual detected and identified
oscillation frequencies and periods, rather than simply using the measured averages of
global patterns based on the asymptotic theory as in Eqs.\,(\ref{freqsep}) and
(\ref{Deltanu}) or Eqs.\,(\ref{periodspacing}) and (\ref{Pi0}).  An intermediate
step between the exploitation of only the average value of the frequency or
period spacing and the full-blown fitting of all  measured individual oscillation
modes is offered by modeling deviations from the expected constant spacings due
to so-called structural glitches.  Sharp features in the sound speed are
called acoustic glitches while those in the Brunt-V\"ais\"al\"a frequency are
termed buoyancy glitches. These glitches may lead to oscillatory deviations in
the patterns of p-mode frequencies or g-mode periods.  Interpretation of such
measured deviations goes beyond the simple use of scaling relations and provides
a good opportunity to derive detailed properties of stellar structure.

\begin{figure}[h!]
\begin{center} 
\rotatebox{270}{\resizebox{6.cm}{!}{\includegraphics{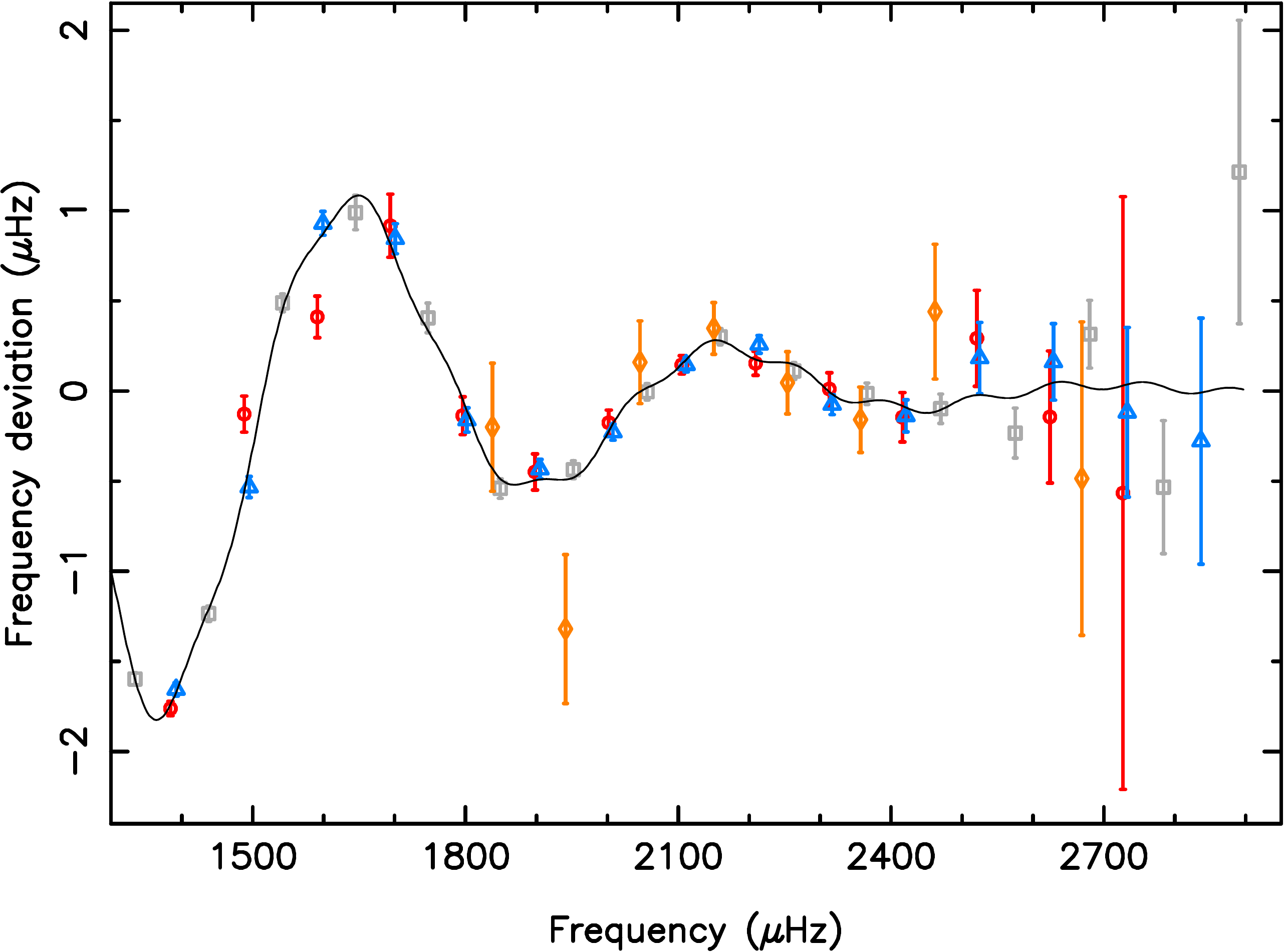}}}
\end{center}
\caption{\label{Kuldeep} Frequency deviations with respect to a fourth-order 
polynomial fit to the measured frequencies for the modes of degree $l=0$
  (blue triangles), $l=1$ (gray squares), $l=2$ (red circles), and $l=3$ (orange
  diamonds) of 16\,Cyg\,A, based on the PD shown in Fig.\,\ref{16cyg}.  The full
  line represents a fit to the oscillatory signal caused by sharp features in
  the star's sound speed in the second  ionization zone of helium. Figure produced from
  data in \citet{Verma2014} kindly made available in electronic format by
  Kuldeep Verma, Aarhus University.}
\end{figure}

Studies of acoustic glitches in the Sun from helioseismology led to the
overshoot properties at the base of the solar convective envelope
\citep{Monteiro1994} and to the capacity to position the second ionization zone
of helium \citep{Monteiro2005}. Following those solar studies, the potential of
exploiting measured oscillatory deviations due to acoustic glitches was
investigated further by \citet{Monteiro2000}, \citet{Basu2004}, and
\citet{Houdek2007} for Sun-like stars covering a mass range of
$M_\star\in [0.85,1.2]\,$M$_\odot$. This led to methods to infer the size of the
convective envelope and to derive properties of the overshoot transition layer
at the bottom of the convective envelope \citep[][for an extensive
discussion]{HekkerJCD2017}.  This methodology was put into practice for {\it
  Kepler\/} data of Sun-like stars by \citet{Mazumdar2014}.
Figure\,\ref{Kuldeep} illustrates the measured oscillatory frequency deviations
due to acoustic glitches in the structure of 16\,Cyg\,A, following its PD in
Fig.\,\ref{16cyg}.  In a series of papers exploiting the measured periodic
deviations from constant spacings, \citet{Verma2014,Verma2017} measured the
locations of the base of the convection zone and of the helium ionization zones
for Sun-like stars observed by {\it Kepler\/}, as well as their helium abundance
\citep{Verma2019a} and the level of helium settling due to atomic diffusion
\citep{Verma2019b}. In these studies, analytical modeling was compared to
numerical asteroseismic modeling following the scheme in Fig.\,\ref{flowcharts}
by estimating
$\bftheta = (M_\star,Y_{\rm ini},{\rm [Fe/H]},\alpha_{\rm MLT}, \alpha_{\rm
  ov})$ for equilibrium models with and without helium settling (i.e., for two
different $\bfpsi$'s in Fig.\,\ref{flowcharts}).  The results of such numerical
modeling for 16\,Cyg\,A are shown as a solid line in Fig.\,\ref{Kuldeep}. This led
to $Y_{\rm ini}\in [0.231,0.251]$, while a similar application to 16\,Cyg\,B
revealed $Y_{\rm ini}\in [0.218,0.266]$. The differences in helium mass
fraction $\Delta Y$ derived from models with and without helium settling were
found to be $\Delta\,Y\!\sim\!0.038$ for both stars.  This result is
representative for an additional $\sim\!30$ Sun-like stars analyzed in the same
way.  As such, frequency deviations due to acoustic glitches offer a unique way
to derive the helium composition of stars too cool to reveal helium spectral
lines.

For the solar analogs 16\,Cyg\,A and B, even more details on the
physics of the envelope could be derived, notably latitudinal differential
rotation due to a dynamo effect. In an updated
study based on the same light curve as the one used by \citet{Davies2015}, the
measured rotational splitting of 16\,Cyg\,A and B led to envelope rotation rates
that are higher at the equator than at the pole.  \citet{Bazot2019} found the
differences of the rotation frequencies between the equator and the pole to be
$320\pm 269$\,nHz and $440\pm 370$\,nHz for 16\,Cyg\,A and B, respectively,
while the equatorial rotation frequencies were $535\pm 75$nHz and
$565\pm 140$\,nHz. This envelope rotation behavior of both binary components is
similar to that of the Sun. The results on turbulence and rotation of this
solar-analog binary and exoplanet system illustrate how modeling of individual
frequencies provides an opportunity to improve the treatment of envelope
convection in cool stars.

Mode trapping due to transition zones with compositional changes is omnipresent
in stars at evolved stages. This phenomenon has already been detected from
ground-based white-dwarf asteroseismology
\citep[e.g.,][]{Winget1991,Winget1994}. It was found in large ensembles
of {\it Kepler\/} $\gamma\,$Dor and SPB stars analyzed by \citet{GangLi2020} and
\citet{Papics2017}, of red giants by \citet{Mosser2015} and of subdwarf B
pulsators by \citet{Reed2011}, \citet{Oestensen2014a}, \citet{Uzundag2017}, and
\citet{Kern2018}, while it was also confirmed in white dwarfs from K2 photometry
\citep{Hermes2017a}. The trapping of g~modes or mixed modes was theoretically
investigated for red giants by \citet{Cunha2015}, for subdwarfs by
\citet{Charpinet2000}, and much earlier for white dwarfs in the seminal paper by
\citet{Brassard1992}. Hence, modeling of mode trapping due to structural
glitches can now be done across the entire HRD \citep{Cunha2019b}.

During the main-sequence phase, buoyancy glitches occur due to the shrinking
convective core of intermediate- and high-mass dwarfs. These glitches lead to
deviations from the constant period spacing of high-order g~modes for stars with
limited chemical mixing in their radiative envelope. Such periodic deviations
were first observed in a SPB observed with CoRoT \citep{Degroote2010}.  The
signature of buoyancy glitches was also found in the period-spacing patterns of
$\gamma\,$Dor stars \citep{VanReeth2015}, as illustrated in Fig.\,\ref{gdor} for
one of them (KIC\,11721304). A sharp deviation from a period-spacing pattern can
also occur when a pure inertial mode in the convective core of a rapid rotator
couples resonantly to a heat-driven gravitoinertial mode in the radiative zone
\citep{Ouazzani2020}. Such resonances appear at specific mode frequencies.  For
evolved dwarfs with a shrinking convective core, the signal due to sharp
features in the structure of the equilibrium model or due to resonances with
inertial modes may be intertwined with mode bumping or avoid crossings
\citep[][for a mathematical description of these phenomena]{Smeyers2010}. A
single resonance between a gravitoinertial mode in the envelope and a pure
inertial mode in the core, or a single avoided crossing leads to just one dip in
the period-spacing pattern, as observed in Fig.\,\ref{rossby}.  A $\mu$-gradient
zone in the near-core region instead gives rise to recurring quasiperiodic
deviations.  Modeling of such regular deviations provides an excellent
opportunity to derive properties of $D_{\rm mix}(r)$ in the deep stellar
interior.  Analytical predictions for the oscillatory deviations due to a
receding convective core were derived for high-order g~modes of nonrotating
$\gamma\,$Dor and SPB stars by \citet{Miglio2008}. \citet{Bouabid2013}
generalized this to a numerical framework for rapidly rotating $\gamma\,$Dor
stars. These theoretical studies have been put into practice for dwarfs to
estimate their levels of chemical mixing from asteroseismic modeling following
Fig.\,\ref{flowcharts}, with adopted $D_{\rm mix}(r)$ profiles as in
Fig.\,\ref{geneva}.  We discuss such recent applications in
Sec.\,\ref{sec-mixing}.

Moving beyond the main sequence, periodic components in the oscillation
frequencies due to sharp features in the convective envelope of the red giant
HR\,7349 were found in its CoRoT data. The signal is due to a local depression
of the sound speed in the second ionization zone of helium
\citep{Miglio2010}. Similar studies from {\it Kepler\/} observations of red
giants were summarized by \citet{HekkerJCD2017}.  A powerful analysis to derive
the helium abundance for red giants was presented by \citet{McKeever2019} for an
ensemble of 27 pulsating red-giant branch (RGB) stars in the metal-rich open cluster NGC\,6791
observed by {\it Kepler\/}.  The helium abundances and ages for each individual
cluster RGB star were combined to create a distribution for the cluster's age
and helium abundance, resulting in $Y_{\rm ini}=0.297\pm 0.003$ and
$\tau=8.2\pm 0.3\,$Gyr.  Such precision for aging and helium abundance
determination is beyond the reach of classical cluster studies involving isochrone
fitting \citep[e.g.,][]{An2007}, even in the Gaia era \citep{Bossini2019}.

Dipole mixed modes in red giants offer additional opportunities. Structural
glitches and mode trapping result in deviations from the asymptotic relations of
period spacings of g~modes as in Eq.\,(\ref{periodspacing}), or of
gravitoinertial modes as in Eq.\,(\ref{spacing-TAR}).  Mixed modes comply with
a more complex asymptotic spacing pattern due to their mixed g- and p-mode
nature. Red giants are slow enough rotators to ignore the Coriolis force for the
derivation of their period-spacing expression, of which Eqs.\,(\ref{freqsep}) and
(\ref{periodspacing}) are the limiting cases for pure pressure and pure gravity
modes, respectively. The asymptotic expression for dipole mixed modes depends on
the evanescent zone between the g-mode cavity determined by buoyancy as the
dominant restoring force and the p-mode cavity where the pressure force is the
dominant restoring force.  The location, shape, and width of this evanescent
zone all play a role in the coupling between these two cavities. The coupling
factor, $q$, reaches extreme values of zero for no coupling and 1 for full
coupling.  The asymptotic expression for mixed dipole modes was
discussed by \citet{Unno1989}. Generalized versions and their use in space
asteroseismology were given by \citet{Mosser2014}, 
\citet{Takata2016a}, \citet{Takata2016b}, and
\citet{Pincon2019}. Observational estimations of $q$ from {\it Kepler\/} data were
made by \citet{Buysschaert2016}, \citet{Mosser2017b}, \citet{Mosser2018},
\citet{JiangJCD2018}, and \citet{Hekker2018}. \citet{Hekker2018} used a
description that explicitly relies on the radial order of the modes and allows
one
to constrain the p-mode and g-mode frequency or period offsets observationally.
This work revealed the g-mode period offsets to correlate with the core boundary
for RGB stars, while the p-mode offsets for core-helium-burning stars require
additional mixing in line with the suggestions by \citet{Constantino2015} and
\citet{Bossini2017}. \citet{Hekker2018} also found that $\ln\,q$ relates
linearly to the width of the evanescent zone normalized by its position. These
findings provide observational guidance to tweak $N(r)$ and deduce the core mass
of red giants in various evolutionary stages. \citet{Pincon2019} demonstated
that $q$ is tightly connected with the width of the evanescent zone and showed how this
zone changes when stars evolve from the subgiant to the RGB and further towards
the red clump. This study highlighted the capacity of $q$ to probe the dynamics
of the zone between the hydrogen-burning shell and the bottom of the convective
envelope. To date, this analytical work has focused on stars with masses below
1.2\,M$_\odot$. It needs to be extended to stars born with a well-developed
convective core, for one to understand the full structural and evolutionary
properties of the evanescent zone for such stars, including those with
$M_\star\simgr2\,$M$_\odot$, which will end up in the secondary clump.

\subsection{Improving the physics of cool-star surface convection}

In the case of low-mass stars, one can go beyond the analytical modeling of
acoustic glitches, which is in itself an improvement over the use of scaling
relations. A major aim of asteroseismology is to let go of the assumption that
low-mass stars adhere to $\bfpsi_{\rm Sun}$. This can be done by comparing the
choice of the solar input physics with any other choice of $\bfpsi$ from
evaluations between predicted and measured values of the individual oscillation
frequencies.  Modeling according to Fig.\,\ref{flowcharts} then requires one to
``overcome'' the surface effects in the case of cool stars with a convective
envelope. Indeed, for those stars, the adiabatic approximation of the
oscillation frequencies and the use of simplified boundary conditions based on
1D atmosphere models are not good enough relative to reality, as such approaches
lead to inappropriate frequency predictions $\omega_{nlm}^{{\rm theo},i}$.
Relying on equilibrium models with model atmospheres based on mlt to
describe the envelope convection rather than taking full account of the
turbulent pressure in the superadiabatic near-surface regions leads to 
p-mode frequencies that are too high \citep{JCD2002}. The offset is larger for higher radial
orders, i.e., for higher mode frequencies. For the solar oscillations, this
leads to offsets for $\Delta\nu_\odot$ of the order 10\,$\mu$Hz.  Similar offsets
are expected to occur for the oscillations of all Sun-like stars with solarlike
oscillations. The values of the offsets are much larger than typical
uncertainties of the measured oscillation frequencies. The surface effects must
hence be assessed to prevent errors in the estimation of $\bftheta$, even for
fixed $\bfpsi_{\rm Sun}$, when going through Fig.\,\ref{flowcharts}.

Methods have been devised to ``correct for'' or ``minimize'' the unknown surface
effects, guided by their properties regarding the Sun. One way to deal with the
surface effects was originally proposed by \citet{Roxburgh2003}, who came up
with combinations of p-mode frequencies (the so-called $r_{02}$ and $r_{01}$
indices) that suppress the sensitivity to the outermost layers, decreasing in
this way the influence of the limitations of the 1D models in the asteroseismic
modeling. More complex indices were subsequently defined with the same aim
\citep[e.g.][]{Roxburgh2005}. Several other methods to fit the surface effects
with a statistical model, with the aim of ``getting rid'' of the differences
between the measured and model frequencies, were developed. These adopted various
levels of sophistication in the fitting following the pioneering paper by
\citet{Kjeldsen2008}. 
\citet{Ball2014,Ball2017}, \citet{Basu2018}, \citet{Compton2018}, and
\citet{Jorgensen2019b} studied this topic.

Even with the simple pragmatic approach by \citet{Kjeldsen2008}, the gain
between modeling based on the scaling relations versus fitting the actual
frequencies is roughly a factor of 2 in the precision of the mass, radius, and
age. Such asteroseismic modeling was done for the CoRoT exoplanet host HD\,52265
by \citet{Lebreton2014}, showcasing how aging (and weighing and sizing) of the
host star by individual oscillation frequency fitting can be done at the level
of 10\%. While this study was applied to an individual CoRoT target star
according to the principles of Fig.\,\ref{flowcharts}, 
\citet{Lund2017} and \citet{SilvaAguirre2017} applied this full scheme to an
ensemble of 66 stars observed in 1-min cadence {\it Kepler\/} photometry
with a time base of up to 4-yr in the so-called legacy sample of the
mission. This study covered the following values for $\bftheta$\,: stellar masses
between 0.8 and 1.6\,M$_\odot$, $Y_{\rm ini}\in [0.2,0.4]$,
$Z_{\rm ini}\in [0.0025,0.05]$, $\alpha_{\rm conv}/\alpha_\odot \in [0.5,1.3]$,
and ages between 1 and 12\,Gyr.  \citet{SilvaAguirre2017} considered seven
different choices for the input physics $\bfpsi$, all of which are nonrotating
nonmagnetic 1D models ignoring radiative levitation in the treatment of the
microscopic atomic diffusion; see their Table\,1. All pulsation computations
were done in the adiabatic approximation and ignored the Coriolis and Lorentz
forces.  The asteroseismic modeling led to average relative uncertainties of 2\%
in radius, 4\% in mass, and 10\% in age, and revealed degeneracies between the
stellar mass $M_\star$ and initial helium abundance $Y_{\rm ini}$.  All seven
adopted $\bfpsi$ led to comparable fit quality when considering calibrations
from the Sun, angular diameter measurements, Gaia parallaxes, and
binarity. An initiative to assess the differences in stellar models computed
with various evolution codes by adopting the same input physics with the aim of
evaluating the level of numerical uncertainties and comparing them to 
asteroseismic uncertainties was provided by \citet{SilvaAguirre2019} and
\citet{JCD2020}, where the results may be found. Such activities are an
important aspect of the model uncertainties at play when one performs modeling via
Fig.\,\ref{flowcharts}.

Currently, the focus in asteroseismology of low-mass stars adopts a true
asteroseismic spirit: rather than shifting the measured frequencies to remove
the surface effects, the latter are considered an ``observational gift'' to
improve the weaknesses in the physics of 1D models. This shift in spirit of the
asteroseismologists rather than in the measured oscillation frequencies offers
great potential. Indeed, the asteroseismic modeling of stars with an outer
convective envelope can be further improved by using the measured surface effect
as an opportunity rather than a nuisance.  A common procedure adopted in 1D
models of low-mass stars is to calibrate an interior model and an atmosphere
model so as to be consistent with the Sun for one single value of
$\alpha_{\rm mlt}$ and do the stitching of the two deep enough in the adiabatic
part of the atmosphere. However, this procedure has limitations for evolved
stars \citep[e.g.,][]{Choi2018}.  Asteroseismology of a variety of stars permits
one to do better. Indeed, the shift between the measured oscillation frequencies and
those predicted by 3D hydrodynamical simulations of convection relevant for the
outer envelopes of low-mass stars is informative when evaluating such simulations
\citep[e.g.,][]{Zhou2019}, assessing nonadiabatic stability analyses
\citep[e.g.,][]{Houdek2019}, and finding out how to ``patch'' 3D atmosphere
models to 1D models of stellar interiors.

Following initial achievement to match a 3D atmosphere model to a 1D solar
interior model by \citet{Rosenthal1999}, detailed studies were conducted to achieve
optimal patching. The measured surface terms from helioseismology by
\citet{Magic2016} and from asteroseismology by \citet{Sonoi2015},
\citet{Ball2016}, and \citet{Trampedach2017} were exploited from the patching of 3D
atmosphere models to 1D interior models.  This led to improved boundary
conditions based on 3D convection simulations.  Guidance from measured surface
effects and adiabatic predictions of solarlike oscillation frequencies
was used to derive the optimal connection depth in the atmosphere.  Although
this approach does not yet take into account nonadiabatic effects, it offers
good potential to come to a better treatment of the 3D simulations of envelope
convection and their use for stellar evolution computations.  
\citet{Jorgensen2018}, \citet{Jorgensen2019a}, \citet{Jorgensen2019b},
and \citet{Mosumgaard2020} presented detailed procedures to include the mean
structure of 3D hydrodynamical simulations as the boundary condition of 1D models to
improve their outer stratification. In these studies of patched models, an
appropriately calibrated solar model with a structure similar to the 
underlying 3D simulations is achieved from helioseismology.  
\citet{Houdek2017} included a full treatment of the interaction between
convection and the oscillations.

The measured oscillation frequencies of {\it Kepler\/} targets across stellar
evolution are now being used to investigate how the convection-oscillation
interaction and the transition between envelope and interior can be achieved
from 3D convection simulations for stellar evolution models.  Such improvements
to stellar structure models partly eliminates the structural contribution to
the surface effect, with discrepancies having decreased from about 10 to some
2\,$\mu$Hz. Hence the patched models do not yet perform to the level of
precision of the asteroseismic data. Moreover, the patching procedures do not
deliver reliable post-main-sequence evolution models when 
performed near the bottom of the convective envelope in currently available 3D
simulations. More refined 3D simulations for deeper convective envelopes are
needed to improve stellar evolution theory of evolved low-mass stars even
further, keeping in mind numerical restrictions
\citep[cf.,][]{SilvaAguirre2019}.

In addition to global frequency shifts due to surface effects, the p-mode
frequencies also undergo time-dependent variability connected with magnetic
activity.  For the Sun this effect was summarized by \citet{JCD2002}.  Magnetic
effects were found for the solarlike p~modes detected in CoRoT data of the F5V
star HD\,49933 \citep{Garcia2010}. The {\it Kepler\/} data allowed such activity
[pulsation connections studied in samples of F-type
stars \citep{Mathur2014} and in the legacy sample \citep{Santos2019}], revealing
that the p-mode frequency shifts increase with increasing chromospheric
activity, increasing metallicity, and increasing effective temperature.  Young
rapid rotators reveal larger frequency shifts than old stars. Moreover, the
nonspherical nature of the magnetic activity in the stellar convective envelope
changes the frequencies of gravitoacoustic modes \citep{Perez2019}.  While
asteroseismic assessments of the physics of stellar activity in terms of its
effect on pulsation-mode behavior has progressed significantly, the improved knowledge
is not at a level at which it can be used to encode temporal magnetic activity in
the theory of stellar evolution. In this sense, the inclusion of the physics of
surface convection via patching of time-averaged 3D stellar atmosphere models to
1D stellar interiors across the evolution of low-mass stars, via calibrations of
solarlike oscillations based on space asteroseismology, advances more steadily
and more targeted than the inclusion of magnetic activity.

\subsection{Improving the theory of angular momentum transport}

A major asset of mixed and g~modes is their probing power of the deep stellar
interior. While this opportunity does not occur for the Sun and Sun-like dwarfs,
we now have thousands of stars with the appropriate g~modes delivering their
interior rotation rates from measured rotational frequency shifts.
Such measurements give quasidirect information about
$\Omega_{\rm core}$ without having to go through Fig.\,\ref{flowcharts}.  In
that sense, the internal rotation of stars has become observational astronomy.
Once the {\it Kepler\/} data reached a duration of 2 yr,
prominent detections of rotational splitting for dipole mixed modes were found
in subgiants by \citet{Deheuvels2012}, \citet{Deheuvels2014}, and
\citet{Deheuvels2020} and in red giants by \citet{Beck2012}, \citet{Mosser2012},
and \citet{Deheuvels2015}. For intermediate-mass dwarfs, both dipole and
quadrupole modes with rotationally split multiplets were detected soon after the
nominal 4-yr {\it Kepler\/} light curves became available, with initial
exploitations by \citet{Kurtz2014}, \citet{Papics2014}, \citet{Papics2015},
\citet{Saio2015}, and \citet{VanReeth2015}.  These  studies
of internal rotation immediately made it clear that the theory of angular
momentum transport as we knew it prior to {\it Kepler\/} failed to explain the
asteroseismic data, with discrepancies of up to 2 orders of magnitude in the
measured $\Omega_{\rm core}$.

Many observational derivations of $\Omega_{\rm core}$ have been done in recent years,
confirming the early findings. For red giants these were summarized by
\citet{Mosser2014} and \citet{Gehan2018}.  The discrepancy between theory and
observations turned out to be independent of the measured rotation rate during
the core-hydrogen burning, i.e., the problem regarding slower than expected
near-core rotation is derived for both the perturbative and TAR regimes of
rotation.  A summary of the asteroseismic results from mixed and g~modes, as
well as ways to improve the theory, was offered in the review by
\citet{Aerts2019} and is not repeated here. One major conclusion of that paper,
which summarized data covering all stages of stellar evolution, is that low- and
intermediate-mass stars are to a good approximation quasirigid rotators during
their core-hydrogen-burning phase, while $\Omega_{\rm core}$ and
$\Omega_{\rm env}$ values differ by less than a factor of 10 during the RGB phase.
The theory of local conservation of angular momentum transport does not explain
this.  The asteroseismic rotation estimates for 1210 stars across stellar
evolution assembled by \citet{Aerts2019} reveal that the CO cores built up
inside red giants and subdwarfs by the end of their core-helium-burning phase
have the same angular momentum as their white-dwarf successors.

Major updates since the summary by \citet{Aerts2019} have become available and
are shown in Fig.\,\ref{araa-update}. A large increase in the sample of dwarfs
was achieved by
\citet{GangLi2019a,GangLi2019b,GangLi2020}, who derived the near-core rotation
frequencies for more than 600 F-type g-mode pulsators.  These are shown in gray
in Fig.\,\ref{araa-update}.  Almost all of the newly included F-type dwarfs
reveal dipole ($l=1$) prograde modes, while about 30\% show
quadrupole ($l=2$) modes, and 16\% of them show retrograde Rossby modes.  Core
rotation rates of 72 core-helium-burning stars have been derived from their dipole
mixed modes \citep[][indicated in blue in
Fig.\,\ref{araa-update}]{Tayar2019}. For both of the new samples, there is no
asteroseismic estimate of $\log\,g$, as is the case for all 1210 stars in
Fig.\,4 given by \citet{Aerts2019}, hence Fig.\,\ref{araa-update} was constructed
differently. For the red giants addressed by \citet{Tayar2019}, $\log\,g$ was derived from
near-IR APOGEE spectroscopy, while \citet{GangLi2020} relied on $T_{\rm eff}$
values from \citet{Mathur2017}, luminosity estimates from Gaia astrometry
computed in \citet{Murphy2019}, and a grid of stellar models to derive the
gravity.  Although this leads to much larger and more systematic uncertainties for
$\log\,g$ than for the asteroseismic $\log\,g$ values used by
\citet{Aerts2019}, with uncertainties between 0.2 and 0.5\,dex for the stars in
Fig.\,\ref{araa-update} (omitted for clarity), the conclusions by
\citet{Aerts2019} are fully confirmed by these additional recent studies,
representing a tenfold increase in the number of dwarfs with
$\Omega_{\rm core}$.

Following \citet{Aerts2019}, two major paths have been followed to try
to fix the theory of stellar rotation, given the prominent results from
asteroseismology. On the one hand, angular momentum transport by IGWs as proposed
by \citet{Kumar1997}, \citet{Rogers2013}, and \citet{Rogers2015}, by mixed modes
as studied by \citet{Belkacem2015a,Belkacem2015b}, and by g~modes as in
\citet{Townsend2018} was considered. On the other hand, instabilities due to
magnetic fields, termed the magnetic Tayler instability, were considered an
explanation by \citet{Fuller2019}, \citet{Goldstein2019}, and
\citet{Eggenberger2019}.  Both physical processes lead to a more efficient
evacuation of angular momentum from the core to the envelope of the star than
any of the processes that were considered in stellar evolution computations
prior to space asteroseismology. While these new theoretical ingredients improve
the discrepancies between the asteroseismic measurements of $\Omega_{\rm core}$
and stellar evolution theory, they still require one to tweak the amount of angular
momentum transported from the core to the surface with a free parameter in order
to achieve compliance with the measured core rotation rates. Currently, none of the
theories are able to explain the quasirigid rotation measured for stars with
ratios of $\Omega/\Omega_{\rm crit}\in [0,75\%]$ during the core-hydrogen
burning phase, as displayed in the right panel of Fig.\,\ref{araa-update}.

\citet{VanReeth2016}, \citet{Ouazzani2017},
\citet{VanReeth2018}, \citet{Christophe2018}, and \citet{GangLi2020} measured
$\Omega_{\rm core}$ for a sample of $\sim\! 650$ $\gamma\,$Dor stars. A
dedicated study by \citet{VanReeth2018} on 37 of those pulsators
with high-precision spectroscopy allowed them to assess whether the stars have differential
envelope rotation while relying on the theoretical formalism derived by
\citet{Mathis2009} as a generalization of the TAR.  \citet{VanReeth2018} combined g-mode
estimation of $\Omega_{\rm core}$ with either a p-mode estimation of
$\Omega_{\rm env}$ or a derivation of $\Omega_{\rm surf}$ from rotational
modulation.  \citet{GangLi2020} added 58 more stars to conclude that the
rotation is almost rigid to within 5\% for all 95 single F-type dwarfs.

Ongoing modeling work considers two more improvements in addition to the TAR
with differential rotation. One generalization takes into account the occurrence
of an axisymmetric magnetic field with poloidal and toroidal components
following the perturbative approach for the magnetism and was elaborated upon by
\citet{Prat2019}. The new dispersion relation derived by \citet{Prat2019} allows
one to
assess how such a field affects the g-mode period-spacing pattern. It was found
that an interior magnetic field with strength above $10^5$\,G leads to
pertinent spiky deviations from the tilted period-spacing patterns
\citep{VanBeeck2020} that are in principle detectable.  These deviating signals
have not yet been identified in {\it Kepler\/} data on g-mode pulsators such as
those shown in Figs.\,\ref{gdor} and \ref{rossby}. Another generalization of the
TAR was derived by \citet{Mathis2019}, who computed a new dispersion relation
for slightly deformed stars including the centrifugal acceleration. The impact
of this inclusion is limited relative to the magnetic effects for rotation rates
up to $\sim\!70\%\,\Omega_{\rm crit}$ \citep{Henneco2021}.

\subsection{\label{sec-mixing} Inference of internal mixing from g~modes}

In this section we concentrate on stars of intermediate mass with a radiative
envelope.  Such stars are much more rapid rotators than low-mass stars with a
convective envelope because they do not experience magnetic braking.  Models of
intermediate-mass stars computed by relying on the Schwarschild or Ledoux
criteria of convection, without extra mixing in the near-core region, have 
convective core masses that are too low.  This is deduced from comparing high-accuracy
model-independent dynamical masses of double-lined eclipsing binaries to those
of stellar evolution models across a wide mass range of
$M_\star\in [1.2;17]\,$M$_\odot$ covered by \citet{Torres2010},
\citet{ClaretTorres2019}, and \citet{Tkachenko2020}. The need for higher
convective core masses for eclipsing binaries stands, irrespective of how 
$D_{\rm ov}(r)$ is treated in the isochrone fitting (see Fig.\,\ref{geneva}),
as discussed by \citet{Constantino2018}, \citet{Costa2019}, and
\citet{Johnston2019a}.  Masses of fully mixed convective cores may also be
derived from a model-dependent isochrone fitting of the observed extended
main-sequence turn-offs (eMSTO). This was done for numerous young open clusters
observed by the Hubble and Gaia space telescopes
\citep{Goudfrooij2018,Li2019}. Interpretation of the shape and diversity of
observed eMSTOs was done mainly by including rotational mixing
\citep[e.g.,][]{Bastian2018,Gossage2018} or magnetism \citep{Georgy2019} in the
stellar evolution models. Other causes of mixing, such as the pulsational or tidal
wave mixing discussed in Sec.\,\ref{models}, are usually ignored.  Inclusion
of $D_{\rm ov}(r)$ and $D_{\rm mix}(r)$ profiles calibrated by
asteroseismology of single field stars can explain some of the eMSTO properties
of young open clusters, thereby impacting their aging. Imposing the asteroseismic
results on cluster isochrone fitting allows for higher convective core masses
for all stars with $M_\star\geq 1.2\,$M$_\odot$ \citep{Johnston2019b}. Cluster
aging is a typical area where asteroseismology can be of interest to
other fields in astrophysics.

Asteroseismology has provided evidence for the need of
higher-than-standard masses of convective cores, in all evolutionary stages and
for a large range of stellar masses covering $M_\star\in [1.1,25]\,$M$_\odot$.
Backtracking the asteroseismic masses of white dwarfs to earlier evolutionary
phases requires more massive helium cores, as discussed by
\citet{Hermes2014}, \citet{Hermes2017a}, and \citet{Hermes2017b}.  The detailed
derivation of the larger-than-expected inferred CO core mass of
$M_{\rm cc}=0.45\,$M$_\odot$ of the pulsating white dwarf KIC\,08626021
\citep{Giammichele2018} is exemplary of the details that can be derived on the
chemical stratification (in this case of oxygen, carbon, and helium) from the
exploitation of g~modes. The core mass of this white dwarf is about 40\% higher
than expected from standard evolution models and points to the need for more CBM
at earlier phases of stellar evolution. The immediate progenitors of the white
dwarfs, i.e., the red-giant and subdwarf stars, also reveal the need for CBM and
higher core masses than those predicted in stellar evolution theory.  This was
quantified for three subdwarf B pulsators by \citet{VanGrootel2010a},
\citet{VanGrootel2010b}, and \citet{Charpinet2011}. These case studies resulted
in constraints on the inner He/C/O core from their g~modes.  Thus, stars not
only transport more angular momentum when they have a convective core, they also
have CBM resulting in more massive mixed cores than anticipated. This need is
most outspoken during the core-hydrogen-burning phase of stellar evolution, so
we focus on dwarfs in the rest of this section.
\begin{table*}[t!]
\tabcolsep8.0pt
\caption{Inferred convective core masses from estimation of 
$D_{\rm ov}(r)$ and $D_{\rm mix}(r)$ via Fig.\,\ref{flowcharts}
for three samples of single dwarf
  pulsators discussed in the text. The level of mixing at the
  bottom of the radiative envelope $D_{\rm  env}$ covers a wide range for B stars.}
\label{coreM}
\begin{center}
\begin{tabular}{|c|c|c|c|c|c|}
\hline
&&&&&\\[-0.3cm]
Sample & Spectral Type & Mass range & $M_{\rm cc}/M_\star$ range & $\Omega/\Omega_{\rm
                                                       crit}$ range & $D_{\rm
                                                                      env}$ range \\[0.1cm]
\hline
&&&&&\\[-0.3cm]
$\sim\!20$ solarlike pulsators & Later than F2 & $[1.1,1.6]$\,M$_\odot$ &
                                                                       $[3,18]\,\%$ & $<10\,\%$ & ? \\[0.1cm]
$\sim\!40$ g-mode pulsators & F0 -- F2 & $[1.3,1.9]$\,M$_\odot$ & $[7,12]\,\%$ &
                                                                                 $[0,70]\,\%$
                                                                  &
                                                                    $<10\,$cm$^2$\,s$^{-1}$ \\ [0.1cm]
$\sim\!30$ g-mode pulsators & B3 -- B9 & $[3.3,8.9]$\,M$_\odot$ & $[6,29]\,\%$ & $[3,96]\,\%$ & 
$[12.0,8.7\times 10^{5}]\,$cm$^2$\,s$^{-1}$\\[0.1cm]
\hline
\end{tabular}
\end{center}
\end{table*}

Prior to space asteroseismology, estimation of $D_{\rm ov}(r)$ assuming
convective penetration in Zahn's prescription \citep{Zahn1991} led to a wide
range of values covering $d_{\rm pen}\in [0.1,0.5]\,H_p$ for $\beta\,$Cep
pulsators \citep{Aerts2015}, but uncertainties from ground-based
asteroseismology remained large ($\simgr 0.1$).  At the low-mass end, an extreme
case requiring a large overshoot is the F5-type $M_\star\simeq\,1.5\,$M$_\odot$
solarlike p-mode pulsator Procyon. \citet{Guenther2014} modeled its p~modes from
equilibrium models with various prescriptions for $D_{\rm ov}(r)$ as of the
ZAMS, considering a radiative or adiabatic temperature gradient and penetration
as well as diffusive overshoot. This led to a fully mixed convective core mass
of $M_{\rm cc}/M_\star=12.4\%$.  Space asteroseismology delivered a better
estimation of $D_{\rm ov}(r)$ and $D_{\rm env}(r)$ from an application of the
method in Fig.\,\ref{flowcharts}.  Most studies had not yet ben able to deduce
the functional form of the profiles for $D_{\rm ov}(r)$ and $D_{\rm mix}(r)$ but
had assessed the global level of internal mixing using a forward method,
adopting parametrized profiles such as those shown in Fig.\,\ref{geneva}.  The
CBM levels and $M_{\rm cc}$ estimates from {\it Kepler\/} g-mode
asteroseismology of a sample of 37 $\gamma\,$Dor stars for which high-resolution
spectroscopy is available revealed equally well explained internal mixing by
convective penetration as by diffusive overshooting when using the observational
trio $\BY^{\rm obs}=(\Pi_0$, $\log\,T_{\rm eff}, \log\,g)$ in the modeling via
Fig.\,\ref{flowcharts}, after estimation of $\Omega_{\rm core}$ as shown in
Fig.\,\ref{araa-update} (gray circles). The results for the asteroseismic
estimation of the stellar parameters
$\bftheta=(M_\star,\Omega_{\rm core},D_{\rm ov},\tau,Z_{\rm ini})$ revealed
$M_{\rm cc}/M_\star\in [8,12]\%$ for the sample, which covers the mass range
$M_\star\in [1.3,1.9]\,$M$_\odot$, rotation rates
$\Omega_{\rm core}\in [0,25]\mu$Hz (i.e.,
$\Omega_{\rm core}/\Omega_{\rm crit}\in [0,70]\%$), and the entire
core-hydrogen-burning phase \citep[Fig.\,7 in][]{Mombarg2019}. On the other
hand, asteroseismology based on solarlike p~modes of nine stars analyzed by
\citet{Deheuvels2016} and \citet{Hjorringgaard2017} covering
$M_\star\in [1.12,1.58]\,$M$_\odot$ resulted in
$M_{\rm cc}/M_\star\in [3,18]\%$, again with equally good results for convective
penetration and exponential diffusive overshooting and for models without and
with atomic diffusion (the latter without radiative
levitation). \citet{Angelou2020} revisited aspects of the methodology to derived
$M_{\rm cc}$ and applied it to 13 stars with solarlike oscillations to arrive at
$M_{\rm cc}/M_\star\leq 14\%$, covering the mass range
$M_\star\in [0.75,1.45]\,$M$_\odot$.

\citet{Pedersen2021} fitted measured dipole g-mode period spacings for a sample
of 26 {\it Kepler\/} SPB pulsators, using eight grids of stellar models with
different CBM and envelope mixing profiles (shown in Fig.\,\ref{geneva}).  This
homogeneous asteroseismic study, via application of the method in
Fig.\,\ref{flowcharts} for ${\cal P}=8$ and ${\cal N}=26$, is the first of its
kind for this mass regime, covering stars with a convective core and radiative
envelope. It 
covers the entire main-sequence phase and allowed
to infer the overall mixing levels calibrated by the detected dipole g~modes for each
of the stars and for each of the eight model grids, limiting the solutions for
each star to its measured spectroscopic and astrometric values of $T{\rm eff}$,
$\log\,g$, and $\log\,(L/L_\odot)$. \citet{Pedersen2021} found that 17 of the 26 stars
were best modeled via convective penetration and 9 out of 26 with exponential diffusive
overshooting; see Fig.\,\ref{geneva}.  
Moreover, stellar models with a stratified envelope mixing
profile (due to vertical shear or IGWs as graphically depicted in
Fig.\,\ref{geneva}) deliver better asteroseismic fits to the data than
unstructured mixing profiles.  This study revealed asteroseismic estimates
$M_{\rm cc}/M_\star\in [6,29]\%$ for the mass range
$M_\star\in [3.3,8.9]\,$M$_\odot$ and rotation rates covering
$\Omega_{\rm core}\in [0.35,21.8]\mu$Hz corresponding to
$\Omega_{\rm core}/\Omega_{\rm crit}\in [3,96]\%$.  The level of envelope mixing
at the bottom of the radiative envelope, where the outer boundary of the
$D_{\rm ov}(r)$ profile occurs (i.e., at the interface of the purple and pink
profiles in Fig.\,\ref{geneva}) reveals a large range for this sample of 26
stars, with values between 12.0 and $8.7\times 10^{5}\,$cm$^2$\,s$^{-1}$. In
contrast to the results for F-type g-mode pulsators, this highlights the need
for considerable envelope mixing in several of these B-type stars. It is found
that the level of mixing at the bottom of the envelope is mildly correlated 
(correlation coefficient of 0.61) with
the rotation frequency in that region. A summary of the asteroseismic inferences
of $M_{\rm cc}/M_\star$, via an estimation of $D_{\rm ov}(r)$ and $D_{\rm env}(r)$
from imposed profiles as in Fig.\,\ref{geneva}, is provided in
Table\,\ref{coreM}.

Further improvements in asteroseismic modeling can come from the inclusion of
microscopic atomic diffusion in the equilibrium models. 
\citet{Deal2017}, \citet{Deal2018}, and \citet{Deal2020} assessed the impact of
adding radiative accelerations to model p-mode pulsators with a convective
envelope. They compared frequency predictions from 1D models based on
atomic diffusion to those from 1D models without diffusion or where it is
treated in a simplified way such as by restricting to gravitational settling (of
helium or heavier elements). Their studies are based on 1D equilibrium models
computed with the {\tt CESTAM} code \citep{Marques2013}, including an advective and
diffusive treatment of rotation.  \citet{Deal2020} found that the inclusion of
radiative levitation is necessary to achieve reliable values for the p~modes of
the F-type stars in the {\it Kepler\/} legacy sample with
$M_\star>1.45\,$M$_\odot$, even in the presence of macroscopic rotational
mixing. The latter was found to be the dominant element transport process in
stars with $M<1.3\,$M$_\odot$, while microscopic and macroscopic mixing are of
equal importance for the mass range $1.3\,$M$_\odot<M<1.45\,$M$_\odot$. The
importance of radiative levitation for g-mode asteroseismology has been
assessed only for two slowly pulsating $\gamma\,$Dor stars so far. This also points
to the need to include this process \citep[][see also
Fig.\,\ref{levitation-gdor}]{Mombarg2020}.

The capacity to infer internal mixing profiles, as well as the thermal structure
in the CBM region of stars with a convective core was assessed for dwarfs by
\citet{Pedersen2018} and \citet{Michielsen2019} and for core-helium-burning
stars by \citet{Constantino2017}.  These studies have yet to be put into
practice.  The full potential of the {\it Kepler\/} data on this front remains
underexploited, given that g-mode asteroseismology of dwarfs only saw its
beginnings since five years and that modeling for ensembles of stars following
the scheme in Fig.\,\ref{flowcharts} is a tedious and time-consuming task.
Nevertheless, Table\,\ref{coreM} reveals a large range of envelope mixing in
stars of similar mass, metallicity, and evolutionary stage during the main
sequence, reflecting the fact that nonlinear interactions between rotation, waves,
microscopic atomic diffusion, and magnetism may be at work.  Refined
calibrations of the mixing due to this multitude of phenomena requires ensemble
modeling of g-mode pulsators for hundreds of stars treated in a homogeneous way,
instead of the few tens addressed thus far.

\subsection{\label{section-tides}The beginnings of tidal asteroseismology}

In all of the previous cases, we considered oscillations based upon 1D equilibrium
models computed under the assumption of a single star. However, a large fraction
of stars occurs in binaries, where tidal forces and tidal interactions come into
play. The binary fraction among stars increases as the stellar birth mass
increases. On average, half of the stars occur in binaries but the occurrence
rate for high-mass stars is much higher than for low-mass stars, as high as
$\sim\!  80\%$ for O-type stars. Their evolution is dominated by binary
interactions \citep{Sana2012}.
\begin{figure*}[t!]
\begin{center} 
\rotatebox{0}{\resizebox{17.cm}{!}{\includegraphics{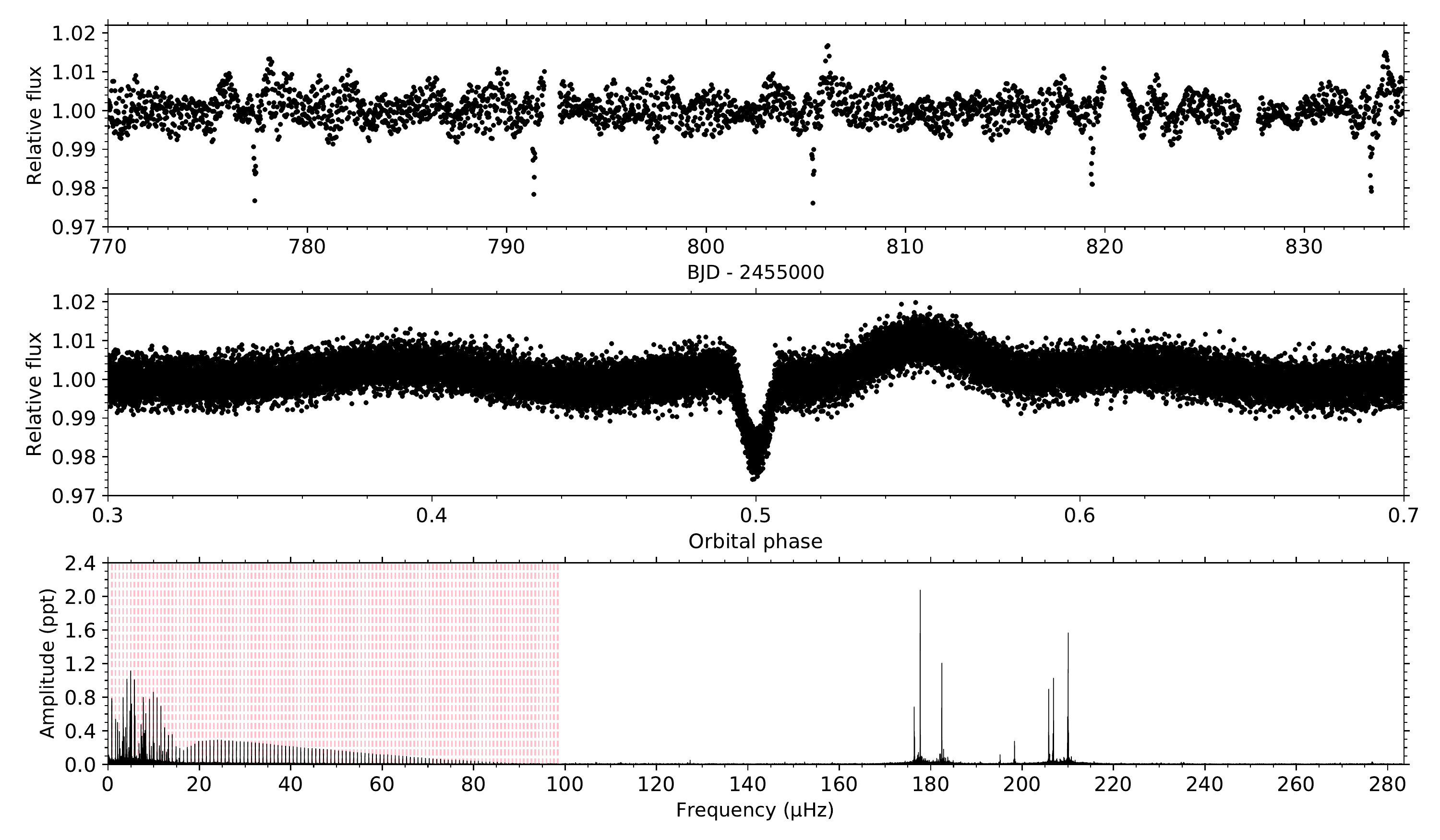}}}
\end{center}
\caption{\label{Zhao} Excerpt of the {\it Kepler\/} light curve (top),
  phase folded according to an orbital period of 14\,d (middle panel), and the
  corresponding LS amplitude spectrum (bottom panel) of KIC\,4142768. This eclipsing
  binary reveals self-excited $\kappa$-driven p~modes and tidally excited
  g~modes occurring at exact multiples of the orbital frequency (indicated in
  red).  Figure based on data in \citet{Guo2019} by courtesy of Timothy Van
  Reeth, KU\,Leuven.}
\end{figure*}

As long as the orbital separation of the two components or the mass ratio is
such that tides can be ignored, the asteroseismic modeling can be done as with
single pulsators.  The orbital motion may offer stringent and model-independent
dynamical masses, particularly for detached double-lined spectroscopic eclipsing
binaries. Some wide binaries reveal two pulsating components in the Fourier transform such that
isochrone fitting offers extra constraints compared to the case of a single
pulsator. The $\alpha\,$Cen system is a prototypical example of this
\citep[e.g.][]{Miglio2005}.  Other binaries with space photometry covering a
variety of pulsating components treated as if concerning a single star were
analyzed by \citet{Telting2012}, \citet{Frandsen2013}, \citet{Maxted2013},
\citet{Beck2014}, \citet{Appourchaux2015}, \citet{Gaulme2016},
\citet{Baran2016}, \citet{Themessl2018}, \citet{Kern2018}, and
\citet{Brogaard2018}. These studies led to stringent constraints
on the global stellar parameters thanks to the binarity.

Tidal asteroseismology treats the case of binaries for which the tide-generating
potential cannot be ignored in the force balance.  For a close binary in which
the tidal forces occur in the term $\boldf$ in Eq.\,(\ref{Eq2}), the tides come
into play at the level of the equilibrium models and for the computation of the
oscillations.  In such a case, the binarity of a pulsating component implies a
complication for an asteroseismic analysis. At the same time, it may offer
unique opportunities to test the effects of tidal forces on stellar structure
and evolution from tidally excited or tidally affected oscillations. The
properties of such oscillations differ from those of self-excited or
stochastically excited oscillations in that they are connected with the orbital
frequency, and they therefore offer additional opportunities to probe stellar
interiors than modes in single stars. Moreover, tidally excited
oscillations may get locked into resonance with the orbit and have a major
effect on the evacuation of orbital energy, efficiently changing the binary
evolution \citep[e.g.,][]{Pap1997,Savonije1997,Witte1999}.  Given that the tides
come into play and that orbital periods are of the order of days, tidally excited
oscillations usually are g~modes. However, tidally affected oscillations may
also occur among p~modes, as found in the close binaries U\,Gru
\citep{Bowman2019c} and V453\,Cyg \citep{Southworth2020}.

Currently, asteroseismic probing to improve the internal structure from tidal
oscillations remains limited.  This is not due to a lack of candidate pulsators,
as systematic searches for oscillations in eclipsing binaries with CoRoT and
{\it Kepler\/} revealed hundreds of cases \citep{GaulmeGuzik2019}. Rather, the
data analysis to deduce the oscillatory properties is extremely
challenging. Iterative schemes have to be devised to unravel the frequencies due
to the orbital motion, synchronous, subsynchronous or supersynchronous rotation,
and pulsations. Once again, CoRoT paved the way to the first proper monitoring
and iterative orbital and pulsational light-curve modeling of close binaries
with g~modes by \citep{Maceroni2009} and \citet{Maceroni2013} and with p~modes
\citep{daSilva2014}.

The real breakthrough in the discovery and analysis of tidal oscillations in
numerous close binaries came for the 4-yr nominal {\it Kepler\/} light
curves. We already showed and discussed the light curve and Fourier transform of
the prototype of high-eccentric binaries with tidally excited modes found by
\citet{Welsh2011} in Sec.\,\ref{excitation} (Fig.\,\ref{KOI-54}). That was a
case where almost all detected frequencies are exact multiples of the orbital
frequency, as expected for dynamical tides.  Another situation occurs for the
binary KIC\,4142768 whose light curve is illustrated in Fig.\,\ref{Zhao} and
discussed by \citet{Guo2019}. This is an eclipsing binary with two evolved
A-type stars in an eccentric orbit with a period of 14\,d. This pulsating binary
reveals low-frequency g~modes, some but not all of which occur at exact
multiples of the orbital frequency (indicated with the red vertical lines in
Fig.\,\ref{Zhao}). The binary also undergoes $\kappa$-driven $\delta\,$Sct-type
p~modes in the frequency range 170--220$\mu$Hz. Spectroscopic follow-up with the
HIRES spectrograph at Keck revealed a surface rotation rate only one-fifth of
the pseudosynchronous rate at periastron. The tidally excited modes were
identified as quadrupole prograde sectoral modes, as anticipated from the theory
of dynamical tides \citep[see,][for a highly didactical paper on tidally excited
oscillations]{Fuller2017a}. The frequency range of the detected self-excited
modes is compliant with theoretical predictions for the fundamental parameters
of the primary, which is a fairly evolved star close to the terminal-age main
sequence. The near-core rotation rate derived from the fitting of the data to
Eq.\,(\ref{spacing-TAR}) corresponds to $\Omega_{\rm core}=0.07\pm 0.03\mu$Hz
and points to an extremely slow rotator, which is in agreement with the spectroscopic surface
velocity projected on the line of sight.

Various binaries discovered from
{\it Kepler\/} photometry have revealed an oscillation signal at
the equilibrium tide, in addition to g~modes triggered by dynamical tides
\citep[e.g., KIC\,8719324, which was addressed by ][]{Thompson2012}. Detailed
observational analyses were made for several systems similar to 
KIC\,4142768 that is shown in Fig.\,\ref{Zhao}. Examples are available in
\citet{Papics2013}, \citet{Murphy2013b}, \citet{Debosscher2013},
\citet{Hambleton2013}, \citet{Borkovits2014}, \citet{Hambleton2016},
\citet{Guo2017a,Guo2017b}, \citet{Fuller2017b}, \citet{Hambleton2018},
\citet{GuoLi2019}, \citet{Bowman2019c}, \citet{Southworth2020},
\citet{Handler2020}, and \citet{Kurtz2020}, where the latter two papers reported
the first cases of tip-tilted oblique binary pulsators.  Hardly any of these
binary systems have yet been modeled asteroseismically, according to
Fig.\,\ref{flowcharts}, with the exception of the double hybrid p- and
g-mode F-type pulsators KIC\,10080943A and KIC\,10080943B. Asteroseismic
isochrone modeling of the two components revealed the need for extra CBM and
higher convective core masses than in standard evolution models for both
components, which is in line with Table\,\ref{coreM} \citep{SchmidAerts2016}.

Space photometry continues to deliver a plethora of close binary pulsators, with
new discoveries by the day. A large diversity of orbital periods,
eccentricities, synchronicities, and oscillation properties have already been
found. The extensive paper by \citet{Fuller2017a} revisited theories of tidal
excitation of nonradial oscillations and provided predictions for flux
variations, mode amplitudes, frequencies, phases, and spin–orbit misalignment
based on a nonadiabatic treatment of the equations, including the Coriolis
force.  Yet, we are only at the beginnings of tidal asteroseismology, because so
few systems have been modeled via the scheme in Fig.\,\ref{flowcharts} and we do
not yet know how common resonance locking is, nor how important it is in
practise for binary evolution and for angular momentum loss. Future modeling
work to understand tidal wave transport phenomena in close binaries is in order.
Searches for oscillation modes in numerous eclipsing binaries in the TESS data
are ongoing. These will undoubtedly uncover objects suitable for better
understanding the evolution of massive binaries, including progenitors of future
gravitational wave emitters. The TESS sample of high-mass stars observed in its
two CVZs holds great potential in this respect.

\section{\label{section-future}Glorious road map for the future} 

The past decade has sparked immense interest in asteroseismology.  Following the
detection of nonradial oscillation modes in ground-based radial-velocity and/or
light curves of a few tens of stars, we have moved on to asteroseismology of
tens of thousands of stars covering all evolutionary phases. This success results
from uninterrupted long-duration space photometric light curves having ppm-level
precision. Asteroseismology is delivering an observational calibration of 
internal rotation and mixing across stellar evolution, as a guide for
improving the theory of angular momentum and element transport inside stars. The
stellar evolution community is currently digesting this flood of asteroseismic
information, given the surprises and challenges that it brought. This
will eventually lead to better stellar evolution models, as important input for
studies of exoplanetary systems and for the chemical evolution of galaxies.

Much more is to come. While past space missions focused on low- and
intermediate-mass stars, the current all-sky TESS mission already delivered data
for high-mass stars in the Milky Way and LMC during its first two years of
operation, covering masses up to $\sim\! 50\,$M$_\odot$.  Prospects are
excellent that we may embark upon asteroseismology of high-mass binaries on their way to
becoming gravitational wave sources, and of blue supergiants nearing their
supernova explosion. The methodology in Fig.\,\ref{flowcharts} is in place but
applications should be generalized to a nonadiabatic framework for the
oscillations. Moreover, dissection of the maximum amount of information present
in the Fourier transforms of the TESS light curves is in order. Asteroseismology based on
stochastically excited GIWs looks appealing now that TESS is delivering proper
data to guide such a development for high-mass stars.  The art will be to
distinguish the signatures of coherent gravitoinertial modes and of stochastic
GIW as input for modeling of the stellar interior.  Similarly, asteroseismology
of pre-main-sequence stars has yet to be put into practice for representative
ensembles. Initial studies were based on just a few short light curves
\citep{Zwintz2014}, but improvements in the physics of accretion, rotational
spin-up, magnetic activity during contraction, and internal mixing of elements
as protostars approach their birth are now within reach for asteroseismic scrutiny.

The ongoing NASA TESS and future ESA PLATO missions lift the probing of 
stellar interiors to all
masses and evolutionary stages. With that glorious prospect,
asteroseismology is entering the big data era. 
Machine-learning methods are advantageous for interpreting the
massive flux of data but must be applied with proper mathematical
modeling, including parameter degeneracies and correlated diagnostics, so as to
ascertain appropriate precision estimation of the stellar parameters, among
them stellar ages. Ensemble asteroseismology will become ever more powerful
when combined with independent and homogeneous nonasteroseismic information
coming from all-sky spectroscopic surveys with multiobject spectrographs such
as SDSS-V \citep[all-sky, near IR,][time frame 2020--2024]{Kollmeier2017}, WEAVE
\citep[Northern Hemisphere, optical,][time frame 2020+]{Dalton2018}, and 4MOST
\citep[Southern Hemisphere, optical,][time frame 2022--2026]{deJong2019}, along
with the final Gaia all-sky space astrometry.

On the theory front, nonlinear asteroseismology has to be redeveloped in this
space era. This has major potential given that a high percentage of pulsators
reveal departures from linearity and evidence for nonlinear resonant mode
coupling. Such coupling occurs in oscillation spectra across all masses and
evolutionary stages. Having been found in CoRoT data of B and Be stars by
\citet{Degroote2009} and \citet{Huat2009}, it has also been detected in BRITE data of
Be stars \citep{Baade2018} and in {\it Kepler\/} photometry of young
intermediate-mass stars \citep{Bowman2016}, of subdwarf pulsators
\citep{Baran2012}, and in various white dwarfs \citep{Hermes2015}. This
observational gold mine is awaiting exploitation in terms of the nonlinear probing
of stellar interiors once a modern theoretical framework gets developed, as in
\citet{Zong2016b}.  Similarly, magnetoasteroseismology is still in its infancy.
Mode predictions for stellar models with strongly magnetic cores have been
triggered to explain mode suppression in red giants by \citet{Loi2020-DeltaP}
and \citet{Bugnet2021}.  Backtracking the results for red giants, their
intermediate-mass main-sequence progenitors should also have strong internal
magnetic fields.  Such fields have a non-negligible effect on g~modes, as shown
by the theoretical developments in \citet{Prat2020} and their magnetic
signatures predicted by \citet{VanBeeck2020}. While these have not yet been
found in g-mode period-spacing patterns of {\it Kepler\/} dwarfs, this might be
because they have not been looked for with dedicated eyes or with
machine-learning artillery. 

Finally, we come back to the use of 1D equilibrium models.  Several of the stars
in Fig.\,\ref{araa-update} rotate faster than 70\% of their critical rotation
frequency. Their asteroseismic modeling will benefit from a 2D treatment.  The
code {\tt ROTORC} by \citet{Deupree2001} delivers 2D equilibrium models and was
used to make pulsation predictions for p~modes of $\beta\,$Cep pulsators (see
Fig.\,\ref{hrd}) by \citet{Lovekin2008} and \citet{Lovekin2009}.  These
predictions were not used in asteroseismic modeling so far.  The public code
{\tt ESTER} \citep[\'Evolution STEllaire en Rotation,][]{Rieutord2016} is under
active development and is advanced in terms of the treatment of transport
processes. This code has great potential given the need for improvements in
models of the fastest rotating intermediate- and high-mass dwarfs.  {\tt ESTER}
delivers 2D axisymmetric static structure models but does not yet treat the
chemical evolution of the star, nor 2D mass loss or envelope convection. Proper
boundary conditions, including a dynamical wind via 2D nonlocal thermodynamic
equilibrium atmosphere models \citep[e.g.,][]{Petrenz2000}, are necessary to
improve pulsation predictions for fast rotators.  While the current limitations
of {\tt ESTER} can be partially circumvented by fixing the hydrogen mass
fraction in the convective core to a seismic estimate of $X_c$ as in a recent
application to the $\delta\,$Sct star Altair \citep[rotating at $\sim\!74\%$ of
its critical velocity,][]{Bouchaud2020}, future developments to turn the code
into a full-blown 2D stellar evolution tool would be highly beneficial.  This
would allow researchers to perform 2D asteroseismic modeling of the most rapid
rotators in Fig.\,\ref{araa-update}, and of the high-mass pulsators discovered
by \citet{Pedersen2019}, \citet{Burssens2020}, \citet{Bowman2020}, and
\citet{Dorn-Wallenstein2020}, in addition to new ones yet to be discovered.

Tremendous progress in our understanding of stellar interiors has been achieved,
thanks to the beauty of nonradial oscillation theory coupled with space photometry
of ppm-level precision for thousands of stars. Asteroseismology turned the study
of stellar interiors into an observational science. Its future is extremely
bright in all aspects of this research field, from instrumentation and
ongoing or planned surveys all the way up to fundamental theory. Major improvements
for stellar evolution theory based on asteroseismology are under way for single
stars, binaries, and star clusters. We thus end with a kind invitation to those
readers whose curiosity might be triggered but who have not yet been active in this
field: it is never too late to become an asteroseismologist\ldots

\begin{acknowledgements}
  \noindent The writing of this review was initiated while I was lecturing as
  the 2019 Oort Professor at Leiden Observatory, the Netherlands, and at the Max
  Planck Institute of Astronomy in Heidelberg, Germany. I am grateful for the
  kind hospitality at both places, and for the genuine interest in my lectures
  expressed by young and not-so-young attendees; this made those stays
  particularly enjoyable.  I am also grateful to many colleagues, too numerous
  to mention, whose inspiring lectures and tutorials educated me in
  astrophysics in general, and in the computation of numerical stellar models
  and their oscillation frequencies in particular.  Trained as a pure
  theoretician at master level, I benefited greatly from practical
  introductions to astronomical observing with various telescopes during my
  PhD and postdoctorate trajectories. Undertaking the full journey from early
  instrument concepts to detailed modeling of stellar interiors, in a
  multicultural and inclusive team spirit, has been crucial for my motivation
  and inspirational for my supervision and training of the next generations of
  astrophysicists.  The PhD students and postdoctoral researchers in the Leuven
  asteroseismology team, colleagues from the community, and three referees are
  thanked for their comments on early versions of the manuscript. I am
  particularly grateful to my local python-artists-in-residence, Dominic Bowman,
  Cole Johnston, Joey Mombarg, P\'eter P\'apics, May Gade Pedersen, and Timothy
  Van Reeth for having produced figures for this review. Appreciation is also
  given   to Ashley Chontos, Philipp Edelmann, Sylvia Ekstr\"om, Gang Li, Tami Rogers,
  Jamie Tayar, and Kuldeep Verma for providing data or figures in
  electronic form.  Those participants of the conference ``{\it Stars and Their
    Variability Observed from Space\/}'' held in Vienna, Austria, in August 2019, who took
  part in the end-of-conference poll on how our research field should progress,
  are thanked for their cooperation; the outcome of the poll shaped the final
  section of this review.

  I acknowledge funding from the KU Leuven Research Council (grant C16/18/005:
  PARADISE).  The funding received from the European Research Council (ERC)
  under the European Union's Seventh Framework (FP7/2007--2013/ERC Grant
  Agreement No.\,227224: PROSPERITY, 2009--2013) and Horizon\,2020 Research and
  Innovation Programme (Grant Agreement No.\,670519: MAMSIE, 2016--2020) has
  been essential for the development of my long-term research goals and
  aspirations in gravity-mode asteroseismology. The ERC has allowed me to
  transform my narrow, bumpy, unconventional path into a broad, royal road.
\end{acknowledgements}

\bibliographystyle{aa}
\bibliography{Aerts-RMP-Proofread}

\end{document}